\documentclass[letterpaper,11pt]{yalephd}
\usepackage{geometry} 
\usepackage{graphicx} 
\usepackage{dcolumn}
\usepackage{bm}
\usepackage{amsmath}
\usepackage{amsfonts}
\usepackage{amssymb}
\usepackage{appendix}
\usepackage{cite}
\usepackage{notoccite}
\usepackage{comment}
\usepackage{siunitx}
\usepackage{wrapfig}
\usepackage{booktabs} 
\usepackage{multirow}
\usepackage{empheq}
\usepackage{cases}
\usepackage{makecell}
\usepackage{blindtext}
\usepackage{tikz}
\usepackage{hyperref} 
\usepackage[all]{hypcap} 
\usepackage{subfiles} 

\hypersetup{
    colorlinks=true,
    linkcolor=blue,
    filecolor=magenta,      
    urlcolor=blue,
    citecolor=blue
    }
\graphicspath{{images/}}

\newcommand{\citeasnoun}[1]{Ref.~\cite{#1}}
\newcommand{\citeasnouns}[1]{Refs.~\cite{#1}}

\newcommand{\Figref}[1]{Figure~\ref{fig:#1}}
\newcommand{\figref}[1]{Fig.~\ref{fig:#1}}
\renewcommand{\eqref}[1]{Eq.~(\ref{eq:#1})}
\newcommand{\Eqref}[1]{Equation~(\ref{eq:#1})}
\newcommand{\eqreftwo}[2]{Eqs.~(\ref{eq:#1},\ref{eq:#2})}
\newcommand{\eqrefrange}[2]{Eqs.~(\ref{eq:#1})--(\ref{eq:#2})}
\newcommand{\vect}[1]{\bm{#1}}

\renewcommand{\Im}{\operatorname{Im}}
\renewcommand{\Re}{\operatorname{Re}}
\newcommand*{\TT}{\mathbb{T}}
\newcommand*{\LL}{\mathbb{L}}
\newcommand*{\DD}{\mathbb{D}}
\newcommand*{\GG}{\mathbb{G}}
\newcommand*{\II}{\mathbb{I}}
\newcommand*{\vP}{\bm{P}} 
\newcommand*{\vx}{\bm{x}}
\newcommand*{\vG}{\bm{G}}
\newcommand*{\vE}{\bm{E}}
\newcommand*{\Vd}{V_d}

\newcommand*{\myforall}{\quad{\rm for}\ \forall}
\newcommand*{\md}{\mathrm{d}}
\newcommand{\hl}[1]{{\color{red}#1}}

\newcommand{\hiddensection}[1]{
    \refstepcounter{section}
    \section*{\Alph{chapter}.\arabic{section}\hspace{1em}{#1}}
}

\newcommand*{\Pscat}{P_{\rm scat}}
\newcommand*{\Pabs}{P_{\rm abs}}
\newcommand*{\Pext}{P_{\rm ext}}
\newcommand*{\ImGO}{\Im \GG_0}
\newcommand*{\Vbb}{\mathbb{V}}
\newcommand*{\cinc}{c_{\rm inc}}
\newcommand*{\BB}{\mathbb{B}}
\newcommand*{\einc}{e_{\rm inc}}

\newcommand*{\SNR}{{\rm SNR}}

\newcommand*{\vr}{\vect{r}}

\newcommand*{\card}{\text{card}}

\newcommand*{\ei}{e_{\rm inc}}
\newcommand*{\es}{e_{\rm scat}}
\newcommand*{\pL}{p^{(\text{L})}}
\newcommand*{\pLd}{p^{(\text{L})\dagger}}
\newcommand*{\pNL}{p^{(\text{NL})}}
\newcommand*{\pNLd}{p^{(\text{NL})\dagger}}

\newcommand*{\chil}{\chi^{-1}}
\newcommand*{\deff}{d_{\rm eff}}

\newcommand*{\vv}{\bm{v}}
\newcommand*{\vw}{\bm{w}}
\newcommand*{\vu}{\bm{u}}
\newcommand*{\Abb}{\mathbb{A}}

\newcommand{\Tr}{\operatorname{Tr}}

\newcommand{\diag}{\operatorname{diag}}

\newcommand{\secref}[1]{Sec.~\ref{sec:#1}}

\newcommand*{\CC}{\mathbb{C}}
 
\newcommand*{\XX}{\mathbb{X}} 
\renewcommand*{\SS}{\mathbb{S}} 
\renewcommand*{\AA}{\mathbb{A}} 
\newcommand*{\kvp}{\vect{k}_\parallel}
\newcommand*{\epsm}{\varepsilon_{\rm m}}

\newcommand*{\dchi}{\Delta\chi}

\newcommand*{\GGd}{\mathbb{G}^{\rm EE\dagger}}
\newcommand*{\WW}{\mathbb{W}}
\newcommand*{\UU}{\mathbb{U}}
\newcommand*{\Lbb}{\mathbb{L}}

\newcommand*{\tpsiinc}{\tilde{\psi}_{\rm inc}}

\newcommand*{\psiadj}{\psi_{\rm adj}}
\newcommand*\widefbox[1]{\fbox{\hspace{1em}#1\hspace{1em}}}

\newcommand*{\ehv}{\vect{\hat{e}}}
\newcommand*{\Nv}{\vect{N}}
\newcommand*{\pv}{\vect{p}}

\newcommand*{\sabs}{\sigma_{\rm abs}}

\newcommand*{\GO}{\Gamma_0}

\newcommand*{\Mv}{\vect{M}}
\newcommand*{\ev}{\vect{e}}
\newcommand*{\kv}{\vect{k}}
\newcommand*{\xv}{\vect{x}}
\newcommand*{\St}{S_{\rm target}}
\newcommand*{\Tt}{T_{\rm target}}
\newcommand*{\td}{\text{d}}
\newcommand*{\tomega}{\tilde{\omega}}
\newcommand*{\Hw}{H_{\omega_0,\Delta\omega}}

\newcommand*{\psiinc}{\psi_{\rm inc}}

\newcommand{\phototm}[1]{%
    \includegraphics[width=3.5cm]{#1}}

\newcommand*{\nv}{\vect{n}}

\newcommand*{\vI}{\mathbf{I}}
\newcommand*{\vV}{\mathbf{V}}
\begin{document}

\title{Fundamental Limits of Nanophotonic Design} 
\author{Zeyu Kuang}
\advisor{Owen D. Miller}
\date{May, 2023} 
\frontmatter

\begin{abstract}
Nanoscale fabrication techniques, computational inverse design, and fields from silicon photonics to metasurface optics are enabling transformative use of an unprecedented number of structural degrees of freedom in nanophotonics.  A critical need is to understand the extreme limits to what is possible by engineering nanophotonic structures. This thesis establishes the first general theoretical framework identifying fundamental limits to light--matter interactions. It derives bounds for applications across nanophotonics, including far-field scattering, optimal wavefront shaping, optical beam switching, and wave communication, as well as the miniaturization of optical components, including perfect absorbers, linear optical analog computing units, resonant optical sensors, multilayered thin films, and high-NA metalenses. The bounds emerge from an infinite set of physical constraints that have to be satisfied by polarization fields in response to an excitation. The constraints encode power conservation in single-scenario scattering and requisite field correlations in multi-scenario scattering. The framework developed in this thesis, encompassing general linear wave scattering dynamics, offers a new way to understand optimal designs and their fundamental limits, in nanophotonics and beyond.
\end{abstract}

\maketitle
\makecopyright{2023} 
\setcounter{tocdepth}{1}
\tableofcontents
\listoffigures 
\listoftables 

\chapter{Acknowledgements}

A year ago, I decided to quit my PhD.

``You seem sad today. Are you alright?'' Owen asked.  

My slides hung on the screen, dry and dull.

So I let everything out: the chain of rolling projects; the weight of creative research; the anxiety that crept upon my back, my neck, and all the way inside my vocal folds; my dream, forfeited; my passion, died.

We talked, about other career choices, the researchers we respect, the high-stake high-reward nature of academic, and in it, his own struggles and burnouts ...

A month later, recuperating in my wife’s apartment in Pennsylvania, I watched the late December snow piling up on the courtyard, burying my life underneath. 
By the time the snow melted and the grass peeked through and the thawing snow streamed across the quadrangle, I started to scribble equations again. 
Not that I've left my depression behind---it is, and always has been, the people that make my life worth living. And Owen, my PhD advisor for the past five years, has taught me not only about research, but also about integrity, mentorship, and courage. The courage to think loud, to speak out, to plow on.

Besides Owen, I also thank my wife Lidan, for the oceans we crossed and distances we fought, my friends Lang, Yaqing, and Bao, for the stars we traced and memories we glazed, and my parents, Minzhi and Bensheng, for our life together and the fine examples you showed me to always cherish the people I love.



\mainmatter

\chapter{Introduction}
\label{chap:intro}

Bounds, or fundamental limits, identify what is possible in a complex design space. 
Besides the famous Shannon's limit~\cite{shannon1948mathematical} that underlies the modern digital communication~\cite{Cover1999}, other well-known bounds include the Carnot efficiency limit~\cite{Callen1985}, the blackbody limit~\cite{planck1914theory}, the Shockley--Queisser bounds in photovoltaics~\cite{Shockley1961}, the Bergman--Milton bounds in the theory of composites~\cite{Bergman1981,Milton1981,Milton2002}, and the Wheeler--Chu bounds on antenna quality factor~\cite{Wheeler1947,Chu1948}, among many more.
Bounds are indispensable tools for designers in many fields and guide research development at three levels.
An individual designer can use bounds to benchmark a device. If its performance is already close to the bound, one stops searching (which saves time and money). 
If its performance falls short, one gets motivated one to find better designs to approach the bound. 
For example, Shannon's bound set off waves of investigations in increasingly more sophisticated coding schemes that eventually came within 99.9\% of  
Shannon's capacity limit~\cite{chung2001design}.
At the next level, bounds distill a complex design process into trade-offs between a few key parameters, such as the bandwidth and size of an antenna in Wheeler-Chu's bound~\cite{Wheeler1947,Chu1948}, as well as the real and imaginary parts of a material susceptibility in Bergman–Milton bound~\cite{Bergman1981,Milton1981,Milton2002}. Bounds of such kinds serve as impartial referees that encompass, compare, and organize different design approaches. At the highest level, bounds guide future research by identifying key constraints in a system. One example is the far-field constraint in Plank's black body limit~\cite{planck1914theory} that later prompted a burst of research in near-field radiative heat transfer when techniques on nano-fabrication and nano-positioning are sufficiently matured~\cite{song2015near, cuevas2018radiative, biehs2021near}. Identifying what is possible under the current technology, as well as its limiting constraints, bounds map out the frontier of science and drives research innovation both in its time and for years to come.  

In nanophotonics, designing the geometry of scatterers has led to numerous impactful applications~\cite{painter1999two, pelton2002efficient, knight2003photonic, eichenfield2009optomechanical, pala2009design, liu2010infrared, men2014robust, piggott2015inverse, piggott2017fabrication}, ranging from solar energy conversion~\cite{pala2009design} to bio-molecule detection~\cite{liu2010infrared} to on-chip integration~\cite{piggott2017fabrication}, but often involves opaque computational optimizations in the complex landscape of Maxwell's equations~\cite{bendsoe2003topology, jensen2011topology, Miller2012a, Molesky2018, pestourie2018inverse, so2020deep,ma2021deep, jiang2021deep,holland1992genetic, kennedy1995particle, schneider2019benchmarking, park2022free}.
Geometry design in nanophotonics is motivated by a simple idea:  different geometries scatter electromagnetic waves differently. Therefore, a designer can alter the geometry of a material to create specific wave interference in the hope of achieving desired scattering effects.  Examples include drilling air holes in the active layer of a solar cell to enhance its absorption efficiency~\cite{pala2009design, Yu2010, brongersma2014light} and judiciously laying out arrays of nano-pillars to create an ultrathin optical lens~\cite{aieta2015multiwavelength,wang2018broadband, chen2018broadband, shrestha2018broadband}.
The state-of-the-art method for geometrical designs is adjoint-gradient-based optimization~\cite{bendsoe2003topology, jensen2011topology, Miller2012a,  Molesky2018, pestourie2018inverse}.
It starts with an initial geometry, iteratively perturbs the geometry in a direction that increases the objective function (via computation of the so-called ``adjoint'' field), and terminates when no perturbation significantly improves the objective function, upon which the design is final.
As the optimization landscape of Maxwell's equations are highly nonconvex and oscillatory, the algorithm often gets stuck in undesirable local optima. Restarting the search multiple times with different initial points helps to discover other local optima~\cite{schneider2019benchmarking}, but it increases the computational runtime and offers no guarantee of finding a better design (let alone the global optimum). 
The same issue arises ubiquitously across all design methods in photonics, ranging from adjoint-based inverse design~\cite{bendsoe2003topology, jensen2011topology, Miller2012a,  Molesky2018, pestourie2018inverse} to evolutionary algorithms~\cite{holland1992genetic, kennedy1995particle, schneider2019benchmarking, park2022free} to neural networks~\cite{so2020deep,ma2021deep, jiang2021deep}, as all of them are essentially finding different local optima in the nonconvex, highly oscillatory landscape of Maxwell's equations.
A natural complement to these ``bottom-up'' design approaches are ``top-down'' theoretical bounds, which provide the necessary benchmarks for what is possible in nanophotonics. 

Bounds to nanophotonic response are few and far between, with each one derived for a specific problem (typically in highly simplfied or asymptotic regimes), with little connective structure between them. 
In the electrostatic limit, established bounds include the Bergman--Milton bounds for composite materials~\cite{Bergman1981,Milton1981,Milton2002} and the Wheeler--Chu bound for small antennas~\cite{Wheeler1947,Chu1948}.
In wave communication, one uses the notion of optimal communication channels to find the best antenna excitation~\cite{Miller2019}. 
For a few analytically well-behaved objectives, there are frequency-integrated sum rules~\cite{gordon_1963, purcell_1969, mckellar_box_bohren_1982,sohl_gustafsson_kristensson_2007}.
Problem-specific bounds include light trapping in solar cells~\cite{Yablonovitch1982, Yu2010}, absorption in graphene~\cite{thongrattanasiri_complete_2012}, radiative heat transfer between nanostructures~\cite{pendry1999radiative}, scattering of antennas and nanoparticles~~\cite{Yaghjian1996,hamam_coupled-mode_2007,kwon2009optimal, ruan2011design, liberal2014least, liberal2014upper, hugonin_fundamental_2015, ivanenko2019optical}, and their maximal optical force and torque~\cite{rahimzadegan2017fundamental,liu2018optimal}, though most assume a few (heuristically determined) well-coupled modes~\cite{Yu2010,hamam_coupled-mode_2007,ruan_superscattering_2010}, resonances~\cite{Yu2010}, or scattering channels~\cite{pendry1999radiative,liberal2014least,liberal2014upper,hugonin_fundamental_2015,ivanenko2019optical,rahimzadegan2017fundamental,liu2018optimal}. 
For general electromagnetic scattering, there is no theory of fundamental limits. Without such theoretical guidance, photonic designers are groping in the dark.

Recently, new bounds emerged in absorptive systems such as plasmonic nanoparticles~\cite{miller_fundamental_2016, ivanenko2019optical, nordebo2019optimal, yang2018maximal}, 2D materials~\cite{miller2017limits, shim2019fundamental}, and metallic metasurfaces~\cite{shim2019fundamental, miller2015shape, michon2019limits, yang2018maximal, yang2017low, dias_fundamental_2019}.
These ``lossy-material bounds'' distinguish themselves  from previous bounds in two ways.
First, all the lossy-material bounds are derived from a single optimization framework: they optimize different objectives but share the same constraint. One can easily apply the framework to a new problem by just swapping the objective without changing the constraint.
Second, the constraint itself captures the key loss mechanism in plasmonic systems, i.e., the material absorption. In particular, the constraint enforces the power absorbed (dissipated) by the polarization currents to be smaller than the power ``extinguished'' (absorbed plus scattered) from the incoming wave. When formulated in a volume-integral form of the optical theorem~\cite{jackson1999classical, Newton1976, bohren2008absorption, carney2004generalized}, the constraint is convex, hence amenable to global bounds.
One constraint, however, is obviously insufficient for general photonic systems. As a result, for common scenarios such as lossless dielectrics~\cite{jahani2016all, kuznetsov2016optically, chen2020flat}, the lossy-material bounds diverge.

In this thesis, I generalize the single constraint in the lossy-material bounds to an infinite number of constraints, and establish  a general framework for computing bounds for almost every electromagnetic scattering scenario. The constraints correspond to power-conservation and field-correlation constraints in photonic systems.  
The power-conservation constraints conserve the complex-valued Poynting flux at each point in space.
Integrating them over a designable region, I retrieve the optical theorem and use it to derive semi-analytical bounds for applications including perfect absorbers and optimal wavefront shaping.
Applying them individually at each point in the designable region, I solve for computational bounds for applications including analog computing and broadband extinction.
The field-correlation constraints, on the other hand, dictate how scatterers behave across multiple scattering scenarios, leading to bounds on multi-functional nanophotonics including two-frequency sensing and beam switching. 
This theory prompts further research in fast algorithms for large-scale meta-optics, optimal wave communications, and optimal photonic designs. All these are described in more detail next, outlining the chapters of this thesis.

To start, Chapter~\ref{chap:analy} generalizes the lossy-material bounds by including both material and radiative losses in the optical theorem.
The optical theorem imposes an power-conservation requirement for any photonic device such that its absorption plus its scattered power equal its extinction.
Lossy-material bounds~\cite{miller_fundamental_2016, shim2019fundamental,miller2015shape,miller2017limits} distill to loosening the optical-theorem constraint to an inequality that absorbed power is bounded above by extinction. 
 Many other previous bounds~\cite{ thongrattanasiri_complete_2012, pendry1999radiative ,Yaghjian1996,hamam_coupled-mode_2007,kwon2009optimal, ruan2011design, liberal2014least, liberal2014upper, hugonin_fundamental_2015, ivanenko2019optical,rahimzadegan2017fundamental,liu2018optimal} distill in essence to loosening this constraint to an inequality that \emph{scattered} power is bounded above by extinction.
 Chapter~\ref{chap:analy} unifies these seemingly disparate bounds, generalizing and tightening all of them, and reveals new insight for optimal nanophotonic design, with applications including far-field scattering, near-field local-density-of-states engineering, optimal wavefront shaping, and the design of perfect absorbers. The
ramifications of our bounds for perfect absorbers are striking: we prove that independent of the geometric patterning, the minimum thickness of perfect or near-perfect absorbers comprising conventional plasmonic materials is typically on the order of 50-100~nm at visible wavelengths, which are roughly 100$\times$ larger than those suggested by previous lossy-material bounds~\cite{miller_fundamental_2016} and sum rules~\cite{sohl_gustafsson_kristensson_2007}.

Chapter~\ref{chap:comp} generalizes the single optical-theorem constraint to an infinite number of local-power-conservation constraints, offering a universal template to compute bounds for nanophotonics.
These local-power-conservation constraints originate from the fact that, for any scattering process, power has to be conserved at every point in space. They can be interpreted as applying the complex-valued Poynting theorem to infinitely small bounding surfaces at every point in space.
They generalize the previously utilized optical theorem from power conservation over an entire design region to power conservation at every point in space, leading to tighter bounds for larger and more complex photonic systems.
An important feature of these local-power-conservation constraints is that they are quadratic forms (of the induced polarization field) which are amenable to convex relaxation~\cite{luo_semidefinite_2010-1, park_general_2017} and hence the computation of bounds.
The resulting bounds lack the intuition of the analytical expressions of the bounds in Chapter~\ref{chap:analy}, but provide a straightforward procedure to derive upper limits for almost every objective in nanophotonics. This procedure is detailed in Chapter~\ref{chap:comp} with two examples: bounds to the minimum size of any
linear optical computing units and bounds on far-field scattering properties over any arbitrary bandwidth.

Chapter~\ref{chap:multi} generalizes the  power-conservation constraints to field-correlation constraints, revealing the fundamental limits of multi-functional nanophotonics.
Multi-functional nanophotonics, ranging from polychromatic mode coupling~\cite{piggott2015inverse, piggott2017fabrication} to tunable beam switching~\cite{Resler1996,He2019, chung2020tunable}, miniaturize the device footprint by condensing multiple functionalities into a single device. However, the added complexity, that a single photonic structure should handle multiple functionalities, cannot be captured by previous bounds.
To this end, Chapter \ref{chap:multi} shows that the single-structure requirement translates to constraints on the correlations between polarization fields induced in different scattering events.
The necessity and utility of these field-correlation constraints is demonstrated in two multi-functional designs: maximal reflectivity contrast for optical sensing and maximum efficiency for optical beam switching.
 
The conservation constraints are reformulated in Chapter~\ref{chap:sparse} with the differential form of Maxwell's equations. This offers a compelling advantage: the resulting optimization problems comprise \emph{sparse} matrices, which can lead to significant computational speedups. 
One limitation of the above bounds is that
their simulation time requirements increase quickly with system size, so they cannot scale to large-area photonic devices. 
In contrast, the computation complexity of the sparsity-accelerated bounds in Chapter~\ref{chap:sparse} increases much slower with the device dimensions, especially when the underlying sparsity pattern is close to a chordal graph~\cite{vandenberghe2015chordal}, which in nanophotonics are designs with one long dimension such as high-aspect-ratio metasurfaces~\cite{she2018large, li2022inverse, zhang2022high}, waveguides, and multilayered thin films~\cite{yeh1990optical, faist1994quantum, xue2021high}. 
In particular, we bound the maximal focusing efficiency of large-area metalenses a hundred wavelength in diameter, paving the way for future large-scale bounds.

Chapter~\ref{chap:comm} transfers the idea of shape-independent bounds from nanophotonic design to wave communication in free space and puts upper limits to fundamental metrics in information science. 
A key problem in electromagnetism is to design the geometries of two communication domains to maximize the number of orthogonal communication channels between them~\cite{miller1998spatial, Miller2000, piestun2000electromagnetic, Miller2019}. 
Instead of designing the geometry of the scatterer (as in photonic design), one must design the geometry of the source and receiver domains which, in practice, often represent antenna arrays of various shapes~\cite{telatar1999capacity,goldsmith2003capacity, tse2005fundamentals}.
Leveraging the monotonic property of the Green's function operator that underlies many nanophotonic bounds, Chapter~\ref{chap:comm} establishes a tight bound on the individual coupling strengths between two domains of any shape, as well as bounds on two fundamental metrics in communication science: the maximal number of non-trivial channels and their information capacities.
With general applicability, the bounds in Chapter \ref{chap:comm} reveal the fundamental limits to what is possible through engineering the communication domains of electromagnetic waves.

\begin{figure} [t]
\centering
\includegraphics[width=1\textwidth]{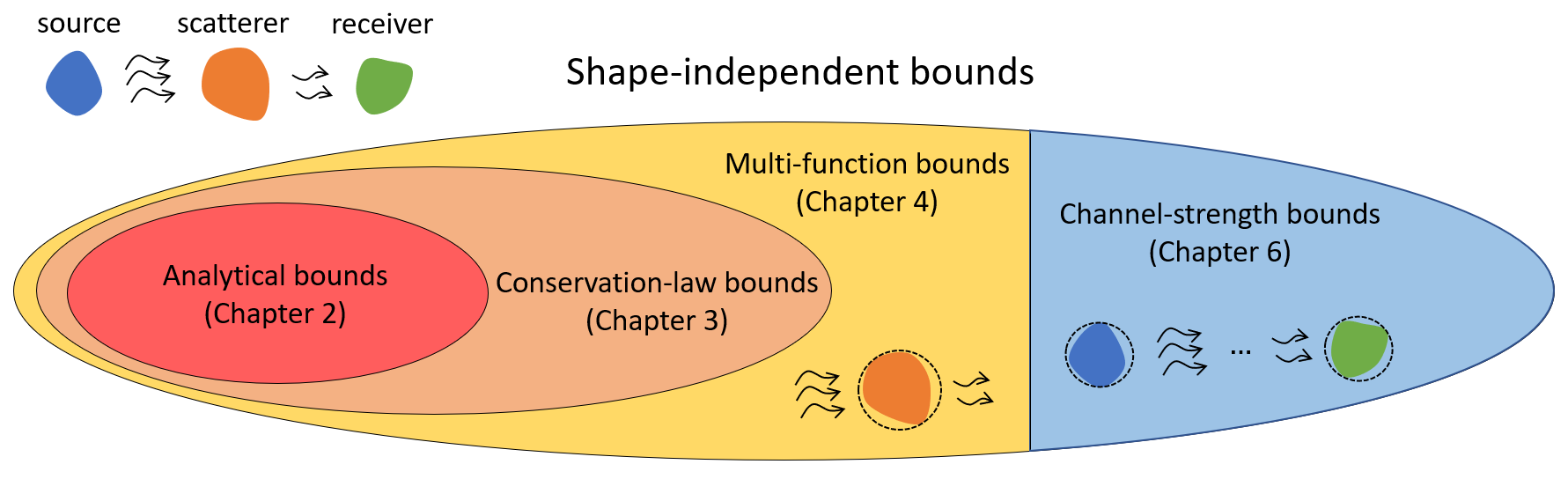}
        \caption{A Venn diagram of the hierarchy of shape-independent bounds on electromagnetic scattering presented in this thesis. A typical scattering event consists of two types of shapes: the shapes of the source and receiver domains, bounded by channel-strengths bounds in Chapter~\ref{chap:comm}, and the shapes of the scatterer, bounded by the techniques of Chapters \ref{chap:analy}, \ref{chap:comp}, and \ref{chap:multi}.} 
	\label{fig:chap1-fig1}
\end{figure} 

Together, the results in this thesis creates a systematic theory of fundamental limits in nanophotonics.
The Venn diagram in Fig.~\ref{fig:chap1-fig1} depicts our hierarchy of shape-independent bounds on electromagnetic scattering.
Electromagnetic scattering comprises a source domain, a scattering region, and a ``receiver'' (which may overlap). 
The geometric design of source and receiver domains is bounded by the channel-strength bounds in Chapter~\ref{chap:comm} on the right side of the Venn diagram. The geometric design of the scatterer is bounded by the bounds on the left of the Venn diagram, which is further categorized into three levels: the multi-functional bounds, the conservation-law bounds, and the analytical bounds. 
The multi-functional bounds in Chapter~\ref{chap:multi} are the most general bounds that apply to devices operating under multiple scattering scenarios. Nested in them, the conservation-law bounds in Chapter~\ref{chap:comp} can be considered as multi-functional bounds applied to a device that operates under only a single scattering scenario. Within those single-scattering devices, the performance of 
certain architectures (e.g., ultrathin perfect absorbers) can be bounded with just the single optical-theorem constraint of Chapter~\ref{chap:analy}.
As discussed in the Conclusion (Chapter~\ref{chap:conclu}), our framework applies to a wide range of emerging design problems. It admits generalizations beyond linear electromagnetic scattering such as nonlinear optics, eigenvalue problems, and quantum controls. It reveals new connections with fast algorithms such as the fast integral-equation solvers~\cite{rokhlin_rapid_1985,harrington_field_1993,johnson_block-iterative_2001}, sparse semidefinite programming~\cite{kim2011exploiting, vandenberghe2015chordal}, and low-rank semidefinite programming~\cite{ burer2003nonlinear, burer2005local, boumal2016non, frank1956algorithm, jaggi2013revisiting, yurtsever2021scalable}. More broadly, our framework bridges designs in nanophotonics with designs in other areas of science~\cite{Boyd1994,Goemans1995,Luo2010,sojoudi2012physics,Candes2013,Horstmeyer2015}, which, in turn, leads to new insights in optimal photonic designs. These tantalizing new directions are  discussed in detail in the Conclusion.

\chapter{Analytical bounds via the optical theorem}
\label{chap:analy}

In this chapter, we derive bounds through a simple, powerful prescription: drop the full constraints of Maxwell's equations, but require the optical theorem~\cite{jackson1999classical, Newton1976, bohren2008absorption, carney2004generalized} to be satisfied by any polarization-current response. A restricted version of this approach was used in the ``lossy-material bounds"~\cite{miller_fundamental_2016, ivanenko2019optical, nordebo2019optimal,  miller2017limits, shim2019fundamental, miller2015shape, michon2019limits, yang2018maximal, yang2017low, dias_fundamental_2019}, in which only the material-loss portion of the optical theorem concerning material loss is imposed. Separately, there had been many ``channel bounds''~\cite{ thongrattanasiri_complete_2012, pendry1999radiative ,Yaghjian1996,hamam_coupled-mode_2007,kwon2009optimal, ruan2011design, liberal2014least, liberal2014upper, hugonin_fundamental_2015, ivanenko2019optical,rahimzadegan2017fundamental,liu2018optimal}, in which electric fields were decomposed into a basis of (heuristically truncated) a few multipolar orders with nontrivial radiative loss~\cite{pendry1999radiative,liberal2014least,liberal2014upper,hugonin_fundamental_2015,ivanenko2019optical,rahimzadegan2017fundamental,liu2018optimal}. The optical theorem we use in this chapter accounts for both material and radiative losses in optical systems, unifying and tightening previous bounds.

This chapter is outlined as follows. We first introduce the key optical-theorem constraint, which enforces the requirement that the sum of absorption and scattered power equals extinction, in terms of the polarization response (Section~\ref{sec:chap2-optical}). 
Channel bounds distill in essence to loosening the optical-theorem constraint to an inequality that scattered power is bounded above by extinction. Lossy-material bounds distill to loosening this constraint to an inequality that absorbed power is bounded above by extinction. Our key innovation is the retention of the full optical theorem. We use Lagrangian duality to solve the resulting optimization problems, ultimately yielding analytical and semi-analytical bounds to arbitrary response functions (Section~\ref{sec:chap2-analytical}). For the important case of plane-wave scattering (Section~\ref{sec:chap2-plane}), we derive explicit bound expressions and identify an important application: perfect absorbers, whose minimal thickness we predict is much tighter than previous bounds. Our bounds explicitly account for  the incident waves; for a given material and designable region, then, we can treat the illumination-field degrees of freedom as the variables and identify the optimal incoming-wave excitation (Section~\ref{sec:chap2-optimal}). As one example, we show that in certain parameter regimes, the extinction of an unpatterned sphere under the optimal illumination field exceeds the upper bound under plane-wave excitation, which means that, as long as the incident field is a plane wave, there is no patterning of any kind that can reach the same power-response level of the optimal illumination. In Section~\ref{sec:chap2-discussion}, we discuss the limitations of the analytical bounds and their resolutions.



\section{Optical theorem}
\label{sec:chap2-optical}

The optical theorem manifests energy conservation: the total power taken from an incident field must equal the sum of the powers absorbed and scattered. The key version of the optical theorem that comprises a meaningful constraint arises from the volume equivalence principle: any scattering problem can be separated into a background material distribution and an additional ``scatterer'' susceptibility. The total fields $\vE(\vx)$ are given by the fields incident within the background, $\vE_{\rm inc}(\vx)$, plus the scattered fields $\vE_{\rm scat}(\vx)$ that arise from polarization fields $\vP(\vx)$ induced in the scatterer volume $V$. 
The volume equivalence principles refers to the idea that the polarization fields effectively replace the scatterer and radiate the scattered field as in free space (or under the background material distribution): $\vE_{\rm scat}(\vx)=\int_{V} \vG_0(\vx,\vx')\cdot\vP(\vx')\md\vx'$, where $\vG_0(\vx,\vx')$ is the background Green's function.
We assume a non-magnetic isotropic material with a scalar susceptibility $\chi$ composing the scatterer, though our framework extends to arbitrary materials with non-local, magnetic, anisotropic response as described in Appendix~\ref{sec:appenG-sec12}.
We use dimensionless units for which the vacuum permittivity and permeability equal 1, $\varepsilon_0 = \mu_0 = 1$.
In addition, we define a variable $\xi$ that is the negative inverse of the susceptibility of the material, $\xi = -\chi^{-1}$. With these conventions, the statement that the total field equals the sum of the incident and scattered fields can be written as 
    \begin{equation}
        -\xi \vP(\vx) = \vE_{\rm inc}(\vx) + \int_{V} \vG_0(\vx,\vx')\cdot\vP(\vx')\md\vx',
        \label{eq:chap2-vie}
    \end{equation}
which is the volume-integral equation that holds for any $\vx\in V$.
\begin{figure} [t]
\centering
\includegraphics[width=0.5\textwidth]{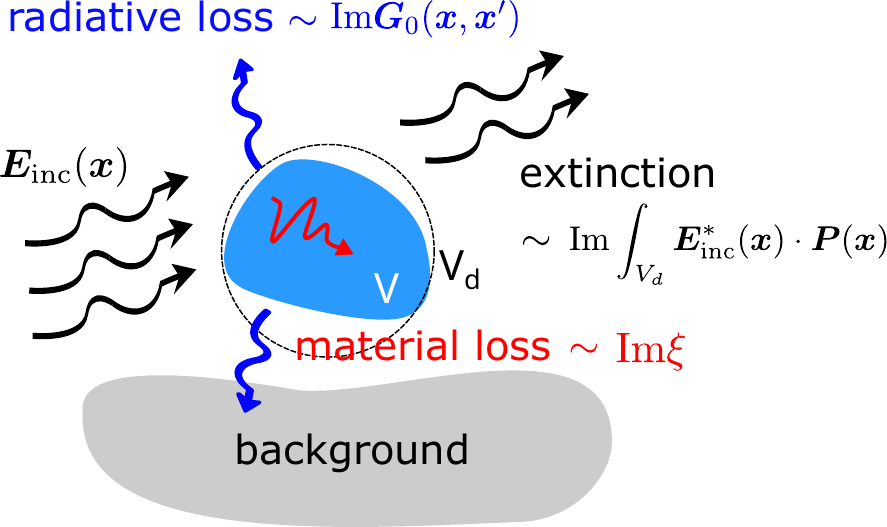}
        \caption{Illustration of the two loss mechanisms in electromagnetic scattering. An incident field $\vE_{\rm inc}(\vx)$ induces polarization fields $\vP(\vx)$ in the scatterer volume $V$ inside the bounding volume $V_d$. Energy dissipated into the material corresponds to material loss, determined by the operator $\Im{\xi}$, which equals $\Im \chi / |\chi|^2$ for susceptibility $\chi$. Energy coupled to the background, into far-field or near-field power exchange, corresponds to radiative loss, determined by the operator $\Im\vG_0(\vx,\vx')$, where $\vG_0(\vx,\vx')$ represents the background (e.g. free-space) Green's function. Total extinction is the sum of the two and is linear in $\vP(\vx)$, as dictated by the optical theorem.} 
	\label{fig:chap2-figure1}
\end{figure} 
The volume-integral-equation optical theorem can be derived from \eqref{chap2-vie} by taking the inner product of \eqref{chap2-vie} with $\vP^*(\vx)$, integrating over the scatter volume $V$, multiplying by $\omega/2$, and taking the imaginary part of both sides of the equation, yielding:
 \begin{align}
        \underbrace{\frac{\omega}{2}\int_V\int_V \vP^*(\vx)\cdot\Im\vG_0(\vx,\vx')\cdot\vP(\vx')\md\vx\md\vx'}_{\Pscat}
        +
        \underbrace{\frac{\omega}{2}\Im\xi\int_V|\vP(\vx)|^2\md\vx}_{\Pabs}  = \underbrace{\frac{\omega}{2}\Im\int_V \vE^*_{\rm inc}(\vx)\cdot\vP(\vx)\md\vx}_{\Pext}
        \label{eq:chap2-optthm_V}
    \end{align}
From left to right, we identify the three terms as scattered, absorbed, and extinguished power of the scatterer $V$~\cite{polimeridis2014computation,kong1972theorems}, illustrated in \figref{chap2-figure1}. 
The operator $\Im\vG_0(\vx,\vx')$ represents power radiated into the background, into near-field or, more typically, far-field scattering channels. For any background materials, $\Im\vG_0(\vx,\vx')$ can be computed by standard volume-integral (or discrete-dipole-approximation) techniques~\cite{chew1995waves,Purcell1973}, and when the background is lossless over the scattering volume it is nonsingular and simpler to compute~\cite{Reid2017subm}. It is a positive semidefinite operator, because the power radiated by any polarization field must be nonnegative in a passive system.  
The second term with $\Im\xi$ represents absorbed power: work done by the polarization field on the total fields. In terms of the susceptibility, $\Im \xi = \Im \chi/|\chi|^2$, which is the inverse of a material ``figure of merit'' that has appeared in many lossy-material bounds~\cite{miller_fundamental_2016, shim2019fundamental, miller2017limits}. The operator $\Im \xi$ is positive definite for any material without gain~\cite{chew1995waves,welters_speed--light_2014}.
Finally, the third term is the imaginary part of the overlap between the incident field and the induced currents, which corresponds to extinction (total power taken from the incident fields). 

A key property of the optical theorem in \eqref{chap2-optthm_V} is that its volume integrals can be extended to a bounding domain that includes all possible scatterer geometries. As the polarization field $\vP(\vx)$ is always zero outside the scatterer,  we can extend all the volume integrals in \eqref{chap2-optthm_V} from the scatterer volume $V$ to a larger bounding volume $V_d$:
 \begin{align}
    \underbrace{\frac{\omega}{2}\int_{V_d}\int_{V_d} \vP^*(\vx)\cdot\Im\vG_0(\vx,\vx')\cdot\vP(\vx')\md\vx\md\vx'}_{\Pscat}
    +
    \underbrace{\frac{\omega}{2}\Im\xi\int_{V_d}|\vP(\vx)|^2\md\vx}_{\Pabs}  = \underbrace{\frac{\omega}{2}\Im\int_{V_d} \vE^*_{\rm inc}(\vx)\cdot\vP(\vx)\md\vx}_{\Pext}
    \label{eq:chap2-optthm}
\end{align}
The three terms now represent the possible scattering, absorption, and extinction of \textit{any} scatterer inside the bounding volume $V_d$. The bounding volume $V_d$ can be a sphere (as in \figref{chap2-figure1}), a planar film, or any highly symmetric geometries where the operator $\Im\vG_0$ can be analytically factorized.
While no simplification of Maxwell's equations will contain every possible constraint, the optical theorem of \eqref{chap2-optthm} has four key features: (1) it contains \emph{both} the powers radiated ($\Pscat$) and absorbed ($\Pabs$) by the polarization fields in a single expression, (2) it is a quadratic constraint that is known to have ``hidden'' convexity for any quadratic objective function~\cite{ben-tal_hidden_1996}, (3) it enforces power conservation in the scattering body, and (4) it incorporates information about the material composition of the scatterer (in the variable $\xi$), the maximal size of the scatterer (defined by the bounding volume $V_d$), while independent of any other patterning details of the scatterer inside the bounding volume.

The optical theorem of \eqref{chap2-optthm} constrains the polarization field $\vP(\vx)$ to lie on the surface of a high-dimensional ellipsoid whose principal axes are the eigenvectors of the operators on the left-hand side of \eqref{chap2-optthm}  and whose radii are constrained by the norm of $\vE_{\rm inc}(\vx)$. In Appendix~\ref{sec:appenG-sec8} we show that all previous channel and lossy-material bounds discussed at the introduction of this chapter can be derived by applying weaker versions of \eqref{chap2-optthm}. Channel bounds can be derived by loosening \eqref{chap2-optthm} to the inequality $\Pscat \leq \Pext$, without the absorption term (but implicitly using the fact that absorbed power is nonnegative). Lossy-material bounds can be derived by loosening \eqref{chap2-optthm} to the inequality $\Pabs \leq \Pext$, without the scattered-power term (but using the fact that scattered power is nonnegative). Of course, including both constraints can only result in equal or tighter bounds, which we derive below.

\section{Analytical bounds}
\label{sec:chap2-analytical}
In this section, we use the optical theorem to constrain the possible distributions of the polarization fields to derive (semi-)analytical bounds on arbitrary scattering responses. 
To start, we discretize the fields in the optical theorem of \eqref{chap2-optthm} in standard basis (e.g., real-space grids, vector spherical waves, plane waves) such that $\vP(\vx)$ and $\vE_{\rm inc}(\vx)$ become vectors $p$ and $\einc$, and the corresponding integrals of $\Im\vG_0(\vx, \vx')$ and $\Im\xi$ become matrices $\Im\GG_0$ and $\Im(\xi)\II$, yielding:
    \begin{equation}
        p^\dagger (\ImGO+\Im\xi \II)p = \Im\left(e^\dagger_{\rm inc} p \right),
        \label{eq:chap2-OptThmDis}
    \end{equation}
where $\dagger$ denotes conjugate transpose. This is the discretized form of the optical theorem that we will use below to constrain the polarization field $p$.

Any electromagnetic power-flow objective function $f$ is either linear or quadratic in the polarization field $p$. In the same basis of \eqref{chap2-OptThmDis}, it can be generically written as $f(p) = p^\dagger \mathbb{A} p + \Im \left(\beta^\dagger p\right)$, where $\mathbb{A}$ is a Hermitian matrix and $\beta$ is a vector, both defined on the bounding domain.  Maximizing $f$ over all possible scatterers amounts to the optimization problem: 
\begin{equation}
\begin{aligned}
& \underset{p}{\text{maximize}}
& & f(p) = p^\dagger \mathbb{A} p +  \Im\left(\beta^\dagger p\right) \\
& \text{subject to}
& & p^\dagger \left\{\Im \xi\II + \ImGO \right\} p = \Im \left(\einc^\dagger p\right).
\end{aligned}
\label{eq:chap2-general-formalism}
\end{equation}
This is a quadratic objective with a single quadratic constraint, which is known to have strong duality~\cite{Boyd2004}. If we follow standard convex-optimization conventions and consider as our ``primal'' problem that of \eqref{chap2-general-formalism}, but instead written as a minimization over the negative of $f(p)$, then strong duality implies that the maximum of the corresponding Lagrangian dual function equals the minimum of the primal problem, and thus the maximum of \eqref{chap2-general-formalism}. By straightforward calculations, the dual function is
\begin{align}
g(\nu)=
\begin{cases}
-\frac{1}{4}(\beta + \nu \einc)^\dagger \BB^{-1}(\nu)(\beta + \nu \einc) & \nu > \nu_0 
\\
-\infty, & \nu< \nu_0
\end{cases}
\label{eq:chap2-gnu}
\end{align}
where $\nu$ is the dual variable, $\BB(\nu) = -\mathbb{A}+\nu(\Im\xi\II+\ImGO)$ and $\nu_0$ is the value of $\nu$ for which the minimum eigenvalue of $\BB(\nu_0)$ is zero. (The definiteness of $\ImGO$ and $\Im(\xi)\II$ ensure there is only one $\nu_0$, cf. Appendix~\ref{sec:appenG-sec1}.) At $\nu=\nu_0$, some care is needed to evaluate $g(\nu_0)$ because the inverse of $\BB(\nu_0)$ does not exist (due to the 0 eigenvalue). If $\beta + \nu_0\einc$ is in the range of $\BB(\nu_0)$, then $g(\nu_0)$ takes the value of the first case in \eqref{chap2-gnu} with the inverse operator replaced by the pseudo-inverse; if not, then $g(\nu_0)\rightarrow -\infty$. (Each scenario arises in the examples below.) By the strong duality of \eqref{chap2-general-formalism}, the optimal value of the dual function, \eqref{chap2-gnu}, gives the optimal value of the ``primal" problem, \eqref{chap2-general-formalism} (accounting for the sign changes in converting the maximization to minimization). In Appendix~\ref{sec:appenG-sec1} we identify the only two possible optimal values of $\nu$: $\nu_0$, defined above, or $\nu_1$, which is the stationary point for $\nu>\nu_0$ at which the derivative of $g(\nu)$ equals zero. Denoting this optimal value $\nu^*$, we can write the maximal response as:
\begin{equation}
f_{\rm max} = \frac{1}{4}(\beta + \nu^* \einc)^\dagger  \left[-\mathbb{A}+\nu^*(\Im\xi\II+\ImGO)\right]^{-1}(\beta + \nu^* \einc).
\label{eq:chap2-gen_bound}
\end{equation}
Although \eqref{chap2-gen_bound} may appear abstract, it is a general bound that applies for any linear or quadratic electromagnetic response function, from which more application-specific specialized results follow.

If one wants to maximize one of the terms already present in the constraint, i.e. absorption, scattered power, or extinction, then the $\mathbb{A}$ and $\beta$ terms take particularly simple forms (cf. Appendix~\ref{sec:appenG-sec2}), leading to the bounds:
\begin{align}
\Pext &\leq \frac{\omega}{2} \einc^\dagger \left(\Im\xi\II+\ImGO\right)^{-1} \einc \label{eq:chap2-ext-gen}\\
\Pabs &\leq \frac{\omega}{2}\frac{\nu^{*2}}{4} \einc^\dagger[(\nu^*-1)\Im\xi\II+\nu^*\ImGO]^{-1} \einc \label{eq:chap2-abs-gen} \\
\Pscat &\leq \frac{\omega}{2}\frac{\nu^{*2}}{4} \einc^\dagger[\nu^*\Im\xi\II + (\nu^* - 1)\ImGO]^{-1} \einc. \label{eq:chap2-sca-gen}
\end{align}
where $\nu^*$ is the dual-variable numerical constant (Appendix~\ref{sec:appenG-sec2}).

Another objective of interest is LDOS, whose bounds represent maximal spontaneous-emission enhancements~\cite{novotny2012principles, liang2013formulation, purcell1946resonance, taflove2013advances,xu2000quantum}. Total (electric) LDOS, $\rho_{\rm tot}$, is proportional to the averaged power emitted by three orthogonally polarized and uncorrelated unit electric dipoles~\cite{wijnands1997green, martin_electromagnetic_1998, d2004electromagnetic, joulain2003definition}. It can be separated into a radiative part, $\rho_{\rm rad}$, for far-field radiation, and a non-radiative part, $\rho_{\rm nr}$, that is absorbed by the scatterer~\cite{jackson1999classical}. As shown in Appendix~\ref{sec:appenG-sec13}, the LDOS bounds for nonmagnetic materials simplify to expressions related to the maximum power quantities given in \eqrefrange{chap2-ext-gen}{chap2-sca-gen}: 
\begin{align}
\rho_{\rm tot} &\leq \frac{2}{\pi\omega^2}\sum_j P^{\rm max}_{\text{ext}, j} + \rho_0 \label{eq:chap2-rhotot_bnd} \\ 
\rho_{\rm nr} &\leq \frac{2}{\pi\omega^2}\sum_j P^{\rm max}_{\text{abs}, j} \\
\rho_{\rm rad} &\leq \frac{2}{\pi\omega^2}\sum_j P^{\rm max}_{\text{sca}, j} + \rho_0,
\label{eq:chap2-rhorad_bnd}
\end{align}
where $\rho_0$ is the electric LDOS of the background material, and equals to $\frac{\omega^2}{2\pi^2c^3}$ for a scatterer in vacuum~\cite{joulain2005surface}. The summation over $j={1,2,3}$ accounts for three orthogonally polarized unit dipoles.  As shown in Appendix~\ref{sec:appenG-sec13}, our bound is tighter than previous bounds on LDOS \cite{miller_fundamental_2016}. In the extreme near field, where material loss dominates, our bound agrees with the known lossy-material bound~\cite{miller_fundamental_2016}.

The bounds of \eqrefrange{chap2-gen_bound}{chap2-rhorad_bnd} can be generalized to arbitrary material composition (e.g., inhomogeneous, nonlocal, and magnetic), which involves Green's function and material susceptibility in six-vector notations. In Appendix~\ref{sec:appenG-sec12}, we provide the most general expressions of bounds and their step-by-step simplifications under the restrictions of the incident field, material, and bounding volumes. In the next section, we consider the important specialization of a plane wave incident upon an isotropic nonmagnetic medium.

\section{Plane-wave scattering}
\label{sec:chap2-plane}
A prototypical scattering problem is that of a plane wave in free space incident upon an isotropic (scalar susceptibility) scatterer. Because $\ImGO$ is positive-definite, we can simplify its eigendecomposition to $\ImGO = \Vbb\Vbb^\dagger$, where the columns of $\Vbb$, which we denote $v_i$, form an orthogonal basis of polarization field in the bounding volume. They are normalized such that the set of $v_i^\dagger v_i$ are the eigenvalues of $\ImGO$ and represent the powers radiated by unit-normalization polarization field. Put simply, the $v_i$ span the space of scattering channels (defined as in Ref.~\cite{Miller2019} in a bounding volume) and the eigenvalues $\rho_i$ represent corresponding radiated powers.  

An incident propagating plane wave (or any wave incident from the far field, cf. Appendix~\ref{sec:appenG-sec13}) can be decomposed in the basis $\Vbb$. We write the expansion as $\einc = \frac{1}{k^{3/2}}\sum_i e_i v_i$, where the $e_i$ are the expansion coefficients and the prefactor (with free-space wavenumber $k$) simplifies the expressions below. Inserting the eigendecomposition of $\ImGO$ and the plane-wave expansion of $\einc$ into \eqrefrange{chap2-ext-gen}{chap2-sca-gen}, we obtain bounds for plane-wave scattering:
\begin{align}
P_{\rm ext} &\leq \frac{\lambda^2}{8\pi^2}\sum_i |e_i|^2\frac{\rho_i}{\Im{\xi} + \rho_i} \label{eq:chap2-ext-cha-bound} \\
P_{\rm abs} &\leq \frac{\lambda^2}{8\pi^2}\frac{\nu^{*2}}{4}\sum_i |e_i|^2\frac{\rho_i}{(\nu^*-1)\Im{\xi}+\nu^*\rho_i} \label{eq:chap2-abs-cha-bound} \\
P_{\rm sca} &\leq \frac{\lambda^2}{8\pi^2}\frac{\nu^{*2}}{4}\sum_i |e_i|^2\frac{\rho_i}{\nu^*\Im{\xi}+(\nu^*-1)\rho_i}. \label{eq:chap2-sca-cha-bound}
\end{align}
The variable $\nu^*$ is the optimal dual variable discussed above; its value can be found computationally via a transcendental equation given in Appendix~\ref{sec:appenG-sec3}. The bounds of \eqrefrange{chap2-ext-cha-bound}{chap2-sca-cha-bound} naturally generalize the channel bounds ($\sim \lambda^2$) and lossy-material bounds ($\sim 1/\Im \xi = |\chi|^2 / \Im \chi$). In Appendix~\ref{sec:appenG-sec8}, we prove that removing either dissipation pathway results in the previous expressions. 

The bounds of \eqrefrange{chap2-ext-cha-bound}{chap2-sca-cha-bound} require knowledge of the eigenvalues of $\ImGO$, and thus the exact shape of the bounding volume, to compute the values of $\rho_i$. However, analytical expressions for $\rho_i$ are known for high-symmetry geometries. 
The following two sub-sections consider two possible scenarios: (a) scattering by finite-sized objects, which are enclosed in spherical bounding volume, and (b) scattering by extended (e.g. periodic) objects, which are enclosed in planar bounding volume.

\begin{figure}[t]
\includegraphics[width=\textwidth]{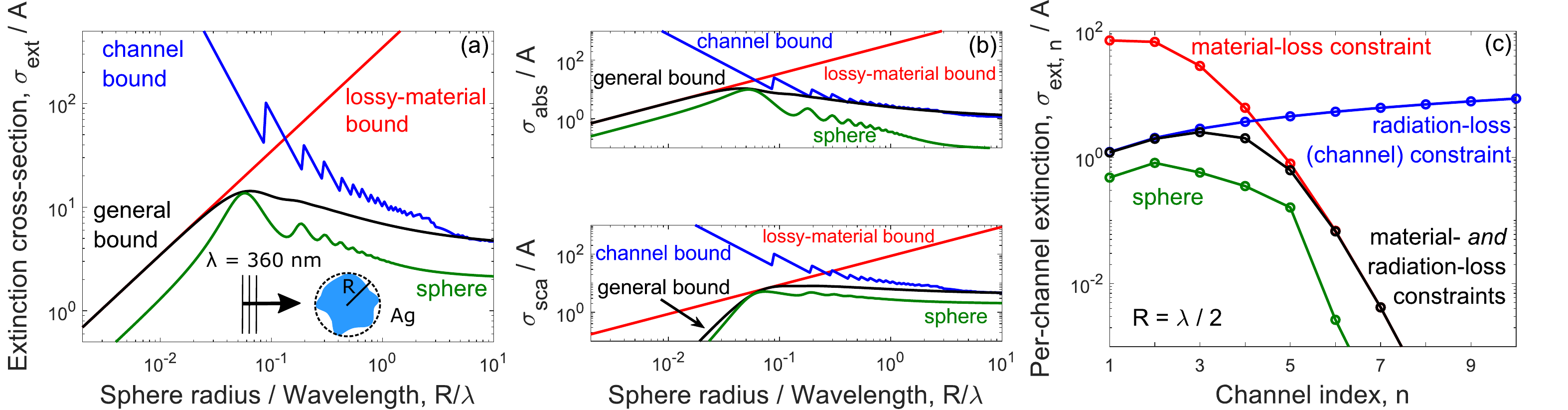}
	\centering
	\caption{Plane wave of wavelength $\lambda=360$ nm scattering from a finite Ag \cite{johnson1972optical} scatterer, enclosed by a spherical bounding volume with radius $R$. The channel bound is heuristically regularized by ignoring small-scattering high-order channels. All cross-sections are normalized by geometric cross-section $A$. (a). Bound of extinction cross-section for different $R$. The general bound regularizes divergence in previous bounds and are tighter for wavelength-scale sizes. (b) Similar behavior is observed in the bounds for scattering and absorption cross-sections. (c) Per-channel extinction cross section $\sigma_{\text{ext},n}$ (defined at the end of Appendix~\ref{sec:appenG-sec3}) for $R=\lambda/2$. Low-order scattering channels are dominated by radiative loss, while high-order scattering channels are dominated by material loss.}
	\label{fig:chap2-figure2}
\end{figure}

\subsection{Finite-sized scatterers}
\label{sec:fss}
Finite-sized scatterers can be enclosed by a minimal bounding sphere with radius $R$, as in the inset of \figref{chap2-figure2}(a). The basis functions $v_i$ are vector spherical waves (VSWs), representing orthogonal scattering channels, with exact expressions given in Appendix~\ref{sec:appenG-sec10}. The state $i$ labels a triplet, $i=\{n,m,j\}$, where $n=1,2,...$ is the total angular momentum, $m=-n,...,n$ is the $z$-directed angular momentum, and $j=1,2$ labels two polarizations. In this basis, the expansion coefficients of a plane wave are $|e_i|^2=\pi(2n+1)\delta_{m,\pm 1}|E_0|^2$, where $E_0$ is the plane-wave amplitude. We show in Appendix~\ref{sec:appenG-sec10} that the value of $\rho_i$ is given by the integrals of spherical Bessel functions. Inserting these expressions into \eqrefrange{chap2-ext-cha-bound}{chap2-sca-cha-bound} and normalizing the power-quantity bounds by plane wave intensity $|E_0|^2/2$, we obtain bounds for extinction, scattering, and absorption cross-sections.

In \figref{chap2-figure2}, we compare cross-section bounds derived from \eqrefrange{chap2-ext-cha-bound}{chap2-sca-cha-bound} to the actual scatterings of a silver sphere (permittivity data from \citeasnoun{johnson1972optical}) at wavelength $\lambda = \SI{360}{nm}$. The \SI{360}{nm} wavelength is close to the surface-plasmon resonance of a silver sphere, simplifying comparisons (instead of requiring inverse design for every data point). We include the previously derived channel~\cite{ruan2011design} and lossy-material \cite{miller_fundamental_2016} bounds for comparison, and in each plots our general bounds are significantly ``tighter'' (smaller) than the previous ones, except in the expected small- and large-size asymptotic limits. At a particular radius, the scattering response even approaches the general bound. In \figref{chap2-figure2}(c), we fix the radius at a half-wavelength and depict the per-channel contributions to the extinction bounds in the  radiation-loss-only, material-loss-only, and tandem-loss constraint cases. Higher-order channels have increasingly smaller radiative losses (causing unphysical divergences discussed below), such that material loss is the dominant dissipation channel. Conversely, material-loss-only constraints are inefficient for lower-order channels where radiative losses dominate. Incorporating both loss mechanisms removes the unphysical divergence, accounts for radiative losses, and sets the tightest bound among the three across all channels. 

Technically, the channel bound diverges for any finite-sized scatterer, and the blue solid line in \figref{chap2-figure2}(a) should be infinitely high. To obtain a reasonable finite value, we only incorporate channels for which the sphere scattering contributions are greater than 1\% of the maximal response. Yet requiring knowledge of the specific scattering structure to compute the upper limit is the key drawback of the channel bounds. This empirical truncation produces two artifacts in the presented channel bounds. First, it results in a step-like behavior which is most prominent at small radii, where only a handful channels contribute. At each radius where a new channel is introduced for consideration (based on this threshold), there is an unphysical increase in the bound due to the suddenly larger power available for scattering, absorption, and extinction. Such behavior is smoothed at large radii, where the contribution from each new channel is subsumed by the large number of existing channels. Second, as we show in Appendix~\ref{sec:appenG-sec9}, there can potentially be large contributions from channels beyond the threshold. The arbitrary cut-off results in inaccurate and unphysical \emph{under}estimates of the cross-sections, which is noticeable in the large size limit of Figs.~\ref{fig:chap2-figure2}(a,b), where the channel bound is slightly smaller than the general bound. The only way to avoid such artifacts would be to include all channels, in which case the channel bounds trivialize to infinite value for any radius.

\subsection{Extended scatterers}
\begin{figure}[t]
\includegraphics[width=1\textwidth]{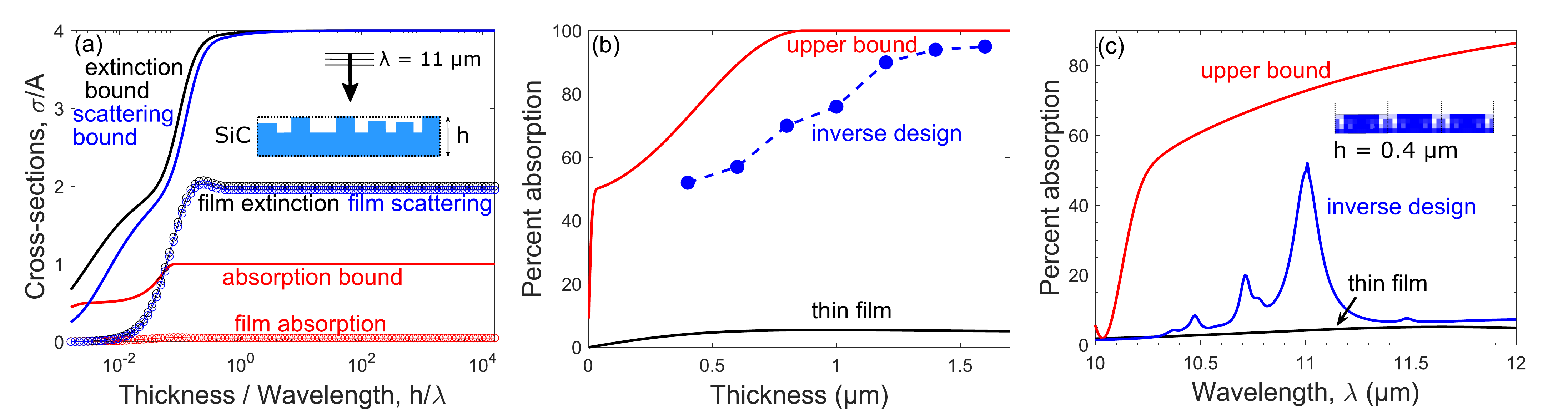}
	\centering
        \caption{Arbitrarily patterned SiC scatterer with maximum thickness $h$ excited by a plane wave at normal incidence and $\lambda=\SI{11}{\micro\meter}$ wavelength, where SiC is polaritonic. (a) Bounds for extinction, scattering, and absorption, compared to their values for a planar SiC \cite{francoeur2010spectral} film. (b) Inverse-designed SiC metasurfaces (blue markers), at varying thicknesses, achieve absorption levels at 64--95\% of the global bounds (red), suggesting the bounds are ``tight'' or nearly so. (c) Absorption spectrum of ultra-thin absorber from (b) with thickness $h=\SI{0.4}{\micro\meter}$. (Inset: inverse-design structure; blue represents SiC, white represents air.) At the target wavelength, the absorption of the inverse-designed structure is more than ten times that of the thin film, and reaches 72\% of the bound.}
	\label{fig:chap2-figure3}
\end{figure}

A second common scenario is scattering from an infinitely extended (e.g. periodic) scatterer. Such scatterers can always be enclosed by a planar ``film'' bounding volume with a minimal thickness $h$, as in the inset of \figref{chap2-figure3}(a). Then the basis functions $v_i$ of $\ImGO$ are the propagating plane waves with wave vectors $\vect{k} = k_x\vect{\hat{x}} + k_y\vect{\hat{y}} + k_z\vect{\hat{z}}$. Now the index $i$ labels the triplet $\{s,p,\vect{k}_\parallel\}$, where $s=\pm$ denotes even and odd modes, $p=M,N$ denotes TE and TM polarizations, and $\vect{k}_\parallel=k_x\vect{\hat{x}} + k_y\vect{\hat{y}}$ denotes the surface-parallel wave vector. In Appendix~\ref{sec:appenG-sec11} we provide the expressions for $v_i$, and show that the eigenvalues $\rho_i$ are given by
\begin{align}
\rho_{\pm, p}(\vect{k}_\parallel) = 
\begin{cases}
\frac{k^2h}{4k_z}(1\pm\frac{\sin(k_zh)}{k_zh}) \quad &p = \textrm{TE}\\
\frac{k^2h}{4k_z}(1\pm\frac{\sin(k_zh)}{k_zh})\mp\frac{\sin(k_zh)}{2}. \quad &p = \textrm{TM}
\end{cases}
\end{align}
The incident wave itself has nonzero expansion coefficients for basis functions with the same parallel wave vector, and is straightforward to expand: $|e_i|^2=2k_zk\delta_{p,p'}|E_0|^2$, where $p'$ is the incident polarization, $E_0$ is the plane wave amplitude, and $k=|\vect{k}|$. The optimal polarization field only comprise waves with parallel wave vector identical to that of the incident wave, simplifying the final bounds. Normalizing the bounds of \eqrefrange{chap2-ext-cha-bound}{chap2-sca-cha-bound} by the $z$-directed plane-wave intensity, $|E_0|^2k_z/2k$, gives cross-sections bounds for extended structures:
\begin{align}
\sigma_{\rm ext} / A &\leq 2\sum_{s=\pm} \frac{\rho_{s, p'}}{\Im{\xi} + \rho_{s, p'}} \label{eq:chap2-ext-cha-extend} \\
\sigma_{\rm abs} / A &\leq \frac{\left(\nu^{*}\right)^2}{2}\sum_{s=\pm}\frac{\rho_{s, p'}}{(\nu^*-1)\Im{\xi}+\nu^*\rho_{s, p'}} \label{eq:chap2-abs-cha-extend} \\
\sigma_{\rm sca} / A &\leq \frac{\left(\nu^{*}\right)^2}{2}\sum_{s=\pm}\frac{\rho_{s, p'}}{\nu^*\Im{\xi}+(\nu^*-1)\rho_{s, p'}}, \label{eq:chap2-sca-cha-extend}
\end{align}
where $A$ is the total surface area and $\rho_{s,p'}$ denotes the radiation loss by a scattering channel with parity $s$, polarization $p'$, and parallel wave vector $\vect{k}_\parallel$. Again, the high-symmetry bounding volume leads to analytical expressions that are easy to compute.

Figure \ref{fig:chap2-figure3}(a) compares the upper bounds of the normalized cross-sections with the cross-sections of SiC thin films at normal incidence and wavelength $\lambda=\SI{11}{\micro\meter}$, where the SiC supports phonon-polariton modes. The bounds indicate that scattering, absorption, and extinction must all be small for an ultrathin metasurface. When its thickness increases, the bounds show possible patterning effect, which plateaus when thickness is roughly one-tenth of the wavelength.

A key question for any bound is whether it is achievable with physical design. In order to test this feasibility, we utilize inverse design~\cite{Jameson1998,Sigmund2003,Lu2011,jensen2011topology,Miller2012a,Lalau-Keraly2013,Ganapati2014,aage2017giga}, a large-scale computational optimization technique for discovering optimal configurations of many design parameters, to pattern SiC films to approach our bounds. We use a standard ``topology-optimization'' approach~\cite{Sigmund2003,Miller2012a} in which the material is represented by a grayscale density function ranging from 0 (air) to 1 (SiC) at every point, and derivatives of the objective function (absorption, in this case) are computed using adjoint sensitivities. We prioritize feasibility tests (i.e., to address whether the bounds are theoretically achievable) over the design of easy-to-fabricate structures. To this end, we use grayscale permittivity distributions, which in theory can be mimicked by highly subwavelength patterns of holes. Recently developed techniques~\cite{Christiansen2019} have identified binary polaritonic structures that come quite close to their grayscale counterparts for many applications, suggesting that binary structures with similar performance levels to those presented here can be discovered. We give algorithmic details for our inverse-design procedure in Appendix~\ref{sec:appenG-sec6}.
    
\Figref{chap2-figure3}(b) depicts the bounds (red solid line) and the performance of thin films (black solid line) as a function of thickness, as well as six different inverse-design structures that bridge most of the gap between the thin films and the bounds. The incident wavelength is $\SI{11}{\micro\meter}$ and the period is $\SI{1.1}{\micro\meter}$, with minimum feature size $\SI{0.1}{\micro\meter}$. For an ultrathin absorber with thickness $\SI{0.4}{\micro\meter}$, the inverse-designed metasurface can reach 72\% of the global bound. In \figref{chap2-figure3}(c) we isolate the design at this smallest thickness and show its spectral absorption percentage, as well as its geometrical design (inset). Detail of the inverse design are given in Appendix~\ref{sec:appenG-sec6}. Since the objective is to compare against the global, we do not impose binarization, lithographic, or other fabrication constraints. The inverse designs approaching the bounds suggests the latter are ``tight'' or nearly so.

\begin{figure}[t]
	\includegraphics[width=\textwidth]{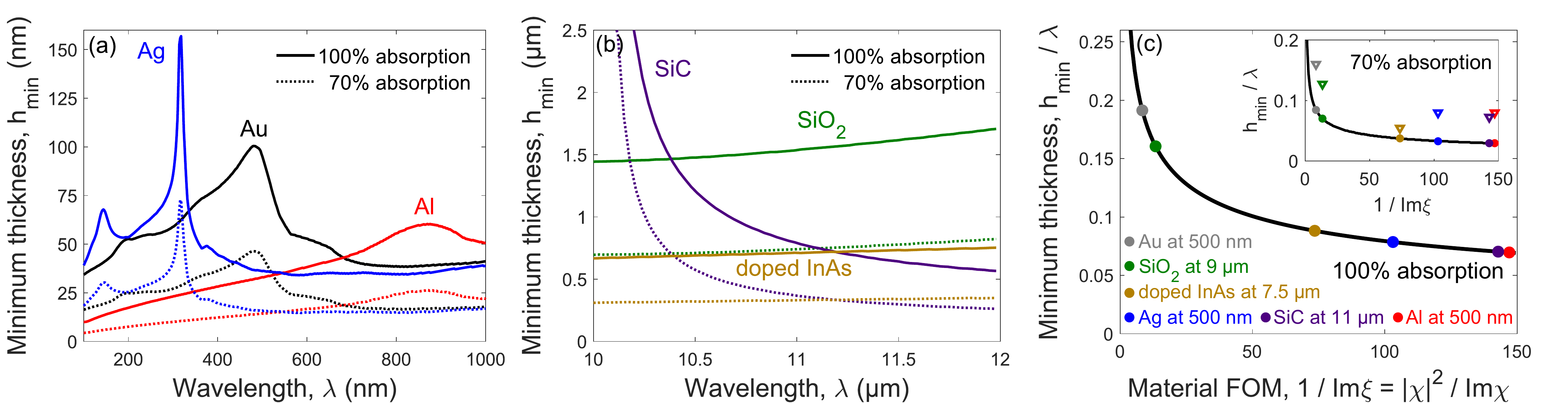}
	\centering
        \caption{Minimum thickness required for a perfect absorber to reach 70\% and 100\% absorption rate under normal incidence for typical materials that are polaritonic at (a) visible~\cite{palik1998handbook} and (b) infrared wavelengths~\cite{law2013all,francoeur2010spectral,popova1972optical}. (c) A universal curve showing minimum possible thicknesses for 100\% absorption as a function of perfect-absorber material figures of merit (FOM), given by $1/\Im\xi = |\chi|^2 / \Im \chi$. The same curve is shown in the inset for 70\% absorption where inverse-designed structures (triangular markers) demonstrate thicknesses within 1.5--2.7$\times$ of the bound.}
	\label{fig:chap2-figure4}
\end{figure}

An important ramification of the bounds of \eqrefrange{chap2-ext-cha-extend}{chap2-sca-cha-extend} is that they determine the minimum thickness of any patterned ``perfect absorber''~\cite{Watts2012,lee2016metamaterials,landy_perfect_2008-1}, achieving 100\% absorption or close to it. Such absorbers are used in sensing applications~\cite{lee2016metamaterials,liu2010infrared} and ultra-thin solar cells~\cite{Yu2010,cui_ultrabroadband_2012,massiot_nanopatterned_2012}. 
Their absorption cross-section per area (i.e., percentage absorption), $\sigma_{\rm abs}/A$, is bounded by \eqref{chap2-abs-cha-extend}. The bound only depends on the incident angle, the absorber thickness (defined as the thickness of its minimum bounding film), and its material susceptibility $\chi(\omega)$. For normally incident waves, we show in Appendix~\ref{sec:appenG-sec5} that the minimum thickness $h_{\rm min}$ to achieve 100\% absorption is given by the self-consistent equation
\begin{align}
h_{\rm min} = \left(\frac{2\lambda}{\pi}\right) \frac{\Im \xi}{1-\operatorname{sinc}^2(kh_{\rm min})}.
\label{eq:chap2-hmin}
\end{align}
\Figref{chap2-figure4}(a,b) shows the minimum thicknesses (solid lines) for 100\% absorption in common metallic and polar-dielectric materials. It is perhaps surprising how large the thicknesses are, averaging on the order of \SI{50}{nm} for metals~\cite{palik1998handbook} at visible wavelengths and \SI{1}{\micro\meter} for polar dielectrics~\cite{law2013all,francoeur2010spectral,popova1972optical} at infrared wavelengths. The only previous bounds that could predict a minimal thickness for perfect absorption are the lossy-material bounds~\cite{miller_fundamental_2016}, which predict minimal thicknesses on the order of \SI{0.5}{nm} and \SI{10}{nm} for the same materials and wavelengths, respectively. Also included in the figures are the minimal thicknesses for 70\% absorption, which are about a factor of two smaller than the 100\%-absorption curves. In Appendix~\ref{sec:appenG-sec6}, we present further analysis suggesting two points: first, that the minimum thickness is typically larger than the skin depth, and can be arbitrarily larger; second, that the nearly linear dependence of Aluminum's minimal thickness relative to wavelength indicates Drude-like permittivity, in contrast to highly non-Drude-like behavior for Ag and Au.  In \figref{chap2-figure4}(c) we present universal curves on which all perfect-absorber materials can be judged, showing the minimum thickness relative to the wavelength as a function of the inverse of material loss, $1/\Im\xi = |\chi|^2 / \Im \chi$, which is a material ``figure of merit'' (FOM) as discussed above~\cite{miller_fundamental_2016}. Using the same inverse-design techniques described above, we discovered ultra-thin absorbers with 70\% absorption rate using both the metals and polar dielectrics presented in \figref{chap2-figure4} (a,b). The grayscale design voxels are specified in Appendix~\ref{sec:appenG-sec7}. As shown in the inset, all of the materials achieve 70\% absorption at thicknesses within a factor of 1.5--2.7 of the bound. In Appendix~\ref{sec:appenG-sec5} we show that in the highly subwavelength limit, the minimum thickness of a perfect-absorber scales with material FOM as $h_{\rm min} / \lambda  \sim (1 / \Im{\xi})^{-1/3}$. The inverse-cubic scaling means that there are diminishing returns to further reductions in loss, and explains the flattening of the curves on the right-hand side of \figref{chap2-figure4}(c).

\section{Optimal Illumination Fields}
\label{sec:chap2-optimal}

In this section, we identify the \emph{incident waves} that maximize the response bounds of \eqrefrange{chap2-ext-gen}{chap2-rhorad_bnd}. There is significant interest in such wavefront shaping~\cite{Popoff2014,Vellekoop2015,Horstmeyer2015,Jang2018}, in particular for the question of identifying optimal illumination fields~\cite{Polin2005,Mazilu2011,Taylor2015,Lee2017,Taylor2017,Fernandez-Corbaton2017,Liu2019}, and yet every current approach identifies optimal fields for a given scatterer. Using the analytical bounds in this chapter, we can instead only specify a designable region, and identify the optimal illumination field that maximizes the bound over all possible scatterers. The resulting bound cannot be exceeded by any geometrical and illumination-field engineerings.

To start, we assume a basis $\Phi$ comprising accessible far-field illumination channels, such as plane waves, vector spherical waves, Bessel beams, or excitations from a spatial light modulator~\cite{Levy2016}. The incident field can be written as
\begin{align}
    \einc = \Phi \cinc,
\end{align}
where $\cinc$ is the vector of basis coefficients to be optimized. The objective is to maximize any of the response bounds, \eqrefrange{chap2-ext-gen}{chap2-rhorad_bnd}, subject to some constraints on the incoming wave. The absorption and scattering bounds, and their near-field counterparts, depend nonlinearly on $\einc$ (due to the presence of a numerically solved dual variable $\nu^*$) and can be optimized locally using gradient-based optimization methods~\cite{Nocedal2006}. Extinction, as well as total near-field local density of states, on the other hand, have analytic forms that lead to simple formulations of \emph{global} bounds over all incident fields. 
Inserting the incident-wave basis into the extinction bound, \eqref{chap2-ext-gen}, the latter becomes
\begin{align}
    P_{\rm ext}^{\rm bound} = \frac{\omega}{2} \cinc^\dagger \Phi^\dagger \left(\Im\xi\II+\ImGO\right)^{-1} \Phi \cinc,
    \label{eq:chap2-Pextbound} 
\end{align}
which is a simple quadratic function of $\cinc$. We maximize this quantity subject to an intensity or power constraint on the fields. Such a constraint has the form $\cinc^\dagger \mathbb{W} \cinc \leq 1$, where $\mathbb{W}$ is a positive-definite Hermitian matrix representing a power-flow measure of $\cinc$. Since the objective and constraint are both positive-definite quadratic forms, the optimal incident-wave coefficients are given by an extremal eigenvector~\cite{Trefethen1997}: the eigenvector(s) corresponding to the largest eigenvalue(s) $\lambda_{\rm max}$ of the generalized eigenproblem
\begin{align}
    \Phi^\dagger \left(\Im\xi\II+\ImGO\right)^{-1} \Phi \cinc = \lambda_{\rm max} \mathbb{W} \cinc.
    \label{eq:chap2-OptIF}
\end{align}
The solution of \eqref{chap2-OptIF} offers the largest upper bound of all possible incident fields.

\begin{figure}[t]
	\includegraphics[width=0.9\textwidth, trim={0 3.5cm 5cm 3.9cm},clip]{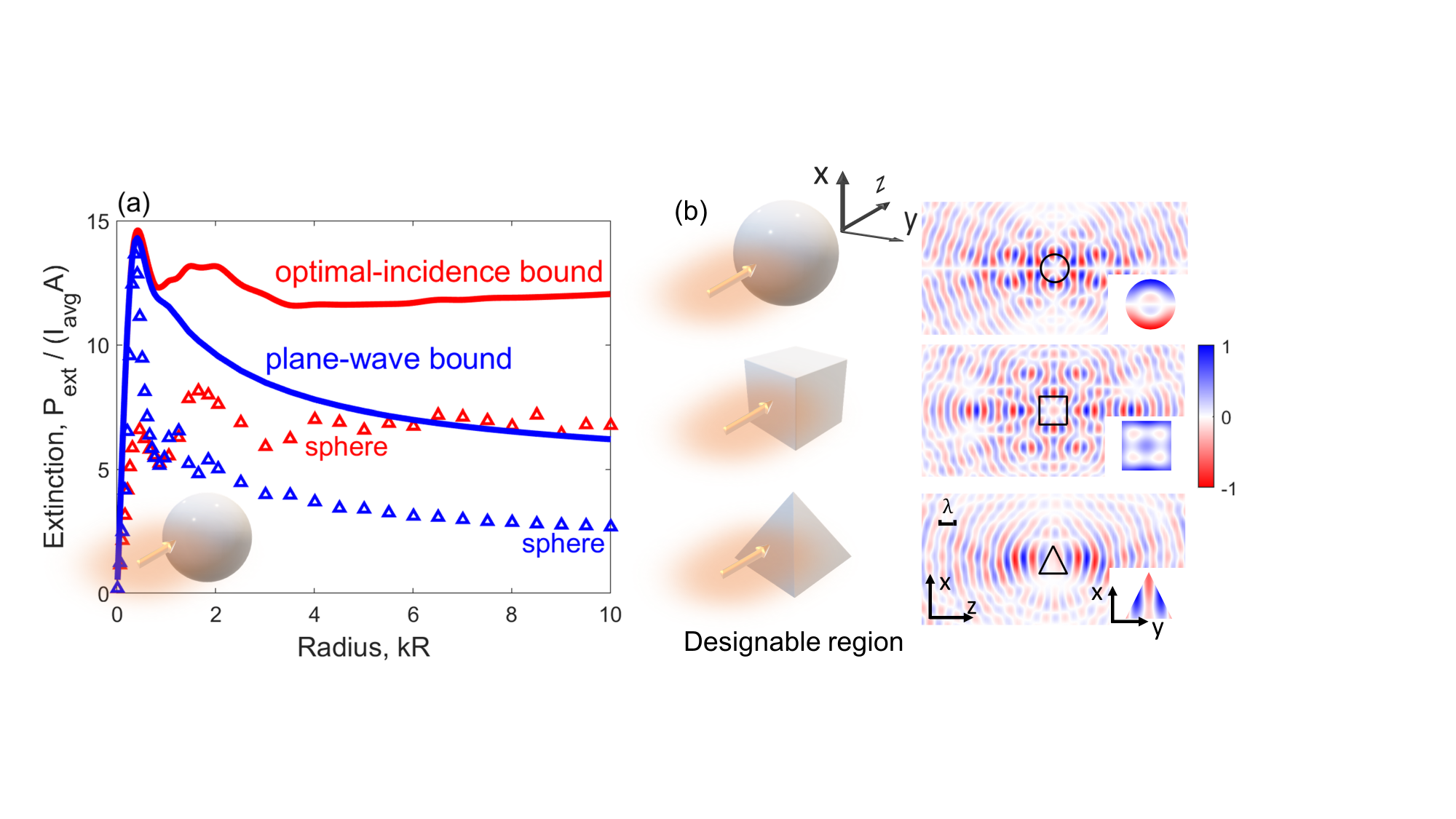}
	\centering
    \caption{Maximum extinction $P_{\rm ext}$ for arbitrary patterning and illumination, normalized by average field intensity $I_{\rm avg}$ and geometric cross section $A$ of the bounding sphere of radius $R$. The solid red line in (a) shows the maximal extinction that can possibly be obtained by the optimal incident field, as compared to the simple plane wave incidence shown by the solid blue line. The triangular markers gives the attained extinction from an unpatterned silver sphere of radius $R$ under either optimal incidence (red triangles) or plane wave illumination (blue triangles). (b) Three possible design regions (sphere, cube, and pyramid) and the corresponding optimal illumination fields ($\Im E_x$) in the x-z plane and x-y plane (inset).}
	\label{fig:chap2-figure5}
\end{figure}

\Figref{chap2-figure5}(a) demonstrates the utility of optimizing over incident fields. We consider incident fields impinging upon a finite silver scatterer within a bounding sphere of radius R at wavelength $\lambda = \SI{360}{nm}$ (as in \figref{chap2-figure2}, near the surface-plasmon resonance). We consider incident fields originating from one half-space, as might be typical in an experimental setup, and use as our basis 441 plane waves with wave vectors $\vect{k}$ whose evenly spaced transverse components range from $-0.8k$ to $0.8k$, where $k =2\pi/\lambda$ is the total wave number. The $0.8$ wave-vector cutoff corresponds to incident-field control over a solid angle of approximately \SI{2.5}{sr}, and can be matched to the specifics of any experimental setup. We impose the constraint that the average intensity over a region  that has twice the radius of the sphere must be equal to that of a unit-amplitude plane wave. \Figref{chap2-figure5}(a) shows the extinction bound evaluated for a plane wave (blue solid), as well as that for the optimal incident field (red solid). As the radius increases, incident-field shaping can have a substantial effect and yield bounds that are almost twice as large as those for plane waves (1.94$\times$ exactly). (Each quantity is normalized by average field intensity $I_{\rm avg}$ and the geometric cross-section $A = \pi R^2$, which is why the extinction bounds may decrease with increasing radius.) Intriguingly, we show that even an unpatterned sphere (red triangles) shows performance trending with that of the bound, and for the larger radii the unpatterned sphere under the optimal illumination field exhibits extinction values larger than the plane-wave bounds. This illustrates a key benefit of bounds: one can now conclude that an unpatterned sphere with optimal illumination fields can achieve extinction values that cannot possibly be achieved by \emph{any} structure under plane-wave illumination.

\Figref{chap2-figure5}(b) further extends the optimal-illumination results, considering three designable regions: a sphere, a cube, and a pyramid. The optimal illumination patterns are shown in two-dimensional cross-sections outside and within the designable regions. The sphere has a radius of one free-space wavelength, while the cube and pyramid have side lengths equal to twice the free-space wavelength. Within each domain the optimal illumination fields exhibit interesting patterns that seem to put field nodes (zeros) in the interior, with the largest field amplitudes around the walls of the domains. This can be explained physically: the optimal incident fields will be those that couple most strongly to the polarization fields that exhibit the smallest radiative losses. The polarization fields that have the smallest radiative losses will tend to have oscillations with far-field radiation patterns that cancel each other, as occurs for oscillating currents along structural boundaries, such as whispering-gallery modes~\cite{Vahala2003,He2013}. This procedure can be implemented for a beam generated by almost any means, e.g., and incident wave passing through a scatterer with a complex structural profile~\cite{Levy2007,Wei2013,KerenZur2016}, precisely controlled spatial light modulators~\cite{Weiner2000,Chattrapiban2003,Guo2007,Zhu2014}, or a light source with a complex spatial emission profile~\cite{Lodahl2004,Ringler2008,Bleuse2011}.

\section{Discussion}
\label{sec:chap2-discussion}
In this chapter, we show that a power-conservation law, in the form of optical theorem, enables the identification of analytical bounds for maximal electromagnetic response. We considered: arbitrary linear and quadratic response functions, \eqref{chap2-gen_bound}, power-flow quantities such as absorption and scattering, \eqreftwo{chap2-abs-gen}{chap2-sca-gen}, and LDOS, \eqrefrange{chap2-rhotot_bnd}{chap2-rhorad_bnd}, more specific scenarios such as plane-wave scattering and perfect absorbers, \eqrefrange{chap2-ext-cha-bound}{chap2-hmin}, and optimal illumination fields, \eqref{chap2-OptIF}. These results demonstrate the utility of our analytical bounds and provide an optimization-based framework for further generalizations and unifications.

One limitation of the analytical bounds is they diverge for lossless dielectrics ($\Re\varepsilon > 0$): the bounds in \eqrefrange{chap2-ext-gen}{chap2-rhorad_bnd} is infinite if $\Im\varepsilon = 0$ (which implies $\Im\xi = 0$). Heuristically, one can truncate the scattering channels, though except in the simplest (e.g., dipolar) systems, it is impossible to predict \emph{a priori} how many channels may contribute in optimal scattering processes.
Another way to resolve the nonphysical divergence is to incorporate bandwidth and causality in the constraint~\cite{shim2019fundamental}, but it only applies to few analytically well-behaved functions such as the local and cross densities of states. 
The most general approach is to incorporate additional constraints in the form of local-power-conservation laws. 
Rooted in the same power-conservation principle behind the optical theorem, the additional local-power-conservation laws regularize the bound for lossless dielectrics, and extend its  applicability to general wave scattering, as demonstrated in the next chapter.

\chapter{Computational bounds via local power conservation}
\label{chap:comp}
The single optical-theorem constraint of the last chapter is not sufficient to capture all the physics in Maxwell's equations. 
In this chapter, we show that local power conservation, a generalized version of the optical theorem, retains all of the information of Maxwell's equations.
Local power conservation imposes a series of quadratic constraints on the polarization fields. Solving for global optima under such constraints is generally NP-hard (i.e., cannot be solved deterministically in polynomial time), but one can relax the optimization problem into a convex semidefinite program whose global optimum can be determined through standard interior-point algorithm.
Computational in nature, such bounds significantly strengthen the previous analytical bounds. They apply to most scattering scenarios, including the case of lossless dielectrics where the analytical bounds diverge.

This chapter is organized as follows.
In Section~\ref{sec:chap3-local}, we derive the local power conservation laws from the complex Poynting theorem applied to each point in the design space.
In Section~\ref{sec:chap3-computational}, we show that optimizing over the local-power-conservation constraints constitutes a nonconvex quadratic constrained quadratic program (QCQP) which can be bounded via semidefinite relaxation. To accelerate the computation of the bound, we develop an algorithm that picks the most important local-power-conservation constraints to impose.
We illustrate the utility of the computational bounds in the two following sections. In Section~\ref{sec:chap3-S}, we identify computational bounds on the minimum size of a scatterer encoding any linear operator, demonstrated for an analog optical discrete Fourier transform (DFT). In Sections~\ref{sec:chap3-far}, we identify bounds on maximum far-field extinction over any bandwidth, resolving a long-standing gap in power–bandwidth limits.
In Section~\ref{sec:chap3-discussion}, we reinterpret the power-conservation laws as binary-material constraints, leading to further generalizations of this computational power-conservation framework.

\section{Local power conservation}
\label{sec:chap3-local}
To start, we generalize the optical theorem in Chapter~\ref{chap:analy} to an infinite number of local-power-conservation laws that must be satisfied by every solution of Maxwell's equations. These conservation laws manifest the complex Poynting theorem~\cite{Jackson1999} over any subdomain of a scatterer, but only when formulated in terms of induced polarization fields do they exhibit properties that enable global bounds. 
We consider a scattering problem comprising arbitrary sources and isotropic electric materials. The formalism below can be easily generalized to anisotropic and magnetic materials. Using the same notations as the last chapter, we denote the electric field by $\vE(\vx)$, the polarization field by $\vP(\vx)$, and the background Green's function by $\vG_0(\vx, \vx')$. We use dimensionless units in which the speed of light is 1.

\begin{figure*}[t]
    \includegraphics[width=0.8\textwidth]{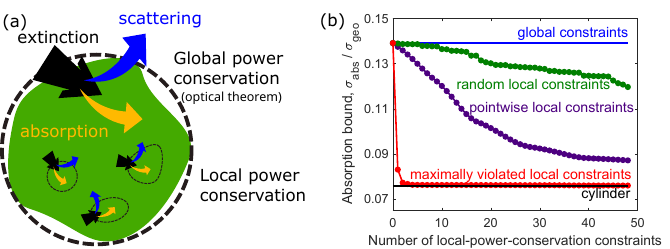}
    \centering
    \caption{(a) A freeform, homogeneous photonic scatterer within a designable region (outer dashed circle). Bounds in Chapter~\ref{chap:analy} utilize the optical theorem that enforce global power conservation. In this Chapter, we introduce a polarization-current-based formulation of local conservation laws that provide an infinite set of constraints to derive upper bounds on light--matter interactions. (b) Example of local constraints (green, purple, red) tightening bounds from global constraints only (blue), for maximum absorption from a material with permittivity $\varepsilon = 12+0.1i$ in a region with diameter $d=0.18\lambda$. Our iterative method of selecting maximally violated constraints rapidly converges.}
    \label{fig:chap3-fig1}
\end{figure*}

The local conservation laws that underlie our bounds arise from the complex Poynting theorem~\cite{Jackson1999}. As depicted in \figref{chap3-fig1}, Poynting's theorem must apply not only globally over an entire scatterer (i.e., the optical theorem), but also locally at at any point within. The usual complex-valued Poynting theorem (a function of the electromagnetic fields) can be rewritten in terms solely of the induced polarization fields $\vP(\vx)$, by taking the inner product between $\vP^*(\vx)$ and the volume-integral equation in Eq.~(\ref{eq:chap2-vie}):  
\begin{equation}
    \xi|\vP(\vx)|^2 + \int_{\Vd} \vP^*(\vx)\cdot\vG_0(\vx,\vx')\cdot\vP(\vx')\md\vx' = -\vP^*(\vx) \cdot \vE_{\rm inc}(\vx),
    \label{eq:chap3-point_consv}
\end{equation}
Again, to eliminate the geometry dependence, we extend the range of integration in \eqref{chap3-point_consv} from the scatterer volume $V$ to a bounding volume $V_d$ (the dashed-circled cylinder in \figref{chap3-fig1}~(a)) as $\vP(\vx) = 0$ outside of $V$.
The first term in Eq.~(\ref{eq:chap3-point_consv}) corresponds to complex-valued Poynting flux into an infinitesimally small spherical surface around $\vx$, the second term to the flux radiated out to the surface, and the last term to the flux extracted from the incident field into the surface. 
The real part of Eq.~(\ref{eq:chap3-point_consv}) corresponds to reactive power conservation at $\vx$.
The imaginary part of Eq.~(\ref{eq:chap3-point_consv}) corresponds to real power conservation at $\vx$, whose three terms correspond to local absorption, local radiation, and local extinction, respectively.
Integrating them over the entire bounding volume gives the optical theorem of Eq.~(\ref{eq:chap2-optthm}) in Chapter~\ref{chap:analy}, which imposes global power conservation as in \figref{chap3-fig1}. In contrast, the local power conservation in Eq.~(\ref{eq:chap3-point_consv}) enforces power conservation at every region in space as shown in \figref{chap3-fig1}, leading to potentially much tighter bounds.

As a generalization of the optical theorem, the local conservation law of \eqref{chap3-point_consv} inherits two key properties that enable global bounds over all possible designs. First, they hold for any scatterer in the bounding domain, as power is conserved at every point in space regardless of the material structuring. Second, it is a quadratic form of the polarization fields, and therefore amenable to semidefinite programming, as we discuss below.

\section{Computational bounds}
\label{sec:chap3-computational}

In this section, we construct a general optimization problem with the local-power-conservation constraint in \eqref{chap3-point_consv}. We discuss how such optimization problems can be bounded above by the standard approach of semidefinite relaxation, leading to computational bounds. We introduce an iterative algorithm to accelerate the computation of bounds by selecting the most important local-power-conservation constraints for a given physics problem.

We first discretize the local-power-conservation law of \eqref{chap3-point_consv} in the real space of the bounding volume. In the same notation as last chapter, where the polarization field, incident field, and Green's function operator are discretized into vector $p$, vector $\einc$, and matrix $\GG_0$, respectively, yielding
\begin{equation}
    p^\dagger\DD_i (\GG_0+\xi \II)p = - p^\dagger\DD_i e_{\rm inc},
    \label{eq:chap3-local-power-discretize}
\end{equation}
where we introduce the diagonal matrix $\DD_i$ to pick only the three polarizations for the $i$-th spatial point in the inner products. The matrix $\DD_i$ is all zero except for the three ones on its diagonal that correspond to the three polarizations of the $i$th point.

The optimization problem is similar to  \eqref{chap2-general-formalism} in the last chapter, the objective function generally expressed as a quadratic function of the polarization field $p$. The only difference is now we constrain the polarization fields by the local power conservation of \eqref{chap3-local-power-discretize} at every point in space:
\begin{equation}
\begin{aligned}
& \underset{p}{\text{maximize}}
& & f(p) = p^\dagger \mathbb{A} p +  \Re\left(\beta^\dagger p\right) \\
& \text{subject to}
& & p^\dagger\DD_i (\GG_0+\xi \II)p = - p^\dagger\DD_i e_{\rm inc},\quad \text{for } i = 1,2,...,n,
\end{aligned}
\label{eq:chap3-local-opt-problem}
\end{equation}
where the index $i$ runs through all the $n$ points we discretize in the bounding volume.
\Eqref{chap3-local-opt-problem} is a key result: it formulates the maximum response, subject to all local-power-conservation laws, as a quadratically-constraint-quadratic-program, i.e., a QCQP optimization problem~\cite{Boyd2004,Luo2010}.
It applies to all possible designs within the bounding volume.
A bound on the solution of \eqref{chap3-local-opt-problem} can be found by standard techniques that relax the original, quadratic program to a higher-dimensional linear program over semidefinite matrices, also known as a semidefinite program~\cite{Laurent2005,Luo2010}, which can be solved by interior-point methods~\cite{Vandenberghe1996,Boyd2004}. Such transformations of QCQP's have led to meaningful bounds in many areas of engineering~\cite{Goemans1995,Vandenberghe1996,Tan2001,Biswas2006,Luo2010,Gershman2010}; we leave the details of the transformation of \eqref{chap3-local-opt-problem} to Appendix~\ref{sec:appenH-sec1}. The final solution represents a global, unsurpassable bound for any electromagnetic scattering response.

It is computationally challenging to solve \eqref{chap3-local-opt-problem} with the local-power-conservation constraints imposed at every point. To reduce the number of constraints, we enforce only $m$ different weighted averages of them in hopes that $m$ can be much smaller than $n$. The optimization problem becomes
\begin{equation}
\begin{aligned}
    & \underset{p}{\text{maximize}}
    & & f(p) = p^\dagger \mathbb{A} p +  \Re\left(\beta^\dagger p\right) \\
    & \text{subject to}
    & & p^\dagger\Re\left[\DD_{\textrm{wei}, j} (\GG_0+\xi \II)\right]p = - \Re\left(p^\dagger\DD_{\textrm{wei}, j} e_{\rm inc}\right),\quad \text{for } j = 1,2,...,m,
\end{aligned}
\label{eq:chap3-opt-weight}
\end{equation}
where each diagonal matrix $\DD_{\textrm{wei},j}$ sums over all matrices in $\{\DD_i\}_{i=1}^n$ with a unique complex-valued weightings.
In \eqref{chap3-opt-weight}, we take only the real part of the power conservation constraint, because $\DD_{\textrm{wei}, j} \rightarrow i\DD_{\textrm{wei}, j}$ accounts for the imaginary part.
The optimization problem of \eqref{chap3-opt-weight} is a relaxation of the original ones in \eqref{chap3-local-opt-problem} as the solution of latter is always feasible in the former, but \eqref{chap3-opt-weight} can be computationally easier to solve if far fewer constraints are needed to reach a meaningful bound.

We propose an iterative algorithm that identifies which weighted-average constraints to use in \eqref{chap3-opt-weight}. We start with the two $\DD_\textrm{wei}$ matrices that correspond to global power conservation. The first is the identity tensor that imposes global reactive-power conservation. The second is the identity tensor multiplied by $i$ that imposes global real-power conservation. The second, in particular, is the optical theorem in \eqref{chap2-OptThmDis}, which leads to a positive semidefinite quadratic form and is crucial to restricting the magnitude of the solutions, as we have seen in the analytical bounds of Chapter~\ref{chap:analy}. As a first iteration, we use only those two $\DD_\textrm{wei}$ matrices to find an initial bound for \eqref{chap3-opt-weight}, as well as the first-iteration optimal polarization fields, $p_{\rm opt,1}$. From those currents, we can identify out of all possible remaining $\DD_\textrm{wei}$-matrix constraints, which ones are ``most violated'' by $p_{\rm opt,1}$, i.e., which constraint has the largest difference between its left- and right-hand sides (measured under the $L_2$ norm). We add the maximally violated $\DD_\textrm{wei}$-matrix constraint into the constraint set, and run a second iteration, identifying new bounds and new optimal polarization fields. This process proceeds iteratively until convergence. Straightforward linear algebra shows (cf. Appendix~\ref{sec:appenH-sec2}) that after iteration $j$, with optimal currents $p_j$, the next constraint to add is the one with  $\DD_\textrm{wei}$ matrix: 
\begin{align}
    \DD_{\textrm{wei}, j+1} = \diag\left[p_{\rm opt,j} p_{\rm opt,j}^\dagger \left(\GG_0+\xi\II\right)^\dagger + p_{\rm opt,j}\einc^\dagger \right],
    \label{eq:chap3-max-vio}
\end{align}
where ``$\diag$'' creates a diagonal matrix with the diagonal of its (matrix) argument. \Figref{chap3-fig1}(b) demonstrates this method for computing bounds on the TE absorption cross-section $\sigma_{\rm abs}$ of a dielectric scatterer in a wavelength-scale cylindrical volume. The designable volume need not be symmetric; in Appendix~\ref{sec:appenH-sec3} we include an example with a triangular region. Whereas the global constraints (blue) are significantly larger than the response of a cylindrical scatterer (black), including local constraints shows that one can clearly identify tighter bounds. Yet both randomly chosen $\DD_\textrm{wei}$ matrices (green) and spatially pointwise, delta-function-based $\DD_\textrm{wei}$ matrices (purple) show slow convergence. 
The iterative method via maximally violated constraints shows rapid convergence, requiring only two local constraints.
The spatial patterns of both the optimal current distribution and local constraints are shown in Appendix~\ref{sec:appenH-sec3}.
With such bound, we can clearly identify the cylinder as a globally optimal structure.

\begin{figure}[t]
    \includegraphics[width=0.9\textwidth]{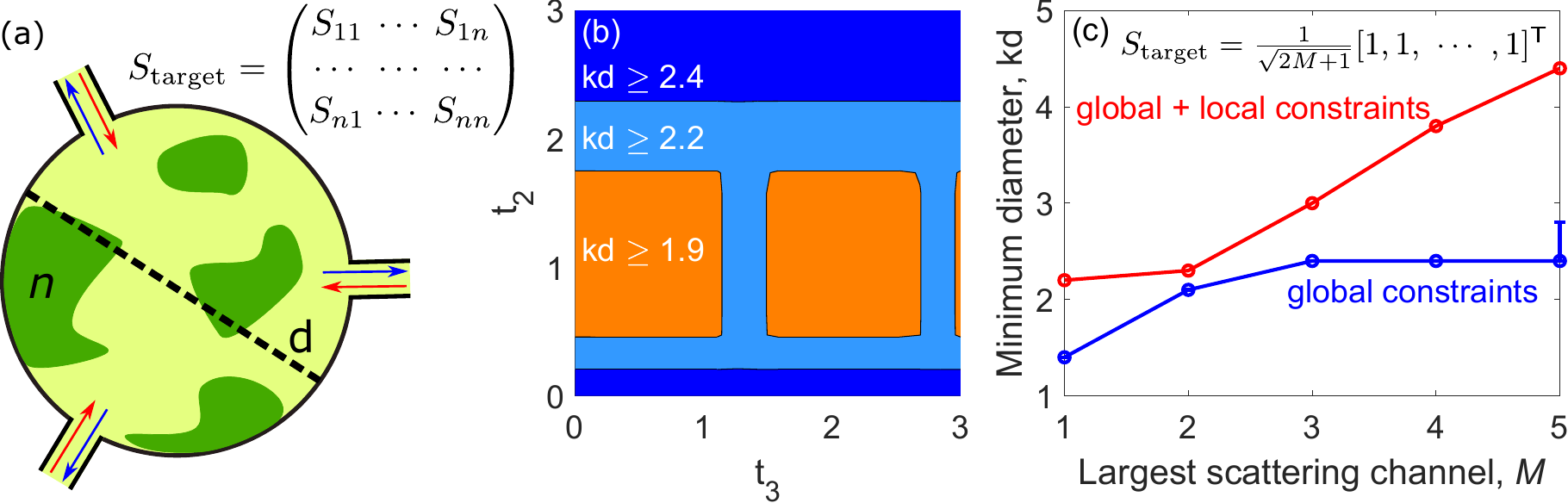}
    \centering
    \caption{(a) Photonic devices are often designed to achieve a specific ``target'' $S$ matrix in a compact form factor. Our bounds enable identification of the minimum diameter $d$ of any such device (relative to wavenumber $k$), for (b) nonuniform DFT-matrix implementation ($t_2$ and $t_3$ are parameters of the DFT matrix) and (c) power splitters for a single input to $2M+1$ outgoing channels. In (b), each point in the image represents a unique DFT matrix, and the colors indicate the minimum diameter for possibly achieving that scattering matrix. In (c) it is evident that local constraints are required to identify feasible design regions as the required functionality increases in complexity. In (b,c) the channels are TE cylindrical waves and the material has refractive index $n=\sqrt{12}$.}
    \label{fig:chap3-fig2}
\end{figure}

\section{S-matrix feasibility}
\label{sec:chap3-S}

To demonstrate the power of this framework, we consider a fundamental question in the fields of analog optical computing~\cite{Silva2014,Pors2015,Zhu2017,Kwon2018,Estakhri2019} and metasurfaces~\cite{yu2014flat,aieta2015multiwavelength,shrestha2018broadband}: what is the minimum size of a scatterer that achieves a desired scattering matrix $\St$? A generic setup is depicted in \figref{chap3-fig2}(a). The target $S$ matrix could manifest lens focusing or meta-optical computing, for example. The objective is to minimize the relative difference between the achievable and target $S$ matrices, i.e., $f_{\rm obj} = \left\| S - \St \right\|^2 / \left\|\St\right\|^2$, where $\|\cdot\|$ denotes the Frobenius norm. It is straightforward to write this objective in the form appearing in \eqref{chap3-opt-weight}, as the $S$ matrix elements are linear in the polarization fields and the objective is a quadratic form (cf. Appendix~\ref{sec:appenH-sec5}). To determine the minimum feasible size for implementing $\St$, we compute bounds on the smallest error between $S$ and $\St$, and define an acceptable-accuracy threshold (1\%) below which the device exhibits the desired functionality with sufficient fidelity.

We apply our framework to two such problems, both of which comprise two-dimensional scatterers with refractive index $n=\sqrt{12}$, discretized by the discrete dipole approximation (DDA)~\cite{Purcell1973,Draine1994}. In the first, we identify the smallest domain within which a scatterer can possibly act as a discrete Fourier transform (DFT) operator over three TE cylindrical-wave channels (cf. Appendix~\ref{sec:appenH-sec4}). The DFT is the foundation for discrete Fourier analysis and many other practical applications~\cite{strang_wavelets_1994}.  With uniform frequencies and nonuniform sample points $t_1$, $t_2$, and $t_3$ (and $t_1$ is fixed as a reference to be $t_1 = 0$), a target $S$ matrix that acts as a DFT can be represented as~\cite{rao_nonuniform_2010}:
\begin{equation}
    S_{\rm target}(t_2,t_3) = \frac{1}{\sqrt{3}}
    \begin{pmatrix}
        1 & 1 & 1\\
        1 & e^{-2\pi it_2/3} & e^{-2\pi it_3/3}\\
        1 & e^{-4\pi it_2/3} & e^{-4\pi it_3/3}
    \end{pmatrix}.
    \label{eq:S_target1}
\end{equation}
\Figref{chap3-fig2}(b) shows the bound-based feasibility map for implementing such an $S$ matrix. Each point in the grid represents a unique DFT matrix (prescribed by the values of $t_2$ and $t_3$), and the color indicates the smallest diameter $d$, relative to wavenumber $k$, of a structure that can possibly exhibit the desired DFT-based scattering matrix (at 99\% fidelity). There is no structure, with any type of patterning, that can act as a DFT matrix if its diameter is smaller than that specified in \figref{chap3-fig2}(b). The bounds dictate the minimal possible size of an optical element implementing specific functionality. A related calculation is shown in \figref{chap3-fig2}(c). In that case we consider a target $S$ matrix for a power splitter, directing a single incident wave equally into outgoing spherical-wave channels index by $m$ (where $m$ is the angular index, $m=-M,\ldots,M$). We depict the minimum diameter as a function of the number of scattering channels, for a bound with only the global constraints (blue) and a bound with both global and local constraints (red). (The error bar indicates a numerical instability in the global-constraint-only approach, cf. Appendix~\ref{sec:appenH-sec6}.) Whereas the bound with only global constraints converges \textit{unphysically} to wavelength scale as the number of channels increases, the bound with the additional local constraints predicts an unavoidable increase in the diameter of the power splitter, representing the first such capability for capturing minimum-size increases with increasing complexity.

\section{Far-field power–bandwidth limits}
\label{sec:chap3-far}

\begin{figure}[t]	\includegraphics[width=0.6\textwidth]{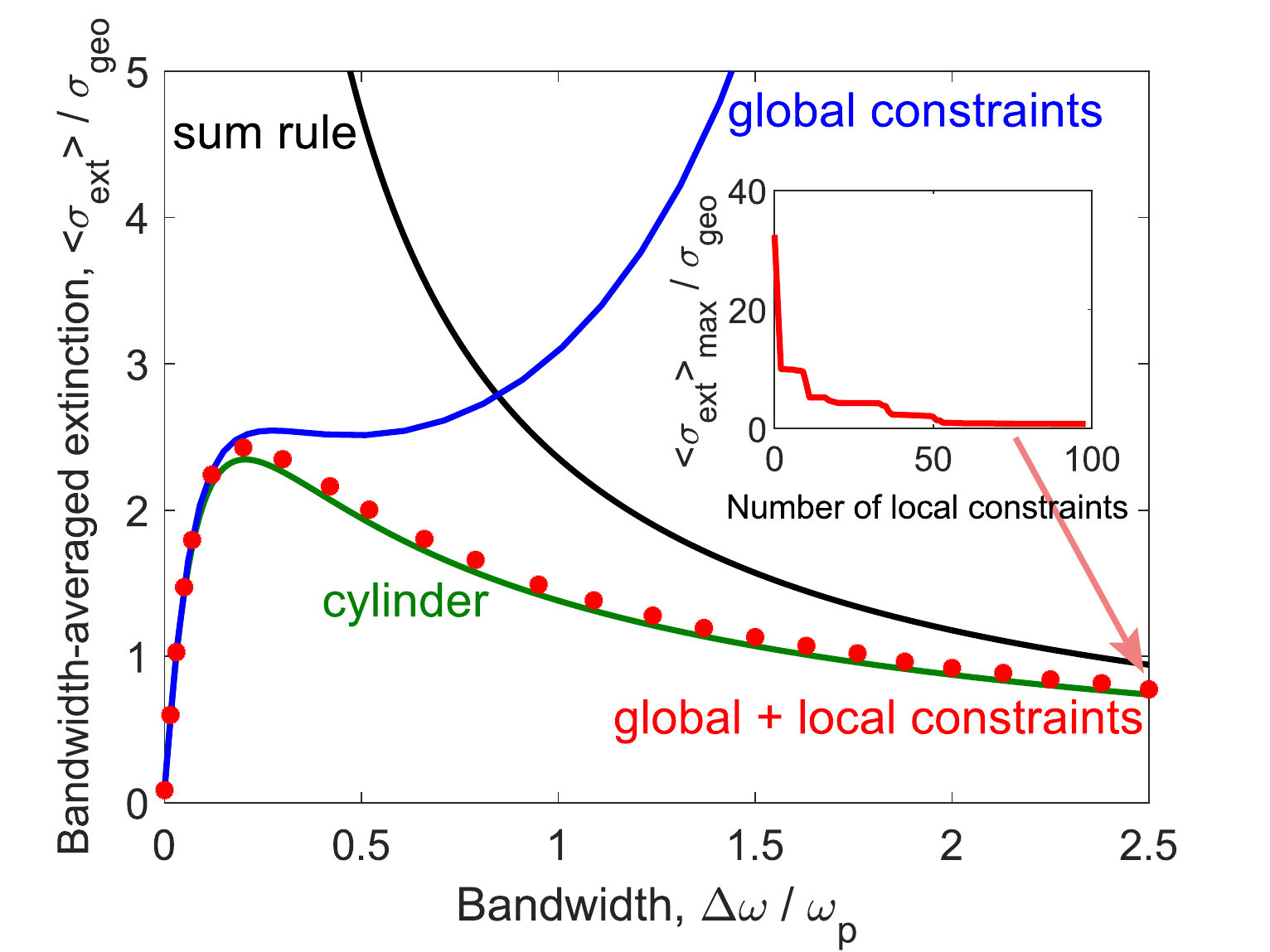}
    \centering
    \caption{Bounds on maximal bandwidth-averaged extinction, $\langle \sigma_{\rm ext}\rangle_{\rm max}$, as a function of bandwidth $\Delta\omega$ for a lossless Lorentz--Drude material with plasma frequency $\omega_p$ and oscillator frequency $\omega_c = 0.3015\omega_p$, which is chosen such that the permittivity is 12 at a center frequency $\omega_0 = 0.05\omega_p$. The bounds are normalized to the geometric cross-section $\sigma_{\rm geo}$ of the designable region, a cylinder with diameter $d=3/\omega_p$. While known sum rules (black) and global-constraint bounds (blue) are loose for many bandwidths, utilizing local constraints (convergence shown in inset) enables apparently tight bounds across all bandwidths.}
    \label{fig:figure3}
\end{figure}

The local-constraint bound framework resolves another outstanding question: how large can far-field scattering be over an arbitrary bandwidth $\Delta \omega$? In \citeasnoun{Shim2019}, bounds for near-field average-bandwidth response were derived using global constraints at a complex frequency; yet it was noted that the same technique diverge under far-field excitation. 
To resolve this, we extend the local conservation law in \eqref{chap3-local-power-discretize} to complex frequencies by multiplying the conjugate frequency $\omega^*$ so that 
\begin{equation}
    p^\dagger\DD_i \omega^*(\GG_0+\xi \II)p = - \omega^*p^\dagger\DD_i e_{\rm inc},
    \label{eq:chap3-complex_con}
\end{equation}
leading to operators that are positive semidefinite over the whole upper half of the complex-frequency plane, by passivity (cf. Appendix~\ref{sec:appenH-sec8}).
Imposing these local-power-conservation constraints regulates divergence of the previous bounds, which we show below.

A prototypical example to consider is the maximum extinction cross-section, $\sigma_{\rm ext}(\omega)$, from a given material over a bandwidth $\Delta\omega$. Using contour-integral techniques from Refs.~(\cite{Hashemi2012,Shim2019}), the average extinction around a center frequency $\omega_0$, over a bandwidth $\Delta\omega$, as measured by integration against a Lorentzian window function, $\Hw(\omega)=\frac{\Delta\omega/\pi}{(\omega-\omega_0)^2+\Delta\omega^2}$, can be written as the evaluation of a single scattering amplitude at a \emph{complex} frequency $\tomega$ (cf. Appendix~\ref{sec:appenH-sec7}):
\begin{align}
    \langle \sigma_{\rm ext}\rangle  &= \int_{-\infty}^{+\infty} \sigma_{\rm ext}(\omega)\Hw(\omega)\td\omega \nonumber \\
	 &=\Im\left[\tomega e^T_{\rm inc}(-\tomega)\phi(\tomega)\right], \label{eq:complex_ext}
\end{align}
where $\tomega = \omega_0 + i\Delta\omega$. \Eqref{complex_ext} is a linear objective function of the form required by \eqref{chap3-opt-weight}, evaluated at a complex frequency. By imposing the global- and local-conservation constraints at the complex frequency $\tomega$, we can identify bounds to the bandwidth-averaged far-field response. \Figref{figure3} shows the results of such a computation for a lossless Lorentz--Drude material (with plasma frequency $\omega_p$) in a designable region with diameter $d = 3/\omega_p$. Included in the figure is a bound on average extinction from a known all-frequency sum rule~\cite{gordon_1963,Yang2015} (black), which is descriptive in the infinite-bandwidth limit, and the global-constraint-only bounds (blue), which are useful in the small-bandwidth limit, but each diverges in the opposite limits. Through the use of global and local constraints (red), we can identify bounds over any bandwidth of interest, and we find that a cylindrical scatterer is nearly globally optimal.

\section{Discussion}
\label{sec:chap3-discussion}

In this chapter, we show that the local-power-conservation laws enable computational bounds to light–matter interactions.
Specifically, the local-power-conservation laws exhibit a mathematical structure that can be relaxed into a convex semidefinite program whose optimum gives a bound to the original problem. 
This computational framework is accelerated by choosing a few maximally violated constraints, and its utility is demonstrated in two examples: optical analog computing and power--bandwidth limits. In the following, we first discuss the challenges and opportunities in scaling up such computational bound to large-scale optical devices, and then introduce an alternative interpretation of the local power conservation that leads to further generalizations of the bounds in Chapters \ref{chap:multi} and \ref{chap:sparse}. 

\subsection{Large-scale computational bound}
The computational bound is  challenging to solve for large photonic structures.
Unlike the analytical bounds that has been applied to three-dimensional structures well beyond the wavelength scale in Chapter~\ref{chap:analy}, the computational bound, because of its additional constraints, is limited to wavelength-scale two-dimensional design regions. Such is the trade off between tightness and computation complexity. 
With local-power-conservation constraints imposed at each point, the computational complexity (i.e., flop counts) of the bound scales as $O(N^4)$ with $N$ being the number of grid points. Choosing only the maximally violated constraints as in \eqref{chap3-max-vio} reduces the number of constraints but the complexity is still above $O(N^3)$, as required by the standard interior-point method for semidefinite programming.
We show in Chapter~\ref{chap:sparse} that in some cases we can reduce the computation complexity to $O(N)$ by leveraging the sparse structure in the Maxwell's equations, which allows us to dramatically accelerate the computations of bounds in certain photonic structures such as multi-layered thin films and high-aspect-ratio metasurfaces.
More general acceleration methods involve replacing the off-the-shelf interior-point solver to algorithms designed specifically for large-scale semidefinite programming. This is discussed in detail in Conclusion.

\subsection{Binary-material interpretation of local power conservation}
\label{subsec:chap3-binary}
In the following, we reinterpret  the local-power-conservation laws as binary-material constraints. This viewpoint connects bounds with binary designs, suggests a new design method, and derives new  conservation laws. The latter leads to further generalizations of the bound to multi-functional devices in Chapter~\ref{chap:multi} and large-scale devices in Chapter~\ref{chap:sparse}.  

We first show the local-power-conservation law in Eq.~(\ref{eq:chap3-point_consv}) is almost equivalent to the binary-material constraint $\chi(\vx) = \{0, \chi\}$ that enforces the susceptiblity at point $\vx$ to be either air or material.
We regard the binary-material constraint as an ``either... or...'' statement, i.e., either $\chi(\vx) = 0$ or $\chi(\vx) = \chi$, and glue the two scenarios together to form a single quadratic constraint that is true in either case:
\begin{equation}
    \chi^*(\vx)[\chi(\vx) - \chi] = 0,
    \label{eq:chap3-binary_material}
\end{equation}
where the complex conjugate $\hphantom{}^*$ is a mathematical convenience to establish the connection that follows.
For scalar fields, one can show Eq.~(\ref{eq:chap3-binary_material}) is exactly equivalent to the local-power-conservation law of Eq.~(\ref{eq:chap3-point_consv}), by substituting $\chi(\vx)$ in Eq.~(\ref{eq:chap3-binary_material}) with $P(\vx) / E(\vx)$ and express the electric field as the sum of incident and radiated field from the polarization field
\begin{equation}
    P^*(\vx)\left[P(\vx) - \chi\left(E_{\rm inc}(\vx) + \int_{\Vd}G_0(\vx,\vx')P(\vx')\md\vx'\right)\right] = 0.
    \label{eq:chap3-local-power_scalar}
\end{equation}
The substitution $\chi(\vx) = P(\vx) / E(\vx)$ is not allowed when $E(\vx)$ is zero, but in that case $P(\vx)$ is also zero, so \eqref{chap3-local-power_scalar} still holds.
For polarized fields with, say, three polarizations, we have three possible substitutions for each $\chi(\vx)$ in Eq.~(\ref{eq:chap3-binary_material}) with nine different combinations in total:
    \begin{equation}
        P_i^*(\vx)\left[\vP(\vx) - \chi\left(\vE_{\rm inc}(\vx) + \int_{\Vd}\vG_0(\vx,\vx')\cdot\vP(\vx')\md\vx'\right)\right]_j = 0,
        \label{eq:chap3-local-power_polarization}
    \end{equation}
    where $i$ and $j$ index the three polarizations. When $i=j$, \eqref{chap3-local-power_polarization} represents local power conservation for each polarization, and summing over three of them gives the local-power-conservation law in Eq.~(\ref{eq:chap3-point_consv}). When $i\neq j$, \eqref{chap3-local-power_polarization} represents a set of constraints on the correlation between polarization fields induced in different polarization states. Such constraints cannot be derived from the Poynting theorem, suggesting the binary-material approach is a more general way of deriving constraints on electromagnetic scattering.

The binary-material approach outlined in \eqrefrange{chap3-binary_material}{chap3-local-power_polarization} provides a straightforward way to construct conservation constraints for a variety of scenarios.
Besides the multi-polarization scattering considered in \eqref{chap3-local-power_polarization}, we show in Chapter~\ref{chap:multi} that one can use the binary-material argument to derive correlation constraints between almost any multi-scattering scenarios, including scattering under multiple incident frequencies and scattering under an actively tuned material.
The same binary-material argument is used in Chapter~\ref{chap:sparse} to derive local-power-conservation laws directly from the differential form of Maxwell's equations (as opposed to the integral form considered in this chapter), where sparsity allows for fast computation of bounds.

The binary-material constraints establish a direct mathematical connection between the binary designs and the bounds.
The standard optimization problem for binary designs is to find the best deposition of air ($\chi(x)=0$) and material ($\chi(\vx)=\chi$) in a design region $V_d$, that maximizes the objective $f(P(x))$ with $P(x)$ solved from Maxwell's equations for each configuration of  $\chi(\vx)$:
\begin{equation}
    \begin{aligned}
    & \underset{\chi(\vx)}{\text{maximize}}
    & & f(P(\vx))  \\
    & \text{subject to}
    & & \chi(\vx) = \{0, \chi\},\quad  \text{ for} \ \forall \  \vx\in \Vd.
    \end{aligned}
    \label{eq:original_design_problem}
    \end{equation} 
This type of problem has been subject to various computational design approaches~\cite{bendsoe2003topology, jensen2011topology, Miller2012a, Molesky2018, pestourie2018inverse, so2020deep,ma2021deep, jiang2021deep,holland1992genetic, kennedy1995particle, schneider2019benchmarking, park2022free}, but due to its nonconvexity, none can only solve globally. 
The binary argument detailed in \eqrefrange{chap3-binary_material}{chap3-local-power_polarization} shows we can at least bound the global optimum by transforming the binary constraint $ \chi(\vx) = \{0, \chi\}$ to a quadratic constraint on the polarization field $\vP(\vx)$, resulting in the QCQP of \eqref{chap3-local-opt-problem} which has a mathematical structure that allows for convex relaxation.
This argument can be extended to designs composing of any two materials, not just air and material, by constructing constraints in the form of $[\chi(\vx)-\chi_1]^*[\chi(\vx) - \chi_2] = 0$ in lieu of \eqref{chap3-binary_material}. 
 Such bound applies to grey-scale structures as the latter can always be mimicked by a binary design given enough spatial resolution~\cite{Bergman1981,Milton1981,Milton2002}.
Previously, we claim the solution of \eqref{chap3-local-opt-problem} is a bound by interpreting the constraints as power-conservation laws that every geometry in scattering has to satisfy.
Now, we show that \eqref{chap3-local-opt-problem} is the exact same optimization problem as the binary-constrained designs in \eqref{original_design_problem}, though instead of searching for local optima as in inverse designs~\cite{bendsoe2003topology, jensen2011topology, Miller2012a, Molesky2018, pestourie2018inverse, so2020deep,ma2021deep, jiang2021deep,holland1992genetic, kennedy1995particle, schneider2019benchmarking, park2022free}, we bound it via convex relaxation for the first time.

The equivalence between the binary design in \eqref{original_design_problem} and the QCQP in \eqref{chap3-local-opt-problem} suggests an alternative pathway for photonic designs:
instead of solving the original design problem as in common design solvers~\cite{bendsoe2003topology, jensen2011topology, Miller2012a, Molesky2018, pestourie2018inverse, so2020deep,ma2021deep, jiang2021deep,holland1992genetic, kennedy1995particle, schneider2019benchmarking, park2022free}, one can solve the global optimum of the QCQP. While the QCQP is still NP-hard, we show in Chapter~\ref{chap:sparse} that in some cases, their solutions can be well approximated by the solutions of the relaxed SDP which are significantly easier to solve and can reach their respective bounds. This perspective on optimal photonic design is further discussed in the Conclusion of the thesis.

The computational framework established in this chapter allows for a number of extensions, one being the bounds to photonic devices with multiple functionalities.
For example, one may design a multi-layered thin film structure (composed of alternating layers between two materials for example) for frequency sensing by adjusting the width of each layer to maximize the reflection of the structure at the frequency of interest and minimize the reflection at a neighboring frequency. 
If the two frequencies are very close, it is perceivable that there is a limit to how much their reflection contrast can be. 
One can directly apply the computational bound developed in this chapter to each frequency separately but the bound would be too loose --- the maximal reflection will be achieved by a certain thin-film layers but the minimal reflection will always be zero as achieved by pure vacuum. 
Clearly, the looseness comes from the fact that we didn't constrain the two structures to be the same when computing the two bounds.
To resolve this, we use the binary-material argument to derive what we call ``field-correlation'' constraints that impose exactly the same-structure requirement for multi-functional devices, and show a natural decrease of reflection contrast between neighboring frequencies in the scenario described above.

\chapter{Bounds on multi-functional nanophotonics}
\label{chap:multi}
Nanophotonic devices that offer multiple functionalities, from liquid-crystal devices for beam steering~\cite{Resler1996,He2019, chung2020tunable}
to polychromatic metasurface lenses~\cite{aieta2015multiwavelength,wang2018broadband, chen2018broadband, shrestha2018broadband}, have tremendous design complexity because the multiple functions ought to be supported by a single photonic structure.
The single-structure constraint  cannot be captured by previous bounds. (See the discussion in Section~\ref{sec:chap3-discussion}.) In this chapter, we introduce ``field-correlation'' constraints between polarization fields induced in different scattering scenarios, which ultimately enforces the required single-structure constraint. 
These field-correlation constraints generalize the power-conservation constraints to scenarios involving multiple scatterings and integrate seamlessly into the framework we developed so far for theoretical bounds.
The bounds presented in this chapter extend the utility of the previous computational bounds to multi-functional photonic devices.

This chapter is outlined as follows.
In Section~\ref{sec:chap4-cross}, we generalize the power-conservation constraints to ``field-correlation'' constraints which enforce a single structure in multiple scattering scenarios. This leads to a general framework for computing bounds for multi-functional devices.
In Sections \ref{sec:chap4-maximum} and \ref{sec:chap4-beam}, we demonstrate the utility of the framework in two multi-functional designs: two-frequency reflection control in optical filters, for applications such as optical sensing (Section~\ref{sec:chap4-maximum}), and optimal beam switching via liquid-crystal-based nanophotonic metagratings (Section~\ref{sec:chap4-beam}). Both examples show the looseness of previous bounds and the necessity of enforcing the field-correlation constraints in multi-functional devices.

\section{Field-correlation constraints}
\label{sec:chap4-cross}
A multi-functional device has limited performance because different functionalities have to be supported by the same structure --- a constraint we introduce in this section as ``field-correlation constraint''. 
Such field-correlation constraints originate from the binary-material constraints discussed in Section~\ref{sec:chap3-discussion} of Chapter~\ref{chap:comp}. In the following, we first review the binary-material constraints for a single scattering, then generalize them to multiple scatterings and show how they enforce a single structure between multiple scenarios. 
These constraints, when expressed in terms of polarization fields, dictate the possible correlations of the fields in a scatterer under different scattering scenarios. These field-correlation constraints are quadratic in the polarization fields, thus amenable to the computation of bounds, as we demonstrate at the end of this section.
For simplicity, we assume scalar fields (e.g., the TE polarization in a 2D domain) scattered by an arbitrarily patterned isotropic materials in free space. The generalization to the scattering of vector fields in a 3D domain under an anisotropic material and an arbitrary background is straightforward.

A flexible way to construct conservation laws is to follow the bindary-material argument introduced in Section~\ref{sec:chap3-discussion}, which we review here for a single scattering.
We index the scattering with letter $i$ and label its associated scattering quantities with subscribe $i$ in preparation for the generalization to multiple scatterings later on.
For a given scattering scenario $i$, its material susceptibility $\chi_i(\xv)$ at each point $\xv$ in space is either zero or given by a constant material susceptibility $\chi_i$ (the constant may change between scenarios as in active modulation, hence the subscript $i$), which can be compactly expressed as
\begin{equation}
\chi_i^*(\xv)\left[ \chi_i(\xv) - \chi_i \right] = 0.
\label{eq:chap4-BinaryConstraint}
\end{equation}
The symbol $\hphantom{}^*$ denotes the complex conjugation, which is a mathematical convenience that maps \eqref{chap4-BinaryConstraint} into local power conservation at point $\xv$~(see Section~\ref{sec:chap3-discussion}).

\begin{figure} [t!]
    \includegraphics[width=1\linewidth]{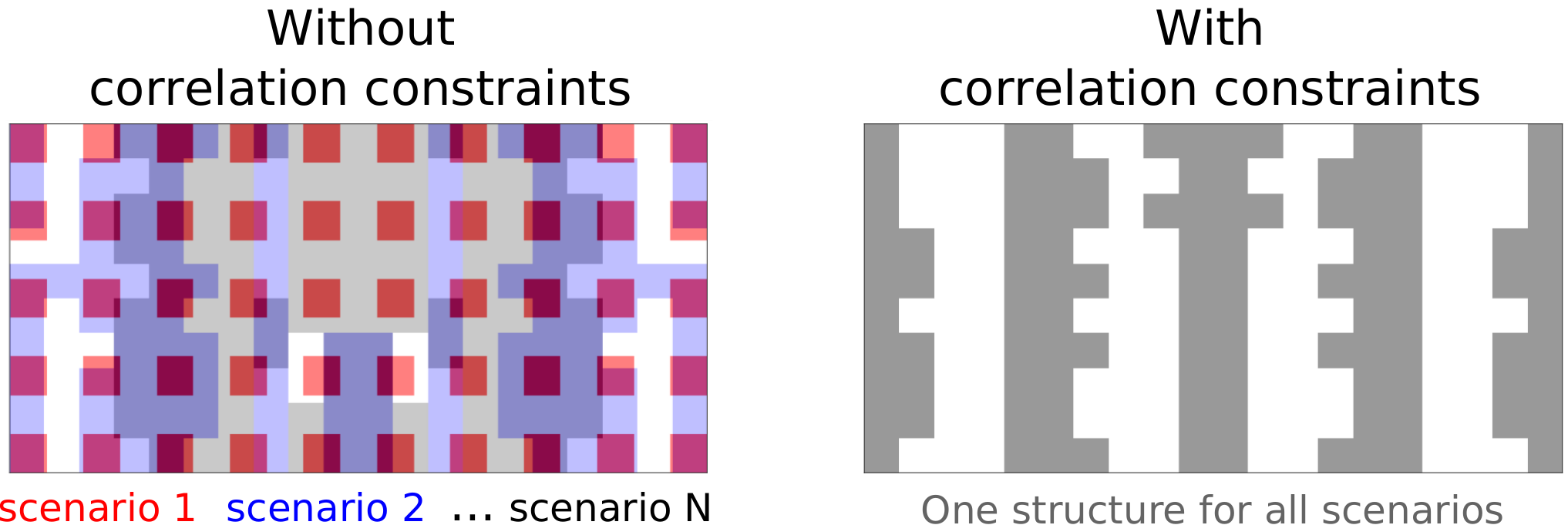} 
    \caption{Approaches in the previous chapters neglect the single-structure constraint in a multi-functional device (left). We introduce correlation constraints in \eqref{chap4-BinaryConstraint_multi} that enforces the same structure across all scattering scenarios (right), yielding meaningful bounds on what is possible.}  
    \label{fig:chap4-crosscorr} 
\end{figure} 

We generalize the binary-material constraint in \eqref{chap4-BinaryConstraint} from a single scenario $i$ to multiple scenarios $I=\{1,2,...,n\}$, by joining two conditions, i.e., placing air ($\chi_i(\vx) = 0$) or placing material ($\chi_j(\vx) - \chi_j = 0$), for every pairs of scenarios:
\begin{equation}
\chi_i^*(\xv)\left[ \chi_j(\xv) - \chi_j \right] = 0, \text{for all } i, j \text{ pair in } I.
\label{eq:chap4-BinaryConstraint_multi}
\end{equation}
At each point $\vx$, the structure is of either air or material, so one of the conditions in \eqref{chap4-BinaryConstraint_multi} has to be zero. Together, their multiplication hold true for every possible structure. \Eqref{chap4-BinaryConstraint_multi} are what we call ``correlation constraints". They capture the key single-structure requirement in multi-functional devices.
To see this, considering a physically impossible situation where scenario $i$ has material present at $\xv$ but scenario $j$ does not, the constraint in \eqref{chap4-BinaryConstraint_multi} is clearly violated. In fact, different scenarios share the same structure if and only if these correlation constraints hold between every pair of scenarios. 
\Figref{chap4-crosscorr} illustrates the difference between imposing the correlation constraints and not: without the correlation constraints, structures are separately optimized for different scattering scenarios; with the correlation constraints, a same structure is enforced across all scattering scenarios.

The correlation constraints in \eqref{chap4-BinaryConstraint_multi}, when expressed in terms of polarization fields $P_i(\xv)$, identify new conservation constraints in photonic devices.
To substitute $\chi_i(\xv)$ for $ P_i(\xv)$ in \eqref{chap4-BinaryConstraint_multi}, we use the constitutive relation $ P_i(\xv) = \chi_i(\xv) E_i(\xv)$ for scenario $i$ (and the same for scenario $j$), and write the total field $ E_i(\xv)$  as the sum of the incident field $ E_{\textrm{inc},i}(\xv)$ and the scattered field which is the convolution between the background Green's function $G_{0,i}(\xv,\xv')$ and $ P_i(\xv')$ in a bounding volume $V_d$, yielding
\begin{align}
 P^*_i(\xv) \cdot \bigg[\chi_j^{-1}  P_j(\xv) &- \int_{V_d} G_{0,j}(\xv,\xv')  P_j(\xv') \,{\rm d}\xv'   -  E_{\textrm{inc},j}(\xv)\bigg] = 0.
    \label{eq:chap4-CCConstr}
\end{align}
Consisting of the inner products between polarization fields induced in different scattering scenarios, \eqref{chap4-CCConstr} corresponds to a new set of ``field-correlation'' constraints that hold regardless of detailed structuring. In the special case of  $i=j$, it degenerates to the power-conservation law of \eqref{chap3-local-power_scalar} in Chapter~\ref{chap:comp}. 
More generally, they encompass not only the power-conservation constraints at each scattering scenario, but all the field-correlation constraints that enforce the single-structure requirement among different scattering scenarios, leading to tighter bounds for multi-functional devices.
     
The field-correlation constraints of \eqref{chap4-CCConstr} limit the optimal response of multi-funcitonal devices. Any typical objective $f$ of interest in linear optics (scattered/absorbed power, mode overlap, switching efficiency, etc.) is a quadratic function of the polarization fields $ \{P_i(\xv)\}_{i=1}^n$ in all $n$ scenarios.  The optimal response is determined by maximizing the objective $f$ over every possible induced polarization field $ \{P_i(\xv)\}_{i=1}^n$ under the field-correlation constraints of \eqref{chap4-CCConstr}:
\begin{equation}
    \begin{aligned}
        & \underset{\{P_i(\xv)\}_{i=1}^n}{\text{max.}} & & f(P_1(\xv), \cdots, P_n(\xv)) \\
        & \text{s.t.} & & \textrm{\Eqref{chap4-CCConstr} satisfied for all $i,j$} \\
        & & & \textrm{ and all $\xv$ in the design region.}
    \end{aligned}
    \label{eq:chap4-QCQP}
\end{equation}
As both the objective and the constraints are quadratic in the optimization variable $ \{P_i(\xv)\}_{i=1}^n$, the optimization problem after numerical discretization is a QCQP, which can be bounded in the same way as the computational bounds in Chapter~\ref{chap:comp} via semidefinite relaxation. Removing the constraints with $i\neq j$ in \eqref{chap4-QCQP} loosens the problem to the power-conservation bounds in \eqref{chap3-local-opt-problem}. Keeping all the field-correlation constraints, the  bound of \eqref{chap4-QCQP} gives a tighter limit on the multi-functional design problem.

\begin{figure} [p]
    \includegraphics[width=1\linewidth]{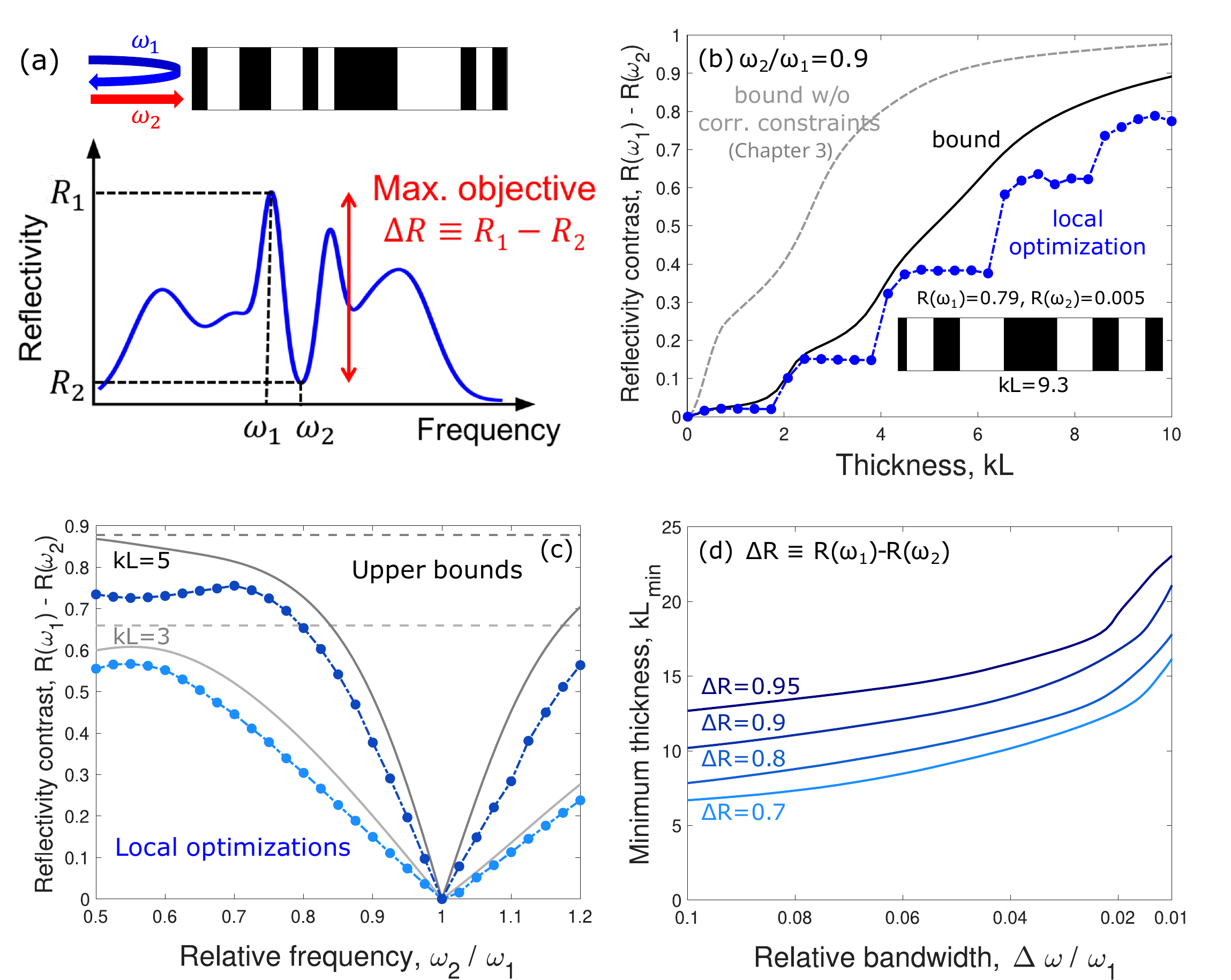} 
    \caption{(a) Maximum reflectivity contrast between two nearby frequencies, $\omega_1$ and $\omega_2$, on a multilayer structure. The bounds capture tradeoffs between the key parameters such as designable-domain size, refractive index, frequency separation, and reflectivity contrast. (b) Comparison of bounds on reflectivity contrast, both with and without field-correlation constraints, to designs achieved via gradient-based local optimization. The relative frequency $\omega_2 / \omega_1$ is set to 0.9, and $kL$ is the normalized thickness of the designable region $k=2\pi / \lambda$. The inset shows a locally-optimized design at $kL=9.3$, with reflectance of 0.79 at $\omega_1$ and 0.005 at $\omega_2$. (c) Similar to (b), but as a function of relative frequency $\omega_2 / \omega_1$, where $kL$ is fixed at 3 and 5. (d) The minimum thickness required for different values of desired reflectivity contrast, as predicted by our bounds with field-correlation constraints.}  
    \label{fig:chap4-RvsL} 
\end{figure}

\section{Maximum reflectivity contrast}
\label{sec:chap4-maximum}

A key problem in sensing applications~\cite{Anker2009b,Chang2010,liu2010infrared,law2013all} is to detect small changes in the incident frequency of light. To do so, the scatterer must be designed to maximize the variation in its response with respect to changes in the incident frequency. We focus here on reflectance, which naturally arises in many applications~\cite{liu2010infrared,law2013all}, and on multilayer films, i.e., optical filters. The problem is depicted schematically in \figref{chap4-RvsL}(a), with any combination of angles and frequencies for the incident, reflected, and transmitted waves, while the primary objective might be the normal-incidence contrast in reflectivity between two nearby frequencies. We take the multilayer film to comprise alternating layers of Al$_2$O$_3$ (as is commonly used for multilayer thin films~\cite{Dobrowolski2008,Poitras2017}) and vacuum. Reflectance is a quadratic function of the polarization fields. Our objective is the contrast in reflectivity at any two frequencies of interest, which we write as $R(\omega_1) - R(\omega_2)$. 
			
The solid black lines in \figref{chap4-RvsL}(b) show the upper bounds for reflectivity contrast as a function of the maximum thickness $L$ of the designable region (normalized to wave number $k = \omega / c$). Also included are the bounds without the field-correlation constraints (i.e., the only constraints in \eqref{chap4-QCQP} are the ones with $i = j$, equivalent to the power-conservation bounds in Chapter~\ref{chap:comp}), as well as the actual reflectivity contrasts achieved by designs identified through a local-optimization routine. For each thickness of the designable region, we use gradient descent~\cite{Nocedal2006}, optimize over many random initial designs, and select the best-performing design. In \figref{chap4-RvsL}(b), the local optimizations approach, though not surpass, the global bounds. The bounds without the field-correlation constraints, shown as the dashed grey line, are loose compared to the local-optimization results. Incorrectly, they always find a vacuum structure that minimizes the reflectance at $R(\omega_2)$, clearly violating the same-structure constraint. For two relatively close frequencies, one must incorporate field-correlation constraints to enforce a single structure and achieve meaningful bounds.

The insufficiency of bounds without field-correlation constraints is further illustrated in \figref{chap4-RvsL}(c), which sweeps across a variety of relative frequency values (holding $\omega_1$ fixed). In each case, the solid line depicts the upper bound with field-correlation constraints, which is tracks closely with the best designs from the local-optimization computations. By contrast, bounds without field-correlation constraints (dashed lines) do not change as functions of relative frequency because they always identify the minimal $R(\omega_2)$ as zero, a trivial result.

\Figref{chap4-RvsL}(d) shows a further extension: the thinnest multilayer film for a desired reflectivity contrast $\Delta R$, as a function of the relative bandwidth of the two frequencies of interest. 
As expected, the minimum thickness increases as a function of the reflectivity difference and the inverse bandwidth. Nonetheless, the scaling lines of \figref{chap4-RvsL}(d) that indicate precise tradeoffs between sensitivity and size is only obtainable by solving the bounds in \eqref{chap4-QCQP}. 

\section{Beam switching via liquid crystals}
\label{sec:chap4-beam}
Another type of multi-functional optical device involves an externally applied voltage that modulates the refractive index of the material and hence its optical response. Here, we consider liquid-crystal-based beam switching, whose target functionality is to direct light to two different directions under two external voltages. 
The design of such two-state switching device can be generalized to many-state switching by stitching together different composing units~\cite{Patel1995,Lee2008}.

\begin{figure} [tb]
    \includegraphics[width=\linewidth]{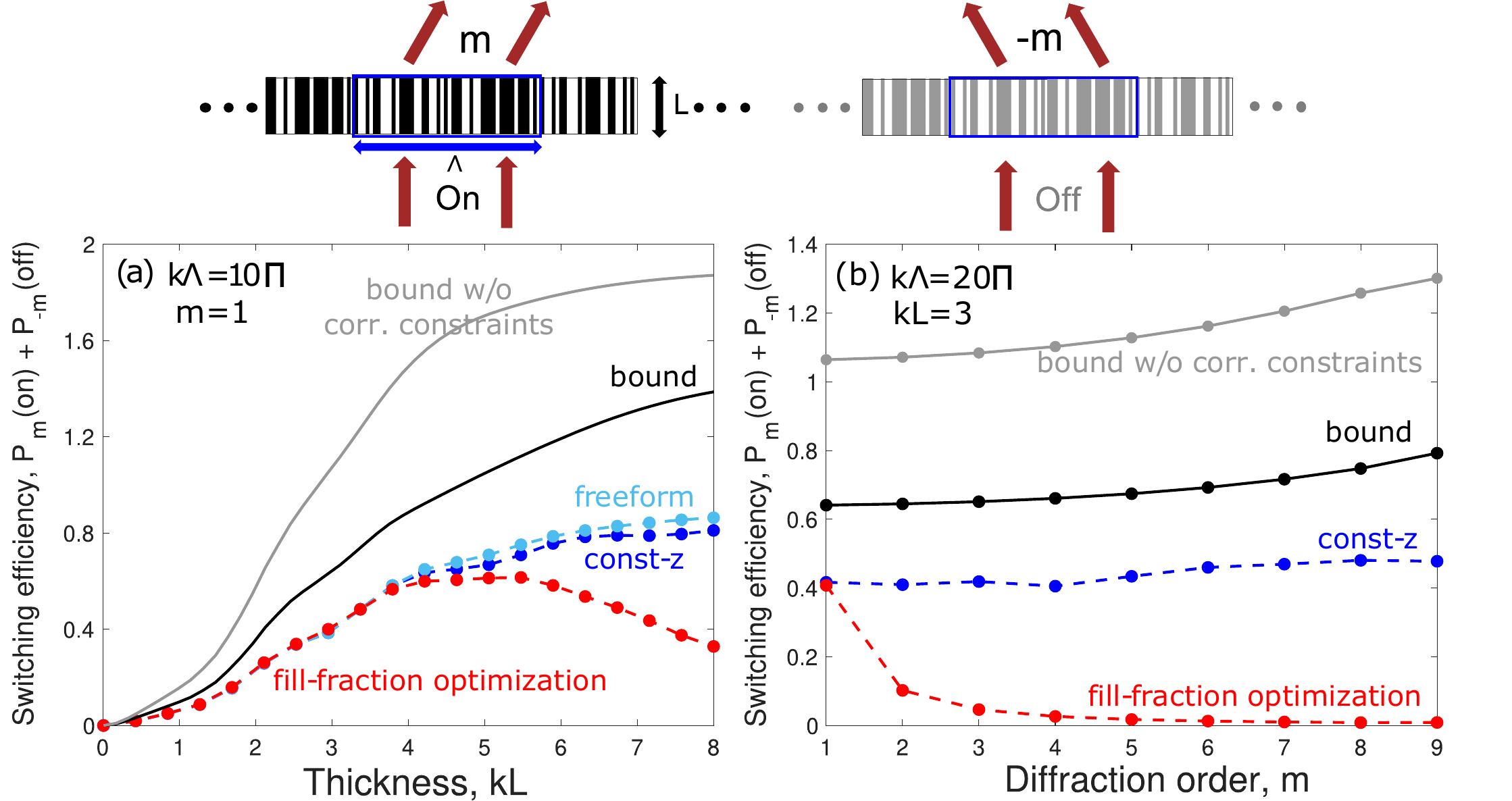} 
    \caption{(Top) Liquid-crystal-based beam beam-switching device. (a,b) Bounds on switching efficiency, both with and without field-correlation constraints, compared to 3 inverse-designed structures (circles). The switching efficiency is the sum of the power into diffraction orders $m$ and $-m$ as illustrated at the top, where the power diffracted into each state is normalized to 1 (so that the maximum switching efficiency is 2).}  
    \label{fig:chap4-2Dswitching} 
\end{figure}

We bound the switching efficiency of a grating structure comprising actively tuned liquid crystal. In the bound, we allow for any possible pattern of the liquid-crystal material within a period $\Lambda$ and grating thickness $L$, as in the top of \figref{chap4-2Dswitching}. The liquid-crystal material, named E7~\cite{Yang2010}, has refractive indices of about $1.7$ and $1.5$ in the voltage-on and off state, respectively (for purposes of demonstration, we introduce a small imaginary part of value 0.1 to the refractive indices in both cases). Given a monochromatic field at normal incidence, we maximize the power switching efficiency, defined to be the sum of the power in target directions for the voltage-on and voltage-off states. The two target directions are chosen to be the diffraction orders $m$ and $-m$, with angles $\theta_m = -\theta_{-m} = \sin^{-1}(2\pi m / k\Lambda)$ with respect to the surface normal. The power diffracted into the two directions, $P_m$ and $P_{-m}$, are quadratic functions of the polarization field, amenable to our framework for theoretical bounds. 

\Figref{chap4-2Dswitching}(a) shows bounds and designs for the liquid-crystal beam-switching problem. The unit cell is about five free-space wavelengths wide ($k\Lambda = 10\pi+\delta$) and the diffraction order $m=1$. (The constant $\delta = 0.1$ is added to avoid the singularities and infinite-$Q$ resonances possible via bound-state-in-continuum modes~\cite{Hsu2016}, which cause numerical instabilities.) The angular deflection is small ($\pm 11.5^\circ$), so one might expect a perfect switching efficiency.
Yet the bound (solid black line) shows one can achieve at most $70\%$ of the maximum possible switching efficiency (which is 2) even when the device is one-wavelength thick.
Compared with the bound are designs optimized under three strategies: a fill-fraction optimization (red), in which the optimal fill fraction of a simple grating structure (fill fraction here specifies the fraction of the material occupying the unit cell) is computed, a ``const-z'' optimization (blue), in which the all air holes must have equal depth to be compatible with lithography, and a freeform optimization, in which the permittivity is allowed to take either material value at any point in the domain (teal). The freeform optimizations have the most degrees of freedom, and come closest to the bounds, while the fill-fraction optimizations are feasible at small thicknesses but deteriorate in quality at larger thicknesses. The freeform and const-z approaches both show similar trendlines to the computed bounds. By contrast, the bounds without field-correlation constraints (i.e., the bounds in Chapter~\ref{chap:comp}) quickly approach 2, a trivial result, because it suggests $100\%$ possible efficiency in both states. \Figref{chap4-2Dswitching}(b) isolates a single thickness $kL = 3$ for a larger unit-cell period $k\Lambda = 20\pi$ and sweeps over the target diffraction order $m$, with angular deflections increasing to $64^\circ$ for $m=9$. Surprisingly, the bound suggests that at this large unit-cell period, the switching efficiency can \emph{increase} as the angular of deflection increases. This is borne out by the const-z optimizations, which show a similar trend, though the noisiness of local optimizations makes the trend less clear than the bounds do. 

\section{Discussion}
\label{sec:chap4-discuss}

In this chapter, we bound the performance limit of multi-functional photonic devices by constraining the correlations between polarization fields in multiple scattering scenarios.
The resulting field-correlation constraints generalize the power-conservation constraints, and ultimately enforce a single binary structure for different functionalities (Section~\ref{sec:chap4-cross}).
The field-correlation constraints are essential for locating tight multi-functional bounds, as we see in the multi-reflectivity bound in Section~\ref{sec:chap4-maximum} and the beam-switching bound in Section~\ref{sec:chap4-beam}. 
The successful implementation of these two examples suggest future applications of this approach to a broad pool of multi-functional devices such as achromatic metalenses~\cite{aieta2015multiwavelength,wang2018broadband, chen2018broadband, shrestha2018broadband}, multi-frequency mode converters~\cite{piggott2015inverse, piggott2017fabrication}, and actively-tuned spectrometers~\cite{yuan2021wavelength}.

The slow computation of the multi-functional bound limits its application to devices with only few functionalities.
This limitation is inherited from the computational bound in Chapter~\ref{chap:comp}, as, algorithmically, the multi-functional bound is the same as the computational bound but with the degree of freedom and constraints multiplied by the number of scattering scenarios considered. Its algorithmic complexity is at least O($N^3$) for $N$ designable degree of freedoms (the number of spatial grids times the number of functionalities).
To reduce the computational complexity, one can minimize the number of constraints by (1). imposing the field-correlation constraints only between neighboring scenarios instead of every pair of them and (2). implementing the algorithm developed in Chapter~\ref{chap:comp} that selects only the maximally-violated local-power-conservation laws.
In certain cases, analytical bounds with one or two ``global'' constraints may be derived to tease out important scaling laws such as the maximal reflectivity contrast in the limit of small frequency separation.
Still, as we demonstrated in Chapter~\ref{chap:analy}, such bounds are only tight if material loss is the major constraining factor as in the plasmonic materials. 
To accelerate the computation of bounds for general materials such as lossless dielectrics, we present in the next chapter an algorithm that leverages the inherent sparsity of Maxwell's equations to dramatically reduce the algorithm complexity to $O(N)$ and demonstrate bounds for metalens (made of lossless dielectrics) for more than a hundred wavelengths in diameter.

\chapter{Accelerating computational bounds with sparsity}
\label{chap:sparse}

The computational bounds we establish in previous chapters are too slow to solve for large-scale photonic devices. 
In this chapter, we accelerate their computation by leveraging sparse structures in Maxwell's equations, leading to bounds for metasurfaces hundreds of wavelengths in diameter.
The bounds in Chapters~\ref{chap:comp} and~\ref{chap:multi} have daunting algorithm complexities, ranging from $O(N^3)$ to $O(N^4)$ for $N$ designable degree of freedoms, and the dense matrices they use permit no direct accelerations.
On the contrary, algorithms that leverage sparse structures in the Maxwell's equations (e.g., the tridiagonal sparsity of the differential operator in 1D problems) can solve an electromagnetic scattering problem in $O(N)$ flops, and are widely used in commercial solvers such as Lumerical and Comsol. 
Those sparsity structures reduce both data storage and calculation flops of an algorithm and, as we demonstrate below, can also accelerate our bounds.

This chapter is outlined as follows. In Section~\ref{sec:chap5-sparse}, we formulate sparse-structured conservation constraints directly from the electromagnetic wave equation. 
The sparsity pattern of the Laplacian operator in the wave equation is often chordal, a specific type of sparsity that allows for fast computation of bounds as discussed in Section~\ref{sec:chap5-fast}.
An important case where chordal sparsity arises is the high-aspect-ratio metasurface, whose maximal focusing efficiency, as we show in Section~\ref{sec:chap5-maximal}, can be bounded in $O(N)$ flops, almost as fast as solving a single scattering problem. 

\section{Sparse-structured conservation constraints}
\label{sec:chap5-sparse}
In this section, we derive conservation constraints that inherit the sparse structures in the Maxwell's equations.
These constraints are formulated in terms of electric fields in the wave equation, involving two regions: a design region where the actual scatterer is, and a perfectly-matched-layer (PML) region that simulates free-space radiation. 
At each point in the design region, either air or material is present, so the conservation law can be constructed via the binary-material argument introduced in previous chapters. 
In the PML region, there is no design, only prespecified PML materials, so the constraint is the wave equation itself.
In both regions, sparsity appears after numerically discretizing the constraints. We discuss these points in detail below.

The sparse-structured conservation constraints originate from the electromagnetic wave equation which involves couplings only between neighboring points.
This ``local coupling'' contrasts with the long-range coupling observed in the volume-integral equation in previous chapters (see \eqref{chap2-vie}), and induces sparsity after numerical discretization.  
Under a certain source excitation $b(\vx)$, a differential operator $\LL(\chi(\vx))$ governs the couplings between electric fields $E(\vx)$ in the wave equation
    \begin{equation}
        \LL(\chi(\vx)) E(\vx) - b(\vx) = 0.
        \label{eq:chap5-wave-eq}
    \end{equation}
The differential operator $\LL(\chi(\vx))$ accounts for the susceptibility $\chi(\vx)$ of the scatterer and is a local operator whose exact form depends on the problem dimension. 
In the following, we consider the scattering of TE polarization in a 2D domain, where the differential operator $\LL(\chi(\vx)) = \nabla^2 + \left[1 + \chi(\vx)\right] \omega^2$, the electric field $E(\vx)$ is a scalar, and the source term $b(\vx)=-i\omega J(\vx)$ for an external current $J(\vx)$.
Given a design with susceptibility $\chi(\vx)$, solving its electric field $E(\vx)$ through \eqref{chap5-wave-eq} requires two simulation regions: a design region where the actual scatterer lies and a PML region where prespecified absorbing materials are placed to truncate the simulation domain.
Constraints in these two domains take different forms, which we discuss separately below.  
    
In the design region, we construct conservation constraints that hold regardless of the shape of the scatterer. 
The binary-material argument put forth in Section~\ref{subsec:chap3-binary} provides a recipe.
At each point $\vx$ in the design region $V_d$, assuming the material susceptibility $\chi(\vx)$ is either $\chi_1$ or $\chi_2$, the wave equation of \eqref{chap5-wave-eq} becomes either $\LL(\chi_1) E(\vx) - b(\vx) = 0$ or $\LL(\chi_2) E(\vx) - b(\vx) = 0$, respectively.
Multiplying these two conditions together, we obtain a single equation that holds in either case:
\begin{equation}
    \left[\LL(\chi_1) E(\vx) - b(\vx)\right]^*\left[\LL(\chi_2) E(\vx) - b(\vx)\right] = 0,\myforall \vx \in V_d,
    \label{eq:chap5-constraint-design}
\end{equation}
where we take the complex conjugation of the first condition so that \eqref{chap5-constraint-design} maps to the local power conservation in \eqref{chap3-local-power_scalar}.
\Eqref{chap5-constraint-design} holds at every point in the design region $V_d$ regardless of its material composition. 
It is also quadratic in the electric fields, suitable for deriving bounds.

Outside the design region are air and PML. The latter, in particular, consists of absorbing materials to truncate the simulation domain by absorbing the scattered field with little reflection as if the waves had radiated into free space. 
For simplicity, we use the phrase ``PML region'' to refer to any space outside of the design region (including both the PML and its air buffer), denoted as $V_{\rm PML}$. Assuming a known material susceptibility distribution $\chi_{\rm PML}(\vx)$ in the PML region $V_{\rm PML}$, the wave equation of \eqref{chap5-wave-eq} becomes
\begin{equation}
    \LL(\chi_{\rm PML}(\vx)) E(\vx) - b(\vx) = 0, \myforall \vx \in V_{\rm PML}.
    \label{eq:chap5-constraint-PML}
\end{equation}
As the PML is chosen independent of the scatterer composition, \eqref{chap5-constraint-PML} holds for every possible design.

\begin{figure}[t]
    \centering
    \includegraphics[width=.95\textwidth]{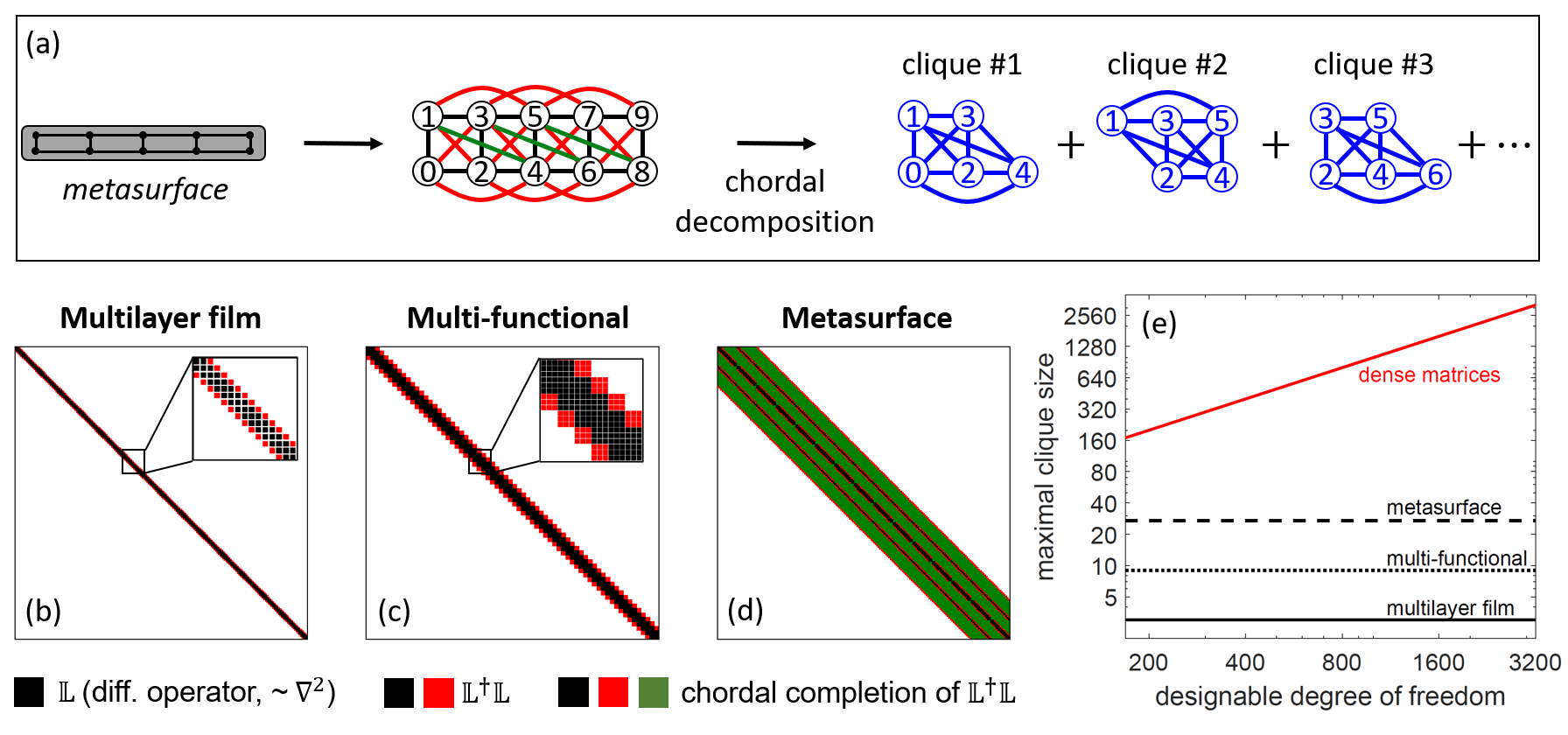}
    \caption{(a) Differential equations on any domain (black grid) have a sparsity pattern that can be represented in an undirected graph (vertices numbered 0--8). The \emph{chordal completion} of that graph leads to decomposition into ``cliques'' that represent smaller units on which the problem can be defined.  (b)-(d) Chordal completions of the sparsity patterns arising in various photonic designs. (e) The solving speed of a semidefinite program depends on  the largest clique size of its sparsity graph. For dense matrices (as arise in integral equations), this is simply proportional to the size of the problem (red). By contrast, the sparsity of differential operators leads to maximum clique sizes that are bounded above by a constant unrelated to the ``long'' dimension of the design problem.}
    \label{fig:chap5-fig1}
\end{figure}

Because electric fields interact locally in the wave equation,  constraints in both \eqref{chap5-constraint-design} and \eqref{chap5-constraint-PML} exhibit sparse matrix structures after being numerically discretized in real space. 
Following our convention, we discretize fields and operators in these two equations into lower-case vectors and  blackboard-bold matrices, respectively, yielding  
\begin{align}
     \left[\LL(\chi_1) e - b\right]^*\DD_i\left[\LL(\chi_2) e - b\right] &= 0,\myforall i \in I_d, \label{eq:chap5-design-discretize}\\
    \left[\LL(\chi_{\rm PML}(\vx_i)) e - b\right]_i &= 0,\myforall i \in I_{\rm PML}, \label{eq:chap5-PML-discretize}
\end{align}
where $I_d$ and $I_{\rm PML}$ are the lists of indices for the points in the design and PML regions, respectively, and the $\DD_i$ matrix is all zero except for its $(i,i)$ entry which has value one.
The matrix $\LL$, in particular, inherits the sparsity pattern of the differential operator $\nabla^2$, which, after the central-difference discretization~\cite{olver2014introduction}, only has few non-zero elements per row. 
We show in \figref{chap5-fig1}(b,d) the sparsity patterns of $\LL$, denoted by black squares, for one- and two-dimensional domains,
which have only three and five nonzero elements per row, respectively. 
This sparsity, as we demonstrate in the next section, enables fast computational bounds.

\section{Fast computational bounds via sparsity}
\label{sec:chap5-fast}
In this section, we show how the sparse-structured conservation constraints accelerate the computation of bounds for photonic devices.
We formulate photonic designs as a QCQP, relax the QCQP into a convex semidefinite program, and solve the semidefinite program for bounds.
Semidefinite programming has high algorithm complexity. 
The key contribution in this chapter is that we accelerate its computation using sparsity, and tackle the key bottleneck, the semidefinite constraint, by decomposing the corresponding chordal graph into small cliques where different parts of the semidefinite constraint can be verified separately. 
The smaller the cliques, the faster the acceleration. To this end, we bound the maximal clique size for a variety of photonic problems, and show, compared to the volume-integral equation, the sparse formulation can lead to much faster bounds.

\subsection{QCQP and its computational bound}
A photonic design can be reformulated as a QCQP under the sparse-structured conservation constraints derived in Eqs.~(\ref{eq:chap5-design-discretize}, \ref{eq:chap5-PML-discretize}).
The key is to directly optimize over the electric fields in the design and PML regions.
In the design region, the electric fields are subject to the sparse-structured conservation constraint in \eqref{chap5-design-discretize}.
In the PML region, the electric fields are subject to the linear wave equation in \eqref{chap5-PML-discretize}. We discretize the entire space into $n$ grid points so that the electric field is a complex-valued vector $e$ of length $n$. 
Most objectives $f(e)$ are quadratic functions in the electric-field vector $e$, maximizing them under the existing constraints amounts to 
\begin{equation}
    \begin{aligned}
    & \underset{e\in\CC^n}{\text{maximize}}
    & & f(e) = e^\dagger \mathbb{A} e +  \Re\left(\beta^\dagger e\right) \\
    & \text{subject to}
    & &  \left[\LL(\chi_1) e - b\right]^*\DD_i\left[\LL(\chi_2) e - b\right] = 0,\myforall i \in I_d, \\
    & & & \left[\LL(\chi_{\rm PML}(\vx_i)) e - b\right]_i = 0,\myforall i \in I_{\rm PML}.
    \end{aligned}
    \label{eq:chap5-QCQP}
\end{equation}
This optimization problem is quadratic in both the objective and the constraints --- a QCQP. 
It is mathematically equivalent to the binary-design problem in \eqref{original_design_problem} but now exhibits a mathematical structure amenable to global bounds, as we show below.

The QCQP of \eqref{chap5-QCQP} can be relaxed to a convex problem via semidefinite relaxation, a standard procedure which we have discussed in both Chapter~\ref{chap:comp} and Appendix~\ref{sec:appenH-sec1}, and review here to highlight its main results. 
Without loss of generality, we express the QCQP of \eqref{chap5-QCQP} in the following form
\begin{equation}
    \begin{aligned}
    & \underset{e\in\CC^n}{\text{maximize}}
    & & e^\dagger \mathbb{A}_0 e +  \Re\left(\beta_0^\dagger e\right) + c_0 \\
    & \text{subject to}
    & &  e^\dagger \mathbb{A}_i e +  \Re\left(\beta_i^\dagger e\right) + c_i = 0, \quad \text{for } i = 1, 2, ..., m,
    \end{aligned}
    \label{eq:chap5-QCQP-standard}
\end{equation}
where $\AA_0$, $\beta_0$, and $c_0$ encapsulate any quadratic objectives, and $\AA_i$, $\beta_i$, and $c_i$ all real-valued quadratic constraints. For each complex-valued constraint in \eqref{chap5-QCQP}, we have two real-valued ones in \eqref{chap5-QCQP-standard}, hence $m=2n$.
Semidefinite relaxation relaxes the QCQP of \eqref{chap5-QCQP-standard} into a convex semidefinite program in two steps: lifting and relaxation.
First, it ``lifts'' the optimization variable $e$ to a higher-dimension matrix space $\XX=e e^\dagger$  where the quadratic function $e^\dagger\AA_i e$ can be written as a linear function $\Tr(\AA_i\XX)$ because $e^\dagger\AA_i e = \Tr(e^\dagger\AA_i e) = \Tr(\AA_i e e^\dagger)$.
The extra linear terms in \eqref{chap5-QCQP-standard} are straightforward to include, leading to modified matrices $\AA_i$ whose expressions are explicitly given in Appendix~\ref{sec:appenH-sec1}.
The matrix $\XX$ cannot vary freely; being the outer product of a vector with itself, $\XX$ have to be positive semidefinite and rank-one. The latter is a nonconvex constraint. The relaxation step removes it, yielding a semidefinite program 
\begin{equation}
    \begin{aligned}
    & \underset{\XX\in\SS^n}{\text{maximize}}
    & & \Tr(\AA_0\XX) \\
    & \text{subject to}
    & &  \Tr(\AA_i\XX) = 0, \quad \text{for } i = 1, 2, ..., m, \\
    & & & \ \XX \geq 0.
    \end{aligned}
    \label{eq:chap5-SDP}
\end{equation}
Both the linear functions, $\Tr(\AA_i\XX)$, and the semidefinite constraint, $\XX\geq 0$, are convex in the matrix variable $\XX$. Hence, the  semidefinite program in \eqref{chap5-SDP} is a convex optimization problem, which can be solved globally with standard solvers such as the interior-point method~\cite{Vandenberghe1996,Boyd2004}. 
Because of the relaxation, the maximum of the semidefinite program bounds the maximum of the QCQP in \eqref{chap5-QCQP-standard}, giving a fundamental limit on the largest photonic response achievable via geometric designs.

We can include additional quadratic constraints in \eqref{chap5-QCQP} to tighten the bound. 
Linear constraints such as $\left[\LL(\chi_{\rm PML}(\vx_i)) e - b\right]_i = 0$ in \eqref{chap5-QCQP} tend to lose its tightness after semidefinite relaxation~\cite{park_general_2017}. 
To remedy this, we create additional quadratic constraints by multiplying the linear constraint with itself:
\begin{equation}
    \left[\LL(\chi_{\rm PML}(\vx_i)) e - b\right]^* \DD_i \left[\LL(\chi_{\rm PML}(\vx_i)) e - b\right] = 0,\myforall i \in I_{\rm PML}.
    \label{eq:chap5-redundant-constraints}
\end{equation}
Those additional quadratic constraints are constructed for every point in the PML region (where the linear constraints lie). 
They are redundant in the QCQP of \eqref{chap5-QCQP} but not in the semidefinite program of \eqref{chap5-SDP}.
They are redundant in the QCQP because the existing linear constraints already imply them.
They are not redundant in the semidefinite program because linear and quadratic constraints undergo different transformations in the semidefinite relaxation~\cite{park_general_2017}. Once after the relaxation, the former cannot derive the latter. 
Thus, incorporating the QCQP in \eqref{chap5-QCQP} with the additional quadratic constraints in \eqref{chap5-redundant-constraints} leads to tighter bounds, sometimes significantly tighter.

\subsection{Fast semidefinite program on chordal graphs}

Semidefinite programs are challenging to solve for large-dimensional problems, unless there is sparsity to be leveraged. 
Sparsity mainly accelerates two operations in the semidefinite program in \eqref{chap5-SDP}: multiplying matrix $\XX$ by matrix $\AA_i$, and verifying the semidefinite constraint, $\XX\geq 0$, is satisfied. 
These two operations are the numerical bottleneck for most  semidefinite programming algorithms~\cite{Boyd2004, Luo2010}, both costing at least $O(n^3)$ per iteration for an $n$-dimensional problem.
To accelerate these two operations, one needs to consider the aggregate sparsity pattern of the semidefinite program in \eqref{chap5-SDP}, which are the places where all the objective and constraint matrices, $\AA_i$, are zero. 
 The aggregate sparsity pattern is given by the sparsity pattern of the $\LL^\dagger \LL$ matrix in the QCQP in \eqref{chap5-QCQP}, which is shown as the black and red squares in \figref{chap5-fig1}(b-d) for a number scenarios.
 Most sparsity can accelerate matrix-matrix multiplications by multiplying only the nonzero entries.
 On the other hand, only a special type of aggregated sparsity based on chordal graphs can accelerate the verification of the semidefinite constraint, which we explain below.

The major bottleneck of large-scale semidefinite programming is the verification of its semidefinite constraint, which is resolvable if the underlying sparsity pattern corresponds to a chordal graph~\cite{Vandenberghe2015}.
There is a unique mapping between sparsity pattern of symmetric matrices and undirectional graphs: an nonzero $(i,j)$ entry of the former maps to a chord between the $i$ and $j$ vertices of the latter. (See Appendix~\ref{appen:graph} for a quick review on graph theories.)
A chordal graph, in particular, is a graph where all cycles of four or more vertices have a chord. 
The sparsity pattern of the $\LL^\dagger\LL$ operator in our photonic problem does not necessarily correspond to a chordal graph (see \figref{chap5-fig1}(a)) but it can be made into one by adding extra chords (green lines in \figref{chap5-fig1}(a)).
Chordal graph is desirable because it allows one to distribute certain complex operations into different ``cliques'', a set of vertices in the graph that are all connected to each other (as in \figref{chap5-fig1}(a)).
In particular, if the aggregate sparsity of the semidefinite program in \eqref{chap5-SDP} can be completed into a chordal graph, then Grone's theorem in Eq.~(\ref{eq:appenJ-Grone}) shows its semidefinite constraint $\XX \geq 0$ can be replaced by ``smaller'' semidefinite constraints on each of its cliques: $\XX[I_j, I_j] \geq 0$, where $I_j$ denoting the indices for the $j$th clique. The optimization problem of \eqref{chap5-SDP} becomes
\begin{equation}
    \begin{aligned}
    & \underset{\XX\in\SS^n}{\text{maximize}}
    & & \Tr(\AA_0\XX) \\
    & \text{subject to}
    & &  \Tr(\AA_i\XX) = 0, \quad \text{for } i = 1, 2, ..., m, \\
    & & & \ \XX[I_j, I_j] \geq 0, \quad \text{for } j = 1, 2, ..., p,
    \end{aligned}
    \label{eq:chap5-SDP-chordal}
\end{equation} 
where $p$ to denotes the total number of cliques in the chordal graph. 
The only difference between the decomposed semidefinite program and the original semidefinite program is the semidefinite constraints are now applied to smaller cliques.
As the semidefinite constraint is the main computational bottleneck that increases as $O(n^3)$ with the matrix size $n$, breaking a large $\XX$ into many smaller ones in \eqref{chap5-SDP-chordal} allows us to significantly accelerate the computational bounds.

The exact amount of acceleration depends on the clique size of the chordal graph.
The smaller the cliques, the faster the verification of the positive semidefiniteness in \eqref{chap5-SDP-chordal}, the faster the bound.
The sparsity patterns of many photonic problems are banded matrices as in \figref{chap5-fig1}(b-d), their clique sizes equal to twice their bandwidths plus one. 
One-dimensional design such as the multilayer film in \figref{chap5-fig1}(b) has a clique size of three. 
Multi-functional bounds in Chapter~\ref{chap:multi} multiply the clique size of three by the number of functionalities; \figref{chap5-fig1}(c) assumes three functionalities, leading to a clique size of nine.
For a two-dimensional domain such as the metasurface in \figref{chap5-fig1}(d), the sparsity pattern acquires side bands, which can be filled into a band matrix with a clique size  proportional to the number of points in the vertical dimension.
In all three examples above, the clique size only depends on the ``short dimension'' (functionalities, metasurface thickness), while independent of the ``long'' dimension (thin-film thickness, metasurface diameter). 
This contrasts with the dense matrices arising from the volume-integral equation whose clique size is the product of the long and short dimensions, as shown in \figref{chap5-fig1}(e).
Here lies an advantage of the sparse framework, for we can bound very large design problem providing that the ``short'' dimension remains short, which we demonstrate for the important case of large-diameter metasurfaces in the next section.

\section{Maximal focusing of large-scale metalenses}
\label{sec:chap5-maximal}

\begin{figure*}[tb]
    \centering    \includegraphics[width=0.9\textwidth]{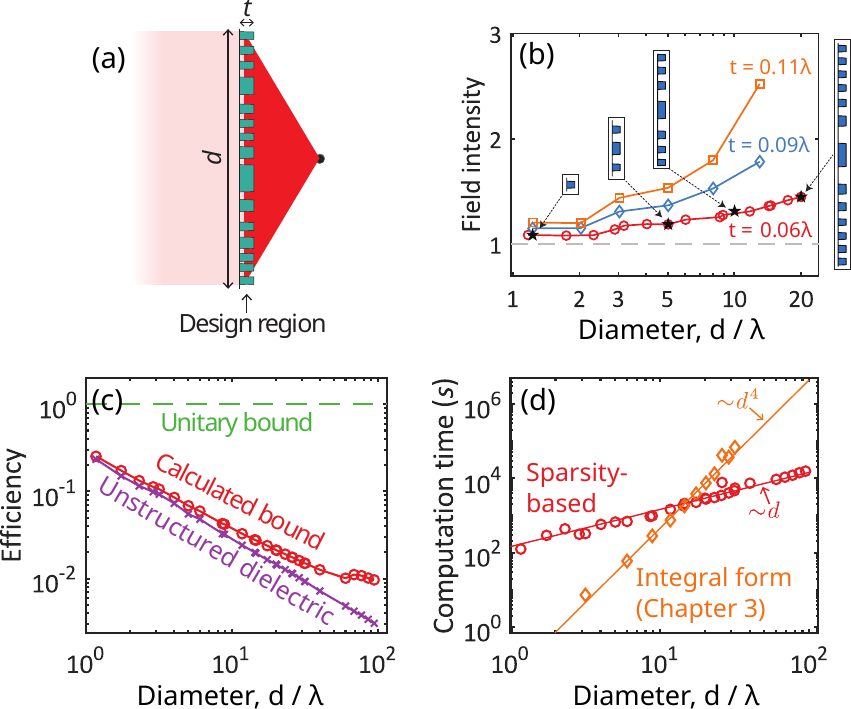}
    \caption{(a) A metalens of diameter $d$ and thickness $t$ that focuses light at wavelength $\lambda$. (b) Maximal field intensities calculated by the bounds for varying diameters and thicknesses, as well as designs that hit the bounds. (c) Bounds on focusing efficiency compared to the unstructured and previous unitary bounds. (d) Compared to the $O(d^4)$ scaling of the integral-form bounds in Chapter~\ref{chap:comp}, the sparsity-based bounds in this Chapter exhibits a much faster, \textit{linear} dependence in computation time.}
    \label{fig:chap5-fig2}
\end{figure*}

In this section, we demonstrate the utility of our sparsity-based fast semidefinite program of \eqref{chap5-SDP-chordal}, for enabling computational bounds of large-scale systems. ``Metasurfaces,'' in which one patterns a wavelength-scale-thickness material for new, compact form factors in optics, offer a compelling example~\cite{yu2014flat,lalanne_metalenses_2017}. A metalens is a metasurface that focuses light to a single focal spot, and one typically wants a maximally efficient metalens with a diameter significantly larger than the free-space optical wavelength. In \figref{chap5-fig2}(a), we consider a two-dimensional metasurface with diameter $d$, thickness $t$, refractive index $n=\sqrt{2}$, and numerical aperture $\textrm{NA} = 0.9$, and pose a fundamental question: what is the maximum possible efficiency any designable pattern could achieve? To answer this question, we formulate the design problem of \eqref{chap5-QCQP} with sparse-structured conservation laws, transform the problem to the QCQP of \eqref{chap5-QCQP-standard}, and relax the problem to the semidefinite program of \eqref{chap5-SDP}. 
We exploit the sparsity of the differential operators in the clique-decomposed semidefinite program of \eqref{chap5-SDP-chordal}, with the open-source software package \texttt{SparseCoLO}\cite{kim2011exploiting}, to dramatically reduce the size of the semidefinite constraint and enable bounds of large-scale metasurfaces.

\Figref{chap5-fig2}(b,c) show bounds on the field intensities as a function of metalens diameter. One can see that the maximum field intensity depends sensitively on the metasurface thickness, and is well below the ``unitary bound'' arising from imposing unitarity on the scattering matrix. A natural question is whether these bounds are achievable, and we include in \figref{chap5-fig2}(b) four data points (black markers): designs, taken from the first singular vectors of the semidefinite-program solutions, directly reach the bounds. Their corresponding structures are shown in the blue inset patterns. Crucially, the sparsity enables computation at very large scale sizes. Whereas the integral-equation formulation of the bounds in Chapters~\ref{chap:comp} and ~\ref{chap:multi} require computational times that scale with the fourth power of the metasurface diameter, the sparsity-based bounds in \figref{chap5-fig2} scale \emph{linearly}, enabling bounds for devices with diameters up to 100 free-space wavelengths in size.

\section{Discussion}
\label{sec:chap5-discussion}
In this chapter, we accelerate the computation of bounds by leveraging the sparsity inherited in Maxwell's equations.
Specifically, we construct a set of sparsity-structured conservation constraints directly from the wave equation, map their sparsity patterns to chordal graphs where a large semidefinite program can be distributively solved, and implement this acceleration technique for large-diameter metasurfaces.
The results illustrate how graph theory, specifically, decomposing a semidefinite program into small cliques as in \eqref{chap5-SDP-chordal}, can facilitate fast computation of bounds.
In the following, we explore additional ramifications of this decomposition technique: its implication for optimal photonic designs, its generalization, and its limitations.

Another consequence of small cliques is that the rank of the semidefinite solution is upper bounded by the maximal clique size of its chordal graph~\cite{madani2014convex}.  As the only difference between the semidefinite program in \eqref{chap5-SDP-chordal} and the QCQP in \eqref{chap5-QCQP} problem is the rank-one constraint, small clique sizes implies the original nonconvex design problem (the QCQP) is almost equivalent to the convex semidefinite problem, and there is little relaxation in between. This suggests our bounds are close to the true global optima when clique sizes are small, and one could retrieve an optimal binary design directly from the solution of the semidefinite program. This is a new way to solve for the optimal photonic design, which we discuss further in Conclusion of the thesis.

Besides the 2D metasurface considered in Section~\ref{sec:chap5-maximal}, the fast computational bound proposed in this chapter can apply to (multi-functional) thin-film designs and optimal quantum controls.
The former design the stacking of materials in a direction; the latter design the laser pulse shape (in the time domain) to maximize certain quantum transition rate. Both are inherently 1D problems with banded sparsity structures similar to the ones in \figref{chap5-fig1}(b,c). Their maximal clique sizes are determined by their short dimensions, i.e., the number of functionalities required for multilayered films and the number of quantum states. As long as these short dimensions are much smaller than the ``long'' dimensions (i.e., the spatial/temporal degree of freedom), the acceleration will be significant.

The algorithm, on the other hand, cannot accelerate bounds on thick metasurfaces, nor any designs with large ``short'' dimensions. 
A thick metasurface, for example, requires many points to discretize in the vertical dimension (the short dimension). Consequently, the sparsity pattern of \Figref{chap5-fig1}(d) carries large cliques, which are detrimental because: 1. each semidefinite constraint in \eqref{chap5-SDP-chordal} is more expensive to verify, and 2. large overlaps between the cliques cause repetitive evaluations. 
For example, consider the following sparsity pattern with only two zeros at the top-right and bottom-left corners:
\begin{equation}
    \XX = \renewcommand{\arraystretch}{0.5}
    \begin{bmatrix}
    * & * & * & * & 0 \\
    * & * & * & * & * \\
    * & * & * & * & * \\
    * & * & * & * & * \\
    0  & * & * & * & * \\
    \end{bmatrix}   
\end{equation}
Matrix $\XX$ has two cliques, $\XX_{1:4, 1:4}$ and $\XX_{2:5, 2:5}$, but obviously splitting it to the two cliques will not accelerate the computation in \eqref{chap5-SDP-chordal}. 
In practice, numerical experiments are usually required to determine when clique decomposition is no longer useful.
When clique sizes are too large (such as in thick metasurfaces), we turn to algorithms that can leverage more more general sparsity patterns in a semidefinite program (i.e., not limited to chordal graphs and their  decomposition), which we discuss in Conclusion of the thesis.

\chapter{Bounds on the coupling strengths of communication channels and their information capacities}
\label{chap:comm}

There are two types of geometries that shall be optimized for optimal electromagnetic communications: the geometry of the scatterer which we consider in previous chapters, and the geometries of the source and receiver domains that are usually represented by different configurations of antenna arrays.  While previous chapters concern the optimal scatterer shape for given source and receiver, they do not address the optimal source and receiver shapes that maximize information flow in wave communications. Such questions have been through intense numerical scrutinization~\cite{solimene2015singular, leone2018inverse, migliore2006role,leone2019comparison,leone2020inverse,pierri1998information,solimene2007number, solimene2013singular,Miller2019,solimene2013role, solimene2014inverse,poon2005degrees,leone2018application, leone2019radiation, suryadharma2017singular,slepian1961prolate, boyd1961confocal, frieden1971viii,bertero1982resolution} but there are no general answers.

In this chapter, we propose shape-independent bounds on source and receiver domains for key quantities in wave communications, including channel strength, number of communication channels, and information capacities. Our bounds suggest that domain sizes and configurations, and not domain shapes, are the key parameters to maximizing these quantities.
It also predicts a much slower, sub-exponential decay of channel strengths than previously thought possible.
In the following, we will introduce the theory of optimal communication channels, the backbone of our analysis, and a number of related studies before we dive into our shape-independent bounds in Section~\ref{sec:chap6-optimal}.

Optimal communication channels represent a unifying framework for optical physics~\cite{Miller2019, fox1961resonant, Miller2007, miller2007fundamental2, Miller2017a} with a wide range of applications in communication sciences~\cite{telatar1999capacity,goldsmith2003capacity, tse2005fundamentals,vellekoop2007focusing,martinsson2008communication,popoff2010measuring,miller2013establishing, miller2013self,li2014space, zhao2015capacity, miller2017better, annoni2017unscrambling,yilmaz2019transverse}. The Green's-function operator that connects a source volume to a receiver volume, while accounting for all possible background scattering, unambigously identifies the optimal channel profiles and their coupling strengths through its singular vectors and singular values, respectively~\cite{miller1998spatial, Miller2000, piestun2000electromagnetic,Miller2019}. Yet identifying the singular-value decomposition is generically an expensive and opaque computation, which has often limited previous work to highly symmetric domains, with little understanding of general properties or scaling laws~\cite{solimene2015singular, leone2018inverse, migliore2006role,leone2019comparison,leone2020inverse,pierri1998information,solimene2007number, solimene2013singular,Miller2019,solimene2013role, solimene2014inverse,poon2005degrees,leone2018application, leone2019radiation, suryadharma2017singular,slepian1961prolate, boyd1961confocal, frieden1971viii,bertero1982resolution} other than overall sum rules~\cite{miller1998spatial, Miller2000, piestun2000electromagnetic, Miller2019}.  

A classical example that is analytically solvable is the communication between two identical rectangular or circular apertures in the paraxial limit, where the optimal communicating channels are prolate spheroidal waves, exhibiting exponentially decaying coupling strengths~\cite{slepian1961prolate, boyd1961confocal, frieden1971viii,bertero1982resolution}. 
Similarly rapid decays of channel strengths are observed across different systems, ranging from simple geometries such as rectangular prisms~\cite{Miller2000, piestun2000electromagnetic}, strip objects~\cite{solimene2013role, solimene2014inverse,solimene2015singular}, and concentric circumferences~\cite{poon2005degrees,leone2018application, leone2018inverse}, to complex geometries involving conformal conic arcs~\cite{migliore2006role, leone2019radiation,leone2019comparison, leone2020inverse} and multiple rectilinear or spherical domains~\cite{pierri1998information, solimene2007number, solimene2013singular,suryadharma2017singular}. Many of these geometries are reexamined in a recent review~\cite{Miller2019}, where numerical observation of apparent exponential decay of coupling strengths past heuristic limits is hypothesized as being possibly universal.

In addition to the channel-strength decay rate, a related open question has been the maximum total number of channels that can be supported between two regions. Identifying bounds on the number of channels has been of interest since the birth of the field~\cite{miller1998spatial, Miller2000, piestun2000electromagnetic, Miller2019}, with partial success: channel sum rules imply upper bounds on the number of ``well-coupled'' channels simply by assumption of a minimum power-measurement threshold and equal division of power among all channels. Yet, as illustrated numerically, for example, in~\citeasnoun{Miller2019}, once we move beyond some simple geometries, such as parallel plane surfaces in a paraxial limit, even well-coupled channels can show substantially different power coupling strengths.

The information capacities of optimal communication channels have been investigated for domains of various shapes including spherical~\cite{poon2005degrees}, cubic~\cite{hanlen2006wireless, lee2016capacity}, and non-symmetrical geometries~\cite{tse2005fundamentals, migliore2006role}.
There are shape-dependent bounds to the information capacities for line-of-sight communications~\cite{chiurtu2000varying, chiurtu2001capacity,wallace2002intrinsic, jensen2004review,jensen2008capacity} and spherical communication domains~\cite{gustafsson2004spectral,glazunov2010physical}.  
A more general computational framework is proposed in Refs.~\cite{ehrenborg2017fundamental,ehrenborg2020physical,ehrenborg2021capacity} which bounds the total information capacity of a communicating domain by optimizing over freely varying currents in a bounding domain under fixed transmitted power, Ohmic loss, and radiation efficiency. In contrast, with a distinct analytical approach, we show in this chapter that the information capacity of any $N$-ports system is also tightly bounded under a fixed energy density in a bounding domain. 

Thus this chapter concerns with three fundamental questions: 
how rapidly must optimal-communication-channel strengths decay, what is the maximum number of usable communication channels, and does this imply bounds on maximum information capacities? We answer each of these questions below.

\section{Optimal communication channels}
\label{sec:chap6-optimal}

Our shape-independent bounds stem from combining two known theorems: (1) the coupling strengths of the optimal communication channels between two domains are the singular values of the corresponding Green's-function operator~\cite{miller1998spatial, Miller2000, piestun2000electromagnetic, Miller2019} and (2) those singular values increase monotonically with the size of the two domains~\cite{hanson2013operator}. The singular value decomposition of the dyadic Green's-function operator $\vG(\vx,\vx')$ from a source region $V_s$ to a receiver region $V_r$ is
\begin{equation}
    \vG(\vx,\vx') = \sum_{q=1}^{\infty} s_q \vu_q(\vx)\vv_q^*(\vx'),
  \label{eq:chap6-SVD_G}
\end{equation}
where $\{\vv_q(\vx)\}_{q=1}^{\infty}$ is a set of orthonormal vector-valued basis functions in the source region, $\{\vu_q(\vx)\}_{q=1}^{\infty}$ is a set of orthonormal vector-valued basis functions in the receiver region, and $\{s_q\}_{q=1}^{\infty}$ is the set of (non-negative) singular values. The tuples $\{(\vv_q,\vu_q,s_q)\}_{q=1}^{\infty}$ represent the optimal communication channels, with the fields radiated from sources $\vv_q(\vx)$ mapping uniquely to fields $\vu_q(\vx)$ in the receiver region with amplitudes $s_q$. The absolute square of the amplitude $|s_q|^2$ is referred to as the coupling strength or channel strength of channel $q$.

The key theorem that enables our shape-independent bounds, and is perhaps less well-known, is that all singular values of a Green's-function operator, as in Eq.~(\ref{eq:chap6-SVD_G}), may not decrease as the source and receiver domains are enlarged~\cite{hanson2013operator}. More precisely: if one domain encloses another, each singular value of the former cannot be smaller than the corresponding singular value of the latter.  We refer to this property of coupling strengths as ``domain monotonicity.'' It can be proven through a recursive argument. To simplify the notation and intuition, we encode the spatial variations of the source amplitude and polarization in a finite-dimensional vector $p$, define the Green's-function operator as a finite-dimensional matrix $\GG$, and use $\dagger$ to denote Hermitian conjugation. The operators $\GG^\dagger \GG$ and $\GG\GG^\dagger$ are necessarily Hermitian operators, which means their eigenvalues are real, and their eigenvalue/eigenfunction pairs can be found variationally via maximization and orthogonalization. (Another important characteristic of these operators is that they are positive semidefinite which implies their eigenvalues are non-negative, though this positive-semidefinite property is not a necessary condition for the variational procedure we discuss below.)
The square of the first singular value is obtained by maximizing the Rayleigh quotient of $\GG^\dagger\GG$: $|s_1|^2 = \max \frac{p^\dagger\GG^\dagger\GG p}{p^\dagger p}$. Clearly this may not decrease as the source domain enlarges, as maximization over a larger space of vectors cannot lead to a smaller optimal value. The second singular value similarly maximizes the Rayleigh quotient, now subject to orthogonality to the first singular vector. Because the first singular vector has changed with the domain, there is not a straightforward comparison to the optimization problem defining the second singular vector of the original domain. Yet the extra freedom given to the first singular vector ultimately only reduces the effect of the orthogonality constraint, such that the second singular value must also increase due to the domain enlargement. (A more precise version of this argument is given in Ref. \cite{molesky2020b}.) The same argument recursively applies to the rest of the singular values, and also for an enlarged receiver domain. Hence we have the key theoretical ingredients: optimal communication channels are defined by the singular-value decomposition of the Green's-function operator between source and receiver domains, and the singular values satisfy domain monotonicity on both domains.

\begin{figure}[t!]
\centering
\includegraphics[width=0.4\textwidth]{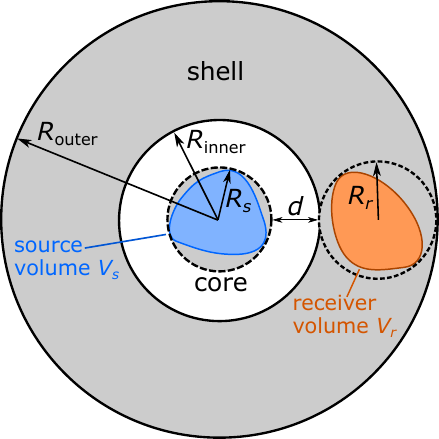}
\caption{The coupling strengths of the communication channels between a source volume $V_s$ and a receiver volume $V_r$ are upper bounded by their counterparts from the core--shell bounding volume (shaded in grey). We can also interchange the roles of the source and receiver volumes, and we obtain tighter bounds by using whichever is smaller as the ``inner'' volume in this figure.}
\label{fig:chap6-BoundingVolume}
\end{figure}

\section{Channel-strength bounds}
\label{sec:chap6-channel}

In this section, we derive shape-independent bounds on total channel strengths, relative channel strengths normalized against a sum rule, and their collective asymptotic decay rates in the many-channel limit. The domain-monotonicity principle discussed above immediately leads to bounds: the coupling strengths $|s_q|^2$ for arbitrary source and receiver domains are individually bounded above by the respective coupling strengths of any enclosing domains. We select an analytically tractable core--shell set of enclosing domains, depicted as the grey shaded region in Fig.~\ref{fig:chap6-BoundingVolume}, which yield the bounds:
\begin{equation}
    |s_{q}|^2 \leq |s_q^{\text{(core--shell)}}|^2,\quad  \text{for}\ q = 1, 2, ...
    \label{eq:chap6-sl_max}
\end{equation} 
In such core--shell configurations we can choose either the source or the receiver to be enclosed in the core; to find the tightest upper bounds, we take the minimum of both possible configurations. The core is a cylinder for 2D and a sphere for 3D. In the following sub-sections, we derive analytical expressions for the bounds in both dimensions.

\begin{figure*}[htb]
\centering
\includegraphics[width=1\textwidth]{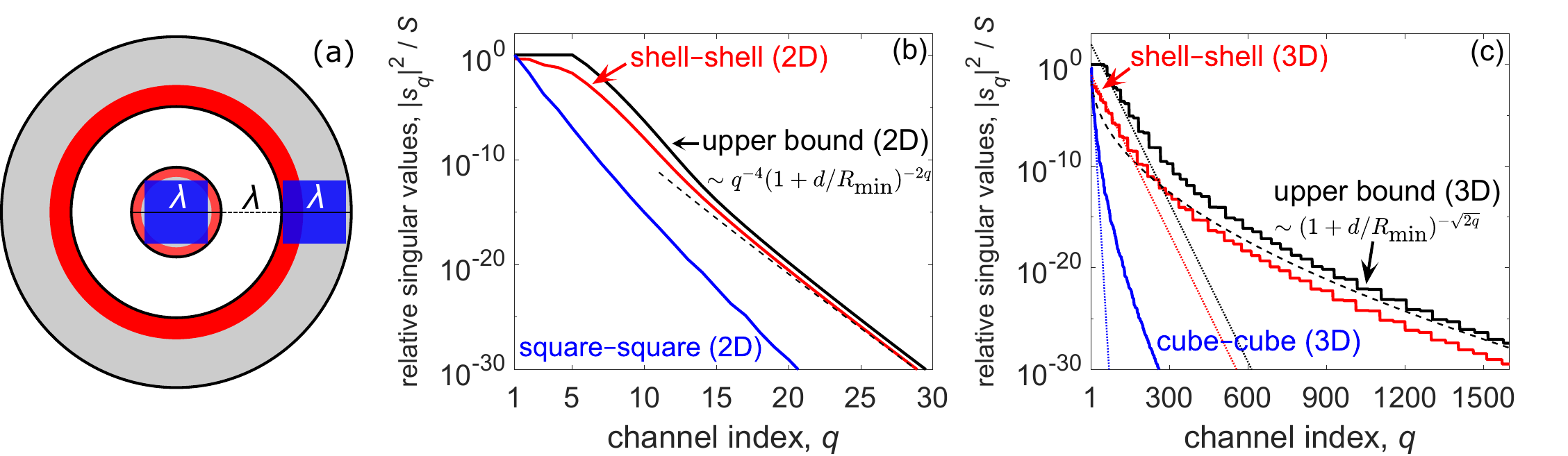}
\caption{Shape-independent upper bounds on the relative coupling strengths $|s_q|^2$ normalized against the total sum rule $S$ in two- and three-dimensional spaces. (a) A grey-shaded concentric core--shell bounding volume enclosing a square--square configuration of sources and receivers in the blue shaded region, as well as a shell--shell configuration in the red shaded region. (b) In two-dimensional (2D) space, the upper bound, calculated for the grey-shaded bounding volume in (a), decays exponentially as in the dashed black line. (c) In three-dimensional (3D) space, the additional azimuthal degeneracy leads to an optimal sub-exponential decay that is achieved by the shell--shell configuration. The sub-exponential decay rate (dashed black line) suggests many more communication channels at the large-channel limit than previously hypothesized exponential decays (dotted lines).}

\label{fig:chap6-Fig2}
\end{figure*}

\subsection{Channel-strength bounds in 2D}
\label{subsec:chap6-2D}
Consider communication in two dimensions between a source domain $V_s$ and a receiver domain $V_r$ as in Fig.~\ref{fig:chap6-BoundingVolume}.  The sources are bounded within a cylindrical core of radius $R_s$ and the receivers sit a minimum distance $d$ and maximum distance $d_{\rm max} = d+2R_r+2R_r$ from the sources. The bounding volumes, comprising an inner cylinder and an outer shell, are shaded in grey in Fig.~\ref{fig:chap6-BoundingVolume}. 
The singular values of the Green's function operator between the concentric cylinder--shell bounding volume can be identified by first performing a separation of variables for the two-dimensional scalar Green's function $G(\vx,\vx') = \frac{ik^2}{4} H^{(1)}_0(k|\vx-\vx'|)$ in polar coordinates~\cite{abramowitz1964handbook}:
\begin{equation}
    G(\vx,\vx') = \frac{ik^2}{4} \sum_{q=-\infty}^{\infty} H_q^{(1)}(k\rho)e^{-iq\phi}J_q(k\rho')e^{iq\phi'},
    \label{eq:chap6-G2D}
\end{equation}
where the functions $H_q^{(1)}(k\rho)e^{-iq\phi}$ and $J_q(k\rho)e^{-iq\phi}$ are the outgoing and regular cylindrical waves, with $H_q^{(1)}(x)$ and $J_q(x)$ being the Hankel function of the first kind and the Bessel function, respectively. Their polar coordinates $(\rho, \phi)$ and $(\rho', \phi')$ are defined on the bounding shell and bounding cylinder, respectively, relative to the center of the cylinder-shell bounding volume. The cylindrical waves $H_q^{(1)}(k\rho)e^{-iq\phi}$ and $J_q(k\rho)e^{-iq\phi}$ are the (unnormalized) left and right singular vectors of the Green's function operator in the cylinder-shell bounding volume.  (The cylindrical symmetry of the bounding volume ensures orthogonality.) 
There are two possible cylinder--shell bounding volumes: one centers around the source domain and one centers around the receiver domain. To tighten the upper bound, we choose the smaller of the two domains as the "inner" volume in Fig.~\ref{fig:chap6-BoundingVolume} because it leads to a smaller coupling strength $|s_{q}^{\text{(cylinder--shell)}}|^2$ which is the product of the norms of the unnormalized singular vectors, $H_q^{(1)}(k\rho)e^{-iq\phi}$ and $J_q(k\rho)e^{-iq\phi}$, in their respective bounding volumes:
\begin{equation}
    |s_{q}^{\text{(cylinder--shell)}}|^2 = \pi^2 k^2  \int_0^{R_{\text{min}}} |J_q(k\rho)|^2 \rho d\rho
    \int_{R_{\rm inner}}^{R_{\rm outer}} |H_q^{(1)}(k\rho)|^2 \rho d\rho.
    \label{eq:chap6-snm_Rmin2D}
\end{equation}
As the inner bounding cylinder is chosen to encompass the smaller domain, its radius is the smaller of the two radii, i.e., $R_{\text{min}} = {\rm min}\{R_s, R_r\}$. Similarly, one can show that the inner and outer radii of the outer bounding shell are $R_{\text{inner}} = d+R_{\rm min}$ and  $R_{\text{outer}} = d+2R_s+2R_r-R_{\text{min}}$, respectively.
The singular values in Eq.~(\ref{eq:chap6-snm_Rmin2D}) are dimensionless quantities because our Green's function, of Eq.~(\ref{eq:chap6-G2D}), differs from the conventional definition \cite{jackson1999classical, Miller2019} by a factor of $k^2$ to be inversely proportional to volume.

The number of non-trivial communication channels is determined by the number of channels whose \textit{relative} channel strengths are above a certain measurement threshold. The relative channel strengths can be normalized either by a total sum rule $S=\sum_{q=-\infty}^\infty|s_q|^2$ or by the largest channel strength~\cite{Miller2019}. Lower bounds on the sum rule can be analytically derived based on the monotonic decay of wave energy in free space, thus leading to bounds on the total number of channels above a certain sum-rule energy fraction. The sum rule $S$ is a double integral of the absolute square of the two-dimensional Green's function over both the source and receiver domains~\cite{Miller2000, Miller2019}
\begin{equation}
    S = \int_{S_s}\int_{S_r} |G(\vx,\vx')|^2 d\vx d\vx' 
      \geq k^4S_sS_r|H_0^{(1)}(kd_{\rm max})|^2/16,
      \label{eq:chap6-sum2D}
\end{equation}
where we further lower bound $S$ by the fact that the magnitude of the Green's function takes its minimal value at the most separated points between the two domains, which is at a distance $d_{\rm max} = d + 2R_r + 2R_s$ for the cylinder--shell bounding volume illustrated in Fig.~\ref{fig:chap6-BoundingVolume}. The variables $S_s$ and $S_r$ in Eq.~(\ref{eq:chap6-sum2D}) denote the total area of the source and receiver domains. Combining Eq.~(\ref{eq:chap6-snm_Rmin2D}) and Eq.~(\ref{eq:chap6-sum2D}), we derive 
\begin{align}
    \frac{|s_{q}|^2}{S} \leq \frac{16|s_{q}^{\text{(cylinder--shell)}}|^2 }{k^4S_sS_r|H_0^{(1)}(kd_{\rm max})|^2},
    \label{eq:chap6-sr_2D}
\end{align}
which is a shape-independent bound on the relative channel strength between domains in two-dimensional space. In the many-channel limit, the bound in Eq.~(\ref{eq:chap6-sr_2D}) simplifies:
\begin{equation}
    \frac{|s_{q}|^2}{S} \leq \frac{R_{\text{min}}^4}{q^4S_sS_r|H_0^{(1)}(kd_{\rm max})|^2(1+d/R_{\text{min}})^{2(q-1)}},\quad \text{as}\ q \rightarrow \infty.
    \label{eq:chap6-decay2D}
\end{equation}
The presence of the exponential factor of $2(q-1)$ indicates that channel strengths in two dimensions must decay at least exponentially fast with channel number, in agreement with the previously hypothesized exponential decay of channel strengths. The exponential decay rate depends only on the separation distance $d$ relative to the smaller radius $R_{\text{min}}$ between the two communication domains.

The upper bound in Eq. (\ref{eq:chap6-sum2D}) and its optimal exponential decay in Eq. (\ref{eq:chap6-decay2D}) applies to any two domains that can be separated by a cylindrical surface. The bound is achieved by concentric communicating domains that fill the bounding volume, while the optimal decay rate can also be achieved with concentric sub-domains. To illustrate the latter point, in Fig. \ref{fig:chap6-Fig2}(a), we arrange a fixed number of sources and receivers in two different configurations inside a bounding volume. The first configuration (blue shaded region) consists of two squares of sources and receivers with the side lengths of $\lambda/\sqrt{2}$. The second configuration (red shaded region) consists of concentric shell-like communicating domains with the same source and receiver areas. Both configurations are enclosed in a concentric cylinder--shell bounding volume of $2R_s=2R_r=d=\lambda$. Inside this bounding volume, the maximal relative coupling strength is given by the solid black line in Fig. \ref{fig:chap6-Fig2}(b), calculated using Eq. (\ref{eq:chap6-sr_2D}).   We observe that, while the square--square configuration (solid blue line) falls far short of the bound, arranging the same number of sources and receivers to cover a wider solid angle in a shell--shell configuration  (solid red line) enables close approach to the upper bound. (The black-line upper bound is clamped to 1; no channel can have strength larger than 1. The looseness of Eq. (\ref{eq:chap6-decay2D}) arises from the dramatic mismatch of the source--receiver volumes to the bounding volumes.)  Moreover, the shell--shell configuration achieves the optimal exponential decay predicted in Eq. (\ref{eq:chap6-decay2D}). This result corroborates previous works~\cite{slepian1961prolate, boyd1961confocal,frieden1971viii,bertero1982resolution,solimene2015singular,leone2018inverse, migliore2006role,leone2019comparison, leone2020inverse,pierri1998information, solimene2007number, solimene2013singular,Miller2019} that predicted exponential decay in wide-ranging scenarios, and hypothesized that exponential decay may be a universal rule. As we show below, however, the three-dimensional behavior is quite different.


\subsection{Channel-strength bounds in 3D}
\label{subsec:chap6-3D}
The derivation of shape-independent bounds on channel strengths in three dimensions is similar to the derivation in two dimensions, with the cylinders replaced by spheres. For this 3D case, we now use a full vector formulation of the problem, as appropriate for a full electromagnetic solution. So, we move to dyadic Green's functions, and we start by expanding the dyadic Green's function as a summation of outer products, now of spherical vector waves: 
\begin{equation}
    \vG(\vx, \vx') = ik^3 \sum_{n=0}^\infty \sum_{m=-n}^n \sum_{j=1,2} \vv_{\text{out}, nmj}(\vx)\vv_{\text{reg}, nmj}^*(\vx'),
    \label{eq:chap6-Gr3D}
\end{equation}
where $\vv_{\text{out}, nmj}$ and $\vv_{\text{reg}, nmj}(\vx')$ are the outgoing and regular spherical vector waves~\cite{tsang2004scattering} defined on the bounding shell and bounding sphere\hl{,} respectively.
The vectors $\vx$ and $\vx'$ are spherical coordinates defined with respect to the center of the concentric bounding volume.
The regular (outgoing) spherical vector waves are formed by combining the angular dependency of vector spherical harmonics with the radial dependency of spherical Bessel (Hankel) functions~\cite{tsang2004scattering}.
Explicit expressions of the vector spherical waves, $\vv_{\text{out}, nmj}$ and $\vv_{\text{reg}, nmj}(\vx)$, and the wave equation we use to define the Green's function are given in Appendix~\ref{sec:appenI-sec1}.
The indices $n$ and $m$ index the underlying spherical harmonics, and $j=1,2$ denotes the two possible polarizations of a transverse vector field.
The orthogonality of the spherical waves in a spherically symmetric domain allows us to identify $\vv_{\text{out}, nmj}$ and $\vv_{\text{reg}, nmj}(\vx)$ as the (unnormalized) left and right singular vectors of the Green's function operator defined on the three-dimensional sphere-shell bounding volumes. 
The corresponding singular values are the products between the norms of functions $\vv_{\text{out}, nmj}$ and $\vv_{\text{reg}, nmj}(\vx)$ in their respective volumes:
\begin{equation}
    |s_{nmj}^{\text{(sphere--shell)}}|^2 = k^6 \int_{V_{\text{shell}}} \left|\vv_{\text{out}, nmj}(\vx)\right|^2 d\vx  \int_{V_{\text{sphere}}} \left|\vv_{\text{reg}, nmj}(\vx)\right|^2 d\vx,
    \label{eq:chap6-snm}
\end{equation}
where $V_{\text{shell}}$ and $V_{\text{sphere}}$ represent the volumes of the bounding shell and bounding sphere.
Explicit expressions of the singular values $|s_{nmj}^{\text{(sphere--shell)}}|^2$ can be found in Appendix~\ref{sec:appenI-sec1}. According to the domain-monotonicity property in Eq.~(\ref{eq:chap6-sl_max}), the $q$-th largest number from the set of all possible $|s_{nmj}^{\text{(sphere--shell)}}|^2$ upper-bounds the $q$-th largest channel strength of any configuration of sources and receivers in the sphere--shell bounding volume. 

Again, the number of non-trivial communication channels is determined by normalizing the channel strengths to the total sum rule. The sum rule is now lower bounded by (cf. Appendix~\ref{sec:appenI-sec3}):
\begin{align}
    S = \int_{V_s}\int_{V_r} ||\vG(\vx,\vx')||_F^2 d\vx d\vx' \geq \frac{k^4V_sV_r}{8\pi^2d_{\rm max}^2} + \mathcal{O}\left(\left(kd_{\rm max}\right)^{-4}\right).
      \label{eq:chap6-sum}
\end{align}
For conciseness, we assume the furthest separated points are in the far field, i.e. $kd_{\rm max} \gg 1$, so  that only the leading term in Eq.~(\ref{eq:chap6-sum}) remains. This can be easily generalized by explicitly including two other higher-order terms, leading to a somewhat more complicated expression but the same asymptotic properties.

By combining the upper bound of channel strengths in Eq.~(\ref{eq:chap6-sl_max}) and the lower bound of the sum rule in Eq.~(\ref{eq:chap6-sum}), we derive a key result for 3D communication domains, a shape-independent upper bound on their relative channel strengths normalized against the total sum rule:
\begin{equation}
    \frac{|s_{nmj}|^2}{S} \leq \frac{8\pi^2d_{\rm max}^2}{k^4V_sV_r} |s_{nmj}^{\text{(sphere--shell)}}|^2,
    \label{eq:chap6-sr}
\end{equation}
where the singular value of the sphere--shell bounding volume, $|s_{nmj}^{\text{(sphere--shell)}}|^2$, is identified in Eq.~(\ref{eq:chap6-snm}), and whose explicit expression can be found in Appendix~\ref{sec:appenI-sec1}.
One immediate prediction of the upper bound in Eq.~(\ref{eq:chap6-sr}) is an optimal sub-exponential decay rate of the channel strengths between two 3D domains, which we now derive. The total number of channels that has $n$-index less or equal to $n$ is $q=2(n+1)^2$. We use this total channel index $q$ as our new index for channel strengths to meaningfully describe their decay rate. When the total number  $q\rightarrow\infty$, Eq.~(\ref{eq:chap6-sr}) can be simplified to (cf. Appendix~\ref{sec:appenI-sec4}):
\begin{equation}
    \frac{|s_{q}|^2}{S} \leq \frac{2\pi^2d_{\rm max}^2}{k^4V_sV_r(1+d/R_{\text{min}})^{\sqrt{2q}+1}}, \quad \text{as}\ q \rightarrow \infty,
    \label{eq:chap6-decay3D}
\end{equation}
where the parameter $R_{\text{min}} = \min\{R_s, R_r\}$ denotes the radius of the smaller domain.
Equation~(\ref{eq:chap6-decay3D}) shows that, regardless of the domain shape, channel strengths $|s_q|^2$ in three-dimensional space have to decay at least as fast as $a^{-\sqrt{q}}$, where $a$ is a bounding-domain-dependent numerical constant ($a=(1+d/R_{\rm min})^{\sqrt{2}}$), and the key new feature is the square root dependence on $q$ in the exponent. 
Such a decay is \emph{sub-exponential}, as its logarithm decays only with the square root of the channel number rather than the (much faster) linear reductions characteristic of exponential decay. 

\begin{figure}[t!]
\centering
\includegraphics[width=0.6\textwidth]{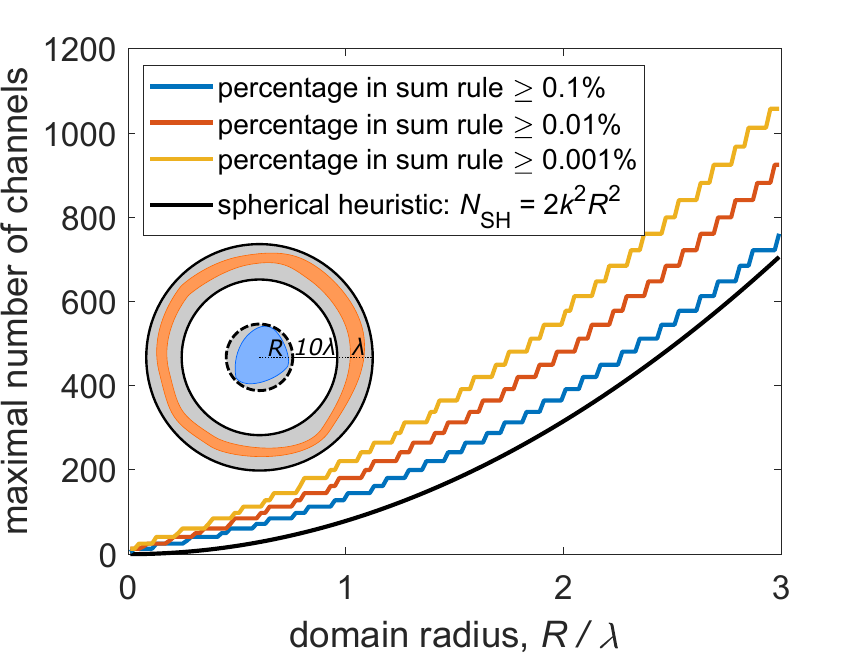}
\caption{ The maximal number of non-trivial communication channels for a domain of maximal radius $R$ and certain measurement thresholds set by their percentage in the total sum rule. The other communication domain is in a bounding shell shown in the inset. 
The quadratic dependence of the bound regards to the domain radius $R$ can be conveniently modelled by a spherical heuristic number, $N_{\text{SH}} = 2k^2R^2$. }
\label{fig:chap6-NoC}
\end{figure}

Figure~\ref{fig:chap6-Fig2}(c) compares the coupling-strengths bound in 3D, with a clearly sub-exponential decay rate, to the coupling strengths of two configurations of sources and receivers (shell--shell and cube--cube) in a sphere--shell bounding volume. Both configurations possess a volume $\lambda^3 / (3\sqrt{3})$ of sources and receivers and follow the same layout as in Fig.~\ref{fig:chap6-Fig2}(a).
Similar to the 2D case, we observe that the shell--shell configuration closely follows the bound while the cube--cube configuration falls short.
Interestingly, both the cube-cube and shell-shell configurations and the upper bound first enter a phase of approximately exponential decay (dotted lines in Fig.~\ref{fig:chap6-Fig2}(b), a phenomenon also observed in Ref.~\cite{Miller2019}) before they exhibit different sub-exponential decays on a larger scale. 
By ``sub-exponential'', we mean that the fall off in the channel strengths is not as fast as exponential; high-index channels have somewhat stronger coupling strength than an exponential fall-off would predict, and, 
in the many-channel limit, the asymptotic sub-exponential decay predicted in Eq.~(\ref{eq:chap6-decay3D}) bounds all geometries and also puts forth the concentric shell-shell configurations as the optimal candidate for achieving the slowest sub-exponential decay.

The sub-exponential decay of the channel-strength bound in 3D is in stark contrast with its exponentially decaying counterpart in 2D.
This point is accentuated by contrasting Fig.~\ref{fig:chap6-Fig2}(c) with Fig.~\ref{fig:chap6-Fig2}(b), where their asymptotic decay rates, shown as the black dashed lines, are fundamentally different. 
This difference originates from the additional azimuthal degeneracy of communication channels in 3D. Such degeneracy manifests through the staircase behavior of the upper bound in Fig.~\ref{fig:chap6-Fig2}(c). It allows one to potentially establish many more useful orthogonal channels in 3D: the bound suggests approximately 145 channels for 3D domains above a threshold of $10^{-4}$ in Fig.~\ref{fig:chap6-Fig2}(c), as compared to only 8 channels in the 2D case above the same threshold.
The difference in the decay rate of upper bounds in two- and three-dimensional spaces underscores the role of dimensionality in channel counting.

\subsection{Bounds on the number of non-trivial channels}
\label{subsec:chap6-channel-number}

The number of non-trivial communication channels is often regarded as the number of ``spatial degrees of freedom'' for communicating between two regions, an idea that generalizes the concept of diffraction limits~\cite{Miller2019} and dictates fundamental response in many wave systems~\cite{Miller2007,miller2015shape, venkataram2020fundamental}. A communication channel is considered non-trivial if its coupling strength is above a certain percentage in the total sum rule~\cite{Miller2019}; the bounds of Eqs.~(\ref{eq:chap6-sr_2D}) and (\ref{eq:chap6-sr}) on relative coupling strengths therefore directly lead to bounds on the number of communication channels.

Figure~\ref{fig:chap6-NoC} shows the maximal number of channels available for any source domain within a three-dimensional sphere of radius $R$, computed from Eq.~(\ref{eq:chap6-sr}). The receiver domain is a shell ten wavelengths away, with a thickness of one wavelength, as shown in the inset of Fig.~\ref{fig:chap6-NoC}. (The source and receiver domains can be transposed.) We also assume both domains occupy at least half of their respective bounding volumes. The bounds are plotted as a function of the maximal domain radius $R$ for a number of measurement thresholds. The bounds are not overly sensitive to the measurement threshold: a hundredfold increase in the sensitivity, as occurs going from the blue line to the yellow line, does not even double the number of available channels. On the other hand, the bounds increase approximately quadratically with the maximal domain radius $R$, suggesting enlarging domain size is the key to gaining more useful channels.

The quadratic increase of the bound with respect to the domain radius $R$ can be understood as arising from the increasing surface area of two sufficiently separated communication domains. At first, one might expect the mode number to increase with the \emph{volume} of the domains, but the waves in the volumes are determined by the waves at the surfaces (by the surface equivalence principle~\cite{Jin2011}), and restrictions on the number of unique wave patterns at the surface will naturally constrain the number of independent volume functions as well. As the domain size increases, we can use the notion of a ``spherical heuristic number,'' denoted $N_{\rm SH}$, to estimate the number of communication channels:
    \begin{equation}
        N_{\rm SH} = 2k^2R^2. \label{eq:chap6-SH}
    \end{equation}
Spherical heuristic numbers were proposed in \citeasnoun{Miller2019}, where the expression $16\pi R^2/\lambda^2$ was suggested. Here we modify the expression, instead assuming one unique spatial mode per $\lambda^2 / \pi$ area on the surface of the spherical bounding domain (instead of $\lambda^2/4$ as previously suggested~\cite{Miller2019}), multiplied by two polarizations, resulting in the expression of Eq.~(\ref{eq:chap6-SH}). The $\lambda^2/\pi$ area expression comes from treating each surface patch on the source and receiver domains as interacting in the paraxial limit--certainly not exactly true, but sufficient to gain intuition. Fig. \ref{fig:chap6-NoC} shows quantitative agreement between the spherical heuristic number and the rigorously calculated bound under a $0.1\%$ threshold on the sum rule, explaining the approximately quadratic increase of the number of channels as a function of domain radius.

The bounds in Fig.~\ref{fig:chap6-NoC} weakly depend on the sum-rule percentage threshold because of the rapid decay of channel strength at large-order channels. Though not shown in this graph, the bound barely depends on the depth of the receivers and their distance from the source (unless in the extreme near-field limit when the separation distance is much less than a wavelength). All these imply that the group of bounds shown in Fig.~\ref{fig:chap6-NoC} represent the intrinsic number of channels one can couple out of any source domain of a given size to the far field.

\section{Bounds on the information capacities of communication channels}
\label{sec:chap6-bounds}

Information capacity, defined as the maximal rate at which the information can be reliably transmitted between two communicating domains, is a notion that has been central to the development of modern communication systems~\cite{shannon1948mathematical,tse2005fundamentals}.
In this section, we show how our coupling-strengths bounds can help one determine the maximal information capacity of communication channels in three-dimensional space.
A key feature of our approach is that it tightly bounds the total information capacity of any given number of channels, which is highly relevant to modern MIMO systems that usually have access to a finite number of antennas.

The information capacity $C$ of $N$ optimal communication channels (per unit time and unit bandwidth) is the sum of the capacity of each channel, each of which logarithmically depends on its input power $P_q$, coupling strength $|s_q|^2$, and noise power $P_{\rm noise}$ \cite{tse2005fundamentals}:
\begin{equation}
    C = \sum_{i=q}^{N}\log_2 \bigg(1+\frac{P_q|s_q|^2}{P_{\rm noise}}\bigg)\quad \text{bits/s/Hz},
    \label{eq:chap6-ShannonMulti}
\end{equation}
where we assume an additive white Gaussian noise background with the same noise power $P_{\rm noise}$ for each channel.

A larger domain size is always favorable to increase the information capacity of the first $n$ optimal communication channels. This is because the capacity $C$ in Eq.~(\ref{eq:chap6-ShannonMulti}) increases monotonically with coupling strength $|s_q|^2$, which in turn increases monotonically with the domain size. Therefore, the capacity of the sphere--shell bounding volume serves as an upper bound for the capacity of all possible sub-domains within:
 \begin{equation}
    C \leq \sum_{q=1}^{N}\log_2 \bigg(1+\frac{P_q|s_q^{(\text{sphere--shell})}|^2}{P_{\rm noise}}\bigg)\quad \text{bits/s/Hz},
    \label{eq:chap6-CapacityBound}
\end{equation}
where the coupling strength $|s_q^{(\text{sphere--shell})}|^2$ of the sphere--shell bounding volume is given in Eq.~(\ref{eq:chap6-snm}). One can solve for the optimal allocation of powers $P_q$ for a fixed total power input $\sum_{q=1}^{N} P_q = P$, by the ``water-filling'' algorithm~\cite{tse2005fundamentals}, with the semi-analytical form 
$P_q = \max\{0,\mu - P_{\rm noise}/|s_q^{(\text{sphere--shell})}|^2\}$, where $\mu$ is the numerical constant for which $\sum_{q=1}^{N}P_q = P$. The signal-to-noise ratio (SNR), defined as the ratio between the total power and noise power, i.e. $\SNR = P / P_{\rm noise}$, is the key external parameter that affects the optimal strategy of the power allocation.

\begin{figure}[t!]
\centering
\includegraphics[width=0.5\textwidth]{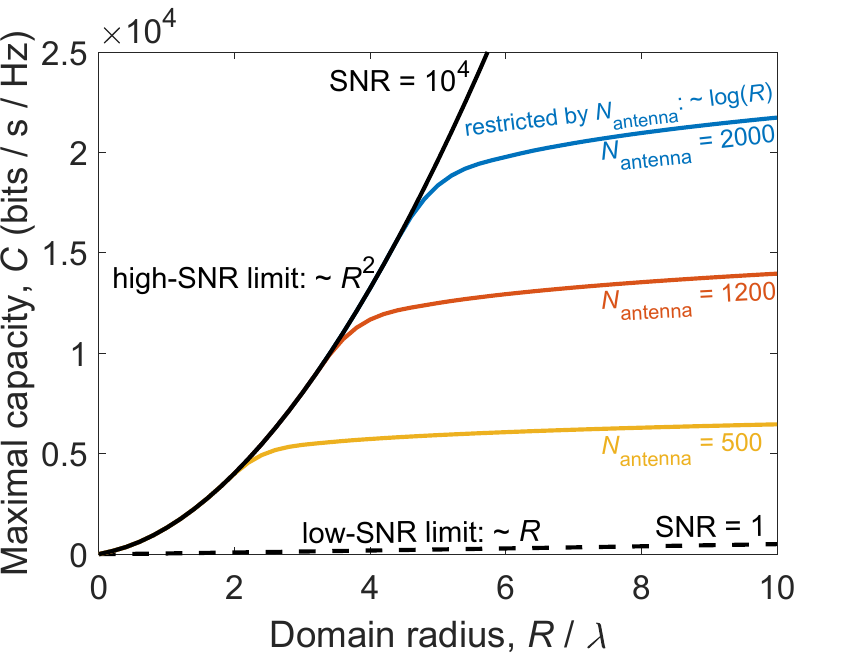}
\caption{\label{fig:chap6-CapacityBound} 
Maximal information capacity $C$ between any two domains that fit within the radius-$R$ sphere and wavelength-thickness bounding shell of Fig.~\ref{fig:chap6-NoC}. In the high SNR limit, the bound increases quadratically with domain size $R$ (solid black), whereas in the low SNR limit, the bound increases only linearly (dashed black). When the number of available channels is restricted  by the number of antennas, $N_{\rm antenna}$, the channel-capacity bounds tail off and increase only logarithmically with domain size (blue, orange, yellow).} 
\end{figure}

Figure~\ref{fig:chap6-CapacityBound} shows the capacity bound for communication between arbitrary domains contained in the sphere--shell bounding volumes in two limits: high SNR (solid black), and low SNR (dashed black). The size dependencies of the capacity bounds are quite different in the two limits. When SNR is very small, the logarithms approximately become linear functions of the power, in which case the optimal allocation puts all of the power in the single channel with the highest coupling strength~\cite{tse2005fundamentals}. The maximum coupling strength scales linearly with the radius $R$:  $\max\left\{|s_{q}^{\text{(sphere--shell)}}|^2\right\} = k^2R_rR$, provided that the radius $R$ of the bounding sphere is much larger than a wavelength and the bounding shell is in the far field of the bounding sphere (cf. Appendix~\ref{sec:appenI-sec2}). Then we have
\begin{equation}
    C \leq \SNR\cdot \log_2(e)  k^2 R_r R,\quad \text{for}\ \ \SNR \rightarrow 0.
\end{equation}
By contrast, in the high-SNR limit, the optimal allocation of power equally divides amongst all channels with nonzero channel strengths~~\cite{tse2005fundamentals}. The information capacity in this case scales with the number of such channels, which, as we established in the Sec.~\ref{subsec:chap6-channel-number}, depends quadratically with the domain radius $R$ (modeled by the spherical heuristic number $N_{\rm SH} = 2k^2R^2$). Hence, the capacity bound increases quadratically with $R$ in the high-SNR limit:
\begin{equation}
    C \leq 2\log_2({\rm SNR})k^2R^2,\quad \text{for}\ \ \SNR \gg 0.
\end{equation}
In many scenarios, the number of communication channels may be restricted well below our electromagnetic limit; one common example may be a MIMO system with antennas spaced more than half a wavelength apart. When the number of communication channels is restricted by the number of antennas, $N_{\text{antenna}}$, the growth in the large-domain limit cannot remain quadratic or even linear; instead, the capacity bound will grow logarithmically at best. This is because for a fixed number of channels, the capacity of each channel increases logarithmically with channel strength, which in turn increases at most linearly with $R$:
\begin{equation}
    C \leq N_{\rm antenna}\log_2\left(\SNR\cdot\frac{ k^2R_rR}{N_{\rm antenna}}\right),\quad \text{for}\ \ \SNR \gg 0.
\end{equation}
The  logarithmic dependence is confirmed by the computations of the blue, orange, and yellow lines in Fig.~\ref{fig:chap6-CapacityBound}, with each having the same SNR as the solid black line, but decreasing $N_{\rm antenna}$. The quadratic increase at the outset of each curve saturates almost exactly at the domain size where the number of electromagnetic channels ($N_{\rm SH} = 2k^2 R^2$) equals $N_{\rm antenna}$. Thus, despite the abundant number of electromagnetic channels in a large domain, antenna restrictions can impose significant constraints on the total information capacity.

\section{Discussion}
\label{sec:chap6-discussion}
The key finding in this chapter is a shape-independent bound on coupling strengths that we derive based on the domain-monotonicity property of the Green's function operator. 
This upper bound leads to two important discoveries. First, the sub-exponential decay in Eq.~(\ref{eq:chap6-decay3D}) identifies the slowest possible decay rate between any two domains in free space, and implies that three-dimensional domains have dramatically more channels available than their two-dimensional counterparts. Second, the ensuing bounds and scaling laws on the maximal number of usable communication channels and their maximal $N$-channel information capacity represent the ultimate limit that no domains can surpass. 
In this section, we briefly touch on other possible extensions of these results.

The bounding volume for the source and receiver domains can be any shape and size. We choose the concentric bounding volume in this article because of its analyticity and generality: its singular values are analytically tractable and the resulting bound is general enough to apply to any two domains that can be separated by a spherical surface.  In practice, if the sources and receivers are constricted to a domain smaller than the concentric bounding volume, one can sacrifice the analyticity by numerically computing the singular values of the largest possible domain for a tighter bound. Another analytical though less general bounding volume arises when the sources and receivers are known to be in the paraxial limit. Then, one can form the bounding volume as two rectangular cuboids whose singular values are known analytically in the paraxial limit~\cite{Miller2000}.  While we mainly focus on the concentric sphere--shell bounding volume in this work, future studies of alternative bounding volumes may reveal the dependence of the bound on the solid angles between the sources and receivers that otherwise cannot be captured by a concentric bounding volume.


Near-field information and power transfer have shown great promise in both wireless communication and fundamental science because of the abundant well-coupled channels in the form of electromagnetic evanescent waves~\cite{courjon1994near,want2011near, kim2015radiative,jawad2017opportunities}. This abundance emerges in our shape-independent bound in Eq.~(\ref{eq:chap6-decay3D}) where the optimal sub-exponential decay $\left(1 + d/R_{\text{min}}\right)^{-\sqrt{2q}}$ tends to unity when the separation distance $d$ goes to zero. Meanwhile, the maximal number of non-trivial communication channels diverges.
While this article mainly focuses on the application of our shape-independent bounds in the far field, it is also interesting to see how this formalism can regulate the maximal information and power transfer for different geometries in the near field.

The $n$-channel capacity bound proposed in this article may have ramifications on the optimal performance of antenna selections in massive multiple-input and multiple-output (MIMO) systems~\cite{molisch2004mimo,molisch2005capacity,gao2017massive,asaad2018massive}.
The technique of antenna selections mitigates the cost and complexity of MIMO systems by judiciously selecting only a fixed-size subset of antennas while maintaining a large total information capacity.  
How large the total information capacity can be among all the possible subsets is a question that falls under the umbrella of our $N$-channel capacity bound, which suggests the possibility to bound the capacity of any $N$-antenna subset by the capacity of the first $N$ optimal channels of the total antenna arrays. 

The presence of external scatterers can strongly affect the scattering amplitude of electromagnetic fields and the information content it carries.
This chapter studies the maximal information transfer in free space. Its technique can be extended to any background scatterer (by modifying the background Green's function) but the shape of the scatterer has to be specified.
For an arbitrary scatterer, there is still a need to understand the maximal information throughput they can support. 
For example, to what degree could an external scatterer alter the sub-exponential decay rate predicted in this chapter? What is the maximal number of non-trivial channels an external scatterer can help to establish and what are the maximal information capacities of those channels? 
Though a few bounds have been identified either in certain physical scenarios~\cite{Miller2007,miller2007fundamental2,molesky2022t,shim2021fundamental} or through heuristic arguments~\cite{miller2022optics, li2022thickness}, those are still open questions that await for general answers. 
In Appendix~\ref{appen:info}, I propose a rigourous framework to answer those questions by iteratively appling the computational bound in Chapter~\ref{chap:comp} to bound the channel strength of each communication channel and find that, in the case of a lossy metal, the presence of a scatterer may alter the exact magnitudes of the coupling strengths through resonance enhancements but cannot change its overall decay rate. This result bounds the number of communication channels through a metal scatterer strictly by its lossiness.
Among various design techniques in search of better scatterer structures or antenna arrays, shape-independent bounds continue to offer a new lens to analyze the fundamental limits of information and power transfer in both fundamental physics and communication science.

\chapter{Conclusion}
\label{chap:conclu}

This thesis builds a general theoretical framework that derives bounds on maximal electromagnetic responses constrained by  fundamental principles of physics. 
We start from the global power conservation (i.e., the optical theorem in Chapter~\ref{chap:analy}), move to local power conservation (Chapter~\ref{chap:comp}), then to field-correlation constraints between different scattering scenarios (Chapter~\ref{chap:multi}), each unlocking more general bounds.
In addition, these physical constraints are reinterpreted as binary-material constraints (Section~\ref{subsec:chap3-binary}), which motivates their sparse reformulation in Chapter~\ref{chap:sparse}, paving ways for fast computations of bounds.
These bounds, concerning the optimal scatterer geometry, are complemented by the channel-strengths bounds in Chapter~\ref{chap:comm} that address the optimal source and receiver geometries. 
We implement this framework for various applications throughout the thesis, ranging from single-frequency perfect absorption to broadband farfield extinction, from multi-frequency sensing to large-scale focusing, from optimal field illumination to optimal wave communication.
Together, these results delineate a general understanding of fundamental limits in nanophotonics.
In the following, we discuss four future research directions: applying the framework to other objectives, extending and augmenting the framework, accelerating the computational bounds, and designing near-optimal photonic structures.

Our framework can apply to many emerging problems in nanophotonics. Besides the ones already demonstrated in this thesis, our framework can also bound the maximal optical forces and torques~\cite{Mazilu2011,rahimzadegan2017fundamental,Lee2017,Liu2019,Horodynski2019}, the maximal field concentration~\cite{Moskovits1985,Nie1997,Stiles2008,michon2019limits}, the maximal cross density of states~\cite{caze_pierrat_carminati_2013, gonzaga-galeana_zurita-sanchez_2013, gonzalez-tudela_martin-cano_2011,lassalle_long-lifetime_2020}, the maximal response of 2D materials~\cite{Novoselov2005,Geim2007,Koppens2011,Basov2016,Low2016} (further discussed in Appendix~\ref{appen:2D}), the maximal response of multi-functional devices such as polychromatic grating coupler~\cite{piggott2015inverse, piggott2017fabrication} and wide field-of-view metalens~\cite{luo2022recent}, and many more. 
Though not obvious, our framework can also apply to the inverse scattering problems~\cite{tarantola2005inverse, chavent2010nonlinear, kern2016numerical} to identify the minimal number of measurements necessary to reconstruct a scatterer of given size, complementing the existing  approaches based on local optimizations~\cite{tarantola2005inverse, chavent2010nonlinear, kern2016numerical}.
In general, our framework applies to any quantity in electromagnetic scattering, as long as it can be written as a linear or quadratic function of the induced polarization fields.  

Future improvements of our framework involve two fronts: practicability and generality.
On the practical side, we can appeal to designers by augmenting the bounds with constraints that impose structural integrity~\cite{augenstein2020inverse, Chung2020}. In cases where device weight is a concern~\cite{miller_fundamental_2016, jin2020inverse}, we can restrict the amount of material utilized in a design and compute its maximal \textit{per-volume} response (detailed in Appendix~\ref{appen:volume}).
On the other front, we can generalize our formalism from linear electromagnetism to more general scattering settings, such as (1) nonlinear optics~\cite{bravo-abad_modeling_2007, smirnova_multipolar_2016, lin_cavity-enhanced_2016, hughes_adjoint_2018, boyd2020nonlinear, gigli_quasinormal-mode_2020,  koshelev_subwavelength_2020} by establishing power conservation between different harmonics (detailed in Appendix~\ref{appen:nonlinear}), (2) gain mediums~\cite{cao2005review, wiersma2008physics, peng2014parity, ashida2020non} by substituting material absorption with external gain, (3) general broadband responses~\cite{ji2017engineering,wang2018broadband, chen2018broadband, shrestha2018broadband} by marrying with the recent $\TT$-matrix bound~\cite{zhang2022all}, (4) eigenvalue problems~\cite{lalanne_light_2018, lu_inverse_2010, raman_upper_2013-1,schab_lower_2018, choi_self-similar_2017, hu_experimental_2018, zhao_minimum_2020, men_bandgap_2010, minkov_inverse_2020} by considering scattering at a complex frequency (detailed in Appendix~\ref{appen:mode}), (5) scattering-assisted information transfer (detailed in Appendix~\ref{appen:info}), and, rather remarkably, (6) other linear differential equations in physics, such as phonon scattering~\cite{venkataram_channel-based_2020, venkataram_mechanical_2020} and quantum control~\cite{mandelstam_energytime_1945, margolus_maximum_1998, taddei_quantum_2013, del_campo_quantum_2013, deffner_quantum_2013, pires_generalized_2016}, where the application of the latter has successfully tightened the previous quantum speed limits~\cite{zhang2021conservation}.
With the proposed advance above, our framework will reach a wider audience and may eventually integrate into every designer's toolbox in nanophotonics.

The wide application of our framework hinges on developing a general acceleration scheme for fast computational bounds.
As demonstrated in Chapter~\ref{chap:sparse},  the standard interior-point method solves the computational bounds in $O(N^4)$ flops for $N$ designable degrees of freedom, unable to scale to photonic devices with $N$ beyond $10^3$.
To resolve this, we have accelerated the bound using two types of sparsity patterns: the chordal sparisty found in thin metasurfaces (Chapter~\ref{chap:sparse}) and the block-diagonal sparsity found in periodic Green's function matrix (Appendix~\ref{appen:fast}), though neither methods is general enough. 
On the other hand, specialized solvers such as the Burer-Monteiror~\cite{burer2003nonlinear, burer2005local, boumal2016non} and Frank-Wolf~\cite{frank1956algorithm, jaggi2013revisiting, yurtsever2021scalable} methods, have solved semidefinite programs with $N = 10^7$ under general sparsity patterns~\cite{yurtsever2021scalable}. 
These specialize solvers circumvent the main bottleneck of the semidefinite programming, the semidefinite constraint in \eqref{chap5-SDP}, and use fast matrix-vector multiplications to calculate the gradient (as opposed to the expensive Hessian in the interior-point method~~\cite{Boyd2004}).
Without Hessian, however, the specialized solvers are sensitive to problem conditioning, requiring us to switch from the ill-conditioned differential formulation in Chapter~\ref{chap:sparse} back to the well-conditioned integral formulation in Chapter~\ref{chap:comp}. 
The dense matrices in the integral formulation exhibit hierarchical-low-rank structures which support fast matrix-vector multiplication, as in the celebrated ``fast integral-equation solvers''~\cite{rokhlin_rapid_1985,harrington_field_1993,johnson_block-iterative_2001}. 
With fast solvers from both communities (i.e., the integral equations and semidefinite programmings), we may finally understand the fundamental limits of large-scale multi-functional photonic designs. 

Our framework, initially developed for computing bounds, also suggests a new way to design near-optimal photonic structures. 
The key is to regard the framework as a series of transformations: one starts with the original design problem in \eqref{original_design_problem}, reformulates it into the QCQP in \eqref{chap3-local-opt-problem}, and then relaxes it to a convex  by dropping the nonconvex rank-one constraint.
If the solution of the semidefinite program is rank-one, then there is no relaxation in the transformation. The solution \textit{is} the optimal binary design.
If the solution is close to rank-one, as suggested in Chapter~\ref{chap:sparse} in many cases, then one can approximate the solution into a rank-one solution by either picking the first eigenvector or other established techniques~\cite{Liu2019b}. The approximated solution is a near-optimal design.
In fact, these are actual design methods in many combinatorial problems involving QCQPs such as optimal power flow~\cite{madani2014convex,sojoudi2012physics}, ptychography~\cite{Horstmeyer2015,yurtsever2021scalable}, VLSI design~\cite{Barahona1988}, Ising problems~\cite{Barahona1988}, and more.
Similarly, by formulating photonic problems as a QCQP, this thesis not only establish bounds, but also suggest photonic structures to reach these bounds.

More broadly, our framework belongs to a recent surge of bounds in nanophotonics~\cite{Angeris2021, chao2022physical} that are all based on rigorous optimizations under various forms of Maxwell's equations. These rigorous formulations contrast with heuristic bounds~\cite{presutti2020focusing, li2022thickness, miller2022optics,shastri2022extent} which are easier to solve but lack the rigorousness and generalizability of the former. 
A future direction for this field is to merge these two distinct approaches by either introducing the heuristic arguments into the optimization frameworks as regularizers~\cite{schab2022upper}, or in turn, justifying the heuristics through rigorous optimizations.
Another important direction in this field is to develop appropriate benchmarks for comparing the generality, tightness, and solving speed of different bounds~\cite{Angeris2021}.
The ultimate goal is to fully delineate what is possible in nanophotonics and have the bounds as a powerful set of tools for practitioners in nanophotonic designs and beyond.

\appendix

\chapter{Fundamental limits of mode engineering in nanophotonics} 
\label{appen:mode}

Modes underlie electromagnetic scattering: illuminating a scatter excites many ``orthogonal current oscillations'', each radiating independently and they adding up coherently into the scattered field.
While their oscillating amplitudes are affected by the external illumination, their spatial patterns are only related to the scatterer itself --- these spatial patterns are the ``modes'' of the scatterer.
Peeling away the facade of external illumination, modes capture the intrinsic property of the scatterer and govern its scattering response.
Therefore, designers tune the quality factor of a mode to control the emission linewidth of an antenna~\cite{capek2017minimization}, minimize the ``volume'' of a mode to enhance the light--matter interaction in a resonator~\cite{albrechtsen2022nanometer}, and alter the band diagram of modes to control the bandgap and dispersion of photonic crystals~\cite{joannopoulos_photonic_2008}.
Most existing designs, however, only rely on intuition or local optimization --- the fundamental limit of mode engineering is yet unknown.

Our local-conservation-based framework can apply to problems in mode engineering.
A mode is a eigensolution of the Maxwell's equations, which self-oscillates without external excitation and losses its energy over time in non-Hermitian systems such as nanoresonators. This damping effect is characterized by a complex-valued oscillating frequency $\omega$.
A fundamental question is whether a mode can oscillate at any complex frequency $\omega$.
This is equivalent to finding a polarization field $p$ that satisfies the complex-frequency local conservation laws of \eqref{chap3-complex_con} at frequency $\omega$ in the absence of the incident field:
\begin{equation}
    \begin{aligned}
        & \underset{p}{\text{maximize}}
        & & 0 \\
        & \text{subject to}
        & & p^\dagger\Re\left\{\DD_i\left[\omega^*\GG_0(\omega)+\omega^*\xi(\omega)\right]\right\}p = 0,\quad i = 1, ...,N \\
        & & & p^\dagger p = 1.
    \end{aligned}
    \label{eq:appenA-opt_prob1}
\end{equation}
The constraints include local power conservation at $N$ points in the design region and a norm constraint on $p$ (arbitrarily chosen to find a ``pattern'', not the exact amplitude).
Is there one $p$ that satisfies all these constraints? To answer this, we relax Problem~(\ref{eq:appenA-opt_prob1}) to a convex semidefinite program which we solve globally. If after the relaxation the problem is still infeasible, then Problem~(\ref{eq:appenA-opt_prob1}) has to be infeasible. In this case, a mode cannot exist at frequency $\omega$, regardless of the material structuring. This is one fundamental limit of mode engineering.

\begin{figure} [t]
\centering
\includegraphics[width=0.5\textwidth]{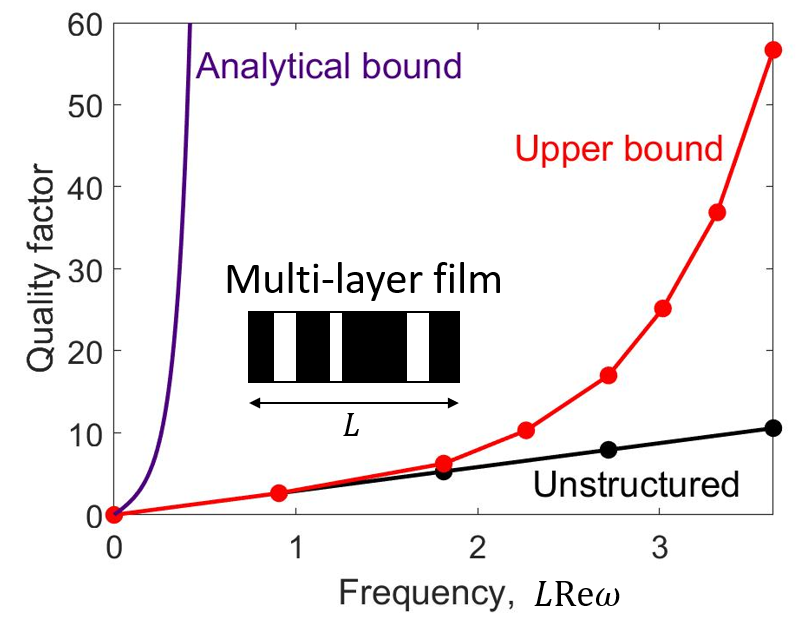}
    \caption{Bounds on the maximal quality factor of a mode in a multilayer film of length $L$. The mode oscillates at a complex frequency with a real part $\Re\omega$.  
    Compared to the previous analytical bound (purple line), our bound (red line) is much closer to the unstructured case (black line), suggesting limited effect of nanostructuring for wavelength-scale multilayer films.} 
	\label{fig:appenA-fig1}
\end{figure}

The feasibility Problem~(\ref{eq:appenA-opt_prob1}) also provides a bound on the maximal quality factor.
The quality factor $Q$ of a mode at a complex frequency $\omega$ is $\Re\omega / (-2\Im\omega)$. For a given $\Re\omega$, the smallest possible $\Im(-\omega)$ where Problem~(\ref{eq:appenA-opt_prob1}) remains feasible translates to an upper bound on the quality factor. This bound is demonstrated in \figref{appenA-fig1} for a multilayer film of length $L$ and composed of alternating materials $\varepsilon=\{1, 12\}$ as shown in the inset of \figref{appenA-fig1}.
When the film is unstructured (filled by $\varepsilon=12$), the modes are the Fabry--Pérot cavity modes~\cite{lalanne_light_2018}, their quality factors shown as the black dots in \figref{appenA-fig1}. The only previous  bound~\cite{osting2013long} is shown in the purple line, increasing exponentially~\cite{osting2013long}, and suggesting extremely large quality factors even for a wavelength-scale structure. In contrast, our bounds (red line) are much tighter, proving nanostructuring is only effective when the scatterer is at least one length of the internal wavelength.

To conclude, our framework applies to mode engineering problems: it identifies whether a mode can exist on a complex-frequency plane, as well as how close it can move towards the real-frequency axis. The latter leads to a bound on the quality factor, which we show is much tighter than the previous one. 
Besides quality factor, the objective 
function could also be mode volume, Purcell factor, bandwidth (inverse of quality factor), or even bandgap between two modes. The last one, especially, relates with the multi-frequency approach developed in Chapter~\ref{chap:multi}.

\chapter{Bounds on channel strengths through arbitrary scatterers}
\label{appen:info}

Chapter~\ref{chap:comm} proves that enlarging the communicating domains always increases the their free-space coupling strengths, but does not show how much an external scatterer can further enhance this coupling.
Existing bounds on maximal scattering-assisted couplings are based on either unconventional assumptions~\cite{Miller2007}, heuristic observations~\cite{li2022thickness} or postulations~\cite{miller2022optics}.
To this end, this appendix presents a rigorous framework to derive such bounds, maximizing the scattering-assisted channel strengths by imposing power conservation constraints on the induced polarization fields in the scatterer. 
In particular, we show lossy materials (such as plasmonics) can enhance the coupling strengths by a factor of $|\chi|^4 / (\Im\chi)^2$ but cannot alter its decay rate, which is just as fast as in free-space.

\hiddensection{Theoretical framework}
In this section, we present a theoretical framework to bound the maximal scattering-assisted channel strengths. We first review the notion of optimal communication channels at the presence of an external scatterer, then cast the problem of interest, i.e., maximizing the coupling strengths, as optimization problems. In the optimization problems, we relax all relevant domains (i.e., source, receiver, and scatterer) into symmetric bounding domains, and impose local-power conservation constraints on the polarization fields induced in the scatterer. Ths leads to bounds on the maximal scattering-assisted channel strengths.

\begin{figure*}[t]
  \centering
  \includegraphics[width=1\linewidth]{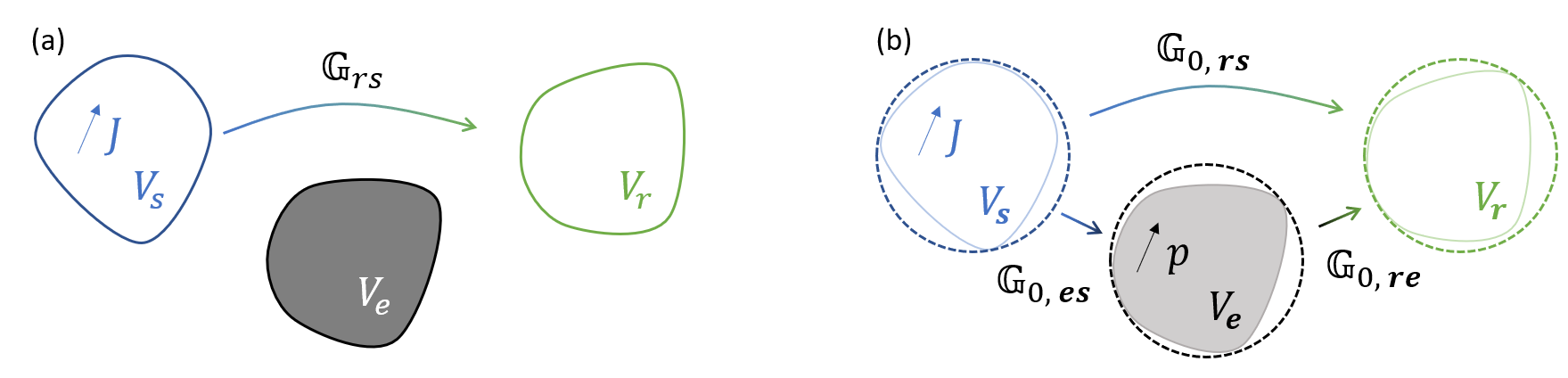}
\caption{Coupling at the presence of an external scatterer. (a) The current $J$ in the source region $V_s$ couples to the field in the receiver region $V_r$ through the Green's function operator $\GG_{rs}$, whose singular values, i.e., the channel strengths, are strongly affected by the presence of the scatterer $V_e$. (b) We bound the maximal possible channel strengths by relaxing all three involved domains to the highly symmetric bounding volumes $V_{\bm{s}}$, $V_{\bm{r}}$, and $V_{\bm{e}}$, denoted with bold subscripts. The scatterer is replaced by polarization fields $p$ whose maximal excitation is constrained by local-power-conservation laws. All involved Green's functions are now free-space Green's functions.}
	\label{fig:appenB-figure1}
\end{figure*}

The optimal communication channels maximize the information transfer between two communication domains. 
These channels are identified by decomposing the Green’s function operator $\GG_{rs}$ into its singular values and singular vectors: 
\begin{equation}
    \GG_{rs} = \sum_{i=1}^n \sigma_i u_i v_i^\dagger,
\end{equation} 
where $v_i$, $u_i$, and $\sigma_i$ are the right singular vector, left singular vector, and singular value, respectively. Together, they define a communication channel whose coupling strength is given by the square of the singular value, $|\sigma_i|^2$. (See Chapter~\ref{chap:comm} for more discussions on the optimal communication channels.)
Placing a scatterer in between the communication domains (as in \figref{appenB-figure1}(a)) affects the Green's function $\GG_{rs}$, as well as its coupling strengths $|\sigma_i|^2$.
An important question is how strong this effect can be, which we address below.

At the presence of a scatterer, the coupling strength $|\sigma_i|^2$ can be determined by solving an optimization problem, which is to maximize the objective  $J^\dagger\GG^\dagger_{rs}\GG_{rs}J$ under the constraints that the optimal currents $J$ of every channel are unitary and orthogonal to each other.
For example, the first two coupling strengths are the maximal objectives of the following two problems:
    
\noindent
\begin{minipage}{.5\linewidth}
\begin{equation}
    \begin{aligned}
    & \underset{J_1}{\text{max.}}
    & & J_1^\dagger\GG^\dagger_{rs}\GG_{rs}J_1  \\
    & \text{s.t. }
    & & J_1^\dagger J_1 = 1. \\
    & & & 
    \end{aligned}
    \label{eq:appenB-opt1-J1}
\end{equation}
\end{minipage}%
\begin{minipage}{.5\linewidth}
\begin{equation}
    \begin{aligned}
    & \underset{J_2}{\text{max.}}
    & & J_2^\dagger\GG^\dagger_{rs}\GG_{rs}J_2  \\
    & \text{s.t. }
    & & J_2^\dagger J_2 = 1, \\
    & & & J_{1, \text{opt}}^\dagger J_2 = 0,
    \end{aligned}
    \label{eq:appenB-opt1-J2}
\end{equation}
\end{minipage}

\noindent
where $J_2$ in Problem~(\ref{eq:appenB-opt1-J2}) is enforced to be orthogonal to $J_{1, \text{opt}}$, the optimal solution of Problem~(\ref{eq:appenB-opt1-J1}). 
Similarly, the coupling strength of the $n$-th channel is obtained by maximizing Problem~(\ref{eq:appenB-opt1-J1}) with $n-1$ orthogonality constraints: $J_{i, \text{opt}}^\dagger J_n = 0$ for  $i=1,2,...,n-1$.
Problems~(\ref{eq:appenB-opt1-J1}, \ref{eq:appenB-opt1-J2}) maximize the same ellipsoid-like objective $J^\dagger\GG^\dagger_{rs}\GG_{rs}J$. The first optimum $J_{1, \text{opt}}$ aligns with its longest axis. The second optimum $J_{2, \text{opt}}$ aligns with its second longest axis. Any deviation of $J_{1, \text{opt}}$ allows $J_{2, \text{opt}}$ to tilt from the second longest to the longest axis, increasing its objective. Thus, if we are interested in obtaining an upper bound of Problem~(\ref{eq:appenB-opt1-J2}), we can replace $J_{1, \text{opt}}$ in its orthogonality constraint with any vector, a fact that we will use below.

Larger channel strengths can always be obtained by enlarging the communicating domains in Problems~(\ref{eq:appenB-opt1-J1}, \ref{eq:appenB-opt1-J2}).
This ``domain monotonic'' property is discussed in Section~\ref{sec:chap6-optimal}, and we review here for clarity. 
As illustrated in \figref{appenB-figure1}(b), we enlarge the domains $V_s$ and $V_r$ to two bounding spheres $V_{\bm{s}}$ and $V_{\bm{r}}$, the bold letters $\bm{s}$ and $\bm{r}$ distinguishing the enlarged bounding domains from the original ones.
Defining current $J_i$ in the bounding volume $V_{\bm{s}}$, and extending the source domain to $V_{\bm{r}}$, Problems~(\ref{eq:appenB-opt1-J1}, \ref{eq:appenB-opt1-J2}) become
 
\noindent
\begin{minipage}{.5\linewidth}
\begin{equation}
    \begin{aligned}
    & \underset{J_1}{\text{max.}}
    & & J_1^\dagger\GG^\dagger_{\bm{rs}}\GG_{\bm{rs}}J_1  \\
    & \text{s.t. }
    & & J_1^\dagger J_1 = 1. \\
    & & & 
    \end{aligned}
    \label{eq:appenB-opt2-J1}
\end{equation}
\end{minipage}%
\begin{minipage}{.5\linewidth}
\begin{equation}
    \begin{aligned}
    & \underset{J_2}{\text{max.}}
    & & J_2^\dagger\GG^\dagger_{\bm{rs}}\GG_{\bm{rs}}J_2  \\
    & \text{s.t. }
    & & J_2^\dagger J_2 = 1, \\
    & & & \beta^\dagger J_2 = 0,
    \end{aligned}
    \label{eq:appenB-opt2-J2}
\end{equation}
\end{minipage}

\noindent
where the Green's function $G_{\bm{rs}}$ connects the two bounding spheres (note the bold subscript in $G_{\bm{rs}}$), and  $\beta$ can be any vector in the bounding sphere $V_{\bm{s}}$. 
Compared to Problems~(\ref{eq:appenB-opt1-J1}, \ref{eq:appenB-opt1-J2}), the additional degree of freedoms introduced in Problems~(\ref{eq:appenB-opt2-J1}, \ref{eq:appenB-opt2-J2}) does not decrease the objective values, and any deviation of $\beta$ from $J_{1, \text{opt}}$ only increases the objective value of Problem~(\ref{eq:appenB-opt2-J2}) as argued above. Thus, the optima of Problems~(\ref{eq:appenB-opt2-J1}, \ref{eq:appenB-opt2-J2}) upper bound the coupling strengths, defined by the optima of Problems~(\ref{eq:appenB-opt1-J1}, \ref{eq:appenB-opt1-J2}), as long as the communicating domains are within the two bounding spheres.

Coupling at the presence of a scatterer includes two parts: a direct coupling between the source and receiver, and an indirect coupling through the scatterer (as in \figref{appenB-figure1}(b)). The latter, in particular, is facilitated by the polarization field $p$ in the scatterer. 
The polarization field $p$ is first excited by an incident field $\GG_{0, \bm{es}}J$ from the source $J$, and then radiates to the receiver, contributing to the coupling. 
Local-power-conservation laws in \eqref{chap3-local-power-discretize} constrain the possible excitation of $p$: $f_{\rm con}(J, p, \DD) \coloneqq p^\dagger\DD(\GG_{0, \bm{ee}} + \xi\II)p + p^\dagger\DD\GG_{0, \bm{es}}J = 0$. 
The excited $p$ contributes to the coupling through the free-space Green's function $\GG_{0, \bm{re}}$, which can be made 
explicit by rewriting the objective in Problems~(\ref{eq:appenB-opt2-J1}, \ref{eq:appenB-opt2-J2}) as $f_{\rm obj}(J, p) \coloneqq (\GG_{0, \bm{rs}}J + \GG_{0, \bm{re}}p)^\dagger (\GG_{0, \bm{rs}}J + \GG_{0, \bm{re}}p)$.
With both $J$ and $p$ as optimization variables, the maximal coupling strengths of the first and second channels can be determined by the following two optimization problems:

\noindent
    \begin{minipage}{.5\linewidth}
    \begin{equation}
        \begin{aligned}
        & \underset{J_1, p}{\text{max.}}
        & & f_{\rm obj}(J_1, p)  \\
        & \text{s.t. }
        & & J_1^\dagger J_1 = 1, \\
        & & & f_{\rm con}(J_1, p, \DD_i) = 0. \\
        & & & 
        \end{aligned}
        \label{eq:appenB-opt3-J1}
    \end{equation}
    \end{minipage}%
    \begin{minipage}{.5\linewidth}
    \begin{equation}
        \begin{aligned}
        & \underset{J_2, p}{\text{max.}}
        & & f_{\rm obj}(J_2, p)  \\
        & \text{s.t. }
        & & J_1^\dagger J_1 = 1, \\
        & & & \beta^\dagger J_2 = 0, \\
        & & & f_{\rm con}(J_2, p, \DD_i) = 0.
        \end{aligned}
        \label{eq:appenB-opt3-J2}
    \end{equation}
    \end{minipage}

\noindent
Problems~(\ref{eq:appenB-opt3-J1}, \ref{eq:appenB-opt3-J2}) involve only free-space Green's functions. Without explicitly specifying an external scatterer, their optima bound the channel strengths of any scatterers. This is in contrast with Problems~(\ref{eq:appenB-opt2-J1}, \ref{eq:appenB-opt2-J2}) which implicitly assume an external scatterer in their Green's function $\GG_{\bm{rs}}$. The parameter $\beta$ in Problem~(\ref{eq:appenB-opt3-J2}) can again be chosen arbitrarily for upper bounds.
Both the objective and constraints in Problems~(\ref{eq:appenB-opt3-J1}, \ref{eq:appenB-opt3-J2}) are quadratic equations of the optimization variables $p$ and $J_i$ --- QCQPs that are amendable to semidefinite relaxation for upper bounds.
Similar to the extension of the source and receiver domains, the polarization fields can also be extended to a highly-symmetric bounding volume in Problems~(\ref{eq:appenB-opt3-J1}, \ref{eq:appenB-opt3-J2}), whose maxima upper bound the scattering-assisted channel strengths of any sources, receivers, and scatterers.


\hiddensection{Analytical bounds for lossy materials}

In the previous section, we demonstrate a rigorous framework to bound the scattering-assisted channel strengths via local-power-conservation constraints. 
While such bounds usually require many conservation constraints, for lossy material, this section shows a single constraint is sufficient. The single constraint captures the key limiting factor of material absorption and leads to analytical bounds on the maximal channel strengths and their optimal decay rate.

The response of lossy material is limited by the material absorption in the system. 
The key material-absorption constraint can be derived by relaxing the global power conservation of \eqref{chap2-OptThmDis} to $p^\dagger\Im(\xi)p \leq \Im\left(J^\dagger\GG_{0, \bm{es}}^\dagger p \right)$. The left-hand side is quadratic in $p$; the right-hand side is linear in $p$. As a quadratic function grows faster than a linear function, this implies a constraint on the magnitude of $p$: $p^\dagger p \leq J^\dagger\GG_{0, \bm{es}}^\dagger\GG_{0, \bm{es}} J / (\Im\xi)^2 $.
Assuming the scattered field in the receiver domain is much stronger than the the incident field, the objective function in Problems~(\ref{eq:appenB-opt3-J1}, \ref{eq:appenB-opt3-J2}) simplifies to $p^\dagger\GG_{0, \bm{re}}^\dagger\GG_{0, \bm{re}}p$.
Optimizing this objective and relaxing the local-power-conservation constraints to the norm constraint of $p$, the maximal coupling strength of the $n$-th channel becomes:
\begin{equation}
    \begin{aligned}
    & \underset{J, p}{\text{max.}}
    & & p^\dagger\GG_{0, \bm{re}}^\dagger\GG_{0, \bm{re}}p   \\
    & \text{s.t. }
    & & J^\dagger J = 1, \\
    & & & p^\dagger p \leq J^\dagger\GG_{0, \bm{es}}^\dagger\GG_{0, \bm{es}} J / (\Im\xi)^2, \\
    & & & J_{i, \text{opt}}^\dagger J = 0, i = 1, 2, \cdots, n-1.
    \end{aligned}
    \label{eq:appenB-opt4-Jn}
\end{equation}
We can eyeball the solution of Problem (\ref{eq:appenB-opt4-Jn}): the optimal $J_{n, \text{opt}}$ is the $n$-th right singular vector of $\GG_{0, \bm{es}}$; the optimal $p_{\rm opt}$ is the first singular vector of $\GG_{0, \bm{re}}$.
The corresponding optimal objective upper bounds the coupling strength $|\sigma_n|^2$: 
\begin{equation}
    |\sigma_n|^2 \leq \left|\sigma^{(\rm vac)}_{n, \bm{es}}\right|^2\left|\sigma^{(\rm vac)}_{1, \bm{re}}\right|^2 / (\Im\xi)^2,
    \label{eq:sigma_n}
\end{equation}
where $\sigma^{(\rm vac)}_{1, \bm{re}}$ is the free-space coupling strength of the first channel between the receiver and the external scatterer, and $\sigma^{(\rm vac)}_{n, \bm{es}}$ is the free-space coupling strength of the $n$-th channel between the external scatterer and the source. Equation~(\ref{eq:sigma_n}) shows that the resonances in lossy materials (e.g., plasmonic resonances) can potentially enhance the coupling strength by a factor of  $(\Im\xi)^2 = |\chi|^4 / (\Im\chi)^2$.

The bound of \eqref{sigma_n} only involves free-space coupling strengths between the source, receiver, and the scatterer. 
It implies $|\sigma_n|^2$, the coupling strength at the presence of a scatterer, decays as fast as $|\sigma_{n, \bm{es}}|^2$, the coupling strength in free space between the source and scatterer.
Exchanging the position of source and receiver in Problem~(\ref{eq:appenB-opt4-Jn}) gives another bound that decays as fast as $|\sigma_{n, \bm{re}}|^2$.
Between the two, i.e., $|\sigma_{n, \bm{es}}|^2$ and $|\sigma_{n, \bm{re}}|^2$, the one that decays faster gives the tighter bound.
The larger the separation between two domains, the faster the decay of their coupling strengths. 
Thus, for the slowest decay rate, one should minimize the largest separation in \figref{appenB-figure1}(b), meaning to put the scatterer in the center of the source and receiver.

One can tighten the analytical bound in \eqref{sigma_n} by including more conservation constraints in Problem~(\ref{eq:appenB-opt4-Jn}).
Additional local-power-conservation constraints as in Problems~(\ref{eq:appenB-opt3-J1}, \ref{eq:appenB-opt3-J2}) can regularize the analytical bound for lossless dielectrics (where \eqref{sigma_n} diverges because $\Im\xi = 0$), as well as its  $|\sigma_{1, \bm{re}}|^2$ scaling. 
The latter is nonphysical, because it requires the optimal current of Problem~(\ref{eq:appenB-opt4-Jn}) to be the first singular vector of the $\GG_{0, \bm{es}}$ for every channel. This cannot be true because their scattered fields are not orthogonal in the receiver region.
Besides the local power conservation, one can add field-correlation constraints of \eqref{chap4-CCConstr} to Problem~(\ref{eq:appenB-opt4-Jn}) to enforce a single optimal scatterer among different channels, further tightening the bound.

\hiddensection{Conclusion} 
Conservation constraints determine the maximal electromagnetic response of any scatterer. Using these constraints, this appendix presents a general framework to compute bounds on scattering-assisted coupling strengths. The bounds are solved analytically for lossy materials, showing scattering-assisted coupling strength can be amplified by resonances but its decay rate remains the same as their free-space counterparts, which are exponential in 2D and sub-exponential in 3D as we proved in Chapter~\ref{chap:comm}. From this result, one can easily bound the maximal number of channels and their information capacities through any scatterers.

\chapter{Analytical bounds for 2D materials}
\label{appen:2D}

How do material choice and illuminating condition affect the maximal response of a photonic structure? One can laboriously apply the computational bounds in Chapter~\ref{chap:comp} to each possible scenario, but its answer is neither intuitive nor scalable. In contrast, the analytical bounds in Chapter~\ref{chap:analy} delineate the tandem effect of material and radiative losses with a single equation. 
Yet, relying only on the real-power-conservation constraint, these analytical bounds are loose for important applications such as lossless dielectrics.

This appendix derives analytical bounds for 2D materials that incorporate both the real- and reactive-power-conservation constraints, regularizing the maximal response of lossless materials.
They reveal how refractive index determines the maximal light--matter interaction, their analyticity owing to the diagonalizability of the background Green's function in an ultrathin domain.
These new bounds significantly tighten the ones in Chapter~\ref{chap:analy}, especially for 2D materials with small negative permittivity or large evanescent couplings, as we show below.

The bound maximizes an arbitrary objective (such as extinction and absorption) under both the real- and reactive-power-conservation constraints (the real and imaginary parts of \eqref{chap3-local-power-discretize} with $\DD_i = \II$):
\begin{equation}
    \begin{aligned}
    & \underset{p}{\text{maximize}}
    & & f(p) = p^\dagger \mathbb{A} p +  \Im\left(\beta^\dagger p\right) \\
    & \text{subject to}
    & & p^\dagger \left\{\Im \xi\II + \ImGO \right\} p + \Im \left(p^\dagger \einc\right) = 0, \\
    & & & p^\dagger \left\{\Re \xi\II + \Re\GG_0 \right\} p + \Re \left(p^\dagger \einc\right) = 0.
    \end{aligned}
\label{eq:appenC-opt}
\end{equation}
Here, we model the 2D material with an effective material susceptiblity $\chi$ (which defines $\xi = -1/\chi$) and an effective thickness $h$.  
The first constraint in Problem~(\ref{eq:appenC-opt}) represents the global-real-power conservation; the second, the global-reactive-power conservation. Without the second, Problem~(\ref{eq:appenC-opt}) relaxes to Problem~(\ref{eq:chap2-general-formalism}). With the second, we retain the information about real part of the material susceptibility in $\Re\xi = -2\frac{\Re\chi}{|\chi|^2}$, leading to more informative bounds.

Problem~(\ref{eq:appenC-opt}) has analytical solutions if the Green's function matrices, $\Im\GG_0$ and $\Re\GG_0$, can be diagonalized simultaneously.
This diagonalization is generally impossible, but for an ultrathin periodic domain (such as patterned 2D materials), it is possible in a Fourier basis, i.e., a Fourier mode only radiates into the same Fourier mode in such a domain.  
For simplicity, we assume the 2D material is periodically patterned in one direction (say $x$) with periodicity $L$, and is homogeneous in the other direction (say $y$). We also assume the incident field is at TE polarization (i.e., polarized along $y$). Within each period, we expand the polarization field in Fourier modes $p(x) = \sum_n p_n e^{ik_nx}$, where $k_n = 2\pi n / L$ is the wave-vector of the $n$-th mode.
In this basis, the free-space Green's function matrix in Problem~(\ref{eq:appenC-opt}) becomes $\GG_{0, nn'} = \frac{ihk}{2\sqrt{1-k^2_n/k^2}}\delta_{nn'}$. When $k_n \leq k$, the mode is a propagating wave that carries real power; $\GG_{0, nn}$ is pure imaginary. Otherwise, the mode is a evanescent wave that carries reactive power; $\GG_{0, nn}$ is real. In either case, the diagonal $\GG$ radiates a current into the same mode.

A diagonalized Green's function matrix allows Problem~(\ref{eq:appenC-opt}) to be solved analytically by Lagrangian duality. 
The derivation follows the one presented in Section~\ref{sec:appenG-sec1}. Here, we just present the results for $\Re\chi \leq 0$ as in common plasmonic materials. Under an incident field at channel $n$, the maximal extinction and absorption cross sections are:
\begin{align}
    \sigma_{\rm ext} / L &= 
    \begin{cases} 
        \frac{hk}{\Im\xi + \rho_n}\left[ 1 + \left(\frac{\Re\xi}{\Im\xi + \rho_n}\right)^2\right]^{-1}\quad  &\text{if}\ \rho_n + \Im\xi \geq \Re\xi \\ 
        \frac{hk}{2\Re\xi}\quad  &\text{if}\ \rho_n + \Im\xi \leq \Re\xi,
    \end{cases} 
    \label{eq:appenC-ext}
    \\
    \sigma_{\rm abs} / L &= 
    \begin{cases} 
       \frac{2hk\Im\xi}{(\Im\xi+\rho_n)^2 + (\Re\xi)^2}\quad  &\text{if}\ (\Im\xi)^2 \geq (\Re\xi)^2 + \rho_n^2 \\ 
        \frac{hk}{\Re\xi}\left(\sqrt{\frac{\rho_n^2}{(\Re\xi)^2}+1} - \frac{\rho_n}{\Re\xi}\right)\quad  &\text{if}\ (\Im\xi)^2 \leq (\Re\xi)^2 + \rho_n^2.
    \end{cases} 
\end{align}
The bound involves three parameters: radiative loss of the $n$-th scattering channel $\rho_n = \frac{ihk}{2\sqrt{1-k^2_n/k^2}}$, material loss $\Im\xi=\Im\chi/|\chi|^2$, and stored (reactive) energy $\Re\xi=\Re\chi/|\chi|^2$. 
When the material or radiative loss dominates, the bounds echo the previous bounds of Eqs. (\ref{eq:chap2-ext-gen}, \ref{eq:chap2-abs-gen}) with a $\Re\xi$-correction. When the stored energy dominates, the bounds depend on $\Re\xi$ but not on $\Im\xi$, which is a new result.

\begin{figure*}[t]
  \centering
  \includegraphics[width=1\linewidth]{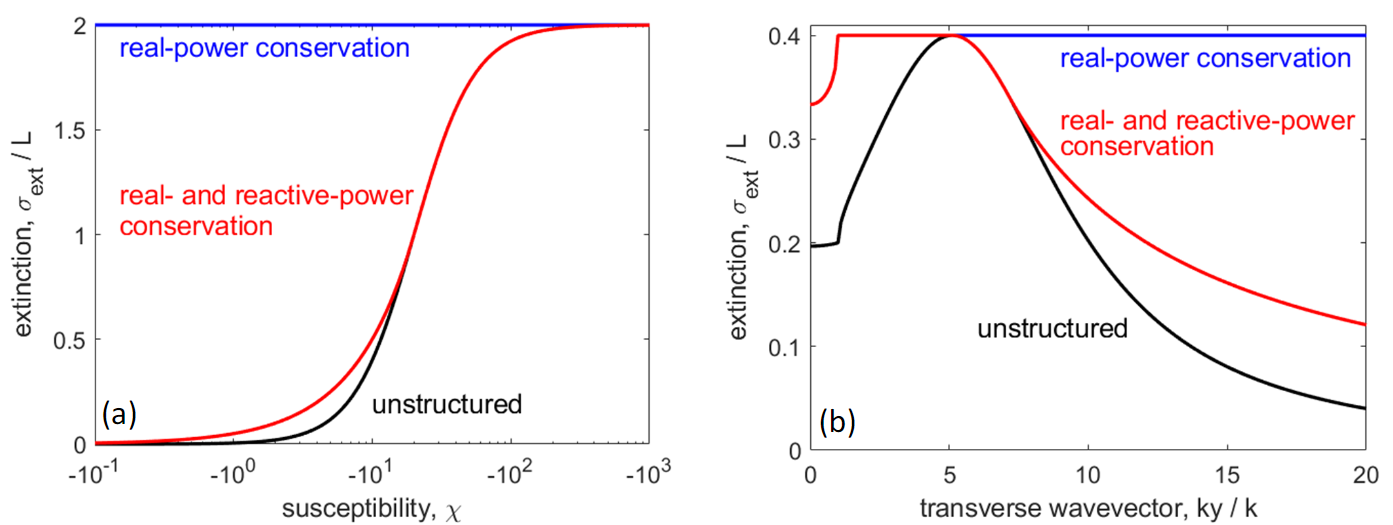}
\caption{Extinction bounds of a 2D material with effective thickness $kh=0.1$ for (a) a normal TE incidence and (b) TM incidences. In both cases, the analytical bounds based on real- and reactive-power conservation (red lines) are much tighter than the previous analytical bounds based on real-power conservation alone (blue lines).}
	\label{fig:appenC-fig1}
\end{figure*}

Figure~(\ref{fig:appenC-fig1}) showcases our analytical bounds with two examples. 
As the first example, \figref{appenC-fig1}(a) considers the maximal extinction as a function of the material susceptibility $\chi$ under a normal TE incidence.
The material is lossless, trivializing the previous real-power-conservation-based bound of \eqref{chap2-ext-gen} to 2.
On the other hand, the bound in \eqref{appenC-ext} considers both the real- and reactive-power conservation, showing maximal extinction decreases with reduced permittivity, tight against the unstructured material (black line).  
As the second example, \figref{appenC-fig1}(b) computes the bound under differently angled incident waves, characterized by their transverse wave vector $k_y$. 
If $k_y \leq k$, the wave is  propagating; otherwise, it is evanescent.
This example assumes a TM polarization (whose analytical bounds are computed via the same apparatus as in Problem~(\ref{eq:appenC-opt})) with susceptibility $\chi = -1 + 2j$, which invokes a plasmonic resonance at $k_y=5k$.
The bound based on real-power conservation (blue line, \eqref{chap2-ext-gen}) is tight at the plasmonic resonance, but loose after that. In contrast, the bound based on real- and reactive-power conservation (red line) is tight throughout. 
Both examples highlight the tightness of the new analytical bounds, delineating the maximal response of a 2D material for varying material permittivity and illumination angles.

\chapter{Fast computational bound via block-diagonal sparsity}
\label{appen:fast}

Applications such as achromatic metalens and mode converters demand large-scale, multi-functional photonic devices.
The fundamental limits of those large-scale devices, however, are unattainable via the computational bound in Chapter~\ref{chap:comp} because its high optimization complexity.
The acceleration method in Chapter~\ref{chap:sparse} scales to metasurfaces with large diameter but the allowed thickness is miniature.  
This appendix presents an acceleration scheme to compute bounds of general metasurfaces, where we use the block diagonal sparsity of the free-space Green's function to break a large semidefinite program into small chunks, each solved swiftly.
The resulting fast bound is demonstrated in the example of metalens beyond 150 wavelengths in diameter. 

We first review our optimization problem and its relaxation method for a computational bound.
Using an $n$-dimensional polarization field $p$ as an optimization variable, we maximize an arbitrary objective function under the real- and reactive-power-conservation constraints:
\begin{equation}
    \begin{aligned}
    & \underset{p\in\CC^n}{\text{maximize}}
    & & f(p) = p^\dagger \mathbb{A} p +  \Im\left(\beta^\dagger p\right) \\
    & \text{subject to}
    & & p^\dagger \left\{\Im \xi\II + \Im\GG^{\rm (vac)} \right\} p + \Im \left(p^\dagger \einc\right) = 0, \\
    & & & p^\dagger \left\{\Re \xi\II + \Re\GG^{\rm (vac)} \right\} p + \Re \left(p^\dagger \einc\right) = 0.
    \end{aligned}
    \label{eq:appenD-opt}
\end{equation}
This QCQP can be relaxed into a semidefinite program
\begin{equation}
    \begin{aligned}
    & \underset{\XX\in\SS^n}{\text{maximize}}
    & & \Tr(\AA_0\XX) \\
    & \text{subject to}
    & &  \Tr(\AA_1\XX) = 0, \\
    & & &  \Tr(\AA_2\XX) = 0, \\
    & & & \ \XX \geq 0.
    \end{aligned}
    \label{eq:appenD-SDP}
\end{equation}
As detailed in Section~\ref{sec:appenH-sec1}, the solution of Problem~(\ref{eq:appenD-SDP}) represents a bound on Problem~(\ref{eq:appenD-opt}).
For simplicity, we ignore the slack variable in Problem~(\ref{eq:appenD-SDP}), which can be included with minimal changes. The major numerical difficulty of Problem~(\ref{eq:appenD-SDP}) is its solution takes $O(n^{3.5})$ flops to compute, hard to scale to large photonic devices.

\begin{figure*}[ht]
  \centering
  \includegraphics[width=1\linewidth]{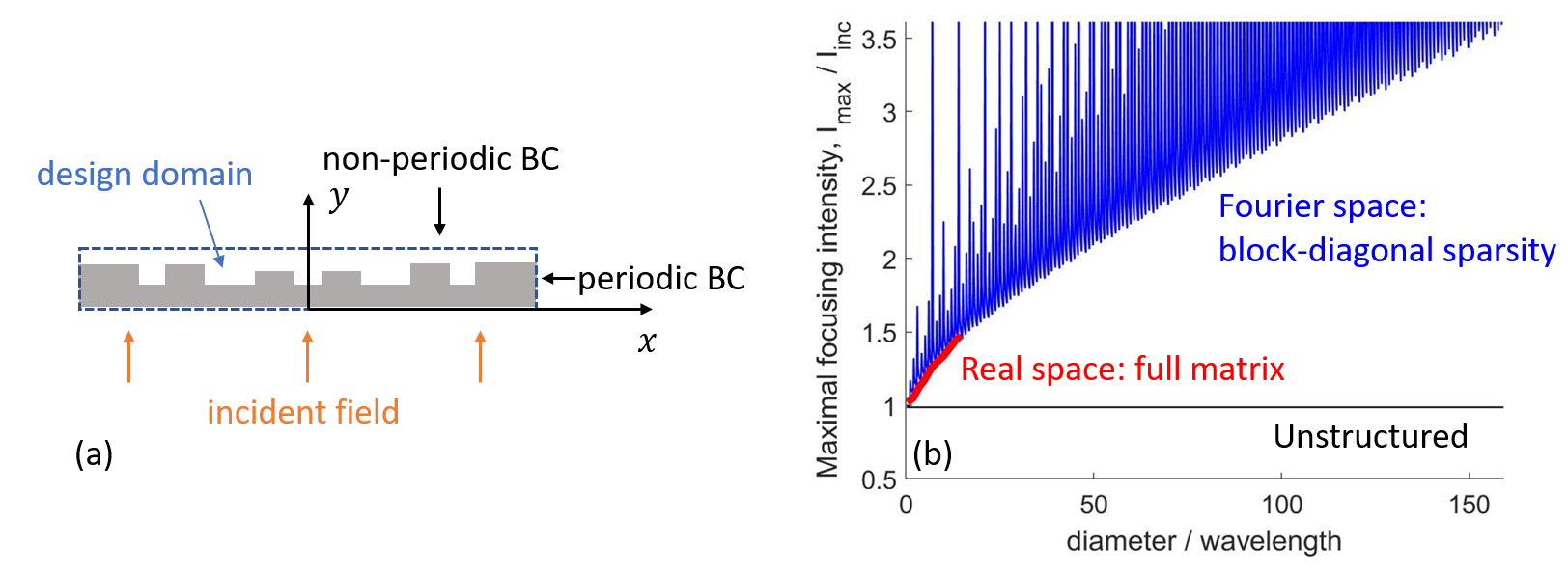}
\caption{(a) A design domain for periodic metasurfaces. (b) Bounds on focusing intensity of a metasurface with focusing distance 1.6$\lambda$, material permittivity 12 + 1i, and thickness 0.0015$\lambda$, where $\lambda$ is the free-space wavelength. Fourier-space-based formulation in Problem~(\ref{eq:appenD-SDP-decomposed}) manifests block-diagonal sparsity, scaling up the previous real-space bound (red line) to larger structure dimensions (blue line). }
	\label{fig:appenD-fig1}
\end{figure*}

The free-space Green's function matrix $\GG^{\rm (vac)}$ in Problem~(\ref{eq:appenD-opt}), a full matrix in real space, can be diagonalized in the Fourier basis. A common design domain is illustrated in \figref{appenD-fig1}(a), which is finite in the $y$ direction with thickness $L_y$ and periodic in the $x$ direction with periodicity $L_x$. We assume the incident field is TE-polarized (perpendicular to the 2D plane).
Any possible polarization field $p(\vr)$ in this domain can be decomposed into a series of Fourier modes 
\begin{equation}
    u_{nm}(\vr) = e^{ik_{x,n}x}e^{ik_{y,m}y},
\end{equation} 
with wavevectors $k_{x,n}=2\pi n / L_x$ and $k_{y,m} = 2\pi m / L_y$ indexed by integers $n$ and $m$ that run from minus infinity to infinity. 
Currents represented by a Fourier mode $u_{nm}(\vr)$ radiate into fields of other Fourier modes $u_{n'm'}(\vr)$, but because of the periodicity, all the radiated mode has $n'=n$. 
Accordingly, in this Fourier basis, the Green's function matrix is block diagonal:
\setcounter{MaxMatrixCols}{20}
\begin{equation}
    \GG^{\rm (vac)} = \renewcommand{\arraystretch}{0.5}
    \begin{bmatrix}
        *   & \cdots &    *   &        &        &        &   &        &        &        &        &   \\
     \vdots & \ddots & \vdots &        &        &        &   &        &        &        &        &   \\
        *   & \cdots &    *   &        &        &        &   &        &        &        &        &   \\
            &        &        &    *   & \cdots &    *   &   &        &        &        &        &   \\
            &        &        & \vdots & \ddots & \vdots &   &        &        &        &        &   \\
            &        &        &    *   & \cdots &    *   &   &        &        &        &        &   \\
            &        &        &        &        &        & \ &    \   & \      &        &        &   \\
            &        &        &        &        &        & \ & \ddots & \      &        &        &   \\
            &        &        &        &        &        & \ &    \   & \      &        &        &   \\
            &        &        &        &        &        &   &        &    *   & \cdots &    *   &   \\
            &        &        &        &        &        &   &        & \vdots & \ddots & \vdots &   \\
            &        &        &        &        &        &   &        &    *   & \cdots &    *   &
    \end{bmatrix},   
\end{equation}
where the blocks run through different $n$ values; within each block, the elements run through different $m$ values.
Explicitly, the $mm'$-th element in the $n$-indexed block is:
\begin{equation}
    \GG^{\rm (vac)}_{n, mm'} = \frac{ik^2}{2k_{y,n}L_y}\int_{0}^{L_y}\int_{0}^{L_y}  e^{i\left(k_{y,n}|y-y'| + k_{y,m'}y' - k_{y,m}y\right)} dy dy'.
\end{equation}
But couldn't a structure induce coupling between Fourier modes with different $n$? The key is to realize the actual scattering problem has been relaxed to the Problem~(\ref{eq:appenD-SDP}). The latter is only constrained by the global power conservations and is always homogeneous in its material parameter $\xi\II$. As a result, the non-sparse scattering problem is bounded by a sparse bound; the latter can even be computed faster than the former.

The block diagonal sparsity in $\GG_0$ can dramatically accelerate the computation of the bound in Problem~(\ref{eq:appenD-SDP}).
This is because Problem~(\ref{eq:appenD-SDP}) inherits the block diagonal sparsity of $\GG_0$ in both its constraint matrices, $\AA_1$ and $\AA_2$, and its objective matrix $\AA_0$. The latter is true for common objectives such as scattering, absorption, and extinction.
As all $\AA_i$ share the same block diagonal sparsity, the only useful entries in $\XX$ are the ones in the corresponding blocks (the others are all multiplied by zero in Problem~(\ref{eq:appenD-SDP})). 
Calling the respective blocks in $\XX$ as $\XX_i$, we decompose Problem~(\ref{eq:appenD-SDP}) into an optimization problem on all the $\XX_i$:
\begin{equation}
    \begin{aligned}
    & 
    \underset{\left\{\XX_i\in\SS^{n_y}\right\}_{i=1}^{n_x}}{\text{maximize}}
    & & \sum_{i=1}^{n_x}\Tr(\AA_0\XX_i) \\
    & \text{subject to}
    & &  \sum_{i=1}^{n_x}\Tr(\AA_1\XX_i)  = 0, \\
    & & &  \sum_{i=1}^{n_x}\Tr(\AA_2\XX_i)  = 0, \\
    & & & \ \XX_i \geq 0,\ \textrm{for}\ i = 1, 2, ..., n_x,
    \end{aligned}
    \label{eq:appenD-SDP-decomposed}
\end{equation}
where we have assumed $n_x$ uncoupled Fourier modes and $n_y$ coupled Fourier modes; the total degree of freedom $n=n_x n_y$. 
The last constraint in Problem~(\ref{eq:appenD-SDP-decomposed}), in particular, decomposes the semidefinite constraint $\XX\geq 0$ into semidefinite constraints on the principle submatrices $\XX_i\geq 0$, reducing the total complexity of Problem~(\ref{eq:appenD-SDP-decomposed}) from $O((n_x+n_y)^{3.5})$ to $O(n_xn_y^3)$. The latter scales only linearly with $n_x$, in proportion to the diameter of the design domain.

Imposing additional local-power-conservation constraints of \eqref{chap3-opt-weight} to Problem~(\ref{eq:appenD-SDP-decomposed}) may break the block diagonal sparsity, though a subset of them do not. This subset is characterized by weighting matrices $\DD_{\rm wei}$ that place different weights for points in the nonperiodic direction but same weights for points in the periodic direction. The latter condition retains the spatial transnational symmetry of the operator $\DD_{\rm wei}\GG_0$ so the Fourier modes remain uncoupled.
Imposing these subset of local-power-conservation constraints to Problem~(\ref{eq:appenD-SDP-decomposed}) can tighten the bound.

This fast algorithm is demonstrated in \figref{appenD-fig1}(b) for a large-scale metalens.
The metalens is contoured by artificial periodic boundary conditions as in \figref{appenD-fig1}(a) so that different horizontal Fourier modes are uncoupled, inducing a block diagonal sparsity. This sparsity allows us to dramatically scale up the problem dimensions, as evident by the comparison in \figref{appenD-fig1}(b) between the fast bounds (blue line) and their real-space counterparts (red line) whose full matrices permit no direct acceleration.
The peaks in the fast bounds are the high-Q BIC modes, an artifact of imposing the periodic boundary conditions for this (non-periodic) problem.
The underside of the blue line traces out the real-space bound and extends to metasurface 150 wavelengths in diameter, seven times larger than before.

Sparsity leads to acceleration. The block diagonal sparsity of the Green's function matrix, in particular, decomposes a large semidefinite program into smaller pieces, whose complexity scales only linearly with the metasurface diameter, as we have demonstrated for a large-scale metalens in this appendix. Furthermore, this fast algorithm can be extended to 1. metasurfaces with layered substrates whose transnational symmetry retains the block diagonal sparsity of the background Green's function matrix, and 2. multi-functional metasurfaces where each block is enlarged by the number of functionalities $n_f$, the total complexity being $O(n_x(n_y+n_f)^3)$.

The block diagonal sparsity generalizes the diagonal sparsity in Appendix~\ref{appen:2D} and is a special case of the chordal sparsity in Chapter~\ref{chap:sparse}. This specialization allows us to bound thick metasurfaces, though to maintain the block diagonal sparsity, one cannot impose every local power conservation constraints, leading to possibly loose bound.
An good future testbed for this method is large-scale, multifunctional plasmonic metasurfaces where few global-power-conservation constraints may suffice, examples including the field concentrators and solar absorbers.

\chapter{Maximal electromagnetic response per volume}
\label{appen:volume}
Space applications such as solar harvesting and light sail demand lightweight photonic devices~\cite{ilic2020nanophotonic}.
Light sail, in particular, propels a nanostructure with a laser beam for high acceleration, which requires maximizing the optical force per material volume. Such per-volume response has no general bound, the only bound being the lossy-material bounds~\cite{miller_fundamental_2016}  which are loose for wavelength-scale devices. Consequently, researchers don't know whether current designs~\cite{jin2020inverse} have reached their fundamental limits.

Building upon our established framework, this appendix bounds the maximal response of nanostructures under a limited material allocation.
Previous bounds in Chapter~\ref{chap:comp} allow material to fill the whole design region, and are loose for lightweight devices. 
To regulate this, this appendix enforces air holes in the design region.
The size of the air hole equals the number of zeros in the polarization-current vector, the latter being our optimization variable.
By enforcing zeros in it, we derive a bound on the maximal per-volume response. Demonstrated in a thin-film design problem, our bounds show the trade-off between the structure weight and their maximal extinction, and predict an optimal filling fraction for maximal per-volume extinction in a thin film.  

As stated, we enforce air holes in a nanostructure through zeros in its polarization field.
Zero polarization field implies zero material susceptibility or zero electric field. Either case, the material is air or can be replaced by air without affecting the eventual response. Therefore, the number of zero (nonzero) in a discretized polarization-current vector $p$ determines the amount of air (material) in a design. 
The number of nonzero elements in $p$ is its cardinality, $\card(p)$. Including it in Problem (\ref{eq:chap3-local-opt-problem}) yields 
\begin{equation}
    \begin{aligned}
    & \underset{p\in\CC^n}{\text{maximize}}
    & & f(p) = p^\dagger \mathbb{A} p +  \Re\left(\beta^\dagger p\right) \\
    & \text{subject to}
    & & p^\dagger\DD_i (\GG_0+\xi \II)p = - p^\dagger\DD_i e_{\rm inc},\quad \text{for } i = 1,2,...,n, \\
    & & & \card(p) \leq q,
    \end{aligned}
    \label{eq:appenE-with-card}
\end{equation}
where we discretize the design region into $n$ points, and assume material occupies at most $q$ of them.
Both the quadratic and cardinality constraints in Problem (\ref{eq:appenE-with-card}) can be relaxed to convex constraints. The former, as we discussed in Section~\ref{sec:appenH-sec1}, can be relaxed to linear constraints on a semidefinite variable $X$; the latter, to a convex $l_1$-constraint on the same $X$:
\begin{equation}
     \sum_{ij} |X_{ij}| \leq (q+1)\Tr(X),
     \label{eq:appenE-X}
\end{equation}
where the left-hand side sums over all matrix elements in $X$~\cite{d2004direct}. 
After the relaxation, Problem (\ref{eq:appenE-with-card}) becomes convex and globally solvable. Its maximum bounds the maximal photonic response under a certain material allocation.

Problem (\ref{eq:appenE-with-card}) also determines the maximal response per volume.
Specifically, we bound Problem (\ref{eq:appenE-with-card}) for every possible material volume, i.e. every possible $q$. The one with the largest bound-by-$q$ ratio bounds the maximal per-volume response, its corresponding $q/n$ being the optimal filling ratio. 
This is the following optimization problem:
\begin{equation}
    \begin{aligned}
    & \underset{p\in\CC^n, q}{\text{maximize}}
    & & f(p) / q \\
    & \text{subject to}
    & & p^\dagger\DD_i (\GG_0+\xi \II)p = - p^\dagger\DD_i e_{\rm inc},\quad \text{for } i = 1,2,...,n, \\
    & & & \card(p) \leq q.
    \end{aligned}
    \label{eq:appenE-per-volume}
\end{equation}
Technically, $\card(p)$ should equal $q$ so that the we divide the actual material volume in the objective. Nonetheless, the optimal $q$ of Problem (\ref{eq:appenE-per-volume}) always satisfies this requirement. If not, decreasing this $q$ can increase the objective, invalidating the premise of $q$ being the optimum.
As already explained, we bound Problem (\ref{eq:appenE-per-volume}) by repeatedly bounding Problem (\ref{eq:appenE-with-card}) for every possible $q$. A bound on the former represents a fundamental limit on the maximal per-volume response.

\begin{figure*}[t]
  \centering
  \includegraphics[width=1\linewidth]{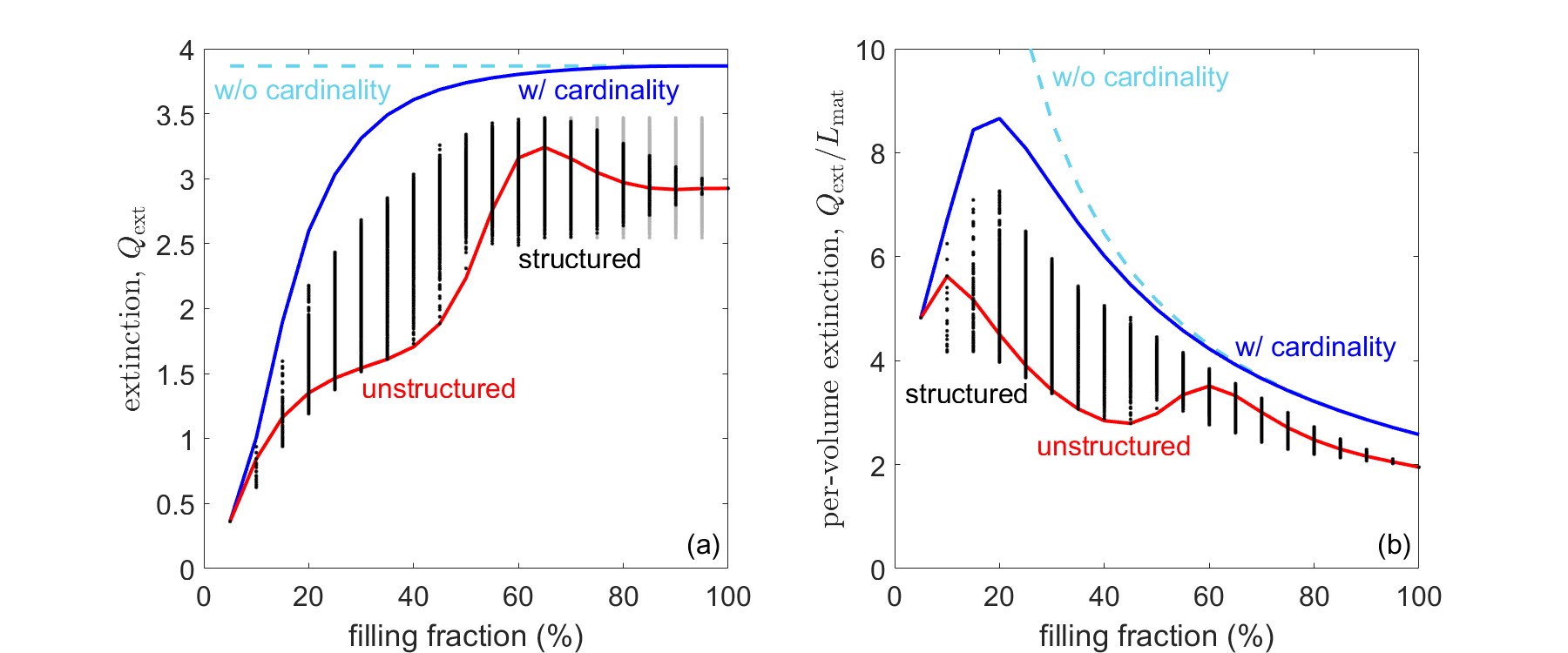}
\caption{Designing multilayer a thin film to maximize its per-volume extinction. (a) Bounds on the maximal extinction for a maximal filling fraction of materials (the blue and cyan lines). Thin-film designs at given filling fractions (the red line and the black dots). For fair comparisons between the designs and bounds, I overlay designs at 65\% filling fraction as the grey dots to designs at larger filling fractions, as in practice one can always choose a smaller filling fraction. (b) Per-volume extinction. The material length $L_{\rm mat}$ is the total design length times filling fraction. Bounds with the cardinality constraint (the blue line) predict optimal filling fraction at 20\%, agreeing with the local designs (the black dots). 
}
	\label{fig:appenE-figure1}
\end{figure*}

We demonstrate our approach by bounding the maximal per-volume extinction of a thin film.
The thin film comprises alternating layers of air and material ($\varepsilon=12 + 1i$) in a design space of length $L = 1.5 / k$. 
For each possible filling fraction of the material, we bound its maximal extinction by imposing the cardinality and two global-power-conservation constraints in Problem (\ref{eq:appenE-with-card}), giving the solid blue line in \figref{appenE-figure1}(a).
Dividing the bounds by its filling fraction gives bounds for per-volume extinction of Problem (\ref{eq:appenE-per-volume}), shown as the blue solid line in \figref{appenE-figure1}(b), its peak predicting the optimal filling ratio is 20\%.
The cardinatity constraint in Problems (\ref{eq:appenE-with-card}, \ref{eq:appenE-per-volume}) is important, without which the bound, the dotted cyan lines in \figref{appenE-figure1}(a, b), can not capture the trade-off between the maximal response and the filling fraction.
Comparing against the bounds are the responses of unstructured thin films (length being $L$ times the filling fraction) and structured designs at each filling fraction in \figref{appenE-figure1}. Red solid line represents the former; black dots, the latter.  
Each black dot is a possible material arrangement, randomly generated for at most 10000 designs for each filling fraction, reaching at least 88\% of the bound. 
The unstructured thin film (red line) suggests the 10\% filling fraction is optimal, but with structuring (black dots), we see 20\% is optimal, a result that is already predicted by the bounds before any actual designs.

To summarize, this appendix bounds the maximal per-volume response by constraining material allocation.
The allocated material is expressed as a cardinality constraint of the polarization field, later relaxed to a convex constraint.
This method is demonstrated in a thin-film design, tightly bounding the maximal per-volume extinction, and correctly predicting the optimal filling ratio of materials before any local designs.

The per-volume bound in this appendix only applies to small-scale systems.
The cardinality constraint, in particular, makes the bound numerically difficult to solve.
This constraint is non-differentiable, and remains so after the convex relaxation in \eqref{appenE-X}, which involves $|X_{ij}|$.
We smooth this absolute value by replacing it with a slack variable $|X_{ij}|\rightarrow t_{ij}$ and constrain $t_{ij}$ with the inequality $-t_{ij}\leq X_{ij} \leq t_{ij}$ (a common method employed in standard convex solvers such as CVX~\cite{cvx}).
 This introduces $n^2$ new variables and $2n^2$ new constraints, making the resulting optimization problem difficult to solve for $n\geq 50$. 
A future direction is to develop solvers (as in Ref.~\cite{d2004direct, beck2019fom}]) to directly optimize $l_1$-constrained semidefinite programs relaxed from Problem (\ref{eq:appenE-with-card}). 
Similar to our discussion in Chapter~\ref{chap:conclu}, once we have formulated the bound, the emphasise shifts to accelerate its computation for large-scale photonic designs.

\chapter{Power conservation laws in nonlinear nanophotonics and possible upper bounds}
\label{appen:nonlinear}
Strong nonlinear light--matter interaction can be induced by structuring materials at the wavelength scale~\cite{smirnova_multipolar_2016, koshelev_subwavelength_2020}, but the fundamental limit of such structuring is unclear. In this appendix, we extend our framework to nonlinear nanophotonics by formulating design problems as optimization problems that are amendable to upper bounds. 
The key is to generalize the power conservation laws to include couplings between different harmonics, which we illustrate in the case of second-harmonic generations. In the end, we point out key numerical challenges for solving the bounds and possible acceleration methods.

\hiddensection{Power conservation laws in nonlinear nanophotonics}
\label{sec:appenF-power}
This section establishes several power conservation laws for a scatterer with nonlinear response. Assuming the incident field is at a single frequency $\omega$, nonlinear interaction generates different harmonics, but at the original frequency $\omega$, its total field $e$ can still be separated into the incident field $\ei$ and a scattered field $\es$, both at frequency $\omega$: 
\begin{equation}
e = \ei + \es.
\label{eq:appenF-vie1}
\end{equation}
The scattered field is radiated by the polarization field $p$ at frequency $\omega$ within the nonlinear scatterer through the background-Green's-function operator $\GG_0$: $\es = \GG_0 p$. The polarization field $p$ includes both the linear polarization $\pL$ and the nonlinear polarization $\pNL$: $p = \pL + \pNL$. The linear polarization $\pL$ follows the usual definition of $\pL=\chi e$. The nonlinear polarization $\pNL$ is defined as the difference between $p$ and $\pL$, incorporating the nonlinear property of the material, and is zero if the material is linear.

\begin{figure*}[t]
  \centering
  \includegraphics[width=1\linewidth]{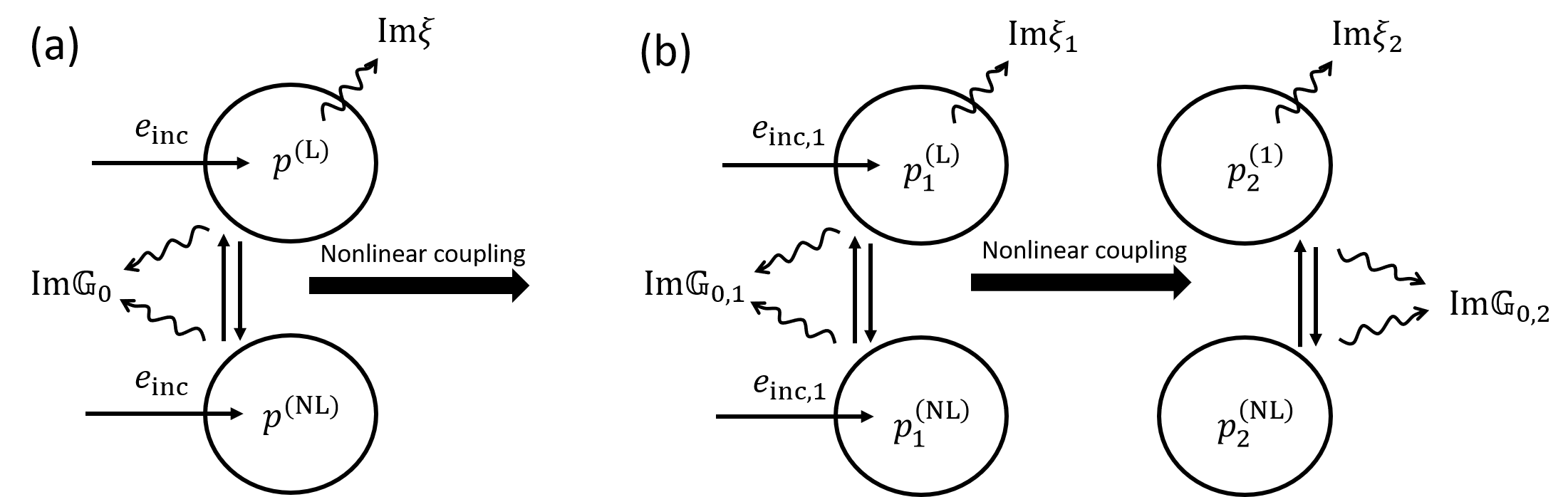}
\caption{(a) Power flow diagram depicting the physical meaning of each term in \eqref{appenF-power_flow2}. Dissipation in the system is marked by squiggle arrows. (b) Power flow in second-harmonic generation, depicting the physical meaning of each term in \eqref{appenF-SHG-power_flow}. Dissipation in the system is marked by squiggle arrows.}
	\label{fig:appenF-figure1}
\end{figure*}

A power conservation law can be derived by taking the inner product of \eqref{appenF-vie1} with the polarization field $p$ and rearrange the terms:
\begin{equation}
-p^\dagger \ei = p^\dagger \es - p^\dagger e.
\end{equation}
Taking the imaginary part of this equation yields
\begin{equation}
\Im(\ei^\dagger p) =\Im(p^\dagger \es) - \Im(p^\dagger e).
\label{eq:appenF-globalconstraint}
\end{equation}
In linear optics, the last term corresponds to material absorption in the scatterer. In nonlinear optics, it also contains the nonlinear coupling from the frequency $\omega$ to other frequencies. To separate these two effect, we insert the relation $p = \pL + \pNL$ into \eqref{appenF-globalconstraint}, yeilding
\begin{equation}
\underbrace{\Im(\ei^\dagger p)}_{\rm extinction} = \underbrace{\Im(p^\dagger \es)}_{\rm scattering} + \underbrace{\Im(-\pLd e)}_{\rm absorption} + \underbrace{\Im(-\pNLd e)}_{\rm nonlinear\ coupling}
\end{equation}
Multiplied by a factor of $\omega / 2$, this equation describes the power conservation in the presence of nonlinear polarization field: extinction = scattering + absorption + nonlinear coupling to other frequencies.

We express the power conservation law only in terms of polarization fields $\pL$ and $\pNL$ (which later will be our optimization variables), in two steps. First, we rewrite the scattered field with the total polarization field $\es = \GG_0 p$, and rewrite the total field by linear polarization field $-e = \xi \pL$ where $\xi = -\chil$ (this is only valid inside the material, but can be extended to air region in the quadratic equation below):
\begin{equation}
\Im(\ei^\dagger p) = \Im(p^\dagger \GG_0 p) + \Im(\pLd \xi \pL) + \Im(\pNLd \xi \pL).
\label{eq:appenF-power_flow1}
\end{equation}
Then, we differentiate the contribution of the nonlinear polarization $\pNL$ from the linear polarization $\pL$:
\begin{align}
    \Im(\ei^\dagger \pL) + \Im(\ei^\dagger \pNL) =&  \Im(\pLd \xi \pL) + \Im(\pNLd \xi \pL) + \Im(\pLd \GG_0 \pL) \label{eq:appenF-power_flow2} \\
    &  + \Im(\pLd \GG_0 \pNL) + \Im(\pNLd \GG_0 \pL) + \Im(\pNLd \GG_0 \pNL). \nonumber 
\end{align}
As illustrated in \figref{appenF-figure1}(a), this equation describes the power flow in and out of the system and between the linear and nonlinear polarization fields. It constraints the possible distributions of the polarization fields, leading to formulations of upper bounds in Section~\ref{sec:appenF-maximal} below.

\hiddensection{Power conservation in second-harmonic generations}
\label{sec:appenF-second}
The nonlinear power conservation law of \eqref{appenF-power_flow1} applies to second-harmonic generations, which has a fundamental frequency $\omega_1$ and a second harmonic frequency $\omega_2=2\omega_1$. Polarization fields at both frequencies satisfy  \eqref{appenF-power_flow1}:
\begin{align}
\omega_1\Im(e_{\text{inc}, 1}^\dagger p_1) = \omega_1\Im(p_1^\dagger \GG_{0,1} p_1)  + \omega_1\Im(\pLd_1 \xi_1 \pL_1) + \omega_1\Im(\pNLd_1 \xi_1 \pL_1),\label{eq:appenF-SHG-1} \\
\omega_2\Im(e_{\text{inc}, 2}^\dagger p_2) = \omega_2\Im(p_2^\dagger \GG_{0,2} p_2)  + \omega_2\Im(\pLd_2 \xi_2 \pL_2) + \omega_2\Im(\pNLd_2 \xi_2 \pL_2).\label{eq:appenF-SHG-2}
\end{align}
The subscript $i$ denotes quantities at frequency $\omega_i$. The frequency prefactor $\omega_i$ maps the quantities in Eqs.~(\ref{eq:appenF-SHG-1}, \ref{eq:appenF-SHG-2}) to true power quantities. The Green's function operators $\GG_{0,1}$ and $\GG_{0,2}$ are the same background Green's function evaluated at frequencies $\omega_1$ and $\omega_2$, respectively.

In second-harmonic generations, the nonlinear polarization fields originate from the electric fields: $\pNL_1(\vx) = 4\deff e^*_1(\vx)e_2(\vx)$, and $\pNL_2(\vx) = 2\deff e^2_1(\vx)$, where we assume the Kleinman symmetry so two equations share the same nonlinear coefficient $\deff$. Further assuming $\deff$ is real, the two nonlinear coupling terms in Eq. (\ref{eq:appenF-SHG-1}) and Eq. (\ref{eq:appenF-SHG-2}) become negative to each other:
\begin{equation}
\omega_1\Im(\pNLd_1 \xi_1 \pL_1) = - \omega_2\Im(\pNLd_2 \xi_2 \pL_2). \label{eq:appenF-SHG-3}
\end{equation}
The left-hand side represents the power lost at frequency $\omega_1$ through nonlinear coupling; the right-hand side power gained at frequency $\omega_2$ through the same coupling. As expected, this nonlinear coupling conserves energy. 

Equating the nonlinear coupling terms in Eqs.~(\ref{eq:appenF-SHG-1}) and (\ref{eq:appenF-SHG-2}) gives a single power-conservation law:
\begin{multline*}
\omega_1\Im(e_{\text{inc}, 1}^\dagger p_1) = \omega_1\Im(p_1^\dagger \GG_{0,1} p_1) + \omega_1\Im(\pLd_1 \xi_1 \pL_1) + \omega_2\Im(p_2^\dagger \GG_{0,2} p_2) + \omega_2\Im(\pLd_2 \xi_2 \pL_2),
\end{multline*}
where we assume a common scenario where one pump is incident at the fundamental frequency $\omega_1$ and no pump at $\omega_2$, so that the incident field at $\omega_2$ is zero.
The left-hand side of the equation is the input power; the right-hand side contain different sources of dissipation. If we explicitly write out the contributions from linear and nonlinear polarization fields, the power-conservation equation becomes:
\begin{align}
&\omega_1\Im(e_{\text{inc}, 1}^\dagger \pL_1)  + \omega_1\Im(e_{\text{inc}, 1}^\dagger \pNL_1) =  \omega_1\Im(\pLd_1 \xi_1 \pL_1)  + \omega_2\Im(\pLd_2 \xi_2 \pL_2) \label{eq:appenF-SHG-power_flow} \\
& +  \omega_1\Im(\pLd_1 \GG_{0,1} \pL_1) + \omega_1\Im(\pLd_1 \GG_{0,1} \pNL_1) + \omega_1\Im(\pNLd_1 \GG_{0,1} \pL_1) + \omega_1\Im(\pNLd_1 \GG_{0,1} \pNL_1)  \nonumber \\
& + \omega_2\Im(\pLd_2 \GG_{0,2} \pL_2) + \omega_2\Im(\pLd_2 \GG_{0,2} \pNL_2) + \omega_2\Im(\pNLd_2 \GG_{0,2} \pL_2) + \omega_2\Im(\pNLd_2 \GG_{0,2} \pNL_2). \nonumber 
\end{align}
This equation describes all the possible power flow in second-harmonic generations, their physical meanings depicted in \figref{appenF-figure1}(b).

\hiddensection{Maximal nonlinear coupling under power-conservation constraints}
\label{sec:appenF-maximal}
The power conservation laws in the last section constrain the possible nonlinear conversions between frequencies. In this section,  we use them to formulate four optimization problems in increasingly more concrete nonlinear scenarios. Those optimizations are amendable to convex relaxation, leading to potential upper bounds on the maximal nonlinear response under material structuring. 

\paragraph{(A) Maximal coupling out of a frequency}
Nonlinearity in the material scatters power out of the incident frequency, the exact amount given by the quantity $\Im(\pNLd \xi \pL)$ in \eqref{appenF-power_flow1}.  Maximizing it under the constraint of \eqref{appenF-power_flow1} equals to:
\begin{equation}
    \begin{aligned}
        & \underset{\pL, \pNL}{\text{maximize}}
        & & \Im(\pNLd \xi \pL) \\
        & \text{subject to}
        & & \Im(\ei^\dagger p) = \Im(p^\dagger \GG_0 p) + \Im(\pLd \xi \pL) + \Im(\pNLd \xi \pL). 
    \end{aligned}
    \label{eq:appenF-opt_prob1}
\end{equation}
With only one constraint, this QCQP is analytical tractable. It is also oblivious to exactly what nonlinearity couples the power out. Therefore, its bound represents the maximal possible coupling out of a frequency regardless of the underlying nonlinear effect.

\paragraph{(B) Maximal coupling between two frequencies}
How much power can be coupled from one frequency to another in a second-harmonic generation? We formulate this question as an optimization problem. The objective is to maximize the radiation from the second harmonic $\omega_2$ under a pump at the fundamental frequency $\omega_1$. The constraints are the power conservation laws at the two frequencies, Eqs.~(\ref{eq:appenF-SHG-1}, \ref{eq:appenF-SHG-2}), as well as the conservation of power flow between the two frequencies, Eq.~(\ref{eq:appenF-SHG-3}). With $\pL_1$, $\pNL_1$, $\pL_2$, $\pNL_2$ as optimization variables, we have
\begin{equation}
    \begin{aligned}
        & \underset{\pL_1, \pNL_1, \pL_2, \pNL_2}{\text{maximize}}
        & & \omega_2\Im(p_2^\dagger \GG_{0,2} p_2) \\
        & \text{subject to}
        & & \omega_1\Im(\pNLd_1 \xi_1 \pL_1) = \omega_1\Im(e_{\text{inc}, 1}^\dagger p_1) - \omega_1\Im(p_1^\dagger \GG_{0,1} p_1) - \omega_1\Im(\pLd_1 \xi_1 \pL_1) \\
        & & -&\omega_2\Im(\pNLd_2 \xi_2 \pL_2) = \omega_2\Im(p_2^\dagger \GG_{0,2} p_2) + \omega_2\Im(\pLd_2 \xi_2 \pL_2) \\
        & & -&\omega_2\Im(\pNLd_2 \xi_2 \pL_2) = \omega_1\Im(\pNLd_1 \xi_1 \pL_1).
    \end{aligned}
    \label{eq:appenF-opt_prob2}
\end{equation}
This is a QCQP with three constraints, which can be upper bounded via semi-definite relaxation~\cite{luo_semidefinite_2010}, as in the computational bound in Chapter~\ref{chap:comp}. The bound represents the maximal possible efficiency of the second-harmonic generation.

\paragraph{(C) Maximal coupling under specific nonlinear susceptibility}
The two previous formulations neglect the underlying nonlinear susceptibility; their bounds are general but may be loose for common nonlinear materials. To include the nonlinear susceptibility $\deff$, we consider the definition of the nonlinear polarization fields:  $\pNL_1(\vx) = 4\deff e^*_1(\vx)e_2(\vx)$ and $\pNL_2(\vx) = 2\deff e^2_1(\vx)$. Replacing the electric fields with polarization fields via the constitutive relation $e_i(\vx) = \xi_i \pL_i(\vx)$, we obtain additional quadratic constraints on the polarization fields. Including these constraints into Problem (\ref{eq:appenF-opt_prob2}) yields
\begin{equation}
    \begin{aligned}
        & \underset{\pL_1, \pNL_1, \pL_2, \pNL_2}{\text{maximize}}
        & & \omega_2\Im(p_2^\dagger \GG_{0,2} p_2) \\
        & \text{subject to}
        & & \omega_1\Im(\pNLd_1 \xi_1 \pL_1) = \omega_1\Im(e_{\text{inc}, 1}^\dagger p_1) - \omega_1\Im(p_1^\dagger \GG_{0,1} p_1) - \omega_1\Im(\pLd_1 \xi_1 \pL_1) \\
        & &  -&\omega_2\Im(\pNLd_2 \xi_2 \pL_2) = \omega_2\Im(p_2^\dagger \GG_{0,2} p_2) + \omega_2\Im(\pLd_2 \xi_2 \pL_2) \\
        & &  -&\omega_2\Im(\pNLd_2 \xi_2 \pL_2) = \omega_1\Im(\pNLd_1 \xi_1 \pL_1). \\
        & & & \pNL_1(\vx) = 4\deff\xi_1^*\xi_2p^{(\text{L})*}_1(\vx)\pL_2(\vx), \quad {\rm for\ every\ } \vx\\
        & & & \pNL_2(\vx) = 2\deff\xi_1^2 p^{(\text{L})2}_1(\vx), \quad {\rm for\ every\ } \vx. 
    \end{aligned}
    \label{eq:appenF-opt_prob3}
\end{equation}
This optimization problem is a QCQP, again amendable to global bounds via semidefinite relaxation. The resulting bound, incorporating the specific nonlinear coefficient of the material, is the tightest bounds for nonlinear materials among all three we have discussed so far.

\paragraph{(D) Perturbation limit}
If the pump is weak, the nonlinear polarization can be treated as a perturbation, so that $p_1(\vx) \approx \pL_1(\vx) \gg \pNL_1(\vx)$.
For simplicity, we choose the objective as $\int_V |\pNL_2(\vx)|^2\text{d}\vx$ to characterize the amount of nonlinear generation at $\omega_2$. Considering the nonlinear coupling relation $\pNL_2(\vx) = 2\deff p^{(\text{L})2}_1(\vx)$ and the relation in the perturbation limit $\pL_1(\vx) \approx p_1(\vx)$, the objective function can be tranformed into a function of $p_1(\vx)$, which reads $4|d_{\rm eff}|^2\int_V |p_1(\vx)|^4\text{d}\vx.$ With this new objective function, problem (\ref{eq:appenF-opt_prob3}) reduces to:
\begin{equation}
    \begin{aligned}
        & \underset{p_1}{\text{maximize}}
        & & \int_V \left|p_1(\vx)\right|^4\text{d}\vx \\
        & \text{subject to}
        & & \Im(e_{\text{inc}, 1}^\dagger p_1) = \omega_1\Im(p^\dagger_1 \GG_{0,1} p_1) + \omega_1\Im(p^\dagger_1 \xi_1 p_1). 
    \end{aligned}
    \label{eq:appenF-opt_prob4}
\end{equation}
In the perturbation limit, the power conservation constraint degenerates to its linear counterpart, with a fourth-order objective function of the polarization field $p_1$. This is a quartic optimization problem, which also has known computational upper bounds~\cite{luo_semidefinite_2010-1}.

\hiddensection{Conclusion}
\label{sec:appenF-conclusion}
In this appendix, we generalize our power conservation laws to nonlinear nanophotonics by considering additional nonlinear polarization fields. The generalized conservation laws promise bounds on maximal nonlinear generations.
In the context of second-harmonic generation, we propose four upper bounds for increasingly more concrete scenarios. First, we show an analytically solvable upper bound in Problem (\ref{eq:appenF-opt_prob1}) that determines the maximal nonlinear coupling out of a given frequency. Then, we establish a tighter computational bound in Problem (\ref{eq:appenF-opt_prob2}) by including potential dissipation at the second frequency. Both Problems (\ref{eq:appenF-opt_prob1}) and (\ref{eq:appenF-opt_prob2}) solve for the maximal nonlinear conversion out of all possible nonlinear materials.
For a given nonlinear material, we formulate the tightest bound in Problem (\ref{eq:appenF-opt_prob3}), and simplify it in the perturbation limit to Problem (\ref{eq:appenF-opt_prob4}). 

Problems (\ref{eq:appenF-opt_prob3}, \ref{eq:appenF-opt_prob4}) are difficult to solve for designs beyond the wavelength scale due to their many constraints. The former, Problem (\ref{eq:appenF-opt_prob3}), is a QCQP with $2N+3$ constraints for $N$ number of descritized spatial points. The latter, Problem (\ref{eq:appenF-opt_prob3}), is a quartic problem which is equivalent to a QCQP with at least $N$ constraints~\cite{luo_semidefinite_2010-1}. Both problems suffer from the same O($N^4$) complexity as the computational bound in Chapter~\ref{chap:comp}. To solve them efficiently, we can replace the constraints with their weighted averages as in Chapter~\ref{chap:comp}. In addition, if the nonlinear design has a ``long'' dimension, the bounds may be susceptible to the sparse-based acceleration in Chapter~\ref{chap:sparse}.
Both acceleration methods may lead to bounds for nonlinear designs beyond the wavelength scale.

\chapter{Analytical bounds via optical theorem: supplementary}
\label{appen:sup_anal}

This appendix provides supplementary information to Chapter~\ref{chap:analy}, “Analytical bounds via optical theorem.”
We (1) establish our general framework through Lagrangian dual function, (2) derive expressions  for bounds on power quantities presented in Section~\ref{sec:chap2-analytical}, (3) derive bounds in the special case for plane wave incidence presented in Section~\ref{sec:chap2-plane}, (4) derive bounds for extended scatters which are greatly simplified for the special case of plane wave incidence, (5) discuss minimum thickness required for perfect absorbers, (6) lay out the inverse design procedure for the ultrathin absorber and (7) the corresponding optimal designs for the data presented in Fig.~\ref{fig:chap2-figure3}(b,c) and Fig.~\ref{fig:chap2-figure4}(c), (8) show how our formalism can incorporate previous predicted limits on nanoparticle scattering, thin film absorption, and thermal absorption. We provide real-space expressions for the eigen-expansions of imaginary part of the electric Green’s function operator for (9) a sphere and (10) a planar film. We (11) summarize expressions for upper bounds on three power quantities at different generality, with additional applications for (12) local density of states manifested from near field interaction and (13) thermal absorption and emission from incoherent sources.

\hiddensection{The optimization problem and its dual function}
\label{sec:appenG-sec1}
The optimization problem is to maximize a response function $f(p) = p^\dagger \mathbb{A} p +  \Im\left[\beta^\dagger p\right]$ under the optical-theorem constraint, where the variable $p$ is polarization field induced in the scatterer. Under a prespecified basis, parameter $\beta$ is a vector, and $\Abb$ is a Hermitian matrix. 
The same basis defines positive semidefinite matrix $\ImGO$ and $\Im\xi\II$, representing radiative and material loss in the system.
Following the standard optimization notation, we rewrite the original maximization problem as a minimization problem by adding a minus sign to the objective function:
\begin{equation}
\begin{aligned}
& \underset{p}{\text{minimize}}
& & -f(p) = -p^\dagger \mathbb{A} p -  \Im\left[\beta^\dagger p\right] \\
& \text{subject to}
& & p^\dagger \left\{\Im \xi\II + \ImGO \right\} p = \Im \left[\einc^\dagger p\right].
\end{aligned}
\label{eq:appenG-gen-form-min}
\end{equation}
The optimization problem stated in \eqref{appenG-gen-form-min} is known to have strong duality~\cite{Boyd2004}, prompting us to find its dual function, which in turn is defined by its Lagrangian:
\begin{equation}
L(p,\nu) = p^\dagger\BB(\nu)p - \Im\left[(\beta + \nu \einc)^\dagger p\right],
\label{eq:appenG-def-L}
\end{equation}
where we introduce dual variable $\nu$ and simplify our notation by introducing matrix
\begin{equation}
\BB(\nu) = -\mathbb{A}+\nu(\Im\xi\II+\ImGO).
\label{eq:appenG-def-B}
\end{equation}

The dual function $g(\nu)$ is defined as the minimum of Lagrangian $L(\phi,\nu)$ over variable $\phi$. We denote $\nu_0$ as the value of $\nu$ when the minimum eigenvalue of $\BB(\nu)$ is zero, leaving $\BB(\nu_0)$ a positive semidefinite matrix with at least one zero eigenvalue. For $\nu<\nu_0$, the positivity of $\Im\xi\II+\ImGO$ implies that $\BB(\nu)=\BB(\nu_0)-(\nu_0-\nu)(\Im\xi\II+\ImGO)$ has negative eigenvalues and $L(\phi,\nu)$ is unbounded below. For $\nu>\nu_0$, $\BB(\nu)=\BB(\nu_0)+(\nu-\nu_0)(\Im\xi\II+\ImGO)$ is positive definite, and $L(\phi,\nu)$ is convex in $\phi$ with a finite minimal value. This minimum is obtained at
\begin{equation}
p(\nu) = \frac{i}{2}\BB^{-1}(\nu)(\beta+\nu\einc),
\label{eq:appenG-phi}
\end{equation}
with the resulting dual function:
\begin{align}
g(\nu)= \min_p L(p, \nu) =
\begin{cases}
-\frac{1}{4}(\beta + \nu \einc)^\dagger \BB^{-1}(\nu)(\beta + \nu \einc) & \nu > \nu_0 
\\
-\infty & \nu< \nu_0.
\end{cases}
\label{eq:appenG-gnu}
\end{align}
Lastly, at $\nu=\nu_0$, if $\beta + \nu_0\einc$ is in the range of $\BB(\nu_0)$, then $L(p,\nu)$ is still convex and $g(\nu_0)$ takes the value of the first case in \eqref{appenG-gnu} with the inverse operator replaced by the pseudo-inverse; if not, then \eqref{appenG-def-L} suggests that $L(p,\nu)$ is unbounded below and  $g(\nu_0)\rightarrow -\infty$.

Due to strong duality, the optimization problem, \eqref{appenG-gen-form-min}, is solved by finding the maximum of the dual function:
\begin{equation}
\begin{aligned}
& \underset{\nu}{\text{maximize}}
& & g(\nu).
\end{aligned}
\end{equation}
According to \eqref{appenG-gnu}, dual function $g(\nu)$ is maximized at a value within range $[\nu_0,+\infty)$, which we denote as $\nu^*$. The maximum response function takes the (negative of the optimal dual) value:
\begin{align}
f_{\rm max} = \frac{1}{4}(\beta + \nu^* \einc)^\dagger \BB^{-1}(\nu^*)(\beta + \nu^* \einc),
\label{eq:appenG-Popt}
\end{align}
and the optimal polarization field $p$ is given by evaluating \eqref{appenG-phi} at $\nu^*$:
\begin{equation}
p^* = \frac{i}{2}\BB^{-1}(\nu^*)(\beta + \nu^* \einc) 
\label{eq:appenG-x},
\end{equation}
except when $\nu^*=\nu_0$, where the $p^*$ can not be uniquely determined due to the presence of zero eigenvalues in $\BB(\nu_0)$.

To solve for the maximum response function $f_{\rm max}$ in \eqref{appenG-Popt}, we need to find the optimal dual variable $\nu^*$, which can only occur either in the interior of the domain $[\nu_0,\infty)$ or its boundary. If $\nu^*$ is in the interior, it has to satisfy the condition:
\begin{align}
\frac{\partial g(\nu)}{\partial \nu}\biggr\rvert_{\nu=\nu^*} = 0.
\end{align}
This can be translated to a transcendental equation that determines the first possible optimum which we denote as $\nu_1$:

\begin{equation}
\begin{aligned}
2\Re\left\{\einc^\dagger\BB^{-1}(\nu_1)(\beta+\nu_1\einc)\right\}& \\
- (\beta+\nu_1\einc)^\dagger\BB^{-1}(\nu_1)&\BB'(\nu_1)\BB^{-1}(\nu_1)(\beta+\nu_1\einc) = 0.
\label{eq:appenG-optlambda}
\end{aligned}
\end{equation}
The concavity of the dual function $g(\nu)$ guarantees the uniqueness of the solution $\nu_1$. The lefthand side of \eqref{appenG-optlambda} is proportional to $-\partial g(\lambda)/\partial\nu$. Its derivative, $-\partial^2 g(\lambda)/\partial\nu^2$, is always non-negative based on the second-order condition of a concave function~\cite{Boyd2004}. Thus, if there is a $\nu_1$ satisfying \eqref{appenG-optlambda}, it can simply be solved by identifying where the sign of the lefthand side changes, using either bisection or Newton’s method.

Based on the concavity of the dual function, we can also argue that if $\nu_1$ exists in the domain $(\nu_0,\infty)$ then it must be the global optimizer of $g(\nu)$. If not, then there is no point in the domain at which the gradient is zero, and $\nu^*$ must be one of the boundary values of $[\nu_0,\infty)$; by the concavity of $g(\nu)$, the maximum must occur at $\nu_0$. Hence we have:
\begin{align}
\nu^* = 
\begin{cases}
\nu_1 & \textrm{if } \nu_1 \in (\nu_0,\infty) \\
\nu_0 & \textrm{else.}
\end{cases}
\label{eq:appenG-det_nu}
\end{align}
The self-consistency implicit in \eqref{appenG-optlambda} for $\nu_1$ can make it to difficult to ascertain whether $\nu_1$ or $\nu_0$ is optimal. Instead, if the derivative of $g(\nu)$ at $\nu_0$ is well-defined, we can check its value to determine whether $g(\nu)$ attains it extremum in the interior of its domain or on its boundary: if and only if it is positive, then $\nu_1$ will be in the interior of the domain $[\nu_0,\infty)$. Hence, if $\beta + \nu_0 \einc$ is in the range of  ${\BB^{-1}(\nu_0)}$, then we can also use the equivalent condition to determine $\nu^*$:
\begin{align}
\nu^* = 
\begin{cases}
\nu_1 & \textrm{if } 2\einc^\dagger\BB^{-1}(\nu_0)(\beta+\nu_0\einc)  \\
 & \quad\quad < (\beta+\nu_0\einc)^\dagger \BB^{-1}(\nu_0)\BB'(\nu_0)\BB^{-1}(\nu_0)(\beta+\nu_0\einc) \\
\nu_0 & \textrm{else.}
\end{cases}
\label{eq:appenG-det-nu-alt}
\end{align}

\hiddensection{Absorbed, scattered, and extinguished power expressions}
\label{sec:appenG-sec2}
We start with extinguished power which is linear in polarization field $p$: $\Pext = \frac{\omega}{2} \Im \left[ \einc^\dagger p \right]$. For simplicity, we take the objective function as $\Im \left[ \einc^\dagger p \right]$, and set $\mathbb{A} = 0$ and $\beta = \einc$ in the optimization problem, \eqref{appenG-gen-form-min}. The matrix defined in \eqref{appenG-def-B} becomes $\BB(\nu) = \nu (\Im\xi\II+\ImGO)$. Its minimum eigenvalue reaches zero when $\nu=\nu_0=0$. Dual function in \eqref{appenG-gnu} takes the form:
\begin{align}
g(\nu) =
\begin{cases}
-\frac{(\nu+1)^2}{4}\einc^\dagger (\Im\xi\II+\ImGO)^{-1}\einc & \nu > 0
\\
-\infty & \nu \leq 0,
\end{cases}
\end{align}
where we identified $g(\nu_0)=-\infty$ since $\beta+\nu_0\einc$ is not in the range of $B(\nu_0)$. Since $g(\nu_0)=-\infty$, the optimal dual variable $\nu^*$ can only be chosen at $\nu_1$. Solving \eqref{appenG-optlambda} gives $\nu_1=1$ and the maximum extinction given by \eqref{appenG-Popt} is (after adding back the $\frac{\omega}{2}$ prefactor):
\begin{align}
\Pext^{\rm max} = \frac{\omega}{2} \einc^\dagger \left(\Im\xi\II+\ImGO\right)^{-1} \einc.
\label{eq:appenG-Pext_bound}
\end{align}
The optimum polarization field $p^*$ is given by \eqref{appenG-x}:
\begin{equation}
p^* = i(\Im\xi\II+\ImGO)^{-1}\einc.
\end{equation}

Absorption has the form $\Pabs = \frac{\omega}{2} p^\dagger (\Im\xi) p$. Taking the objective function as $p^\dagger (\Im\xi) p$, we have $\mathbb{A} = \Im \xi\II$ and $\beta = 0$ in the optimization problem, \eqref{appenG-gen-form-min}. The matrix defined in \eqref{appenG-def-B} becomes $\BB(\nu) = (\nu-1)\Im\xi\II+\nu\ImGO$. Dual function takes the form of \eqref{appenG-gnu}:
\begin{align}
g(\nu) =
\begin{cases}
-\frac{\nu^2}{4}\einc^\dagger [(\nu-1)\Im\xi\II+\nu\ImGO]^{-1}\einc & \nu > \nu_0
\\
-\infty & \nu < \nu_0,
\end{cases}
\label{eq:gnu_abs}
\end{align}
At $\nu=\nu_0$, the value of $g(\nu_0)\rightarrow-\infty$ if $\einc$ is not in the range of $\BB(\nu_0)$, otherwise $g(\nu_0)$ takes the form of the first case in \eqref{gnu_abs} with the inverse replaced by pseudo-inverse.
As in \eqref{appenG-det_nu}, the optimal dual variable $\nu^*$ is obtained either at the  interval $(\nu_0,\infty)$ or its boundary $\nu_0$. The value of $\nu_0$ depends on the nature of both $\Im\xi\II$ and $\ImGO$. The value of $\nu_1$ is given by \eqref{appenG-optlambda}:
\begin{equation}
\einc^\dagger\left[ 2\BB^{-1}(\nu_1) - \nu_1\BB^{-1}(\nu_1) (\Im\xi\II+\ImGO) \BB^{-1}(\nu_1) \right]\einc = 0.
\label{eq:appenG-nu1_abs}
\end{equation}
Using \eqref{appenG-Popt} and adding back the $\frac{\omega}{2}$ prefactor, we have maximum absorption:
\begin{align}
\Pabs^{\rm max} = \frac{\omega}{2}\frac{\nu^{*2}}{4} \einc^\dagger[(\nu^*-1)\Im\xi\II+\nu^*\ImGO]^{-1} \einc.
\label{eq:appenG-Pabs_bound}
\end{align}
The optimal current can be determined by \eqref{appenG-x} in the case of $\nu^*=\nu_1$:
\begin{equation}
p^* = i\frac{\nu^*}{2}[(\nu^*-1)\Im\xi\II+\nu^*\ImGO]^{-1}\einc.
\end{equation}

Scattering power has the form $\Pscat = \frac{\omega}{2} p^\dagger (\ImGO) p$, such that $\mathbb{A} = \ImGO$ and $\beta=0$ after suppressing the $\frac{\omega}{2}$ prefactor. Following a similar procedure as absorption, we have maximum scattering as:
\begin{align}
\Pscat^{\rm max} = \frac{\omega}{2}\frac{\nu^{*2}}{4} \einc^\dagger[\nu^*\Im\xi\II + (\nu^*-1)\ImGO]^{-1} \einc.
\label{eq:appenG-Psca_bound}
\end{align}
Again, $\nu^*$ takes two possible values: $\nu_1$ and $\nu_0$, as dictated by \eqref{appenG-det_nu}. The determinant equation for $\nu_1$ takes the same form as \eqref{appenG-nu1_abs} with $\BB(\nu) = \nu\Im\xi\II+(\nu-1)\ImGO$.

An equivalent formulation for all three power quantities is to write them as the difference (or sum) of the other two. For example, scattering power can be written as the difference between extinction and absorption: $\Pscat = \frac{\omega}{2} \Im \left[ \einc^\dagger p \right] - \frac{\omega}{2} p^\dagger (\Im\xi) p$. With $\Abb=-\Im\xi\II$ and $\beta=\einc$ after suppressing the $\frac{\omega}{2}$ prefactor, this gives the same optima as in \eqref{appenG-Psca_bound} but with a different form:
\begin{align}
\Pscat^{\rm max} = \frac{\omega}{2} \frac{(1+\nu^*)^2}{4} \einc^\dagger \left[(\nu^*+1)\Im\xi\II + \nu^*\ImGO \right]^{-1}\einc,
\label{eq:appenG-Psca_bound_2nd}
\end{align}
where the optimal dual variable $\nu^*$ is determined by \eqref{appenG-det_nu}.

\hiddensection{Bounds for a nonmagnetic scalar material under plane wave incidence}
\label{sec:appenG-sec3}
Let us consider a typical case where the incident field is a plane wave and the scatterer is composed of nonmagnetic scalar material. 
Because $\Im \GG_0$ is positive-semidefinite, we can simplify its eigendecomposition to write $\Im\GG_0 = \Vbb\Vbb^\dagger$, where the columns of $\Vbb$, which we denote $v_i$, form an orthogonal basis of polarization fields. They are normalized such that the set $\rho_i=   v_i^\dagger    v_i$ are the eigenvalues of $\Im \GG_0$ and represent the powers radiated by unit-normalization polarization fields. The expansion of incident plane wave, $\einc$, in these channels is assumed to be: $\einc = \frac{1}{k^{3/2}}\sum_i e_i    v_i$, where the exact value of $|e_i|^2$ depends on the choice of $   v_i$. 

We decompose general bounds given by Eq. (\ref{eq:appenG-Pext_bound}, \ref{eq:appenG-Pabs_bound}, \ref{eq:appenG-Psca_bound}) into contributions from these channels: 
\begin{align}
P_{\rm ext} &\leq \frac{\omega}{2}\frac{1}{k^3}\sum_i |e_i|^2\frac{\rho_i}{\Im{\xi} + \rho_i} \label{eq:appenG-pw_ext}\\
P_{\rm abs} &\leq \frac{\omega}{2} \frac{\nu^{*2}}{4}\frac{1}{k^3}\sum_i |e_i|^2\frac{\rho_i}{(\nu^*-1)\Im{\xi}+\nu^*\rho_i} \label{eq:appenG-pw_abs} \\
P_{\rm scat} &\leq \frac{\omega}{2} \frac{\nu^{*2}}{4}\frac{1}{k^3}\sum_i |e_i|^2\frac{\rho_i}{\nu^*\Im{\xi}+(\nu^*-1)\rho_i} \label{eq:appenG-pw_sca}.
\end{align}
Taking $\omega=k$ in our unitless convention and write $k=2\pi/\lambda$ gives the expressions presented in \eqrefrange{chap2-ext-gen}{chap2-sca-gen} in Chapter~\ref{chap:analy}.
Bounds for both absorption and scattering contain $\nu^*$, which is determined by \eqref{appenG-det_nu}. For absorption, $\nu_0 = 1$, and $\nu_1$ is computationally evaluated by solving the following equation:
\begin{equation}
\sum_i |e_i|^2 \frac{(\nu_1-2)\Im{\xi} + \nu_1\rho_i}{[(\nu_1-1)\Im{\xi}+\nu_1\rho_i]^2} = 0.
\end{equation}
For scattering bound, $\nu_0=\rho_{\rm max}/(\rho_{\rm max}+\Im{\xi})$, where $\rho_{\rm max}$ is the largest $\rho_i$. The other potential optimum, $\nu_1$, is solved computationally through equation:
\begin{equation}
\sum_i |e_i|^2\rho_i \frac{(\nu_1-2)\rho_i + \nu_1\Im{\xi}}{[(\nu_1-1)\rho_i+\nu_1\Im{\xi}]^2} = 0.
\end{equation}

Bounds on maximal cross sections for a finite-size scatterer is obtained by normalizing \eqrefrange{appenG-pw_ext}{appenG-pw_sca} by plane wave intensity $|E_0|^2/2$ (the vacuum resistance $Z_0=1$):
\begin{align}
\sigma_{\rm ext} &\leq \frac{\lambda^2}{4\pi^2|E_0|^2}\sum_i |e_i|^2\frac{\rho_i}{\Im{\xi} + \rho_i}  \\
\sigma_{\rm abs} &\leq \frac{\lambda^2}{4\pi^2|E_0|^2}\frac{\nu^{*2}}{4}\sum_i |e_i|^2\frac{\rho_i}{(\nu^*-1)\Im{\xi}+\nu^*\rho_i}  \\
\sigma_{\rm scat} &\leq \frac{\lambda^2}{4\pi^2|E_0|^2}\frac{\nu^{*2}}{4}\sum_i |e_i|^2\frac{\rho_i}{\nu^*\Im{\xi}+(\nu^*-1)\rho_i}. 
\end{align}

For a plane wave incidence with $|e_i|^2=\pi(2n+1)\delta_{m,\pm 1}|E_0|^2$, we can simplify the above expression by summing over index $m$ within $i=\{m,n,j\}$, leaving contributions indexed only by total angular momentum $n$ and polarization state $j$:
\begin{align}
\sigma_{\rm ext} &\leq \frac{\lambda^2}{2\pi}\sum_{n,j} (2n+1)\frac{\rho_{n,1,j}}{\Im{\xi} + \rho_{n,1,j}}  \label{eq:appenG-sigma_ext_n}\\
\sigma_{\rm abs} &\leq \frac{\lambda^2}{2\pi}\frac{\nu^{*2}}{4}\sum_{n,j} (2n+1)\frac{\rho_{n,1,j}}{(\nu^*-1)\Im{\xi}+\nu^*\rho_{n,1,j}}  \\
\sigma_{\rm scat} &\leq \frac{\lambda^2}{2\pi}\frac{\nu^{*2}}{4}\sum_{n,j} (2n+1)\frac{\rho_{n,1,j}}{\nu^*\Im{\xi}+(\nu^*-1)\rho_{n,1,j}}. 
\end{align}
In Fig.\ref{fig:chap2-figure2}(c) of Chapter~\ref{chap:analy}, we use the notation $\sigma_{\rm ext,n}$ to denote the contribution from the n-th channel in the summation of \eqref{appenG-sigma_ext_n}.

\hiddensection{General bound for extended scatterers}
\label{sec:appenG-sec4}
In a planar bounding volume for extended scatterers, the most general far-field incidence has the expansion:
\begin{equation}
\einc = \frac{1}{k^{3/2}} \sum_i \int_{k_\parallel\leq k} e_i(\kv_\parallel)    v_i(\kv_\parallel)\frac{\text{d}\kv_\parallel}{(2\pi)^2},
\label{eq:appenG-cont_inc}
\end{equation}
where index $i=\{s,p\}$. Plugging the expansion of $\einc$ in Eq. (\ref{eq:appenG-Pext_bound}, \ref{eq:appenG-Pabs_bound}, \ref{eq:appenG-Psca_bound}) gives the integral form of cross-sections bounds after normalization by the z-directed plane wave intensity $|E_0|^2k_z/2k$:
\begin{align}
\sigma_{\rm ext} \leq& \frac{1}{kk_z|E_0|^2}\sum_i\int_{k_\parallel \leq k} |e_i(\kvp)|^2\frac{\rho_i(\kvp)}{\Im\xi + \rho_i(\kvp)}\frac{\text{d}\kv_\parallel}{(2\pi)^2} \label{eq:appenG-ext-cont} \\
\sigma_{\rm abs}\leq& \frac{1}{kk_z|E_0|^2}\frac{\nu^{*2}}{4} 
\sum_i\int_{k_\parallel \leq k} |e_i(\kvp)|^2\frac{\rho_i(\kvp)}{(\nu^*-1)\Im\xi + \nu^*\rho_i(\kvp)}\frac{\text{d}\kv_\parallel}{(2\pi)^2} \label{eq:appenG-abs-cont} \\
\sigma_{\rm scat} \leq& \frac{1}{kk_z|E_0|^2}\frac{\nu^{*2}}{4} \sum_i\int_{k_\parallel \leq k} |e_i(\kvp)|^2\frac{\rho_i(\kvp)}{\nu^*\Im\xi + (\nu^*-1)\rho_i(\kvp)}\frac{\text{d}\kv_\parallel}{(2\pi)^2}. \label{eq:appenG-sca-cont}
\end{align}

Now we restrict our scope to a plane wave incidence with total wave vector $\vect{k}=k_x\vect{\hat{e}}_x+k_y\vect{\hat{e}}_y+k_z\vect{\hat{e}}_z$  and polarization $p'$. We denote its parallel wave vector as $\kvp'=k_x\vect{\hat{e}}_x+k_y\vect{\hat{e}}_y$ where the $'$ symbol differentiates $\kvp'$ from $\kvp$ that is used to label different channels in \eqref{appenG-cont_inc}. The plane wave has the expression:
\begin{equation}
\einc = E_0\hat{\ev}e^{i\kvp'\cdot\vect{r}_\parallel}e^{ik_zz},
\label{eq:appenG-pw_in_pw}
\end{equation}
where $\hat{\ev}$ is a unit vector denotes incident polarization, taking the form $(k_y\vect{\hat{e}}_x-k_x\vect{\hat{e}}_y)/k_\parallel'$ for $p'=M$, and $(-k_z\vect{\hat{k}'_\parallel}+k_\parallel'\vect{\hat{e}}_z)/k$ for $p'=N$. Equating \eqref{appenG-pw_in_pw} with \eqref{appenG-cont_inc} gives the expansion coefficients, $e_i(\kvp)$. 
Plugging its absolute value $|e_i(\kvp)| = |E_0|\sqrt{2k_zk}(2\pi)^2\delta(\kvp'-\kvp)\delta_{p,p'}$ into \eqrefrange{appenG-ext-cont}{appenG-sca-cont} gives bounds for plane wave incidence:
\begin{align}
\sigma_{\rm ext} / A & \leq 2\sum_{s=\pm} \frac{\rho_{s,p'}(\kvp')}{\Im\xi+\rho_{s,p'}(\kvp')} \\
\sigma_{\rm abs} / A &\leq \frac{\nu^{*2}}{2}\sum_{s=\pm}\frac{\rho_{s, p'}(\kvp')}{(\nu^*-1)\Im{\xi}+\nu^*\rho_{s, p'}(\kvp')} \label{eq:appenG-abs-cha-extend} \\
\sigma_{\rm scat} / A &\leq \frac{\nu^{*2}}{2}\sum_{s=\pm}\frac{\rho_{s, p'}(\kvp')}{\nu^*\Im{\xi}+(\nu^*-1)\rho_{s, p'}(\kvp')}, 
\end{align}
where we identified factor $A=(2\pi)^2\delta^2(0)$ corresponding to total surface area.

Bounds for both absorption and scattering contain $\nu^*$, which is determined by \eqref{appenG-det_nu}. For absorption, $\nu_0 = 1$, and $\nu_1$ is computationally evaluated by solving the following equation:
\begin{equation}
\sum_{s=\pm} \frac{(\nu_1-2)\Im\xi + \nu_1\rho_{s,p'}(\kvp')}{\left[(\nu_1-1)\Im\xi+\nu_1\rho_{s,p'}(\kvp')\right]^2}=0.
\label{eq:appenG-ext-det-nu}
\end{equation}	
For scattering, $\nu_0=\rho_{\rm max}/(\rho_{\rm max}+\Im{\xi})$, where $\rho_{\rm max}$ is the largest $\rho_i$. The other potential optimum, $\nu_1$, is solved computationally through equation:
\begin{equation}
\sum_{s=\pm} \rho_{s,p'}(\kvp') \frac{(\nu_1-2)\rho_{s,p'}(\kvp') + \nu_1\Im\xi}{\left[(\nu_1-1)\rho_{s,p'}(\kvp')+\nu_1\Im\xi\right]^2}=0.
\label{eq:appenG-ext-det-nu-sca}
\end{equation}	

\begin{figure}[t!]
	\includegraphics[width=0.45\textwidth]{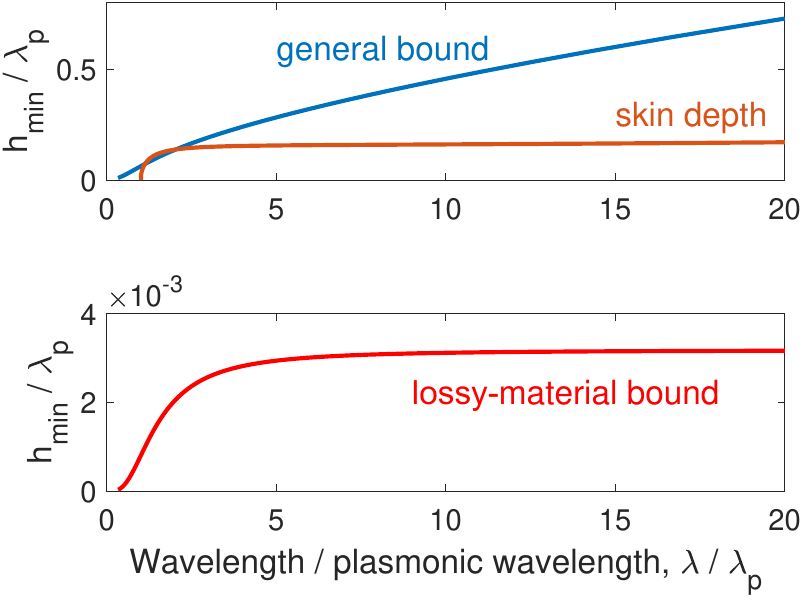}
	\centering
	\caption{Comparison between skin depth and minimum thickness, $h_{\rm min}$, for a perfect absorber as a function of wavelength $\lambda$ for a Drude metal. General bound and lossy-material bound are shown seperately in two plots. The former one gives a much more modest prediction.
		All quantities are normalized by the plasmonic wavelength, $\lambda_p$, of the Drude metal.}
	\label{fig:appenG-figure7}
\end{figure}

\hiddensection{Minimum thickness for perfect absorbers}
\label{sec:appenG-sec5}
Following Section~\ref{sec:appenG-sec4}, this section studies minimum thickness required for a perfect absorber that has 100\% absorption. Usually, one determines the optimal $\nu^*$ in \eqref{appenG-abs-cha-extend} by comparing the values of $\nu_0$ and $\nu_1$. Here, we take an alternative approach introduced through \eqref{appenG-det-nu-alt}, where the derivative of the dual function at $\nu_0$ is used as a threshold, giving a explicit expression for maximum absorption cross section:
\begin{align}
&\sigma_{\rm abs} / A  \leq \begin{cases}
\frac{\nu_1^2}{2}\sum_{s=\pm}\frac{\rho_{s, p'}(\kvp')}{(\nu_1-1)\Im{\xi}+\nu_1\rho_{s, p'}(\kvp')}, &\frac{1}{\Im{\xi}} > \frac{1}{2}\left[\frac{1}{\rho_{+,p'}(\vect{k}_\parallel)} + \frac{1}{\rho_{-,p'}(\vect{k}_\parallel)}\right] \\
1, &  \text{otherwise.}
\end{cases} 
\label{eq:appenG-film_abs}
\end{align}
Threshold for maximum absorption (at a given incident angle) corresponds to the condition:
\begin{equation}
\frac{1}{\Im{\xi}} = \frac{1}{2}\left[\frac{1}{\rho_{+,p'}(\vect{k}_\parallel)} + \frac{1}{\rho_{-,p'}(\vect{k}_\parallel)}\right],
\end{equation}
where we can solve for its required minimum thickness:
\begin{equation}
h_{\rm min} = \frac{k_z}{k^2}\frac{4\Im{\xi}}{1-\text{sinc}^2(k_zh_{\rm min})}.
\end{equation}
Under normal incidence ($k_z=k$), when the absorber is much thinner than the wavelength, $kh_{\rm min}\rightarrow 0$, it can be shown that:
\begin{equation}
kh_{\rm min} = (24\Im{\xi})^{1/3}.
\end{equation}
This prodicts a much more modest improvement over reduced material loss, compared with previous lossy-material bound~\cite{miller_fundamental_2016} where the expression for minimum thickness under normal incidence is $kh_{\rm min} = \Im \xi$. \figref{appenG-figure7} shows that this contrast is on the order of $10^2$ for a Drude metal modeled by permittivity
\begin{equation}
\varepsilon(\omega) = - \frac{\omega_p^2}{\omega^2+i\gamma\omega},
\end{equation}
with loss rate $\gamma = 0.02\omega_p$. Plasmonic wavelength is  $\lambda_p = 2\pi c/\omega_p$, with $c=1$ being the speed of light in our unitless convention. 

It is also shown in \figref{appenG-figure7} that, minimum thickness predicted by the general bound is on the same length scale as skin depth in the metal~\cite{novotny2012principles} near plasmonic wavelength, $\lambda_p$. For $\lambda<\lambda_p$, there is no surface plasmonic mode inside a Drude metal and skin depth is ill-defined, though it is still possible to realize a perfect absorber according to the general bound. For $\lambda>\lambda_p$, while both skin depth and lossy-material bound reach a plateau at large wavelength limit, general bound has $h_{\text{min}}$ increases proportionally to wavelength. This comes from the effectively thinner material under large wavelength incidence and explains the behavior of Al in Fig.~\ref{fig:chap2-figure4}(a) of Chapter~\ref{chap:analy}.

\hiddensection{Inverse design procedures for perfect absorbers}
\label{sec:appenG-sec6}
In Fig.~\ref{fig:chap2-figure3}(b,c) and Fig.~\ref{fig:chap2-figure4}(c) of Chapter~\ref{chap:analy}, we showed examples of maximum absorption of topology-optimized metasurfaces with subwavelength periodicity, which are generally within $70\%$ of the bounds, and therefore confirming our bounds to be tight or nearly so. Here we present the details of the topology optimization procedures. Given the permittivity of the material $\epsm$, using a material density function $\alpha_i$, with the subscript $i$ standing for its spatial coordinate, $\alpha_i=1$ meaning material and $\alpha_i=0$ meaning air at pixel $i$, then the design problem of perfect absorbers is formulated as a maximization of the absorption cross section over all permissible choice of $\alpha_i$ at each pixel $i$:
\begin{equation}
\begin{aligned}
& \underset{\varepsilon_i}{\text{maximize}}
& &\sabs \\
& \text{subject to}
& &\varepsilon_i = 1 + \alpha_i(\epsm-1),\\
& & &\alpha_i \in [0,1],
\end{aligned}
\end{equation}
and the absorption cross section as function of the six-vector field $\psi$ is given by $\sabs(\psi)=\frac{1}{A}\int_{V}\frac{\omega}{2}\psi^{\dagger}\frac{\Im\chi}{|\chi|^2}\psi$, where $A$ is the unit cell area. The Maxwell constraint, i.e. that all solutions satisfy Maxwell's equations is implied. 

Global optimization methods tend not to provide reasonable convergences with such large dimensionality of the problem. Hence local optimizations with random initial starting points were tested to approach the global bounds. Fast calculations of the gradients $\partial \sabs / \partial \alpha_i$ are facilitated with the adjoint method~\cite{Miller2012a}. Following the volume-integral formalism, one can take the variation of any generic figure of merit $f(\psi)$ due to changes in the susceptibility $\dchi$:
\begin{equation}
\begin{aligned}
\delta f & = 2\Re\int_{V}\left(\delta \psi\right)^{\rm T}\frac{\partial f}{\partial \psi}.
\end{aligned}
\end{equation}
Considering that the perturbed field, expressed with six-vector Green's function $\Gamma_0(x,x')$:
\begin{equation}
\begin{aligned}
\delta \psi(x) = \int_{V}\Gamma_0(x,x')\dchi(x')\psi(x').
\end{aligned}
\end{equation}
The total variation can be written as:
\begin{equation}
\begin{aligned}
\delta f & = 2\Re\int_{V}\int_{V}\psi(x')^{\rm T}\dchi(x')\Gamma_0^{\rm T}(x,x')\frac{\partial f}{\partial \psi(x)}.
\end{aligned}
\end{equation}
Using reciprocity relations, $\Gamma_0^{\rm T}(x,x')=Q^{\rm T}\Gamma_0(x',x)Q$, where $Q=\begin{pmatrix}
\II & 0 \\
0 & -\II 
\end{pmatrix}$ is the parity operator. Then by rearranging, the variation in the figure of merit is given by
\begin{equation}
\begin{aligned}
\delta f & = 2\Re\int_{V}\psi(x')^{\rm T}\dchi(x')Q\int_{V}\Gamma(x',x)Q\frac{\partial f}{\partial \psi(x)}.
\end{aligned}
\end{equation}
Now one can define the adjoint field $\psiadj(x')=\int_{V}\Gamma(x',x)Q\frac{\partial f}{\partial \psi(x)}$, which is essentially fields resulting from the six-vector current sources, the so-called adjoint sources, $\phi_{\rm adj}=Q\frac{\partial f}{\partial \psi}$.
In the case of absorption cross-section $\sabs$, the adjoint sources are given by
\begin{equation}
\phi_{\rm adj}=\frac{1}{A}\frac{\omega}{2}\frac{\Im\chi}{|\chi|^2}Q\psi^{\star},
\end{equation}
and so the variation in $\sabs$ is
\begin{equation}
\begin{aligned}
\delta \sabs = 2\Re\int_{V}\psi(x')^{\rm T}\dchi(x')Q\psiadj(x').
\end{aligned}
\end{equation}
Hence the fields $\psi$ from the prescribed structure with direct incidence plus the adjoint fields $\psi_{\rm adj}$ provide the gradients with respect to any number of design variables. Numerically, in each iteration of the topology optimizations, one direct simulation to compute $\psi$ and another simulation with $\phi_{\rm adj}$ as sources to compute $\psi_{\rm adj}$ are need.

\begin{figure}[t]
\centering
  \includegraphics[width=0.5\textwidth]{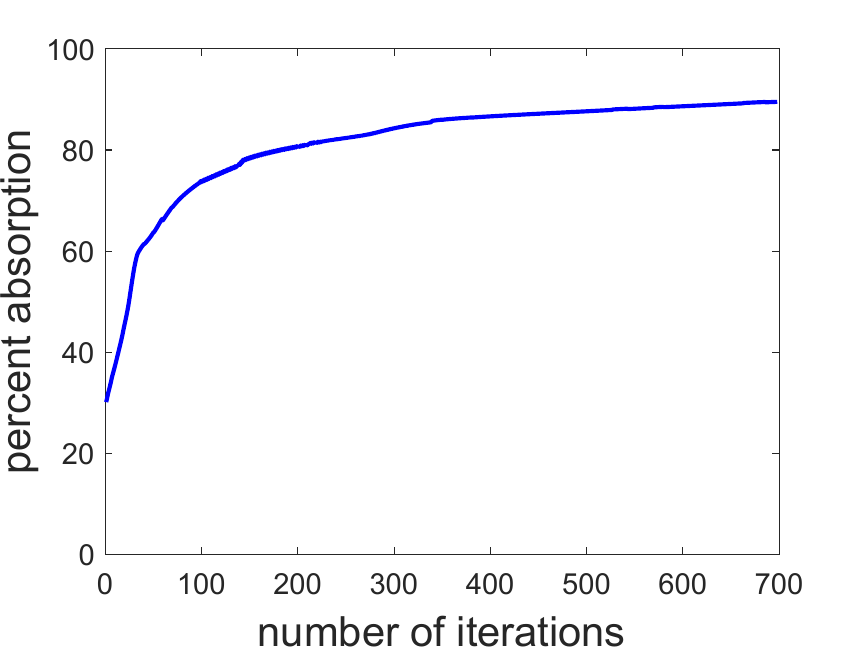}
  \caption{Percent absorption as a function the number of iterations for the best design optimization of $\SI{1.2}{\micro\meter}$-thick SiC absorber.}
  \label{fig:appenG-evol}
\end{figure}

The simulations are performed with a finite-difference time-domain~\cite{taflove2013advances} open-source solver~\cite{MEEP}. In all design figures below, periodic conditions are imposed in the horizontal direction, and light is incident from below and propagating upward. For all sets of hyper-parameters, including material permittivities and thicknesses, we test at least 10 initial starting points, and run simulations with resolutions up to 110 grids per wavelength. Almost all optimizations converge within 700 iterations, and we show in \figref{appenG-evol} the evolution of $\sabs$ in $\SI{1.2}{\micro\meter}$-thick SiC absorber optimization.

\hiddensection{Optimized designs}
\label{sec:appenG-sec7}
\subsection{Different thicknesses of SiC absorbers at$\SI{11}{\micro\meter}$ wavelength}
As an example, we investigated absorber inverse designs with SiC at $\SI{11}{\micro\meter}$ wavelength and a range of thicknesses. The resolution is $\SI{0.1}{\micro\meter}$, and unit cell period is $\SI{1.1}{\micro\meter}$. Their percent absorption and designs are presented in Table \ref{tab:appenG-id1}.

\subsection{Minimum thicknesses of 70\% absorbers for different materials}
\label{sec:appenG-sec8}
Thinnest perfect absorbers are designed for different types of materials, such as metals, doped semiconductors and polar dielectrics. In Table \ref{tab:appenG-id2}, we demonstrate designs of six representative materials at different wavelengths where 70\% absorption is achieved with minimum thicknesses of the metasurfaces.

\hiddensection{Deriving previous bounds from general bound formalism}
Different derivations of upper bounds can be formulated as optimization problems with same objective functions but different constraints. In this section, we showed that how the analytical bound, developed in Chapter~\ref{chap:analy}, can incoorperate previous bounds by either relaxing the energy equality constraint, or taking the result of the general bound in certain limit.

\begin{table*}[t]
	\centering
	\begin{tabular}{|l|c|c|c|}
		\hline
		& Extinction & Absorption & Scattering   \\ \hline
		General	bound & \makecell{max. $P_{\rm ext}$ \\ s.t. $P_{\rm scat}+P_{\rm abs}=P_{\rm ext}$}   & \makecell{max. $P_{\rm abs}$ \\ s.t. $P_{\rm scat}+P_{\rm abs}=P_{\rm ext}$} & \makecell{max. $P_{\rm scat}$ \\ s.t. $P_{\rm scat}+P_{\rm abs}=P_{\rm ext}$}   \\ \hline
		Material bound~\cite{miller_fundamental_2016} & \makecell{max. $P_{\rm ext}$ \\ s.t. $P_{\rm abs}\leq P_{\rm ext}$}   & \makecell{max. $P_{\rm abs}$ \\ s.t. $P_{\rm abs}\leq P_{\rm ext}$} & \makecell{max. $P_{\rm ext}-P_{\rm abs}$}    \\ \hline
		Channel	bound~\cite{hugonin_fundamental_2015} & \makecell{max. $P_{\rm ext}$ \\ s.t. $P_{\rm scat}\leq P_{\rm ext}$}   & \makecell{max. $P_{\rm ext}-P_{\rm scat}$}  & \makecell{max. $P_{\rm scat}$ \\ s.t. $P_{\rm scat}\leq P_{\rm ext}$}   \\ \hline
	\end{tabular}
	\caption{Upper bounds as optimization problems with the same objective function but different constraints. General bound uses an equality energy constraint, while the other two relax it to inequality constraints (or even unconstrained), resulting looser bounds.}
	\label{tab:appenG-opt-for}
\end{table*}

\begin{table*}[t]
	\centering
	\makebox[\textwidth]{%
		\begin{tabular}{|l|c|c|c|}
			\hline
			& Extinction & Absorption & Scattering   \\ \hline
			General	bound & $\einc^\dagger \left(\Im\xi\II+\ImGO\right)^{-1} \einc$  & \makecell{$\frac{\nu^{*2}}{4} \einc^\dagger[(\nu^*-1)\Im\xi\II$ \quad\quad\quad\quad \\ \quad\quad\quad\quad $+\nu^*\ImGO]^{-1} \einc$} & \makecell{$\frac{\nu^{*2}}{4} \einc^\dagger[\nu^*\Im\xi\II$ \quad\quad\quad\quad\quad\quad\quad \\ \quad\quad\quad $+ (\nu^*-1)\ImGO]^{-1} \einc$} \\ \hline
			Material bound~\cite{miller_fundamental_2016} &  $\einc^\dagger (\Im\xi)^{-1} \einc$  & $\einc^\dagger (\Im\xi)^{-1} \einc$ & $\frac{1}{4}$ $\einc^\dagger (\Im\xi)^{-1} \einc$    \\ \hline
			Channel	bound~\cite{hugonin_fundamental_2015} & $\einc^\dagger (\ImGO)^{-1} \einc$  & $\frac{1}{4}\einc^\dagger (\ImGO)^{-1} \einc$ & $\einc^\dagger (\ImGO)^{-1} \einc$    \\ \hline
	\end{tabular}}
	\caption{Optimum of different upper-bound formulations presented in Table \ref{tab:appenG-opt-for}. Optimal dual variable of the general bound is determined by \eqref{appenG-det_nu}.}
	\label{tab:appenG-opt-sol}
\end{table*}

\paragraph{Channel and lossy-material bounds}
Table \ref{tab:appenG-opt-for} compares general bound with lossy-material bound (material bound) and channel bound.
General bound purposed in Chapter~\ref{chap:analy} utilizes the \textit{equality} energy conservation constraint: $P_{\rm scat}+P_{\rm abs}=P_{\rm ext}$. Throwing away either $P_{\rm scat}$ or $P_{\rm abs}$ gives the \textit{inequality} energy conservation constraint used in previous material bound~\cite{miller_fundamental_2016} or channel bound~\cite{hugonin_fundamental_2015}. In both formalisms, the disregarded term itself is treated by an unconstrained optimization.

All optimization problems in Table \ref{tab:appenG-opt-for} have strong duality, thus their optimums can be analytically determined by the optimal of their dual functions, given in Table \ref{tab:appenG-opt-sol} (with prefactor $\omega/2$ suppressed in every expression). Results for material bound appears in~\cite{miller_fundamental_2016}. Results for channel bound appears in~\cite{hugonin_fundamental_2015}. Moreover, expanding channel bound into VSWs for a spherical scatterer gives the expressions in~\cite{hamam_coupled-mode_2007,kwon2009optimal,ruan2011design,liberal2014least} (after adding back prefactor $\omega/2$): $P^{\rm max}_{\rm scat}=P^{\rm max}_{\rm ext}=4P^{\rm max}_{\rm abs}=\frac{|E_0|^2}{k^2}\sum_{n=1}^{+\infty}\pi(2n+1)$, where $E_0$ is the plane wave amplitude, $k$ is the amplitude of the wave vector, and $n$ is total angular momentum.

\paragraph{$\mathbb{T}$-operator bound}
As discussed in Section \ref{sec:appenG-sec14}, our bound is tighter than $\mathbb{T}$-operator bound~\cite{Molesky2019a} for maximum absorption from a thermal incident field. Though using different approaches, the general bound can reproduce the same result as in $\mathbb{T}$-operator bound by relaxing the energy constraint to $\Pabs\leq\Pext$ and replacing objective function $\Pabs$ with $\Pext-\Pscat$:
\begin{equation}
\begin{aligned}
& \text{maximize} & \Pext-\Pscat \\
& \text{subject to} & \Pabs\leq\Pext.
\end{aligned}
\label{eq:appenG-dual-T}
\end{equation}
Similar to Section~\ref{sec:appenG-sec1}, we solve \eqref{appenG-dual-T} by its dual function:
\begin{align}
g(\lambda)=\frac{(\lambda+1)^2}{4}\einc^\dagger\left[\ImGO+\lambda\Im\xi\II\right]^{-1}\einc,
\label{eq:appenG-dual-func-T}
\end{align}
where $\lambda$ is the notation used in~\cite{Boyd2004} to denote dual variable for an inequality constraint. The range for $\lambda$ is $[0,+\infty)$. When $\lambda=0$, the inverse operator in \eqref{appenG-dual-func-T} is ill-defined and we replace it with pseudo inverse if $\einc\in\text{Range}\{\ImGO\}$, otherwise $g(0)\rightarrow -\infty$ 

Following assumptions made in $\TT$-operator bound, we assume far-field thermal incidence and nonmagnetic material, where $\einc$ and $\Gamma_0$ is replaced by $\einc$ and $\GG_0$. As discussed in Section~\ref{sec:appenG-sec14}, thermal incident field can be expanded by a set of uncorrelated orthogonal fields. We choose it to be $\{   v_i\}$, the eigenvectors of $\Im\GG_0$, with expansion coefficients given by $|e_i|^2=\frac{4}{\pi\omega}\Theta(T)$ and $\Theta(T)=\hbar\omega/(e^{\hbar\omega/k_BT}-1)$ is the Planck energy of a harmonic oscillator at temperature $T$.

Maximizing $g(\lambda)$ gives the expression for optimal absorption of thermal incident fields in \cite{Molesky2019a}:
\begin{equation}
P_\text{abs} \leq \frac{2}{\pi}\Theta(T) \sum_i
\begin{cases}
\frac{\rho_i}{\Im{\xi}} & \text{for } 2\rho_i \leq \Im{\xi} \\  
\frac{1}{4} & \text{for } 2\rho_i \geq \Im{\xi},
\end{cases}
\label{eq:appenG-thermal_bound}
\end{equation} 
where two cases correspond to optimal dual variable taking the value of either $\nu_1 \in (0,+\infty)$ or $\nu_0=0$. Such a bound is looser than the general bound presented in Section~\ref{sec:appenG-sec14}, as a result of its inequality energy constraint in \eqref{appenG-dual-T}, rather than the equality energy constraint.

\paragraph{Patterned thin film bound}
It is predicted that within a vacuum background, a patterned thin film with thickness much smaller than the incident wavelength has a maximum absorption of 50\%~\cite{thongrattanasiri_complete_2012}. To validate this, we take the limit $k_zh\rightarrow 0$ in \eqref{appenG-film_abs} and obtain:
\begin{equation}
\sigma_{\rm abs}/A \leq \frac{2(\Im{\xi})\rho_{+,p'}}{(\Im{\xi}+\rho_{+,p'})^2}.
\label{eq:appenG-thin-film_abs}
\end{equation}
Because a thin film only has dipole radiation that is symmetric respect to the $z=0$ plane, only mode with index $s=+$ survived in \eqref{appenG-thin-film_abs}. 

When $\Im\xi=\rho_{+,M}$, the absorption rate $\sigma_{\rm abs}/A$ in \eqref{appenG-thin-film_abs} reaches its maximum of 50\%, agreeing with the prediction made in~\cite{thongrattanasiri_complete_2012}. The advantage of our formalism is that we can also predict the minimum thickness for the patterned thin film to reach $50\%$ absorption: $h_{\rm min} = 2\Im\xi/k$, as solved from the optimal condition $\Im\xi=\rho_{+,M}$.


\hiddensection{Underestimation of the channel bounds from cutoff channels}
\label{sec:appenG-sec9}
The channel bounds shown in Table \ref{tab:appenG-opt-sol} are in fact infinite for a plane wave incident. Physically, this is due to the negligible radiative loss in high-order VSW channels, corresponding to the eigenvectors of $\ImGO$ with near-zero eigenvalues. To regularize such divergence, one needs to truncate its radiation channels to a finite number based on certain threshold. Such an empirical truncation is certainly a disadvantage of the channel bound, moreover, as we will show below, it also introduces unwanted underestimation of the channel bound itself.

\begin{figure}[t!]
	\includegraphics[width=0.5\textwidth]{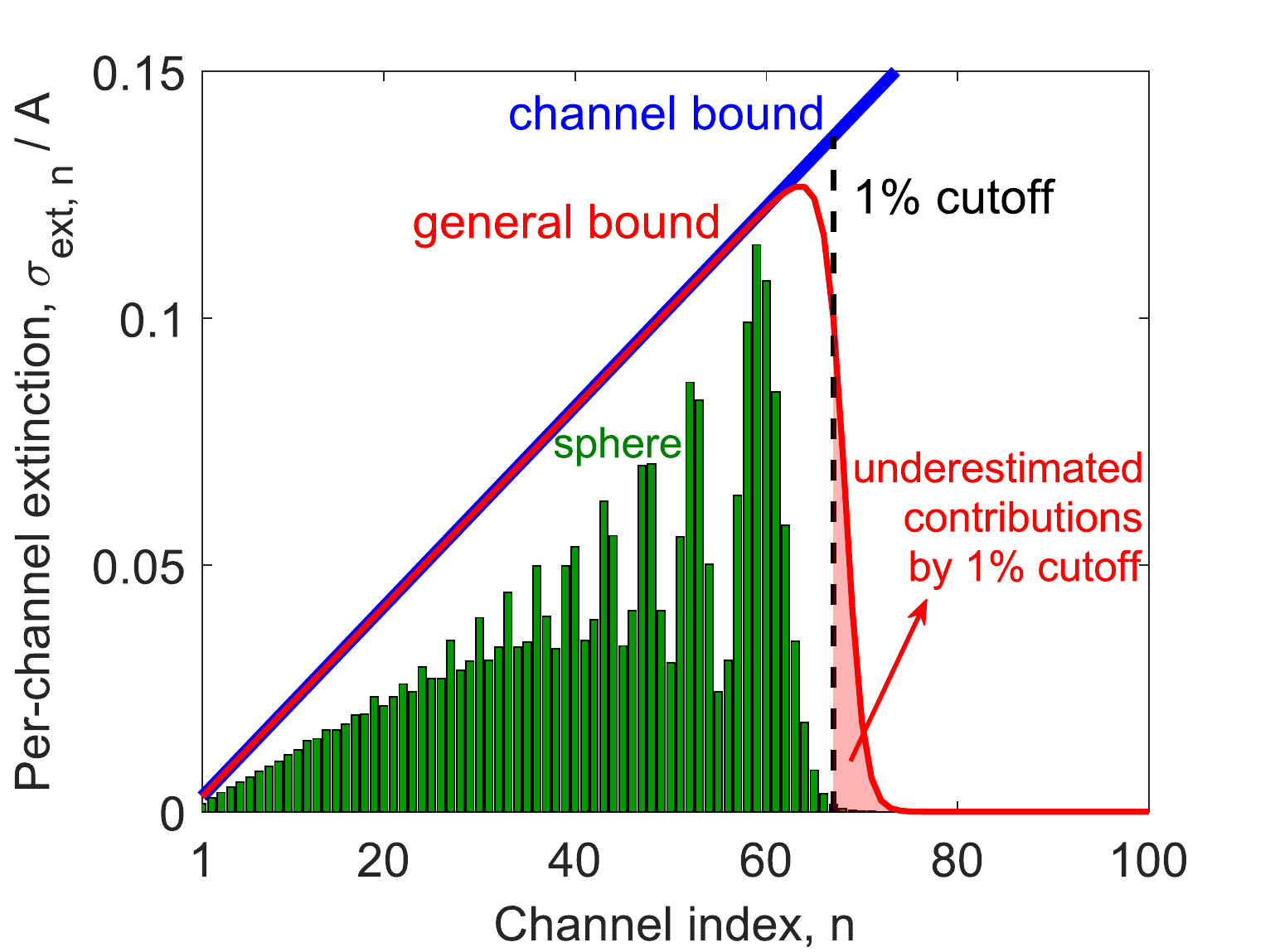}
	\centering
	\caption{Per-channel extinction $\sigma_{\text{ext},n}$ for material Ag in a spherical bounding volume with radius $R = 10\lambda$ at wavelength $\lambda = 360$ nm. Compared to the general bound, the 1\% cutoff threshold excludes channels whose potential contributions are marked by the red shaded region, resulting in an underestimated channel bound.}
	\label{fig:appenG-fig3}
\end{figure}

As an example, Fig. \ref{fig:appenG-fig3} shows channel bounds for per-channel extinction $\sigma_{\text{ext},n}$ within a bounding volume of radius $R=10\lambda$. The material is Ag and incident wavelength $\lambda=360$ nm. Also shown in the same figure are the general bound and spherical scattering. As expected, the channel bound diverges at high-order radiative channels, and is regularized by a 1\% cutoff line, which excludes channels for which the sphere scattering contributions are less than 1\% of the channel bound. 

Compared with the general bound, we see that the potential contribution of those excluded channels (red shaded region), are ignored by the 1\% threshold. Such an underestimation results in a seemly tighter bound in Fig.~\ref{fig:chap2-figure2} in Chapter~\ref{chap:analy} at large radius limit. Of course, the 1\% threshold is empirical. One could reduce the threshold to eliminate the unwanted underestimation, but that usually results in an overall overestimation of the channel bound since more channels are now included without the inhibition of material loss. We found 1\% is a good empirical threshold for  estimating the channel bound.

\hiddensection{The imaginary part of the Green's function operator for a sphere}
\label{sec:appenG-sec10}
The expressions of $\Im\GG_0$ is given in~\cite{tsang2004scattering}, whose imaginary part is Hermitian and can be decomposed as:
\begin{equation}
\Im{\GG_0}=\frac{1}{2i}(\GG_0 - \GGd_0) = \sum_{n,m,j} v_{n,m,j}v_{n,m,j}^\dagger,
\end{equation}
where $n=1,2, ...$, $m=-n,...,n$, $j=1,2$ represents two polarizations. $v_{n,m,j}$ are reguarized VSWs whose definition can be found in~\citeasnoun{tsang2004scattering}: 
\begin{align}
\vect{v}_{n,m,1}(\vect{x}) &= k^{\frac{3}{2}} Rg \vect{M}_{n,m}(kr,\theta,\phi) \\
\vect{v}_{n,m,2}(\vect{x}) &= k^{\frac{3}{2}} Rg \vect{N}_{n,m}(kr,\theta,\phi).
\end{align}
The inner product of $v_{n,m,j}$ with itself gives the eigenvalue of $\Im{\GG_0}$:
\begin{align}
\rho_{n,m,j} &= v_{n,m,j}^\dagger v_{n,m,j} \\
&= \int_V \vect{v}_{n,m,j}^*(\vect{x})\cdot\vect{v}_{n,m,j}(\vect{x}) \text{d}V. 
\end{align}
Integrating over angular coordinates gives the expression:
\begin{align}
\rho_{n,m,1} &= \int_0^{kR} x^2j_n^2(x)\text{d}x \\
\rho_{n,m,2} &= n(n+1) \int_0^{kR} j_n^2(x)\text{d}x + \int_0^{kR}[xj_n(x)]^{\prime 2}\text{d}x,
\end{align}
which can be computationally evaluated or even reduced to simpler analytical forms~\cite{bloomfield2017indefinite}.

\hiddensection{The imaginary part of the Green's function operator for a film}
\label{sec:appenG-sec11}
As in~\citeasnouns{tsang2004scattering,kruger2012trace}, $\Im{\GG_0}$ in Cartesian coordinate can be decomposed into a complete set of plane waves:
\begin{equation}
\Im{\GG_0} = \sum_{s,p} \int_{k_{\parallel}\leq k} \vect{v}_{s,p}(\vect{k_{\parallel}})\vect{v}^\dagger_{s,p}(\vect{k_{\parallel}})\frac{\text{d}\vect{k_{\parallel}}}{(2\pi)^2}.
\end{equation}
Index $s=\{-1,+1\}$ represents odd and even parity, index $p=M,N$ represents different polarization, $\vect{k_{\parallel}}$ are in-plane wave vector whose integration only runs through propergating modes. 
Real-space expressions of $v_{s,p}(\vect{k}_{\parallel})$ are:
\begin{align}
\vect{v}_{+, M}(\vect{k_{\parallel}},\vect{x}) &= ik\frac{e^{i\vect{k}_\parallel\cdot\vect{r}_\parallel}}{\sqrt{2k_z}k_\parallel}(k_y\vect{\hat{e}}_x-k_x\vect{\hat{e}}_y)\cos(k_zz)  \\
\vect{v}_{-, M}(\vect{k_{\parallel}},\vect{x}) &= -ik\frac{e^{i\vect{k}_\parallel\cdot\vect{r}_\parallel}}{\sqrt{2k_z}k_\parallel}(k_y\vect{\hat{e}}_x-k_x\vect{\hat{e}}_y)\sin(k_zz)\\
\vect{v}_{+, N}(\vect{k_{\parallel}},\vect{x}) &= \frac{e^{i\vect{k}_\parallel\cdot\vect{r}_\parallel}}{\sqrt{2k_z}}\left[k_\parallel\cos(k_zz)\vect{\hat{z}} - ik_z\sin(k_zz)\vect{\hat{k}_\parallel}\right]\\
\vect{v}_{-, N}(\vect{k_{\parallel}},\vect{x}) &= \frac{e^{i\vect{k}_\parallel\cdot\vect{r}_\parallel}}{\sqrt{2k_z}}\left[k_\parallel\sin(k_zz)\vect{\hat{z}} + ik_z\cos(k_zz)\vect{\hat{k}_\parallel}\right].
\end{align}

Inner products of $v_{s, p}(\vect{k}_{\parallel})$ in a thin film (thickness $h$, centered at $z=0$) is~\cite{Molesky2019a}:
\begin{align}
v_{s, p}^\dagger(\vect{k}_{\parallel}) v_{s', p'}(\vect{k}_{\parallel}') &= \int_V \vect{v}_{s, p}^*(\vect{k}_{\parallel},\vect{x}) \cdot \vect{v}_{s', p'}(\vect{k}_{\parallel}',\vect{x})\text{d}V \\
&= \rho_{s,p}(\vect{k}_\parallel)(2\pi)^2\delta(\vect{k}_\parallel-\vect{k}_\parallel')\delta_{s,s'}\delta_{p,p'},
\end{align}
where the eigenvalues are:
\begin{align}
\rho_{\pm, M}(\vect{k}_\parallel) &= \frac{k^2h}{4k_z}(1\pm\frac{\sin(k_zh)}{k_zh})\\
\rho_{\pm, N}(\vect{k}_\parallel) &= \frac{k^2h}{4k_z}(1\pm\frac{\sin(k_zh)}{k_zh})\mp\frac{\sin(k_zh)}{2}.
\end{align}

\hiddensection{Upper bounds at different generality}
\label{sec:appenG-sec12}  
For generality, we first express the bounds with six-vector notations for the incident field $\psiinc$, Green's function $\Gamma_0$, and implicitly for $\Im\xi$ which is now a tensor operator for arbitrary material susceptibility.
\begin{enumerate}
	\item Most general form (include non-local, magnetic, inhomogeneous materials, any incident field, any geometry of the scatterer):
	\begin{align}
		\Pext &\leq \frac{\omega}{2} \psiinc^\dagger \left(\Im\xi+\Im\Gamma_0\right)^{-1} \psiinc\\
		\Pabs &\leq \frac{\omega}{2}\frac{\nu^{*2}}{4} \psiinc^\dagger[(\nu^*-1)\Im\xi+\nu^*\Im\Gamma_0]^{-1} \psiinc  \\
		\Pscat &\leq \frac{\omega}{2}\frac{\nu^{*2}}{4} \psiinc^\dagger[\nu^*\Im\xi + (\nu^* - 1)\Im\Gamma_0]^{-1} \psiinc, 
	\end{align}
	where $\Im\xi$ and $\Im\Gamma_0$ are matrices that depends on the exact shape and material compositions of the scatterer. 
	\item Scalar material (electric or magnetic scalar material, any incident field, any geometry of the homogeneous scatterer):
	\begin{align}
	P_{\rm ext} &\leq \frac{\omega}{2}\frac{1}{\Im{\xi}}\Big[\psiinc^\dagger \psiinc - \psiinc^\dagger \Vbb (\Im{\xi}+\Vbb^\dagger \Vbb)^{-1} \Vbb^\dagger\psiinc\Big] \label{eq:appenG-scalar_bound1} \\
	P_{\rm abs} &\leq \frac{\omega}{2}\frac{\nu^{*2}}{4}\frac{\nu^*}{\nu^*-1}\frac{1}{\Im{\xi}}\Big\{ \frac{1}{\nu^*}\psiinc^\dagger\psiinc \nonumber \\
	& - \psiinc^\dagger \Vbb [(\nu^*-1)\Im{\xi}+\nu^*\Vbb^\dagger \Vbb]^{-1}\Vbb^\dagger \psiinc \Big\} \\
	P_{\rm scat} &\leq \frac{\omega}{2}\frac{\nu^{*2}}{4}\frac{\nu^*-1}{\nu^*}\frac{1}{\Im{\xi}}\Big\{ \frac{1}{\nu^*-1}\psiinc^\dagger\psiinc \nonumber \\
	& - \psiinc^\dagger \Vbb [\nu^*\Im{\xi}+(\nu^*-1)\Vbb^\dagger \Vbb]^{-1}\Vbb^\dagger \psiinc \Big\}, \label{eq:appenG-scalar_bound3}
	\end{align}
	where $\Im\xi$ is a scalar represents either the isotropic electric or magnetic susceptibility and we write the eigendecomposition of $\Im\Gamma_0$ as $\Im\Gamma_0=\Vbb\Vbb^\dagger$.  
	\item Isotropic \emph{electric} material (electric scalar material, any incident field, any geometry): same form as \eqrefrange{appenG-scalar_bound1}{appenG-scalar_bound3} with $\psi_{\rm inc}$ replaced by $e_{\rm inc}$, and $\Gamma_0$ replaced by $\GG_0$. Eigenbasis $\Vbb$ is now defined by the eigendecomposition: $\Im \GG_0 = \Vbb \Vbb^\dagger$, with $   v_i$ being the i-th column of $\Vbb$.
	\begin{enumerate}
		\item For far field scattering, where the incident electric field $e_{\rm inc}$ is characterized by the property $e_{\rm inc} \in \text{Range}\{\Im \GG_0\}$, bounds in \eqrefrange{appenG-scalar_bound1}{appenG-scalar_bound3} can be dramatically simplified:
		\begin{align}
		P_{\rm ext} &\leq \frac{\omega}{2}e_{\rm inc}^\dagger (\Im\xi+\Im \GG_0)^{-1}e_{\rm inc} \\
		P_{\rm abs} &\leq \frac{\omega}{2}\frac{\nu^{*2}}{4}e_{\rm inc}^\dagger [(\nu^*-1)\Im\xi+\nu^*\Im \GG_0]^{-1}e_{\rm inc} \\
		P_{\rm scat} &\leq \frac{\omega}{2}\frac{\nu^{*2}}{4}e_{\rm inc}^\dagger [\nu^*\Im\xi+(\nu^*-1)\Im \GG_0]^{-1}e_{\rm inc}.
		\end{align}
		\begin{itemize}
			\item Plane wave incidence (applies to both finite and extended scatterers) with $e_{\rm inc} = \sum_i e_i   v_i$. Explicitly written out contributions from different channels:
			\begin{align}
			P_{\rm ext} &\leq \frac{\omega}{2}\sum_i |e_i|^2\frac{\rho_i}{\Im{\xi} + \rho_i} \label{eq:ext-cha-bound} \\
			P_{\rm abs} &\leq \frac{\omega}{2} \frac{\nu^{*2}}{4}\sum_i |e_i|^2\frac{\rho_i}{(\nu^*-1)\Im{\xi}+\nu^*\rho_i} \label{eq:appenG-abs-cha-bound} \\
			P_{\rm scat} &\leq \frac{\omega}{2} \frac{\nu^{*2}}{4}\sum_i |e_i|^2\frac{\rho_i}{\nu^*\Im{\xi}+(\nu^*-1)\rho_i}, 
			\end{align}
			where $\rho_i =    v_i^\dagger   v_i$ is analytically known for highly symmetric bounding volumes.
			
			\item VSW incidence (applies to finite scatterers). Now the incident field is one specific VSW: $e_{\rm inc} = e_i    v_i$, under which:
			\begin{align}
			P_{\rm ext} &\leq \frac{\omega}{2} \frac{|e_i|^2}{\Im\xi + \rho_i} \\
			P_{\rm abs} &\leq \frac{\omega}{2} \frac{\nu^{*2}}{4}\frac{|e_i|^2}{(\nu^*-1)\Im\xi + \nu^*\rho_i} 
			 = \begin{cases}
			\frac{\Im\xi}{(\Im\xi+\rho_i)^2}|e_i|^2 & \text{if\ \ } \rho_i\leq\Im\xi \\
			\frac{1}{4\rho_i}|e_i|^2 & \rm{else} 
			\end{cases} \\
			P_{\rm scat} &\leq \frac{\omega}{2} \frac{\nu^{*2}}{4}\frac{|e_i|^2}{\nu^*\Im\xi + (\nu^*-1)\rho_i}  = 
			\begin{cases}
			\frac{\rho_i}{(\Im\xi+\rho_{\rm max})^2}|e_i|^2 & \hspace{-55pt} \text{if\ \ } \rho_i\geq\frac{\rho_{\rm max}\Im\xi}{2\Im\xi+\rho_{\rm max}} \\
			\frac{1}{4}\frac{\rho_{\rm max}^2}{\Im\xi(\rho_{\rm max}+\Im\xi)(\rho_{\rm max}-\rho_i)}|e_i|^2 & \rm{else},
			\end{cases}
			\end{align}
			where the choice of $\nu^*$ is simple enough that we can write out explicit two possible solutions of $P_{\rm abs}$ and $P_{\rm sca}$. We denote the maximum in $\{\rho_i\}$ as $\rho_{\rm max}$.
		\end{itemize}
		\item Incident field in near-field scattering is not necessarily in the range of $\Im \GG_0$ as evanescent waves may contribute (for an extended scatter). Expression for its bound takes the most general form as \eqrefrange{appenG-scalar_bound1}{appenG-scalar_bound3} with $\psi_{\rm inc}$ replaced by $e_{\rm inc}$, and $\Gamma_0$ replaced by $\GG_0$.
		For arbitrary dipole sources $ p$, the incident field can be written as $\ev_{\rm inc}=\GG_{0,  p\rightarrow V} p$ where $\GG_{0,  p\rightarrow V}$ is an integral Green's function mapped from the region of dipole source $ p$ to the scatterer $V$. Taking the singular vector decomposition of $\GG_{0,  p\rightarrow V}=\UU\WW^\dagger$, bounds for near field scattering can be written as:
	\end{enumerate}
\end{enumerate}
	\begin{align}
	P_{\rm ext} &\leq \frac{\omega}{2}\frac{1}{\Im\xi} p^\dagger\WW[\UU^\dagger\UU - \UU^\dagger\Vbb(\Im\xi + \Vbb^\dagger\Vbb)^{-1}\Vbb^\dagger\UU]\WW^\dagger p \nonumber \\
	&= \frac{\omega}{2}\frac{1}{\Im\xi} \sum_i | p^\dagger  w_i|^2 \left(  u_i^\dagger u_i - \frac{| u_i^\dagger   v_i|^2}{\Im\xi+   v_i^\dagger   v_i} \right) \label{eq:appenG-nf_ext_bnd} \\
	P_{\rm abs} &\leq \frac{\omega}{2}\frac{\nu^{*2}}{4(\nu^*-1)}\frac{1}{\Im\xi} p^\dagger\WW\{\UU^\dagger\UU - \UU^\dagger\Vbb[(\nu^*-1)\Im\xi/\nu^* + \Vbb^\dagger\Vbb]^{-1}\Vbb^\dagger\UU\}\WW^\dagger p \nonumber \\
	&= \frac{\omega}{2}\frac{\nu^{*2}}{4(\nu^*-1)}\frac{1}{\Im\xi}   \sum_i | p^\dagger  w_i|^2 \left(  u_i^\dagger u_i - \frac{| u_i^\dagger   v_i|^2}{(\nu^*-1)\Im\xi/\nu^*+   v_i^\dagger   v_i} \right) \label{eq:appenG-nf_abs_bnd} \\
	P_{\rm scat} &\leq \frac{\omega}{2}\frac{\nu^*}{4}\frac{1}{\Im\xi} p^\dagger\WW\{\UU^\dagger\UU  - \UU^\dagger\Vbb[\nu^*\Im\xi/(\nu^*-1) + \Vbb^\dagger\Vbb]^{-1}\Vbb^\dagger\UU\}\WW^\dagger p \nonumber \\
	&= \frac{\omega}{2}\frac{\nu^*}{4}\frac{1}{\Im\xi} \sum_i | p^\dagger  w_i|^2 \left(  u_i^\dagger u_i - \frac{| u_i^\dagger   v_i|^2}{\nu^*\Im\xi/(\nu^*-1)+   v_i^\dagger   v_i} \right). \label{eq:appenG-nf_sca_bnd}
	\end{align}

\hiddensection{Bound for local density of states (LDOS)}
\label{sec:appenG-sec13}
In this section, we first derive general bounds of LDOS quantities for arbitrary materials in six-vector notations (defined in Section~\ref{sec:appenG-sec12}), then narrow the scope down to non-magnetic materials to simplify the expressions of the bounds. 

\paragraph{General bounds for LDOS}
We start with the expressions of total, non-radiative, radiative electric LDOS in six-vector volume-integral form~\cite{miller_fundamental_2016}: 	
\begin{align}
\rho_{\rm tot} &= \rho_0 + \frac{1}{\pi\omega}\sum_j\Im \left(\tilde{\psi}_{\rm inc, j}^\dagger\phi_j\right)  \\
\rho_{\rm nr} &= \frac{1}{\pi\omega}\sum_j \phi_j^\dagger(\Im\xi)\phi_j  \\
\rho_{\rm rad} &= \rho_0 + \frac{1}{\pi\omega}\sum_j \left[\Im \left(\tilde{\psi}_{\rm inc, j}^\dagger\phi_j\right) - \phi_j^\dagger(\Im\xi)\phi_j \right], \label{eq:appenG-rad_LDOS_gen_bound}
\end{align}
where $\rho_0$ is the electric LDOS of the background material, and takes the value of $\frac{\omega^2}{2\pi^2c^3}$ for a scatterer in the vacuum~\cite{joulain2005surface}.
Summation $j=1,2,3$ denotes power quantities from three orthogonally polarized unit dipoles. Incident field from dipole $j$ is denoted by $\psi_{\rm inc, j}=(e_{\rm inc, j}, h_{\rm inc, j})$. Here we use lowercase notations for both electric and magnetic fields to emphasis their vector nature, as opposed to capitalized characters that are usually reserved for operators and matrices. Such incident field excites polarization field $\phi_j$ in the scatterer. Complex conjugate of $\psi_{\rm inc, j}$ (with a minus sign in front of magnetic fields) is denoted by $\tilde{\psi}_{\rm inc, j} = (e^*_{\rm inc, j}, -h^*_{\rm inc, j})$.

Because three dipoles are uncorrelated, we can first solve the bound for one unit dipole. For simplicity, we omit its index $j$ and write its incident field as $\psiinc$, which excites polarization field $\phi$ in the body. For this dipole, its non-radiative LDOS can be bounded by maximum absorption in \eqref{appenG-Pabs_bound} by identifying the objective function as $\phi^\dagger(\Im\xi)\phi$. Bounds on total and radiative LDOS are less straightforward and are discussed below.

Objective function for total LDOS is $\Im \left(\tilde{\psi}_{\rm inc}^\dagger\phi\right)$ with energy conservation constraint $\phi^\dagger \left(\Im \xi + \Im\Gamma_0 \right) \phi = \Im \left(\psiinc^\dagger \phi\right)$. This six-vector form is generalized from \eqref{appenG-gen-form-min} with $\Abb=0$ and $\beta=\tpsiinc$. Its maximum is given by \eqref{appenG-Popt}:
\begin{equation}
\begin{aligned}
\max_{\phi}
\left\{
\Im \left(\tilde{\psi}_{\rm inc}^\dagger\phi\right)\right\} = \frac{1}{4\nu^*}(\tpsiinc+\nu^*\psiinc)^\dagger \left(\Im \xi + \Im\Gamma_0 \right)^{-1} (\tpsiinc+\nu^*\psiinc),
\end{aligned}
\end{equation}
where the optimal dual variable $\nu^*$ is always chosen at $\nu_1>\nu_0=0$ similar to \eqref{appenG-optlambda}:
\begin{equation}
\nu^* =\nu_1= \left[\frac{\tpsiinc^\dagger\left(\Im \xi + \Im\Gamma_0 \right)^{-1}\tpsiinc}{\psiinc^\dagger\left(\Im \xi + \Im\Gamma_0 \right)^{-1}\psiinc}\right]^{\frac{1}{2}}.
\label{eq:appenG-nu_tot_LDOS}
\end{equation}
For non-magnetic scatterer, the above expression can be significantly simplified. No magnetic current can be excited in the non-magnetic scatterer such that $\vect{M}(\xv)=0$. Examining the object function $\Im \left(\tilde{\psi}_{\rm inc}^\dagger\phi\right)$, we can find that it is equivalent to set $h_{\rm inc}=0$. \Eqref{appenG-nu_tot_LDOS} gives $\nu^*=1$ and the maximum objective function for non-magnetic scatterer can be simplified to:
\begin{align}
\max_{\phi}
\left\{
\Im \left(\tilde{\psi}_{\rm inc}^\dagger\phi\right)\right\} &= \left[\Re e_{\rm inc}\right]^\dagger \left(\Im \xi + \Im\GG_0 \right)^{-1} \left[\Re e_{\rm inc}\right] \\
&\leq e_{\rm inc}^\dagger \left(\Im \xi + \Im\GG_0 \right)^{-1} e_{\rm inc} \\
&= \frac{2}{\omega}P^{\rm max}_{\rm ext}.
\end{align}
where in the last two lines, we relax the bound to the maximum-extinction bound given in \eqref{appenG-Pext_bound} with the same assumption of non-magnetic scatterer. 

Objective function for radiative LDOS defined in \eqref{appenG-rad_LDOS_gen_bound} can be chosen as $\Im \left(\tilde{\psi}_{\rm inc}^\dagger\phi\right) - \phi^\dagger(\Im\xi)\phi$. Thus, $\Abb=-\Im\xi$, $\beta=\tpsiinc$. Maximal objective function given be \eqref{appenG-Popt} can be written as:
\begin{equation}
\begin{aligned}
\max_\phi & \left\{\Im \left(\tilde{\psi}_{\rm inc}^\dagger\phi\right) - \phi^\dagger(\Im\xi)\phi\right\}
= \\
&\frac{1}{4} (\tpsiinc+\nu^*\psiinc)^\dagger \left[(\nu^*+1)\Im\xi + \nu^*\Im\Gamma_0 \right]^{-1} (\tpsiinc+\nu^*\psiinc)
\end{aligned}
\label{eq:appenG-fmax_radiative_LDOS}
\end{equation}
with optimal dual variable $\nu^*$ given by \eqref{appenG-det_nu}. For non-magnetic scatterer (effectively $h_{\rm inc}=0$ in \eqref{appenG-fmax_radiative_LDOS}), bound in \eqref{appenG-fmax_radiative_LDOS} reduces to:
\begin{equation}
\begin{aligned}
\max_\phi& \left\{\Im \left(\tilde{\psi}_{\rm inc}^\dagger\phi\right) - \phi^\dagger(\Im\xi)\phi\right\} \\
&= \frac{1}{4} (\tilde{e}_{\rm inc}+\nu^*\einc)^\dagger \left[(\nu^*+1)\Im\xi + \nu^*\Im\GG_0 \right]^{-1} (\tilde{e}_{\rm inc}+\nu^*\einc).
\end{aligned}
\label{eq:appenG-fmax_radiative_LDOS_2}
\end{equation}
Radiative LDOS bound in \eqref{appenG-fmax_radiative_LDOS_2} can be relaxed to scattering bound in \eqref{appenG-Psca_bound_2nd} by observing that the dual function of the former, $g_1(\nu)$, is always greater than or equal to the latter (after suppressing its $\frac{\omega}{2}$ factor), $g_2(\nu)$, for any $\nu\geq\nu_0$:
\begin{align}
g_1(\nu) &= -\frac{1}{4} (\tilde{e}_{\rm inc}+\nu\einc)^\dagger \left[(\nu+1)\Im\xi + \nu\Im\GG_0 \right]^{-1} (\tilde{e}_{\rm inc}+\nu\einc) \\
&\geq -\frac{(1+\nu)^2}{4}\einc^\dagger \left[(\nu+1)\Im\xi + \nu\Im\GG_0 \right]^{-1} \einc = g_2(\nu). \label{eq:appenG-g1_g2}
\end{align}
The last inequality can be proved by performing Cholesky decomposition on the Hermitian matrix $\left[(\nu+1)\Im\xi + \nu\Im\GG_0 \right]^{-1} = \Lbb^\dagger \Lbb$ and using Cauchy–Schwarz inequality to relax the cross term:
\begin{equation}
\begin{aligned}
\Re\{\einc^\dagger \Lbb^\dagger \Lbb  \ev^*_{\rm inc}\} &\leq \left|\einc^\dagger \Lbb^\dagger \Lbb  \ev^*_{\rm inc}\right| \\
&\leq  \left\Vert \Lbb\einc \right\Vert\cdot \left\Vert \Lbb\ev^*_{\rm inc} \right\Vert = \left\Vert\Lbb \einc\right\Vert^2 = \einc^\dagger \Lbb^\dagger \Lbb  \einc.
\end{aligned}
\end{equation}
It follows from \eqref{appenG-g1_g2} that the maximum of $g_1(\nu)$ is greater than the maximum of $g_2(\nu)$. The optimum of a primal function is given by the negative of the maximum of a dual function, so the optimal objective function considered here is smaller than the optimal scattering bound in \eqref{appenG-Psca_bound_2nd}, and equivalently \eqref{appenG-Psca_bound}:
\begin{equation}
\max_\phi \left\{\Im \left(\tilde{\psi}_{\rm inc}^\dagger\phi\right) - \phi^\dagger(\Im\xi)\phi\right\} \leq \frac{2}{\omega}P_{\rm sca}^{\rm max}.
\end{equation}
To summarize, we derive general LDOS bounds for any material. For non-magnetic material specifically, LDOS can be directly bounded by maximum power response in Eqs. (\ref{eq:appenG-Pext_bound}), (\ref{eq:appenG-Pabs_bound}), and (\ref{eq:appenG-Psca_bound}):
\begin{align}
\rho_{\rm tot} &\leq \frac{2}{\pi\omega^2}\sum_j P^{\rm max}_{\text{ext}, j} + \rho_0 \label{eq:appenG-rhotot_bnd} \\ 
\rho_{\rm nr} &\leq \frac{2}{\pi\omega^2}\sum_j P^{\rm max}_{\text{abs}, j} \\
\rho_{\rm rad} &\leq \frac{2}{\pi\omega^2}\sum_j P^{\rm max}_{\text{sca}, j} + \rho_0,
\label{eq:appenG-rhorad_bnd}
\end{align}
where $j=1,2,3$ denotes the summation of maximum power quantities from three orthogonally polarized unit dipoles.

\begin{figure}[t!]
	\includegraphics[width=0.5\textwidth]{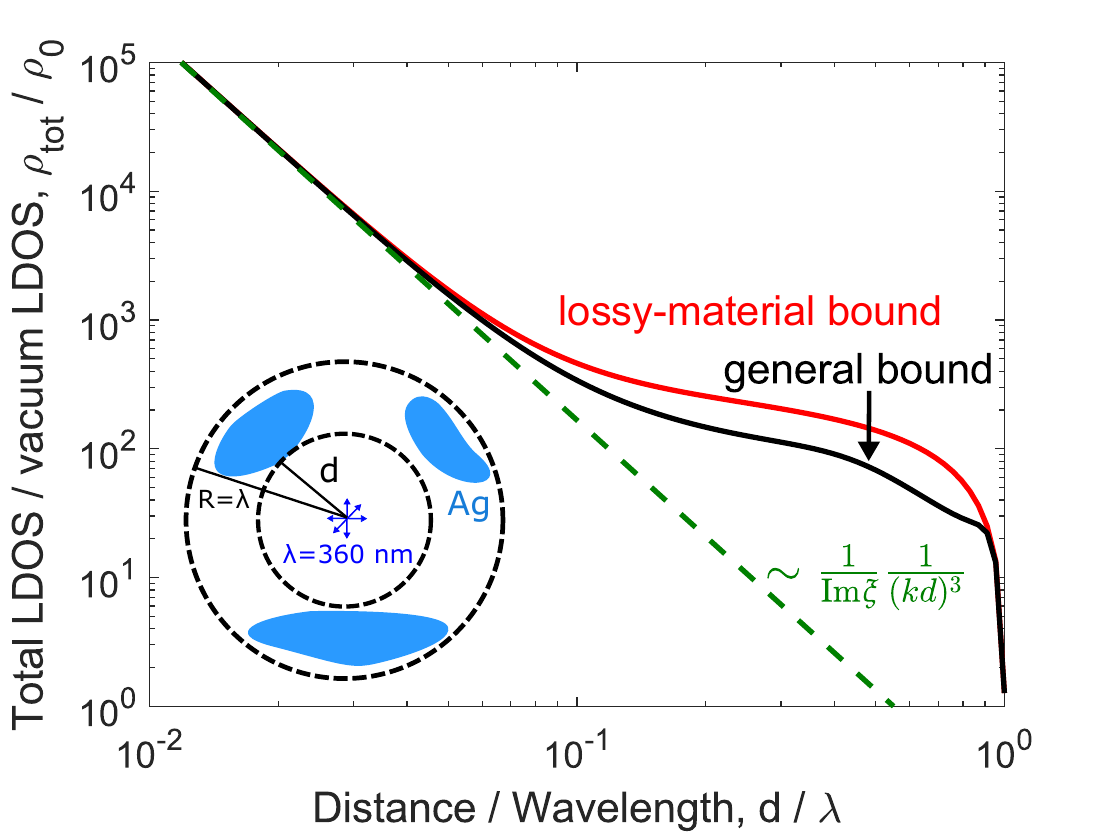}
	\centering
	\caption{Bounding volume for the LDOS problem is chosen to be a spherical shell with three randomly oriented dipoles in the center, radiating at wavelength $\lambda=360$ nm. Inner radius of the shell is determined by the minimum distance $d$ to the scatterer comprising only Ag. Outer radius $R$ of the shell covers the far end of the scatterer and is assumed to be one wavelength in the figure. In the far field, general bound is tighter than lossy-material bound. In the near field, general bound converge to lossy-material bound, and both follow the same divergence as $\frac{1}{\Im\xi}\frac{1}{k^3d^3}$.}
	\label{fig:appenG-figure5}
\end{figure}

\paragraph{LDOS bounds for a finite non-magnetic scatterer}
In the following, we assume the scatterer is non-magnetic and finite, embedded in the vacuum. The non-magnetic nature of the scatterer allows us to use \eqrefrange{appenG-rhotot_bnd}{appenG-rhorad_bnd} to decompose LDOS bounds to previous power bounds for three orthogonally polarized unit dipoles. In \eqrefrange{appenG-nf_ext_bnd}{appenG-nf_sca_bnd}, we presented power bounds for arbitrary dipole distributions $\pv(\xv)$. Here, we start with a point dipole oriented along $\ehv_j$ at origin $\pv(\xv) = \pv_j(\xv) = p_0 \delta(\xv)\ehv_j$ with $p_0=1$, and later sum up the contributions from three orthogonal polarizations.
We also assume the scatterer is finite, thus can be enclosed by a spherical shell (see Fig. \ref{fig:appenG-figure5} inset).
A shell-like bounding volume has spherical symmetry, so $   v_i$ and $  w_i$ in \eqrefrange{appenG-nf_ext_bnd}{appenG-nf_sca_bnd} are regular VSWs:
\begin{align}
   \vv_{mn1}(\xv) &=   \vw_{mn1}(\xv) = k^{\frac{3}{2}}Rg\Mv_{mn}(kr,\theta,\phi) \\
   \vv_{mn2}(\xv) &=   \vw_{mn2}(\xv) = k^{\frac{3}{2}}Rg\Nv_{mn}(kr,\theta,\phi),
\end{align}
$ u_i$ are outgoing VSWs:
\begin{align}
 \vu_{mn1}(\xv) &= k^{\frac{3}{2}}\Mv_{mn}(kr,\theta,\phi) \\
 \vu_{mn2}(\xv) &= k^{\frac{3}{2}}\Nv_{mn}(kr,\theta,\phi).
\end{align} 

Power bounds in \eqrefrange{appenG-nf_ext_bnd}{appenG-nf_sca_bnd} require us to evaluate four overlap integrals: $p_j^\dagger  w_i,  u_i^\dagger u_i,  u_i^\dagger   v_i,    v_i^\dagger   v_i$.
We first evaluate overlap integral between the point dipole and regular VSWs in the source volume $V_s$:
\begin{align}
p_j^\dagger  w_i &= \int_{V_s}\pv_j^*(\xv)\cdot  \vw_i(\xv)\text{d}\xv \\
&= p_0 \ehv_j\cdot  \vw_i(\xv=0) \\
&= k^{\frac{3}{2}}p_0
\begin{cases}
\ehv_j\cdot Rg\Nv_{m,1}(0,\theta,\phi) & \text{if  } j=2, n=1 \\
0 & \text{else,}
\end{cases}
\end{align}
where we used the fact that only $Rg\Nv_{m,1}$ has nonzero value at the origin. Exact value of the dot product $\ev_j\cdot Rg\Nv_{m,1}(0,\theta,\phi)$ depends on the orientation of the dipole:
\begin{equation}
\ehv_j\cdot Rg\Nv_{m,1}(0,\theta,\phi) = 
\begin{cases}
\pm \frac{1}{2\sqrt{3\pi}}\delta_{m,\pm 1} & \text{if\ } \ehv_j = \vect{\hat{x}} \\
\frac{1}{2i\sqrt{3\pi}}\delta_{m,\pm 1} & \text{if\ } \ehv_j = \vect{\hat{y}} \\
-\frac{1}{\sqrt{6\pi}}\delta_{m,0} & \text{if\ } \ehv_j = \vect{\hat{z}}. 
\end{cases}
\end{equation}
Later for LDOS, we will need to evaluate averaged power from three randomly oriented dipoles, which is related to the quantity:
\begin{equation}
\frac{1}{3}\sum_j |p_j^\dagger  w_i|^2 = k^3 \frac{p_0^3}{18\pi}\delta_{n,1}\delta_{j,2},
\label{eq:appenG-random_dipole}
\end{equation}
where $\ehv_j$ runs through directions $\vect{\hat{x}}$, $\vect{\hat{y}}$, and $\vect{\hat{z}}$.
We now evaluate overlap integrals between different VSWs within the bounding volume $V$:
\begin{align}
   v_i^\dagger   v_i &= \int_V    \vv_{mnj}^*(\xv)\cdot   \vv_{mnj}(\xv)\text{d}\xv = I_j\left(j_n^{(1)}(x), j_n^{(1)}(x)\right) \\
 u_i^\dagger u_i &= \int_V  \vu_{mnj}^*(\xv)\cdot \vu_{mnj}(\xv)\text{d}\xv = I_j\left(h_n^{(1)*}(x), h_n^{(1)}(x)\right) \\
 u_i^\dagger   v_i &= \int_V  \vu_{mnj}^*(\xv)\cdot   \vv_{mnj}(\xv)\text{d}\xv = I_j\left(h_n^{(1)*}(x), j_n^{(1)}(x)\right), 
\end{align}
where we defined function:
\begin{equation}
I_j\left(z_n^{(1)}(x), z_n^{(2)}(x)\right) =  
\begin{cases}
\int_{kR_1}^{kR_2}x^2z_n^{(1)}(x)z_n^{(2)}(x)\text{d}x &\text{if } j=1 \\
n(n+1)\int_{kR_1}^{kR_2}z_n^{(1)}(x)z_n^{(2)}(x)\text{d}x \\
\quad + \int_{kR_1}^{kR_2}[xz_n^{(1)}(x)]'[xz_n^{(2)}(x)]'\text{d}x &\text{if } j=2.
\end{cases}
\end{equation}

Bound for total extinction from three randomly oriented dipoles is bounded by:
\begin{align}
\frac{1}{3}\sum_j P_{\text{ext}, j} \leq \frac{1}{3}\sum_j \sum_i \frac{\omega}{2} |p_j^\dagger  w_i|^2 \underbrace{\frac{1}{\Im\xi}\left(  u_i^\dagger u_i - \frac{| u_i^\dagger   v_i|^2}{\Im\xi+   v_i^\dagger   v_i} \right)}_{f_i},
\label{eq:appenG-def_f}
\end{align}
where we defined enhancement factor $f_i$ (depends only on $n$ and $j$). Using \eqref{appenG-random_dipole}, we can show that:
\begin{equation}
\frac{\frac{1}{3}\sum_j P_{\text{ext}, j}}{P_0} \leq f_{n=1,j=2},
\end{equation}
where $P_0=\omega k^3p_0^3/12\pi$ is the power radiated by a dipole with amplitude $p_0$ in vacuum. Similarly, one can show that:
\begin{equation}
\frac{\rho_{\rm tot}}{\rho_0} \leq 1+f_{n=1,j=2}.
\end{equation}
The enhancement factor $f_{n=1,j=2}$ shows how large the light extinction of three uncorrelated dipoles can be, compared to the vacuum. While the first term in \eqref{appenG-def_f} appears in previous lossy-material bound~\cite{miller_fundamental_2016}, the second term comes from radiation coupling between the bounding volume and the vacuum.
In near field when material loss dominates, $f_{n=1,j=2}$ can be simplified to the lossy-material bound:
\begin{align}
f_{n=1,j=2} &= \frac{1}{\Im\xi} u_i^\dagger u_i \\
&= \frac{1}{\Im\xi}\left(x-\frac{1}{x}-\frac{1}{x^3}\right)\bigg|_{kR_1}^{kR_2} \\
&\rightarrow \frac{1}{\Im\xi}\frac{1}{k^3R_1^3} \label{eq:appenG-nf_exp},
\end{align}
where, in the last line, we take the limit of extreme near field where $kR_1\ll 1, kR_2$. 
In Fig. \ref{fig:appenG-figure5}, we showed the general bound and lossy-material bound for LDOS enhancement at wavelength $360$ nm by Ag surroundings. It is clear that both bounds follow \eqref{appenG-nf_exp} in near field limit. In far field, general bound is slightly tighter than the lossy-material bound due to the consideration of additional radiative loss.

Absorption and scattering bounds can also be written through an enhancement factor over the vacuum radiation:
\begin{align}
\frac{\frac{1}{3}\sum_j P_{\text{abs}, p_j}}{P_0} \leq f^{\rm abs}_{n=1,j=2}(\nu^*) \\
\frac{\frac{1}{3}\sum_j P_{\text{sca}, p_j}}{P_0} \leq f^{\rm sca}_{n=1,j=2}(\nu^*).
\end{align}
Though they are more complicated in the sense that both enhancement factors (defined below) are functions of $\nu^*$, the optimal dual variable. Similarly, for non-radiative and radiative LDOS we can write:
\begin{align}
\frac{\rho_{\rm nr}}{\rho_0} &\leq f^{\rm abs}_{n=1,j=2}(\nu^*) \\
\frac{\rho_{\rm rad}}{\rho_0} &\leq 1+f^{\rm sca}_{n=1,j=2}(\nu^*).
\end{align}

Lastly, we present the explicit expressions of absorptive and scattering enhancement factors.
For absorption, the enhancement factor is:
\begin{equation*}
f^{\rm abs}_{n=1,j=2}(\nu^*) = \frac{\nu^{*2}}{4(\nu^*-1)}\frac{1}{\Im\xi}\left( u_i^\dagger u_i-\frac{| u_i^\dagger   v_i|^2}{(\nu^*-1)\Im\xi/\nu^* +    v_i^\dagger   v_i}\right),
\end{equation*}
where $\nu^*$ is determined by solving $a = (\nu^*-1)\Im\xi/\nu^*$ in the following equation:
\begin{align*}
& 2a\left( u_i^\dagger u_i-\frac{| u_i^\dagger   v_i|^2}{a+   v_i^\dagger   v_i}\right)= \left\{ u_i^\dagger u_i\Im\xi + | u_i^\dagger   v_i|^2\left[1 - \frac{(\Im\xi+   v_i^\dagger   v_i)(2a+   v_i^\dagger   v_i)}{(a+   v_i^\dagger   v_i)^2}\right]\right\}.
\end{align*}
For scattering, the enhancement factor is:
\begin{equation*}
f^{\rm sca}_{n=1,j=2}(\nu^*) = \frac{\nu^*}{4}\frac{1}{\Im\xi}\left( u_i^\dagger u_i-\frac{| u_i^\dagger   v_i|^2}{\nu^*\Im\xi/(\nu^*-1) +    v_i^\dagger   v_i}\right),
\end{equation*}
where $\nu^*$ is determined by solving $a = \nu^*\Im\xi/(\nu^*-1)$ in the following equation:
\begin{align*}
& 2\Im\xi\left( u_i^\dagger u_i-\frac{| u_i^\dagger   v_i|^2}{a+   v_i^\dagger   v_i}\right) = \left\{ u_i^\dagger u_i\Im\xi + | u_i^\dagger   v_i|^2\left[1 - \frac{(\Im\xi+   v_i^\dagger   v_i)(2a+   v_i^\dagger   v_i)}{(a+   v_i^\dagger   v_i)^2}\right]\right\}.
\end{align*}

\hiddensection{Thermal absorption and emission} 
\label{sec:appenG-sec14}
Our formalism applies equally to thermal absorption and emission. By Kirchhoff's Law (reciprocity), or its nonreciprocal generalization~\cite{Miller2017a}, total thermal absorption and emission are equivalent and can be found by considering a weighted average of incoherent, orthogonal incoming fields $e_{\textrm{inc},i}$:
\begin{equation}
\langle \left| e_{\rm inc} \right|^2 \rangle = \sum_i w_i \left| e_{\text{inc}, i} \right|^2,
\end{equation}
where $w_i$ is a weighting factor. For a continuum of incoming fields the sum is instead an integral with a differential weight. A direct consequence of the incoherent averaging is that an upper bound to the average absorptivity/emissivity is given by the average of the bounds for each independent incident field. Surprisingly, the bounds computed by this averaging procedure varies depending on which basis is used for the incoming fields. If the incident field is treated as an incoherent sum of plane waves, over all propagation angles, for example, then the absorptivity/emissivity cross-section bounds would simply be a scalar multiple of \eqref{appenG-abs-cha-bound}. However, the bound can be tightened (decreased) if the incident fields are instead decomposed in vector spherical waves, for which the weight function $w_i$ is determined by the fluctuation-dissipation theorem~\cite{kruger2012trace}: $w_i = \frac{4}{\pi\omega}\Theta(T)$, where $\Theta(T)=\hbar\omega/(e^{\hbar\omega/k_BT}-1)$ is the Planck energy of a harmonic oscillator at temperature $T$ without the zero-point energy. The resulting bound is a sum over all VSW channels $i$:
\begin{equation}
P_\text{abs} \leq \frac{2}{\pi}\Theta(T) \sum_i
\begin{cases}
\frac{\rho_i \Im\xi}{(\Im{\xi}+\rho_i)^2} & \text{for }\rho_i \leq \Im{\xi} \\  
\frac{1}{4} & \text{for } \rho_i \geq \Im{\xi}
\end{cases}
\label{eq:appenG-thermal_bound2}
\end{equation}  
where $i=\{n,m,j\}$ includes all VSW channels: $n=1,2,...$, $m = -n,...,n$, $j=1,2$, and the sum converges for any nonzero $\Im \xi$. \eqref{appenG-thermal_bound} shows a distinct threshold behavior within each VSW channel. In the asymptotic limits of radiation-dominant ($\rho_i \gg \Im\xi$) or material-loss-dominant ($\rho_i \ll \Im \xi$) scenarios, \eqref{appenG-thermal_bound} simplifies to the known channel-~\cite{pendry1999radiative} and lossy-material bounds~\cite{miller_fundamental_2016}. In tandem, accounting for both mechanisms yields a significantly tighter bound than any previous approach.

\begin{figure}[t!]
	\includegraphics[width=0.5\textwidth]{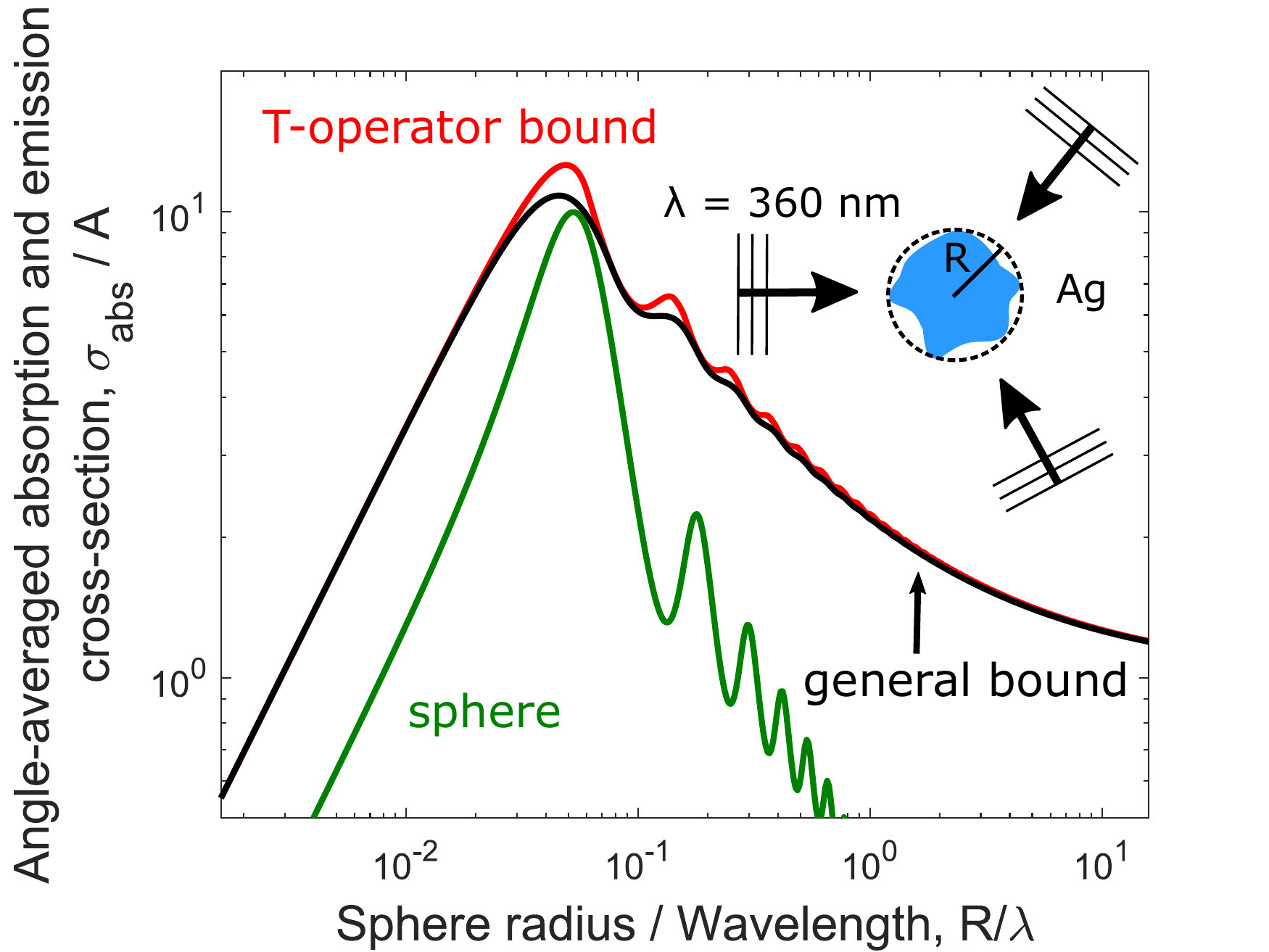}
	\centering
	\caption{General bound for maximum thermal absorption and emission, compared with $\mathbb{T}$-operator bound and thermal absorption of an actual Ag \cite{johnson1972optical} sphere with radius $R$.}
	\label{fig:appenG-figure6}
\end{figure}

Taking the same approach as in Sec.~\ref{sec:chap2-plane} in Chapter~\ref{chap:analy}, we compute the bound for finite-sized thermal absorbers with a spherical bounding volume. \Figref{appenG-figure6} shows the thermal absorption/emission cross-section as a function of the size of a spherical silver~\cite{johnson1972optical} nanoparticle at wavelength $\lambda = \SI{360}{nm}$. Included is the bound of \eqref{appenG-thermal_bound}, which is nearly achieved by the sphere at its ideal resonant size. We also include the recently published $\mathbb{T}$-operator bound of \citeasnoun{Molesky2019a}, which considered the effect of radiation and material losses separately for thermal sources. As shown in \figref{appenG-figure6}, by incorporating both losses in one optical theorem constraint, even for thermal fields the new bounds are slightly tighter.

\clearpage
%

\begin{table*}[t]
\centering
\begin{tabular}{ | m{1.5cm} | m{1.6cm}| m{4cm} | m{1.5cm} | m{1.6cm}| m{4cm} | }
\hline
    thickness ($\SI{}{\micro\meter}$)  & absorption (\%) &   \hspace{14mm} design & thickness ($\SI{}{\micro\meter}$)  & absorption (\%)   &  \hspace{14mm} design\\
    \hline
    \hspace{15mm} 0.4 &   \hspace{15mm} 52 & \vspace{1mm}\hspace{1.5mm} \phototm{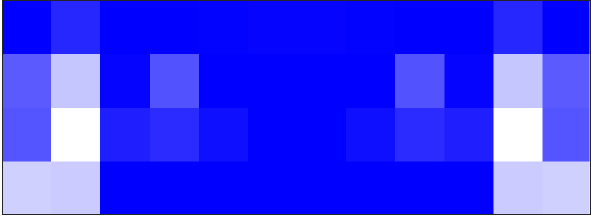} & \hspace{15mm} 0.6 &   \hspace{15mm} 57 & \vspace{1mm}\hspace{1.5mm} \phototm{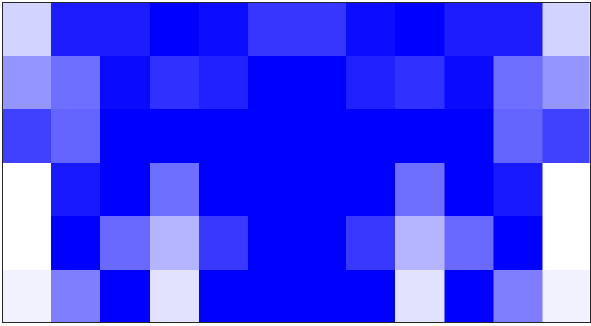}\\
    \hspace{15mm} 0.8 &   \hspace{15mm} 70 & \vspace{1mm}\hspace{1.5mm} \phototm{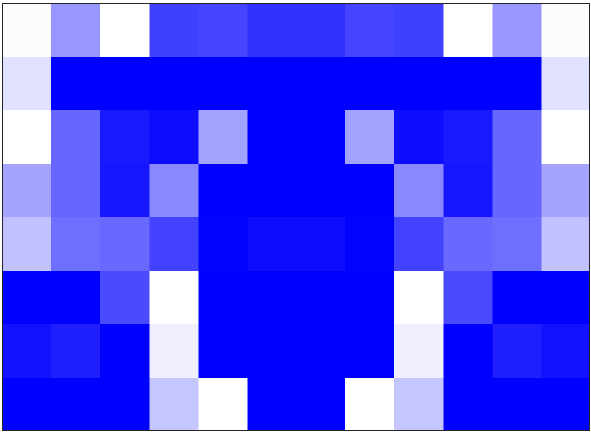} &
    \hspace{15mm} 1.0 &   \hspace{15mm} 76 & \vspace{1mm}\hspace{1.5mm} \phototm{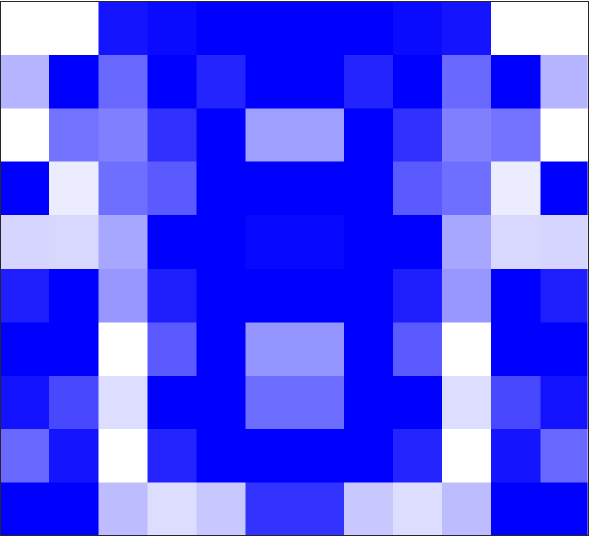} \\
    \hspace{15mm} 1.2 &   \hspace{15mm} 90 & \vspace{1mm}\hspace{1.5mm} \phototm{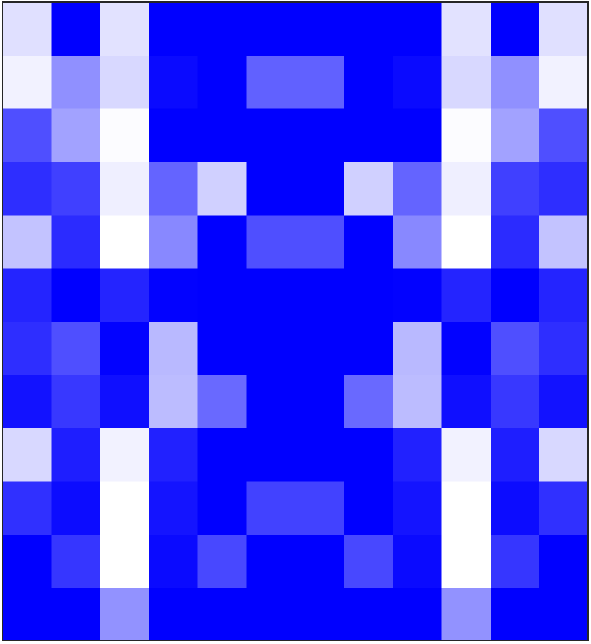} &
    \hspace{15mm} 1.4 &   \hspace{15mm} 94 & \vspace{1mm}\hspace{1.5mm} \phototm{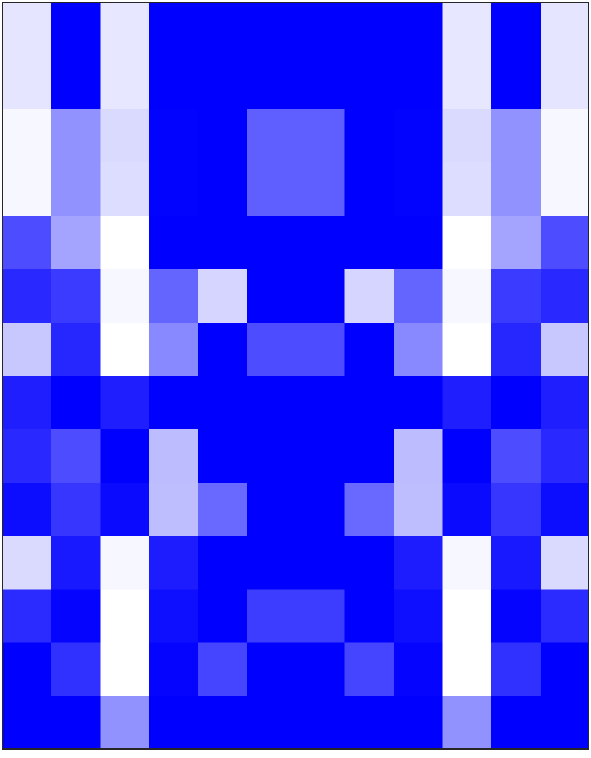} \\
    \hspace{15mm} 1.6 &   \hspace{15mm} 95 & \vspace{1mm}\hspace{1.5mm} \phototm{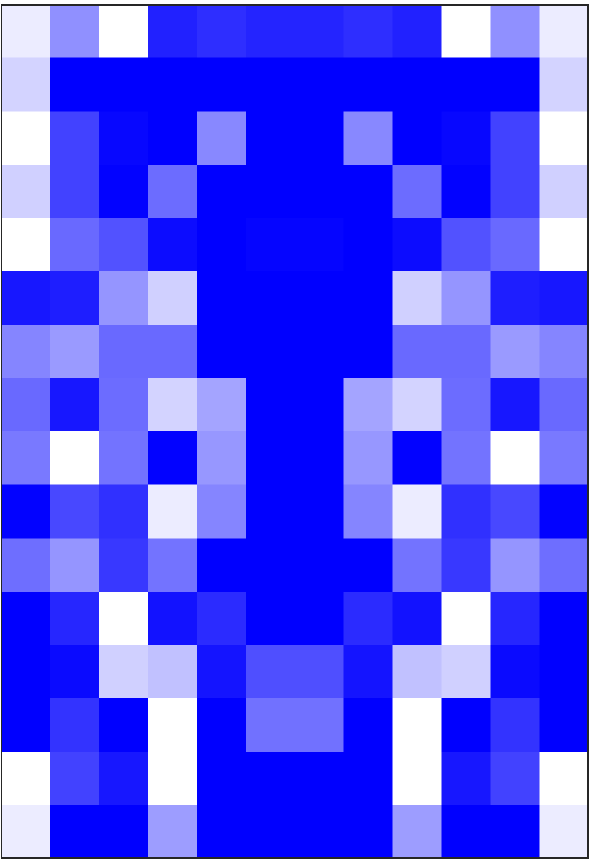} & & & \\
\hline
\end{tabular}
\caption{Inverse-designed SiC ultra-thin absorbers at $\SI{11}{\micro\meter}$. These are grey-scale designs with material ranges from pure air (purely white) to pure SiC~\cite{francoeur2010spectral} (dark blue).}
\label{tab:appenG-id1}
\end{table*}

\clearpage

\begin{table*}[t]
	\centering
	\begin{tabular}{ | m{1.5cm} | m{2cm} | m{1.5cm} | m{2cm} | m{3cm}| m{4cm} | }
\hline
    material  & wavelength  &\hspace{1.5mm} period &\hspace{1.8mm} thickness &\hspace{13.5mm} $\varepsilon$ &\hspace{14mm} design\\
    \hline
    \hspace{11.5mm} Au~\cite{palik1998handbook}  & 500 nm  & 55 nm &\hspace{5mm} 80 nm  &\hspace{5mm} -2.99+2.93i &\vspace{1mm}\hspace{1.5mm} \phototm{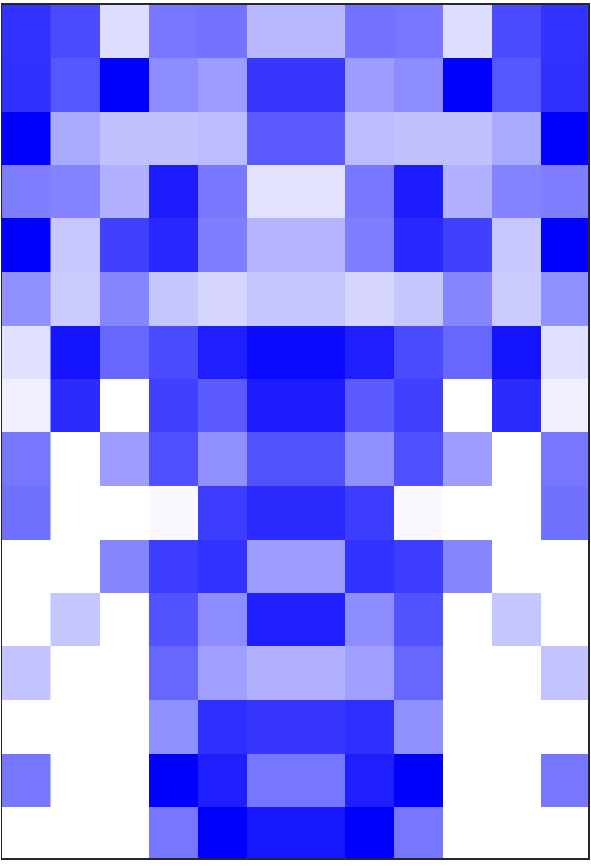}\\
    \hspace{11.5mm} Ag~\cite{palik1998handbook}  & 500 nm  & 55 nm &\hspace{5mm} 40 nm &\hspace{5mm} -7.63+0.73i &\vspace{1mm}\hspace{1.5mm} \phototm{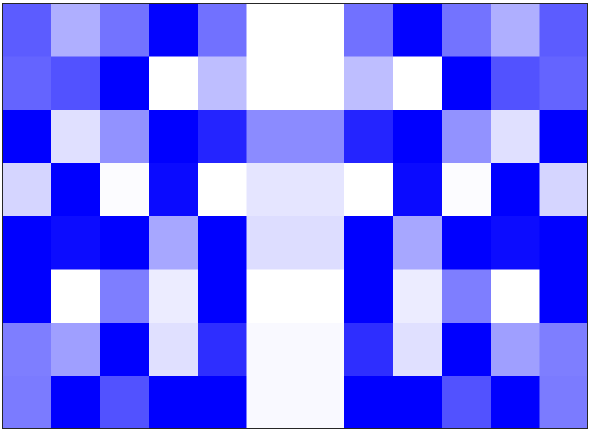}\\
    \hspace{11.5mm} Al~\cite{palik1998handbook}  & 500 nm  & 55 nm &\hspace{5mm} 40 nm &\hspace{5mm} -34.23+8.98i &\vspace{1mm}\hspace{1.5mm} \phototm{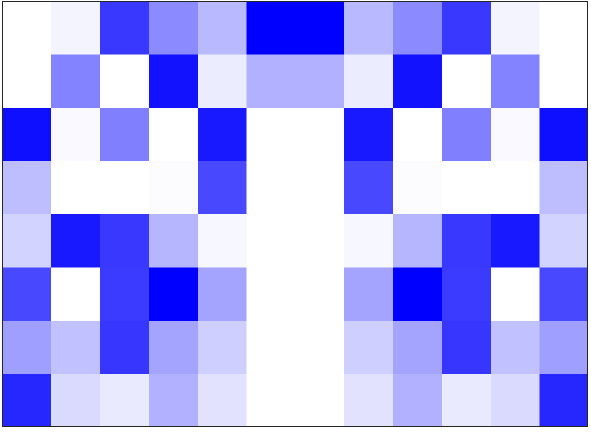}\\
    
    \hspace{11.5mm} SiO$_2$~\cite{popova1972optical}  & $\SI{9}{\micro\meter}$  &\hspace{4mm} $\SI{1.1}{\micro\meter}$ &\hspace{4.5mm} $\SI{1.4}{\micro\meter}$ &\hspace{5mm} -4.71+3.20i &\vspace{1mm}\hspace{1.5mm} \phototm{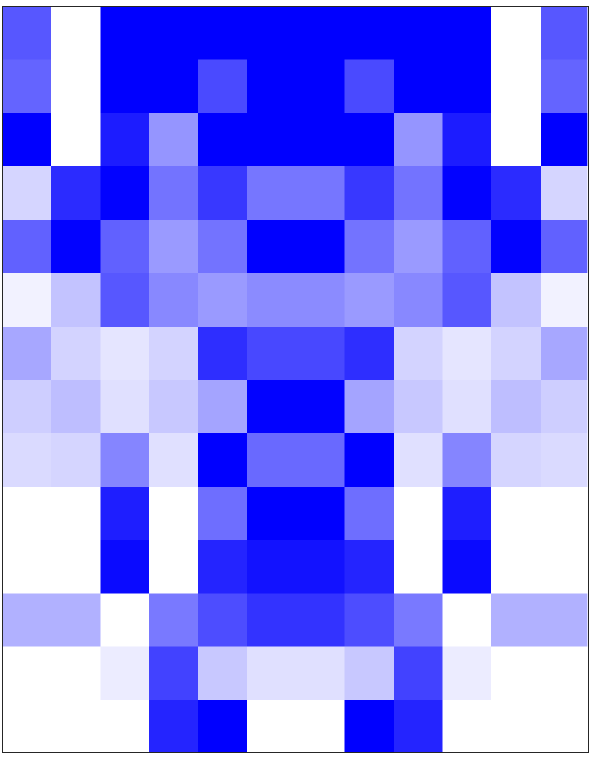}\\
    \hspace{6.5mm} doped InAs~\cite{law2013all}  &\hspace{10mm} $\SI{7.5}{\micro\meter}$  &\hspace{4mm} $\SI{1.1}{\micro\meter}$ &\hspace{4.5mm} $\SI{0.6}{\micro\meter}$ &\hspace{5mm} -10.39+1.80i &\vspace{1mm}\hspace{1.5mm} \phototm{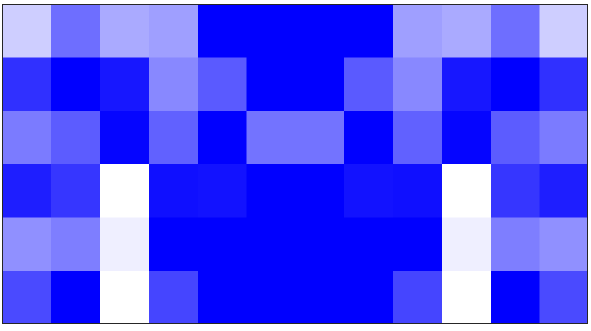}\\
    \hspace{11.5mm} SiC~\cite{francoeur2010spectral}  &\hspace{10mm} $\SI{11}{\micro\meter}$  &\hspace{4mm} $\SI{1.1}{\micro\meter}$ &\hspace{4.5mm} $\SI{0.8}{\micro\meter}$ &\hspace{5mm} -3.81+0.23i &\vspace{1mm}\hspace{1.5mm} \phototm{appenG-fig5-SiC_h08um}\\
\hline
\end{tabular}
\caption{Inverse-designed ultra-thin absorbers with 70\% absorption rate. These are grey-scale designs with material ranges from pure air (purely white) to pure material (dark blue).}
\label{tab:appenG-id2}
\end{table*}

\chapter{Computational bounds via local power conservation: supplementary }
\label{appen:sup_comp}

This Appendix provides supplementary materials to support the arguments in Chapter~\ref{chap:comp}. We (1) show the standard semidefinite relaxation process of the QCQP that leads to computational bounds, (2) derive the maximally violated local constraints that are used in our iterative algorithm, (3) show the detailed evolution of the iterative algorithm in the exemplar case of the maximal absorption cross-section in cylindrical and triangle domains, (4) derive the expressions of the $T$ matrix in cylindrical wave basis, (5) translate the expression from $T$ matrix to $S$ matrix and formulate $S$-matrix-feasibility objective as a quadratic function of polarization fields, (6) show the numerical details of the computational bounds that determine the minimum diameter of a power splitter, (7) formulate bandwidth-averaged extinction as a quadratic function of the polarization field at a complex frequency via contour integrals, and (8) prove the positive semidefiniteness of operators $\Im\xi$ and $\Im\GG_0$ at complex frequencies.  

\hiddensection{Semidefinite relaxation of the QCQP problem}
\label{sec:appenH-sec1}
In \eqref{chap3-local-opt-problem} of Chapter~\ref{chap:comp}, we show that one can formulate the bound problem with a quadratic-form objective and the conservation-law constraints:
\begin{equation}
    \begin{aligned}
        & \underset{p}{\text{max.}}
        & & f(p) = p^\dagger \mathbb{A}  p +  \Re\left(\beta^\dagger  p\right)+c \\
        & \text{s.t.}
        & &   p^\dagger \Re\left\{\DD_j(\xi+\GG_0)\right\}  p = -\Re \left( p^\dagger\DD_j \einc\right),\quad \text{for } i = 1,2,...,2n.
    \end{aligned}
    \label{eq:appenH-opt_prob}
\end{equation}
Here we separate the real and imaginary parts of the complex-valued constraints in \eqref{chap3-local-opt-problem} by taking the real part of the original constraints in \eqref{appenH-opt_prob} and appending $n$ more $\DD$-matrix constraints with $\DD_{n+j} = i\DD_{j}$ constraints to account for the imaginary part.
This type of problem with quadratic objective and quadratic constraints is well studied in the optimization literature~\cite{ben-tal_hidden_1996,Luo2010,park_general_2017}.
In this section we describe how it is translated to a semidefinite program using standard techniques: each step below is also clearly explained in \citeasnoun{Luo2010}. 
The first step of the transformation is to \emph{homogenize} the quadratic forms on \eqref{appenH-opt_prob}, which means introducing an additional variable in order to have purely quadratic and scalar terms without any linear term. To do this, in the objective function we introduce a complex-valued scalar variable $s$ into the linear term:
\begin{align}
    f( p) =  p^\dagger \mathbb{A}  p +  \Re\left(s^\dagger \beta^\dagger  p\right)+c.
    \label{eq:appenH-SDR1}
\end{align}
The key advantage of introducing this variable is that now one can write $f$ as a homogeneous quadratic form:
\begin{align}
    f\left(
        \begin{bmatrix}
             p \\
            s
        \end{bmatrix}
    \right)
    = 
    \begin{pmatrix}
         p \\
        s
    \end{pmatrix}^\dagger 
    \begin{pmatrix}
        \mathbb{A} & \frac{1}{2}\beta \\
        \frac{1}{2}\beta^\dagger & 0
    \end{pmatrix}
    \begin{pmatrix}
         p \\
        s
    \end{pmatrix}
    + c.
\end{align}
We can do this for each of the $j$ constraints as well, introducing the dummy variable s for each constraint, which then takes the form:
\begin{align}
    \begin{pmatrix}
         p \\
        s
    \end{pmatrix}^\dagger 
    \begin{pmatrix}
        \Re\left\{\DD_j(\xi+\GG_0)\right\} & \frac{1}{2} \DD_j  \einc \\
        \frac{1}{2}\omega  \einc^\dagger \DD_j^\dagger & 0
    \end{pmatrix}
    \begin{pmatrix}
         p \\
        s
    \end{pmatrix}
    = 0.
\end{align}
One cannot allow $s$ to take arbitrary values or else it will modify the initial problem. Instead, it should be required to have modulus one, i.e., $|s|^2 = 1$, which is itself a quadratic form in the degrees of freedom $ p$ and $s$. Finally, we can lump all degrees of freedom into a single vector $v$:
\begin{align}
    v = 
    \begin{pmatrix}
         p \\
        s
    \end{pmatrix}.
\end{align}
With this notation, the objective, the $2n$ conservation-law constraints, and the modulus constraint of $s$ are all written in the form 
\begin{align}
    v^\dagger \mathbb{F} v.
\end{align}
The way to optimize over such quadratic forms is to ``lift'' them to a higher-dimensional space where they become linear forms. The first step is to use the trace operator to rewrite the quadratic form:
\begin{align}
    v^\dagger \mathbb{F} v = \Tr{\mathbb{F} v v^\dagger}.
\end{align}
Then one defines a rank-one \emph{matrix variable} $\mathbb{X}$ given by $vv^\dagger$, in which case we now have a linear form:
\begin{align}
    \Tr{\mathbb{F} vv^\dagger} = \Tr{\mathbb{F} \mathbb{X}}.
    \label{eq:appenH-SDR2}
\end{align}
One cannot optimize arbitrarily over $\mathbb{X}$ and have an equivalent problem; one must additionally impose constraints that $\mathbb{X}$ be a rank-one, positive-definite matrix. The rank-one constraint is nonconvex; the ``relaxation'' in semidefinite relaxation (SDR) refers to dropping this rank-one constraint. Once that constraint has been removed, one is left with a linear objective function (in $\mathbb{X}$), and $2n+1$ linear constraints, over the space of positive-definite matrices. The transformation to a semidefinite program is complete.

The same transformation applies to the weighted average of the $\DD_j$-matrix constraints as in \eqref{appenH-opt-weight} of Chapter~\ref{chap:comp}:
\begin{equation}
\begin{aligned}
    & \underset{p}{\text{maximize}}
    & & f(p) = p^\dagger \mathbb{A} p +  \Im\left(\beta^\dagger p\right) \\
    & \text{subject to}
    & & p^\dagger\Re\left[\DD_{\textrm{wei}, j} (\GG_0+\xi \II)\right]p = - \Re\left(p^\dagger\DD_{\textrm{wei}, j} e_{\rm inc}\right),\quad \text{for } j = 1,2,...,m,
\end{aligned}
\label{eq:appenH-opt-weight}
\end{equation}
which essentially has the same form as \eqref{appenH-opt_prob}. If there is only one (global) constraint in the optimization problem \eqref{appenH-opt-weight}, the semidefinite relaxation mentioned in \eqrefrange{appenH-SDR1}{appenH-SDR2} does not introduce any actual relaxation~\cite{Luo2010}. Furthermore, even though a certain degree of relaxation may be triggered by additional (local) constraints, it is straightforward to show that the additional constraints can \emph{only} tighten the bound. This can also be seen in our examples for both absorption cross-section (Fig.~\ref{fig:chap3-fig1}(b) in Chapter~\ref{chap:comp}) and broadband extinction (Fig.~\ref{fig:chap3-fig2} in Chapter~\ref{chap:comp}), where the bounds are always monotonically decreasing with the additional local constraints.

\hiddensection{Algorithm: Maximally violated local constraints}
\label{sec:appenH-sec2}
In this section we derive the optimal new $\DD_{\rm wei}$ matrix, and corresponding conservation-law constraint, that should be added to a given set of constraints by our principle of maximum violation. The conservation-law constraints as given in \eqref{appenH-opt-weight} are of the form
\begin{align}
    p^\dagger\Re\left[\DD_{\textrm{wei}, j} (\GG_0+\xi \II)\right]p = - \Re\left(p^\dagger\DD_{\textrm{wei}, j} e_{\rm inc}\right),
    \label{eq:appenH-loc_law}
\end{align}
where $j$ runs from 1 to $m$, where $m$ is the current number of constraints that have been imposed. The key remaining question, then, is how to select the $(m+1)^{\rm th}$ constraint? From the first $m$ constraints, one can identify a potentially optimal polarization field $ p_{\rm opt}$ as the first singular vector of the optimal matrix solution of the SDP (as discussed in \secref{appenH-sec1}). Given this polarization field, then, a sensible approach to selecting a new constraint is to identify the constraint whose residual is largest when evaluated for polarization field $ p_{\rm opt}$. In other words, we want the $\DD_{\textrm{wei}, m+1}$ that maximizes the quantity
\begin{equation}
    \begin{aligned}
        & \underset{\DD}{\text{maximize}}
        & & \left| \Re\left\{  p_{\rm opt}^\dagger \DD(\xi+\GG_0)  p_{\rm opt} +  p_{\rm opt}^\dagger\DD \einc \right\} \right|.
    \end{aligned}
\end{equation}
By the cyclic property of the matrix trace, we can rewrite this expression as
\begin{align}
    \Re\Tr\left\{ \DD\left[ (\xi+\GG_0)  p_{\rm opt}  p_{\rm opt}^\dagger +   \einc  p_{\rm opt}^\dagger \right] \right\},
\end{align}
where we dropped the absolute value since any optimal negative value can be reversed through $\DD \rightarrow -\DD$. Let us denote the matrix in square brackets as $\mathbb{C}$. Expanding the real (Hermitian) part, we have
\begin{align}
    \frac{1}{2} \left[ \Tr\left(\DD\mathbb{C} \right) + \Tr\left(\DD^\dagger \mathbb{C}^\dagger\right) \right].
\end{align}
Clearly one can maximize the residual by allowing the norm of $\DD$ to be arbitrarily large, but that would not give insight into which spatial pattern $\DD$ should take. As a normalization we can take the Frobenius norm of $\DD$ to be 1, i.e. $\Tr\left(\DD^\dagger \DD\right) = 1$. Then, straightforward variational calculus yields an optimal $\DD$ matrix given by $\DD = \mathbb{C}^\dagger$; since $\DD$ must be (spatially) diagonal, we take $\DD$ to comprise the diagonal elements of $\mathbb{C}^\dagger$:
\begin{align}
    \DD_{\textrm{wei}, m+1} &= \diag\left[\mathbb{C}^\dagger\right] \nonumber \\
              &= \omega \diag\left[ p_{\rm opt}  p_{\rm opt}^\dagger \left(\xi+\GG_0\right)^\dagger +  p_{\rm opt} \einc \right],
              \label{eq:appenH-D_iter}
\end{align}
where now ``$\diag$'' strips its matrix argument of all elements except along the (spatial) diagonal, as in Chapter~\ref{chap:comp}. This is the optimal selection of the $\DD_{\rm wei}$ matrix as presented in \eqref{chap3-max-vio} of Chapter~\ref{chap:comp}, which significantly accelerates convergence of the bound computation.

\begin{figure}[!htp]
	\includegraphics[width=1\textwidth]{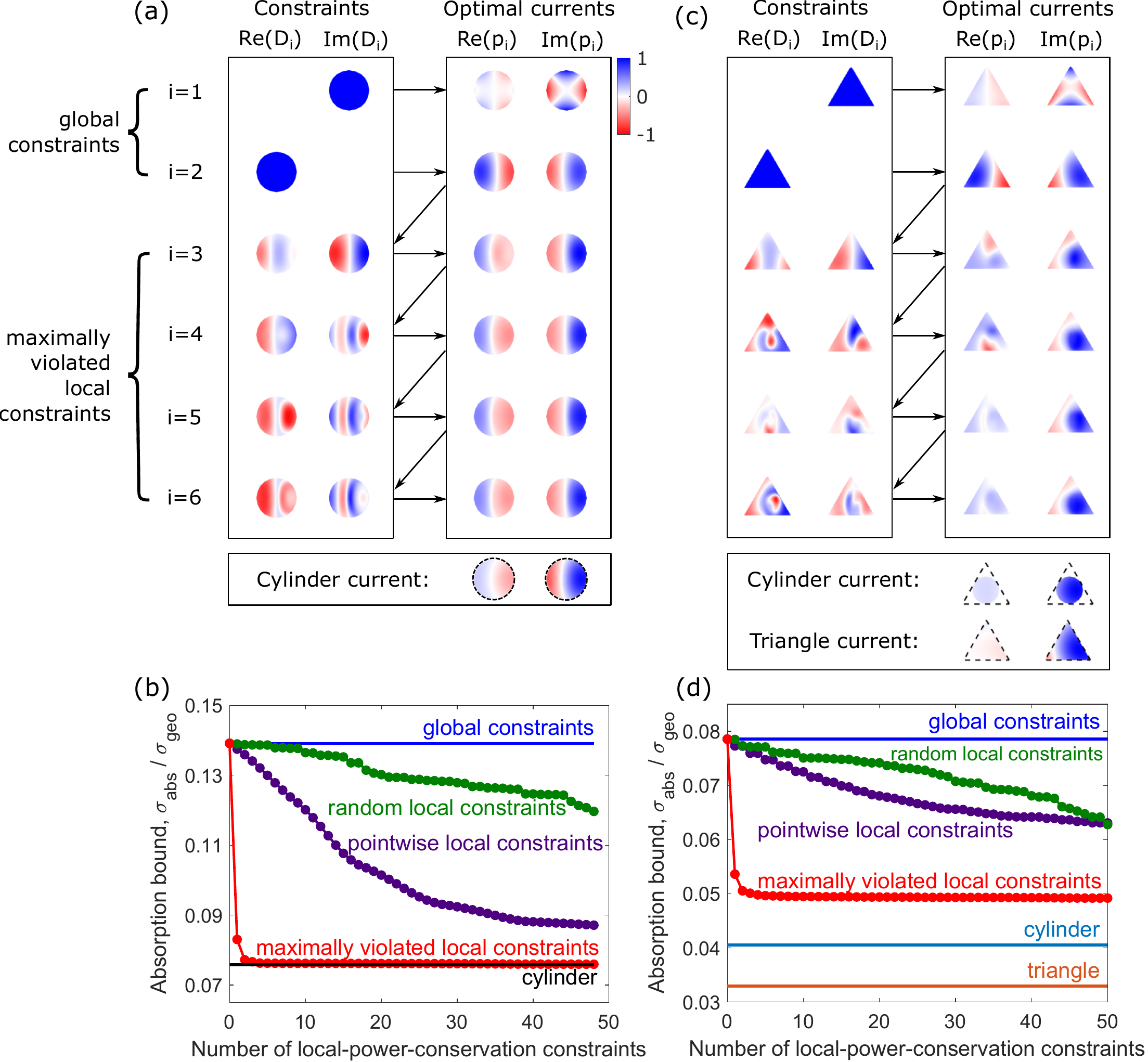}
    \centering
    \caption{(a) The left panel shows spatial profiles of the global and the first four maximally-violated local constraints. The design region is a cylinder with diameter $0.18\lambda$ and the presented profiles are its 2D cross sections. D$_i$ denotes the $E_z$ component of $\DD_{\textrm{wei}, i}$ in space, normalized so its maximal magnitude is one. Each iteration generates an optimal current, as indicated by the direction of the arrows. The normalized $E_z$ component of the optimal currents are shown on the right panel, which in turn generate the next maximally-violated local constraints.   (b) Convergence of the upper bound on absorption cross-section, $\sigma_{\rm abs}$, under maximally-violated local constraints (red line) as compared to the ones from only the global constraints (blue line) or other type of local constraints (green and purple lines). Figures (c,d) are the same as (a,b) but the designableregion is now an equilateral triangle with side length 0.18$\lambda$. }
    \label{fig:appenH-SM_figure1}
\end{figure}

\hiddensection{Maximal absorption cross-section under local constraints}
\label{sec:appenH-sec3}
In Fig.\ref{fig:chap3-fig1}(b) of Chapter~\ref{chap:comp}, we provide an example of maximizing absorption cross-section under local constraints. In this section, we provide detail on the formulation of the optimization problem and the iteration process involved in identifying the maximally-violated local conservation laws. The main result in this section is summarized in \figref{appenH-SM_figure1}, where we consider not only a cylindrical design region, but also a triangular design region to showcase the generality of this computational approach.

The exact expression of absorption cross-section in terms of the polarization field $ p$ can be identified from the global power-conservation law~\cite{Lytle2005,Newton1976,Jackson1999}, which can be derived from \eqref{appenH-loc_law} by choosing $\DD_{\textrm{wei},j}$ as an identity multiplied by the unit imaginary number and multiplying \eqref{appenH-loc_law} by a factor of $\omega/2$:
\begin{align}
\underbrace{\frac{\omega}{2}  p^\dagger \left(\ImGO\right)  p}_{\Pscat} + \underbrace{\frac{\omega}{2}  p^\dagger \left(\Im \xi \right)  p}_{\Pabs}  = \underbrace{\frac{\omega}{2} \Im \left(\psi_{\rm inc}^\dagger  p\right)}_{\Pext},
\label{eq:appenH-optthm}
\end{align}
where each term from left to right represents scattered power, absorption, and extinction, respectively. If the incident wave is a plane wave, the expression of its intensity in our dimensionless unit is $I_0 = |E_0|^2 / 2$, where $E_0$ is the plane-wave amplitude. The absorption cross-section is defined as the ratio between absorption and plane-wave intensity $\sigma_{\rm abs} = P_{\rm abs} / I_0 = \omega p^\dagger \left(\Im \xi \right) p/|E_0|^2$. 

Maximizing absorption cross-section under local conservation laws is equivalent to the optimization problem:
\begin{equation}
    \begin{aligned}
        & \underset{ p}{\text{max.}}
        & & \sigma_{\rm abs} = \omega p^\dagger \left(\Im \xi \right) p/|E_0|^2 \\
        & \text{s.t.}
        & &  p^\dagger \Re\left\{\DD_{\textrm{wei}, j}(\xi+\GG_0)\right\}  p = -\Re \left( p^\dagger\DD_{\textrm{wei}, j} \einc\right).\quad\quad j = 1, 2, ..., m
    \end{aligned}
    \label{eq:appenH-opt_prob_abs}
\end{equation}
Given designable region, incident field, and material properties as inputs, one solves this optimization problem via semidefinite relaxation discussed in Sec. \ref{sec:appenH-sec1}. The rest of this section considers a specific example where the incident wave is a TE-polarized plane wave and the material is nonmagnetic with susceptibility $\varepsilon = 12 + 0.1i$. We consider two designable regions: a cylinder and a equilateral triangle. The cylindrical design region has diameter $d = 0.18\lambda$, total length $h$ in its translational invariant direction ($h \rightarrow \infty$, so we can solve the 2D simplification), and a geometric cross section $\sigma_{\rm geo} = dh$. The equilateral triangle has side length $l=0.18\lambda$, total length $h$ in its translational invariant direction, and a geometric cross section $\sigma_{\rm geo} = \sqrt{3}/2lh$.

As mentioned in Sec. \ref{sec:appenH-sec2}, the algorithm for generating the maximally-violated local conservation constraints is built up from the existing global conservation constraints.
Thus, we first solve the optimization problem (\ref{eq:appenH-opt_prob_abs}) with only the global real-power conservation constraint where $m = 1$ and $\DD_{\textrm{wei}, 1} = iI$. For a cylindrical design region, the optimization program returns an optimal polarization field $p_1$ shown in the right panel of \figref{appenH-SM_figure1}(a), and an upper bound $\sigma_{\rm abs} / \sigma_{\rm geo} = 4.12$, too loose to be shown in \figref{appenH-SM_figure1}(b). Adding an additional global reactive-power conservation constraint ($m = 2$, $\DD_{\textrm{wei}, 1} = iI$, and $\DD_{\textrm{wei}, 2} = I$) gives us a dipole-like optimal current $p_2$ shown in \figref{appenH-SM_figure1}(a), and an upper bound $\sigma_{\rm abs} / \sigma_{\rm geo} = 0.139$, marked by the blue line in \figref{appenH-SM_figure1}(b). Next, we include extra local conservation laws in the optimization problem (\ref{eq:appenH-opt_prob_abs}) to tighten the global bound (result shown in \figref{appenH-SM_figure1}(b)). In particular, we use the algorithm derived in Sec. \ref{sec:appenH-sec2} of this Appendix to generate maximally-violated local conservation constraints. For example, we use \eqref{appenH-D_iter} to find out a local constraint, $\DD_{\textrm{wei}, 3}$, that is maximally violated by the optimal current $p_2$. The spatial profile of its diagonal components (denoted by D$_3$) are shown in the left panel of \figref{appenH-SM_figure1}(a). This additional constraint reduces the upper bound (the second red marker from the left in \figref{appenH-SM_figure1}(b)), and together with global constraints $\DD_{\textrm{wei}, 1}$ and $\DD_{\textrm{wei}, 2}$, predicts an optimal current $p_3$ which resembles the polarization field in an unpatterned cylinder.
We continue this iteration for 50 more times in \figref{appenH-SM_figure1}(b) and show the spatial profile of the first six in \figref{appenH-SM_figure1}(a). After the fourth iteration, both the upper bound (red line \figref{appenH-SM_figure1}(b)), and the optimal currents have converged to the solution of an unstructured cylinder, suggesting the ineffectiveness of structuring in this particular case. 

The same algorithm is applied to a equilateral triangular design region shown in \figref{appenH-SM_figure1}(c,d). In this example, we consider two possible scattering structures: an unpatterned triangle with a dimension the same as the design region, and the largest unpatterned cylinder that can fit in the design region (bottom panel of \figref{appenH-SM_figure1}(c)). Neither structure generates the optimal current distribution predicted in the right panel of \figref{appenH-SM_figure1}(c), and consistently, neither reach the predicted upper bound in \figref{appenH-SM_figure1}(d). Unlike a cylindrical design region where an unpatterned cylinder is already the optimum, a triangular design region may benefit from a more complex structure. From a computational perspective, the asymmetry of the triangular region has no effect on the speed or convergence of the bound computations.

\hiddensection{Volume integral form of T-matrix}
\label{sec:appenH-sec4}
In \secref{chap3-S} of Chapter~\ref{chap:comp}, one of the examples considered is whether a specific scattering-matrix can be targeted by some designable region, an example that we discuss more in the next section. In this section, in preparation for that, we derive the transition-matrix ($T$-matrix) elements for waves impinging upon and exiting from a 2D circular bounding region. The $T$-matrix calculation is simpler than a direct $S$-matrix calculation, and the two are related in a simple way, as noted in the next section.

We first derive the volume integral form of $T$-matrix elements as a linear function of the polarization field in arbitrary basis functions in six-vector notations (i.e., $\psi$ for fields, $\phi$ for sources, $\Gamma_0$ for the background Green's function operator). Then, specifically for a 2D circular bounding region, we derive the $T$-matrix expression in the basis of vector cylindrical waves.

Given arbitrary bounding volume $V$, a set of incoming basis $\{\psi_{\rm in,n}\}$ is defined on its surface $\partial V$ through the orthogonal relation:
\begin{equation}
-\frac{1}{4}\int_{\partial V} \psi_{\rm in, i}(\xv_s)^\dagger P(\xv_s) \psi_{\rm in, j}(\xv_s) = \delta_{ij}, \quad \quad
P = 
\begin{pmatrix}
0 & \hat{\nv}\times \\
-\hat{\nv}\times & 0
\end{pmatrix},
\label{eq:appenH-orth_relation}  
\end{equation}
with $\hat{\nv}$ being the unit normal vector. When $i=j$, the right hand side of the orthogonality relation measures the power flow of state $ \psi_{\rm in, i}$ through the surface $\partial V$. (We choose the convention pointing outward for outgoing states and inward for incoming states.)
Outgoing states can be defined as the time reverse of the incoming states:
\begin{equation}
\psi_{\rm out, i}(\xv_s) = Q\psi^*_{\rm in, i}(\xv_s),  \quad \quad
Q = 
\begin{pmatrix}
I & 0 \\
0 & -I
\end{pmatrix},
\end{equation}
where the operator $Q$ flips the sign of the magnetic field, as required by time reversing.
The incident basis $\{\psi_{\rm inc, i}\}$ is defined by a linear combination of the incoming and outgoing basis:
$\psi_{\rm inc, i} = \alpha \psi_{\rm out, i} + \beta \psi_{\rm in, i}.$
Coefficients $\alpha$ and $\beta$ depend on the exact basis one choose. For example, for vector cylindrical waves, they are both $\frac{1}{2}$. 

Incident field $\psi_{\rm inc}$ can be expanded by the incident basis with coefficients $c_{\rm inc, i}$.
Similarly, scattered field $\psi_{\rm out}$ can be expanded by the outgoing basis with coefficients $c_{\rm out, i}$. These two sets of coefficients are connected by $T$-matrix.
\begin{equation}
\psi_{\rm inc} = \sum_i c_{\rm inc, i}\psi_{\rm inc, i}, \quad\quad \psi_{\rm scat} = \sum_i c_{\rm out, i}\psi_{\rm out, i},  \quad\quad 
\begin{pmatrix}
  \\
 c_{\rm out} \\
 \
\end{pmatrix}
=
\begin{pmatrix}
  \\
 T \\
 \quad\quad\quad\quad\quad
\end{pmatrix}
\begin{pmatrix}
  \\
 c_{\rm inc} \\
 \
\end{pmatrix}.
\end{equation}
Thus, the entry $T_{ij}$ measures the ratio $c_{\rm out, i} / c_{\rm inc, j}$. In other words, when the incident field $\psi_{\rm inc} = \psi_{\rm inc, i}$, $T_{ij}$ takes the value of $c_{\rm out, i}$. Using this definition, we can express $T_{ij}$ as a linear function of polarization field $\phi$ after some mathematical manipulation:
\begin{align}
T_{ij} &= -\frac{1}{4}\int_{\partial V} \psi_{\rm out, i}^\dagger(\xv_s) P(\xv_s) \psi_{\rm scat}(\xv_s) \\
 &= -\frac{1}{4\alpha}\int_{\partial V} \psi_{\rm inc, i}^\dagger(\xv_s) P(\xv_s) \psi_{\rm scat}(\xv_s), \\
 &= -\frac{1}{4\alpha}\int_{\partial V} \int_{V} \psi_{\rm inc, i}^\dagger(\xv_s)P(\xv_s) \Gamma_0(\xv_s,\xv_v)\phi(\xv_v)
\end{align}
where we used the fact that $\psi_{\rm out,i} = \frac{1}{\alpha}\psi_{\rm inc,i} - \frac{\beta}{\alpha}\psi_{\rm in}$, and the incoming and outgoing fields are orthogonal in this inner product. To further simplify this equation, we first take its transpose, and then use the properties $P^T(\xv_s) = P(\xv_s)$ and $\Gamma_0^T(\xv_s,\xv_v)=\Gamma_0(\xv_v,\xv_s)$ to write $T_{ij}$ as:
\begin{align}
T_{ij} &= -\frac{1}{4\alpha}\int_{\partial V} \int_{V} \phi^T(\xv_v)\Gamma_0(\xv_v,\xv_s)P(\xv_s)\psi_{\rm inc, i}^*(\xv_s) \\
 &= \frac{1}{4\alpha}\int_{\partial V} \int_{V} \phi^T(\xv_v)\Gamma_0(\xv_v,\xv_s)P(\xv_s)\psi_{\rm inc, i}(\xv_s),
\end{align}
where we use the properties  $\psi_{\rm inc, i}^*(\xv_s) = Q\psi_{\rm inc, i}(\xv_s)$ and $-P(\xv_s)Q= P(\xv_s)$ to derive the second equality. Lastly, we identify that the product $P(\xv_s)\psi_{\rm inc, i}(\xv_s)$ gives the surface equivalent current $\xi_{\rm inc, i}(\xv_s)$ on the surface $\partial V$, which can be propagated back to the volume through the Green's function:
\begin{align}
T_{ij} &=\frac{1}{4\alpha}\int_{\partial V} \int_{V} \phi^T(\xv_v)\Gamma_0(\xv_v,\xv_s)\xi_{\rm inc, i}(\xv_s)    \\
 &= \frac{1}{4\alpha}\int_{V} \phi^T(\xv_v)\psi_{\rm inc, i}(\xv_v) \\
 &= \frac{1}{4\alpha}\phi^T\psi_{\rm inc, i}. \label{eq:appenH-Tij0}
\end{align}
The key result, \eqref{appenH-Tij0}, identifies $T_{ij}$ as a overlap integral between incident channel $\psi_{\rm inc, i}$ and polarization field $\phi$ that is induced by incident field $\psi_{\rm inc, j}$.

For a highly symmetric bounding volume, the derivation of the volume integral form of $T$-matrix can be greatly simplified. In the example provided in \secref{chap3-S} of Chapter~\ref{chap:comp}, we assume nonmagnetic material with a 2D bounding area and TE incidence. The basis for outgoing and incident field can be chosen as the set of vector cylindrical waves:
\begin{align}
v_{\rm inc, n}(\xv) &= \frac{1}{2}\hat{z}J_n(k\rho)e^{in\phi} \label{eq:appenH-vcw_1} \\
v_{\rm out, n}(\xv) &= \frac{1}{2}\hat{z}H^{(1)}_n(k\rho)e^{in\phi}, \label{eq:appenH-vcw_2}
\end{align}
where $J_n(x)$ is the Bessel function of order $n$, and $H^{(1)}_n(x)$ is the Hankel function of the first kind of order $n$. Conventionally, these two basis written here do not include magnetic field, and are not normalized based on \eqref{appenH-orth_relation}, so we use different notations other than $\psi_{\rm inc, n}$ and $\psi_{\rm out, n}$.

As discussed before, when $\psi_{\rm inc} = v_{\rm out, j}$, the entry $T_{ij}=c_{\text{scat}, i}$. By virtue of the Green's function expansion $\GO(\xv, \xv') = i \sum_n v_{\rm out, n}(\xv) v_{\rm inc, n}^\dagger(\xv')$ for $\rho>\rho'$, we can easily derive the volume integral form of $T_{ij}$:
\begin{align} 
	T_{ij}	&= \frac{\int_{\partial V} v^\dagger_{\rm out, i}(\xv_s)\psi_{\rm scat}(\xv_s)\td S}{\int_{\partial V} v^\dagger_{\rm out, i}(\xv_s)v_{\rm out, i}(\xv_s)\td S} \\
			&= \frac{\int_{\partial V} v^\dagger_{\rm out, i}(\xv_s)  \left[\int_V \GO(\xv, \xv')\phi(\xv') \td V\right]     \td S}{\int_{\partial V} v^\dagger_{\rm out, i}(\xv_s)v_{\rm out, i}(\xv_s)\td S} \\
			&= \frac{\int_{\partial V} v^\dagger_{\rm out, i}(\xv_s)  \left[\int_V i\left(\sum_n v_{\rm out, n}(\xv) v_{\rm inc, n}^\dagger(\xv') \right)\phi(\xv) \td V\right]     \td S}{\int_{\partial V} v^\dagger_{\rm out, i}(\xv_s)v_{\rm out, i}(\xv_s)\td S} \\
			&= i\int_V v_{\rm inc, i}^\dagger(\xv')\phi(\xv')\td V \\
			&= i v_{\rm inc, i}^\dagger \phi.
\end{align}
Similar as before, the result suggests that $T_{ij}$ is the projection of $\phi$ into the given incident basis $v_{\rm inc, i}$ with an additional phase delay, under the incident field $\psi_{\rm inc} = v_{\rm inc, j}$. There is slight difference between this and the more general result in \eqref{appenH-Tij0} because the vector cylindrical waves defined in  equations (\ref{eq:appenH-vcw_1}) and (\ref{eq:appenH-vcw_2}) do not include magnetic field components and are not normalized based on \eqref{appenH-orth_relation}.

\hiddensection{Formulation of S-matrix feasibility bound}
\label{sec:appenH-sec5}
The objective for the S-matrix feasibility problem is to minimize the relative difference between the achievable and target $S$ matrices:
\begin{equation}
\text{Min\ \ } f_{\rm obj} = \left\| S - \St \right\|^2 / \left\|\St\right\|^2,
\label{eq:appenH-fobj}
\end{equation}
where we choose $\|\cdot\|$ to denote Frobenius norm. 

It is simpler to translate the scattering matrix $S$, which relates \emph{incoming} waves to \emph{outgoing} waves, into the transition matrix $T$, which relates \emph{incident} waves to \emph{scattered} waves. One can typically choose a basis (such as the cylindrical-wave basis) for which $S = I+2T$. Inserting this relation into \eqref{appenH-fobj}, we have:
\begin{align}
f_{\rm obj} &= 4\left\| T - \Tt \right\| / \left\|\St\right\|^2 \\
	 &= \frac{4}{||\St||^2}\sum_{ij}|T_{ij} - T_{\text{target}, ij}|^2 \\
	 &= \frac{4}{||\St||^2}\sum_j f_{\text{obj}, j},
\end{align}
where in the last equality we separate out the objective into contributions from different incident fields:
\begin{align}
	 f_{\rm obj,j} = \sum_i |T_{ij} - T_{\text{target}, ij}|^2. \label{eq:appenH-fobj_j} 
\end{align}	
Each $f_{\rm obj,j}$ corresponds to the scattering from incident field indexed by $j$, so we bound them separately and later add up their contributions. As we proved in \secref{appenH-sec4}, $T_{ij}$ can be written as a linear function of $\phi$, which is the induced polarization field under the incident field $\psi_{\rm inc} = \psi_{\rm inc, j}$. Assume this linear relation is $T_{ij}=w_i^\dagger \phi$. We can plug it in \eqref{appenH-fobj_j} to express each $f_{\rm obj,j}$ as a quadratic function of $\phi$:
\begin{align}
	 f_{\rm obj,j} = \phi^\dagger\left(\sum_iw_i w_i^\dagger\right)\phi + \Re\left[\left(-2\sum_i T_{\text{target}, ij}w_i\right)^\dagger\phi\right] + \sum_i|T_{\text{target}, ij}|^2, \label{eq:appenH-Sobj}
\end{align}	
This can be written in the form of the objective in \eqref{chap3-local-opt-problem} of Chapter~\ref{chap:comp} (after adding a minus sign to the objective to turn minimization into maximization) with $\mathbb{A} = -\sum_iw_iw_i^\dagger$, $\beta = 2\sum_i T_{\text{target}, ij}w_i$, $c = -\sum_i|T_{\text{target}, ij}|^2$, and $\psi_{\rm inc} = \psi_{\rm inc, j}$. 

For the general case where the incident basis $\psi_{\rm inc, i}$ is defined through \eqref{appenH-orth_relation}, we substitute $\omega_i$ in \eqref{appenH-Sobj} with $\psi_{\rm inc, i}^* / (4\alpha^*)$. For the specific case where the we assume nonmagnetic material with a 2D bounding area and TE incidence, the incident basis $\psi_{\rm inc, i}$ is vector cylindrical waves $v_{\rm inc, i}$ defined in \eqref{appenH-vcw_1}, and we substitute $\omega_i$ with $-iv_{\rm inc, i}$.




\begin{figure}[t]
	\includegraphics[width=1\textwidth]{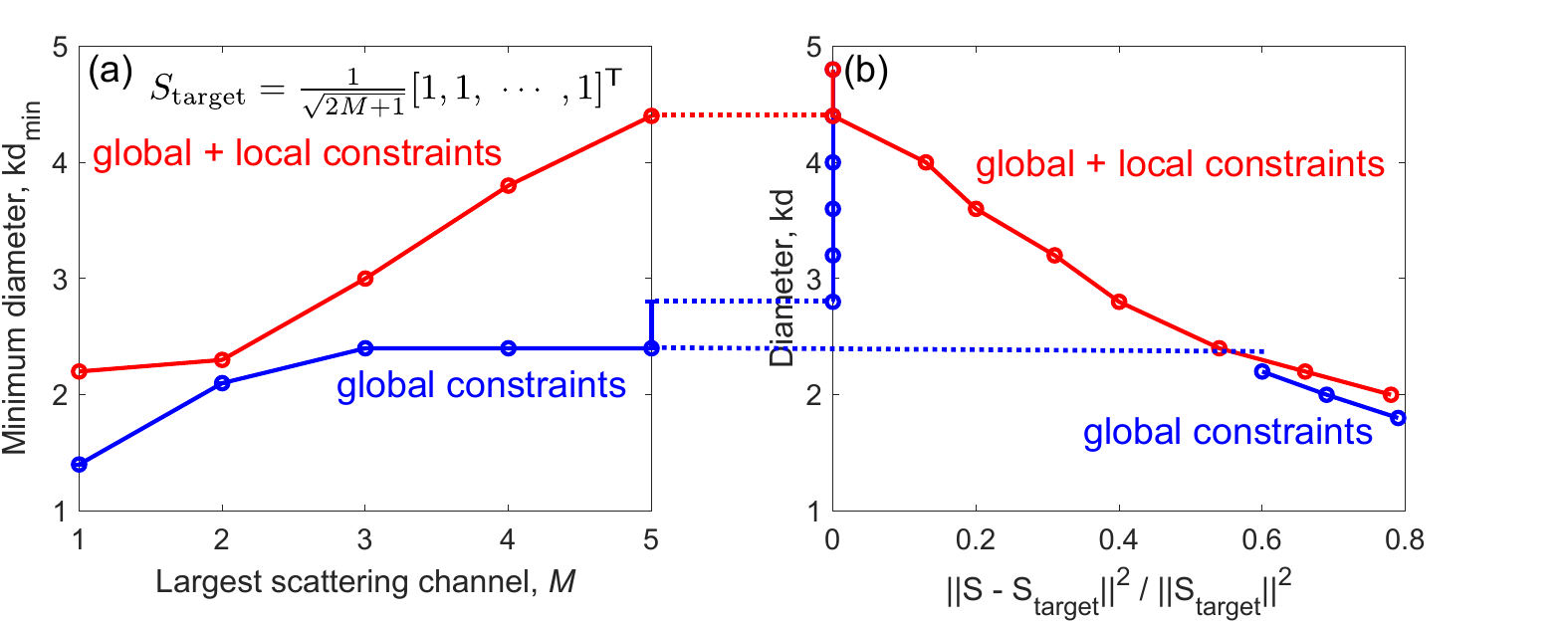}
    \centering
    \caption{(a) Minimum diameter required for a power splitter for a single input to 2M + 1 outgoing channels. (b) Lower bounds on the objective $||S-S_{\rm target}||^2 / ||\St||^2$ at each diameter when the largest scattering channel $M=5$. Uncertainties rising from the numerical instabilities in global constraints are marked by the dotted blue lines.}
    \label{fig:appenH-SM_figure4}
\end{figure}

\hiddensection{Minimum diameter of a power splitter}
\label{sec:appenH-sec6}
In \secref{chap3-S} of Chapter~\ref{chap:comp}, we show the minimum diameter required for a power splitter for a single input to 2M + 1 outgoing channels in the cylindrical-wave basis. The way we determine the minimum diameter for each $M$ is to minimize the objective function $\left\| S - \St \right\|^2 / \left\|\St\right\|^2$ for every diameter $d$, and choose the smallest one that satisfies $\left\| S - \St \right\|^2 / \left\|\St\right\|^2 < 1\%$. This process is shown in \figref{appenH-SM_figure4}(b) for the case with $M=5$.

The gap between two blue lines in \figref{appenH-SM_figure4}(b) originates from a numerical instability in the global-constraint-only approach. Higher orders of the cylindrical waves yield widely separated numerical scales in the corresponding matrices, such that with only global constraints the optimization does not terminate successfully for some diameters. The two dashed blue lines indicate the uncertainty region for determining the minimum diameter. The lower bound of this uncertainty region is estimated from the asymptotic limit of the global-constraint-only approach in \figref{appenH-SM_figure4}(a). The minimum diameter can be lower bounded by the lower dashed line of the uncertainty region, which explains the location of the circular point with the errorbar.

\hiddensection{Formulation of the bandwidth-averaged extinction bound}
\label{sec:appenH-sec7}
In this section, we transform the bandwidth-averaged extinction to a single scattering amplitude at a complex frequency by Cauchy's residue theorem, using a similar technique to that which has been demonstrated in Refs.~\cite{Hashemi2012,Shim2019}.
We start with the expression of single-frequency extinction cross section at a real frequency:
\begin{align}
	\sigma_{\rm ext}(\omega) = \Im\left[\omega  e^\dagger_{\rm inc}(\omega)  p(\omega) \right] \label{eq:appenH-ecs-1}
\end{align}
Incident field $ e_{\rm inc}(\omega)$ in far-field scattering is often approximated as a plane wave. Without loss of generality, we assume it has unit intensity and is propagating along the $x$ direction. We use dimensionless quantities with $c=1$, so the plane-wave frequency dependence can be written as $e^{i\omega x}$. In anticipation of an analytic continuation into the complex plane, we use the general relation $ e_{\rm inc}^*(\omega) =  e_{\rm inc}(-\omega)$ for real-valued frequencies~\cite{Landau1960} to remove the complex conjugation (which cannot be analytically continued):
\begin{align}
	\sigma_{\rm ext}(\omega) &= \Im\left[\omega  e^T_{\rm inc}(-\omega)  p(\omega) \right] \label{eq:appenH-ecs-2} \\
							&= \Im s(\omega). \label{eq:appenH-ecs-3}
\end{align}
Here, we define a new term $s(\omega)=\omega  e^T_{\rm inc}(-\omega)  p(\omega)$ that we identify as the far-field scattering amplitude.
Since the incident plane wave $ e_{\rm inc}(\omega)$ has the frequency dependence $e^{i\omega x}$ (analytic everywhere), and the polarization field $ p(\omega)$ is a causal linear-response function~\cite{Nussenzveig1972}, the amplitude $s(\omega)$ is analytic in the upper half of the complex-frequency plane (UHP).

The average extinction cross section $\langle \sigma_{\rm ext}\rangle$ in a bandwidth $\Delta\omega$  around a center frequency $\omega_0$ can be defined as the integral of the product of $\sigma_{\rm ext}(\omega)$ and a Lorentzian window function $\Hw(\omega)=\frac{\Delta\omega/\pi}{(\omega-\omega_0)^2+\Delta\omega^2}$: 
\begin{align}
	\langle \sigma_{\rm ext}\rangle &= \int_{-\infty}^{+\infty} \sigma_{\rm ext}(\omega)H(\omega)\td\omega  \label{eq:appenH-int_1}\\
	&= \Im\int_{-\infty}^{+\infty} s(\omega)H(\omega)\td\omega \label{eq:appenH-int_2}  
\end{align}
The integrand $s(\omega)H(\omega)$ has two properties that allows us to use Cauchy’s residue theorem to equate
the all-frequency integral to a single pole in the UHP. The first property is that the $s(\omega)H(\omega)$ only has one pole $\tomega = \omega_0 + i\Delta\omega$ from the window function in the UHP, since $s(\omega)$ is complex analytic in the UHP as discussed above. The second property is the magnitude of  $s(\omega)H(\omega)$ decays faster than $1/|\omega|$ when $|\omega|\rightarrow +\infty$. In this asymptotic limit, the window function $H(\omega)$ decays at a rate of $1/|\omega|$, and the amplitude $s(\omega)$ decay at the rate of $1 / |\omega|$, which can be proved as follows.

In the high-frequency limit, the polarization field must decay towards zero (the bound charges cannot respond to
such high frequencies), and on physical grounds~\cite{Landau1960} the decay must occur in proportion to $1/|\omega|$. Conventionally, the decay constant is chosen to be a ``plasma frequency'' $\omega_p$ that is physically meaningful for metals but applies to dielectrics as well. Because the scatterer becomes transparent at high frequencies, the Born approximation applies and the polarization field will be directly proportional to the incident field: $ p(|\omega|\rightarrow\infty) = -\frac{\omega_p^2}{\omega^2} e_{\rm inc}(\omega)$, so that $s(|\omega|\rightarrow\infty) = -\frac{\omega_p^2}{\omega} e_{\rm inc}^T(-\omega) e_{\rm inc}(\omega) \sim 1/\omega$. Note that the inner product $ e_{\rm inc}^T(-\omega) e_{\rm inc}(\omega)$ does not dependent on frequency as the frequency dependence of the incident plane wave is just $e^{i\omega x}$.

Taking these two properties into account, we can connect the upper and lower limit of the integral in \eqref{appenH-int_1} by a half circle in the UHP, which does not actually contribute to the integral due to the fast decay rate of the $s(\omega)H(\omega)$. Integration of this closed loop can be transformed into the single pole of $s(\omega)H(\omega)$ at $\tomega = \omega_0 + i\Delta\omega$ by Cauchy’s residue theorem, giving the expression in Chapter~\ref{chap:comp}:
\begin{equation}
\langle \sigma_{\rm ext}\rangle = \Im\left[\tomega  e^T_{\rm inc}(-\tomega) p(\tomega)\right].
\end{equation}
In the case of TE incidence in a 2D geometry with nonmagnetic material, we only need to consider the $z$ polarization component of the electric incident field, which is a scalar quantity. If we still use notation $ e_{\rm inc}$ to denote this quantity, we can solve for the maximum $\langle \sigma_{\rm ext}\rangle$ by the optimization problem with $\beta = i\tomega^*e^{i\omega_0x+\Delta\omega x}$ and incident field $ e_{\rm inc}(\tomega)=e^{i\omega_0x-\Delta\omega x}$.

\hiddensection{Positive semidefinite property of scattering and absorption operators}
\label{sec:appenH-sec8}
The power-bandwidth limit discussed in \secref{chap3-far} of Chapter~\ref{chap:comp} relies on the fact that the local power-conservation laws can be extended to complex frequency $\omega$. Explicitly writing out the frequency dependency of the operators, we have:
\begin{align}
\frac{\omega^*}{2}  p^\dagger \DD_\textrm{wei}\GG_0(\omega)  p + \frac{\omega^*}{2}  p^\dagger \DD_\textrm{wei}\xi(\omega)  p  = -\frac{\omega^*}{2}  p^\dagger\DD e_{\rm inc}.
\label{eq:appenH-optthm2}
\end{align} 
Among all the possible local conservation laws we can impose, the most important one is the global power-conservation law. It constrains the optimization variable $ p$ to the boundary of a high-dimensional ellipsoid, and can be derived by assigning $\DD_\textrm{wei}$ an identity tensor and take the imaginary part of \eqref{appenH-optthm2}: 
\begin{align}
\frac{1}{2}  p^\dagger \Im\left[\omega^*\GG_0(\omega)\right]  p + \frac{1}{2}  p^\dagger \Im\left[\omega^*\xi(\omega)\right]  p  = -\frac{1}{2} \Im(\omega^* p^\dagger e_{\rm inc}).
\label{eq:appenH-global_power}
\end{align} 
In this section, we prove the positive semidefinite property of the two involving operators, $\Im\{\omega^*\GG_0(\omega)\}$ and $\Im\{\omega^*\xi(\omega)\}$, in the UHP, using a similar technique to that which has been used in Ref.~\cite{zemanian_hilbert_1970, zemanian_realizability_1995, welters_speed--light_2014}.

We first prove the positive semidefinite property of the operator $\Im\{\omega^*\xi(\omega)\}$ in a passive scattering problem. Passivity requires that the polarization fields $ p$ in the material do not do work. The total work they do up to a time $t$ must be greater than or equal to zero:
\begin{equation}
    \Re\int {\rm d}x \, \int_{-\infty}^{t} {\rm d}t'\,  e^\dagger(x,t') \frac{\td p(x,t')}{\td t'} \geq 0.
\label{eq:appenH-pvt1}
\end{equation}
In a scattering problem where $ e$ is the total field, we can interpret $ p$ as the polarization fields, which are the convolution of the susceptibility in time and space (we allow for spatial nonlocality):
\begin{align}
     p(x,t') = \int {\rm d}x' \int_{-\infty}^{t'} {\rm d}t'' \, \chi(x,x',t'-t'')  e(x',t'') = \int {\rm d}x' \int_0^\infty \chi(x,x',\tau)  e(x',t'-\tau) \,{\rm d}\tau,
\end{align}
where in the second expression the variable $\tau$ can be interpreted as the delay since the excitation that is creating a response. Inserting the latter expression into \eqref{appenH-pvt1} we have:
\begin{align}
    \Re\int \int {\rm d}x' {\rm d}x \, \int_{-\infty}^{t} {\rm d}t'\,  e^\dagger(x,t') \int_0^{\infty} {\rm d}\tau \, \chi(x,x',\tau)  e'(x',t'-\tau) \geq 0,
    \label{eq:appenH-TI1}
\end{align}
where $ e'$ denotes the derivative of $ e$. The expression of \eqref{appenH-TI1} must be valid for all $ e$. We can choose a simple time-dependence for $ e$, following Refs.~\cite{zemanian_realizability_1995, welters_speed--light_2014}:
\begin{align}
     e(x,t') = 
    \begin{cases}
         e(x) e^{-i\omega t'} & \text{ for } t' < T \\
        0 & \text{ for } t' \geq T,
    \end{cases}
\end{align}
where $\omega$ is a complex-valued frequency, i.e. $\omega = \omega_0 + i \Im \omega$, and $T$ is simply a shut-off time that we will always choose larger than $t$ and which assures technical conditions are satisfied in rigorous proofs~\cite{zemanian_realizability_1995, welters_speed--light_2014}. Given this form, \eqref{appenH-TI1} becomes:
\begin{align}
    \Re\int \int {\rm d}x' {\rm d}x \,  e^{\dagger}(x) \int_{-\infty}^{t} {\rm d}t'\, e^{i\omega_0 t'} e^{(\Im \omega) t'} \int_0^{\infty} {\rm d}\tau \, \chi(x,x',\tau) \left(-i\omega\right) e^{-i\omega_0 (t'-\tau)} e^{(\Im \omega) (t'-\tau)}  e(x') \geq 0,
    \label{eq:appenH-TI2}
\end{align}
Re-arranging terms then gives
\begin{align}
    \Re \left[ \left(-i\omega\right) \int \int {\rm d} x' {\rm d}x \,  e^{\dagger}(x) \int_{-\infty}^{t} {\rm d}t'\, e^{2(\Im \omega) t'} \left\{ \int_0^{\infty} {\rm d}\tau \, \chi(x,x',\tau) e^{i\omega \tau} \right\}  e(x') \right] \geq 0.
    \label{eq:appenH-TI3}
\end{align}
The term in curly brackets is proportional to the Fourier transform of $\chi$, i.e. $\chi(x,x',\omega)$ at complex frequency $\omega$, and we can drop the constants related to $2\pi$. The integral over $t'$ is easily evaluated. Finally, noting that $\Re(-iz) = \Im(z)$, we have the expression
\begin{align}
    \frac{e^{2(\Im\omega)t}}{2 \Im \omega} \Im \int \int {\rm d} x' {\rm d}x \,  e^{\dagger}(x) \left[ \omega \chi(x,x',\omega) \right]  e(x') \geq 0.
    \label{eq:appenH-TI4}
\end{align}
This expression must be valid for all $ e(x)$ distributions. We can remove the spatial dependence in $\chi(x,x',\omega)$ and instead treat it as a square matrix $\chi(\omega)$ (as in any standard discretization), in which case we can simply write that
\begin{align}
    \Im \left[ \omega \chi(\omega) \right] \geq 0 \quad \text{ for } \Im \omega > 0,
    \label{eq:appenH-chipos}
\end{align}
where the imaginary part of the matrix argument refers to its anti-Hermitian part; e.g., $\Im A = (A - A^\dagger)/2i$. 

To convert \eqref{appenH-chipos} to an inequality for $\xi(\omega)$, we use the fact that $\chi(\omega) = -\xi^{-1}(\omega)$ to rewrite \eqref{appenH-chipos} as
\begin{align}
    \Im \left[ \omega \chi(\omega) \right] &= \Im \left[-\omega \xi^{-1}(\omega) \right] \\
           &= \Im \left[-\omega \left(\xi^\dagger(\omega)\xi(\omega)\right)^{-1} \xi^\dagger(\omega)\right] \\
           &= \left(\xi^\dagger(\omega)\xi(\omega)\right)^{-1} \left( \Im \left[ \omega^* \xi(\omega) \right] \right),
\end{align}
which implies that
\begin{empheq}[box=\widefbox]{align}
    \Im \left[ \omega^* \xi(\omega) \right] \geq 0 \quad \text{ for } \Im \omega > 0.
\end{empheq}

Thus we have our proof for the positive semidefinite property of the first of our two operators. Now we can follow similar logic for the second one. We start with an expression similar to \eqref{appenH-pvt1}, but now we change our interpretation: we will take the $ p$ to be free currents, $ e$ to be the fields radiated by them, and the quantity in \eqref{appenH-pvt1} then represents the \emph{negative} of the work done by those currents on the outgoing field (which again must be positive). Thus our starting point is the negative of \eqref{appenH-pvt1}:
\begin{equation}
    \Re\left[ - \int {\rm d}x \, \int_{-\infty}^{t} {\rm d}t'\, \frac{\td p(x,t')}{\td t'}^\dagger  e(x,t')  \right] \geq 0,
\label{eq:appenH-pvt2}
\end{equation}
where we also reversed the order of our arguments in the integrand for simplicity below. (That is allowed because $\Re z = \Re z^*$.)

Now our convolution relation will connect the fields at a time $t'$ to the polarization fields at an earlier time through the background Green's function:
\begin{align}
   e(x,t') = \int {\rm d}x' \int_{-\infty}^{t'} {\rm d}t'' \, \GG_0(x,x',t'-t'')  p(x',t'') = \int {\rm d}x' \int_0^\infty \GG_0(x,x',\tau)  p(x',t'-\tau) \,{\rm d}\tau.
\end{align}
We are going to insert this convolution relation into \eqref{appenH-pvt2}, analogous to what we did before. We can combine this with the step of specifying a time-dependence for the function $ p(x,t)$:
\begin{align}
     p(x,t') = 
    \begin{cases}
         p(x) e^{-i\omega t'} & \text{ for } t' < T \\
        0 & \text{ for } t' \geq T.
    \end{cases}
\end{align}
Performing these two steps in \eqref{appenH-pvt2} we have:
\begin{equation}
    \Re\int\int {\rm d}x {\rm d}x' \, \int_{-\infty}^{t} {\rm d}t'\, (-i\omega^*) e^{i\omega_0 t'} e^{(\Im \omega)t'}  p^\dagger(x) \int {\rm d}\tau \, \GG_0(x,x',\tau) e^{-i\omega(t'-\tau)}  p(x') \geq 0.
\label{eq:appenH-TI1b}
\end{equation}
As before, the oscillatory terms cancel, the integral over $\tau$ is proportional to the $\GG_0(x,x',\omega)$, i.e. the Fourier transform of $\GG_0(x,x',\tau)$, and the integral over $t'$ is simple to do. We are left with:
\begin{align}
    \frac{e^{2(\Im\omega)t}}{2 \Im \omega} \Im \int \int {\rm d}x {\rm d}x' \,  p^{\dagger}(x) \left[ \omega^* \GG_0(x,x',\omega) \right]  p(x') \geq 0.
    \label{eq:appenH-TI4b}
\end{align}
If we again treat $\GG_0(x,x',\omega)$ in space as a square matrix $\GG_0(\omega)$, we have
\begin{empheq}[box=\widefbox]{align}
    \Im \left[ \omega^* \GG_0(\omega) \right] \geq 0 \quad \text{ for } \Im \omega > 0,
\end{empheq}
where again the imaginary part of the matrix refers to its anti-Hermitian part.

\chapter{Bounds on the coupling strengths of communication channels and their information capacities: supplementary }
\label{appen:sup_comm}

This appendix provides supplementary information to Chapter~\ref{chap:comm} “Bounds on the Coupling Strengths of Communication Channels and Their Information Capacities.”
We (1) present explicit expressions for the singular vectors of the Green's function operator in the sphere--shell bounding volume and derive their corresponding singular values, (2) derive asymptotic form of the largest coupling strength in the limits of large bounding sphere and far-field bounding shell, (3) derive a lower bound on the total sum rule, and (4) derive an upper bound on the relative coupling strengths in the large-channel limit.

\hiddensection{Singular values of the Green's function operator in the sphere--shell bounding volume}
\label{sec:appenI-sec1}

In this section, starting from the full-electromagnetic wave equation, we define the dyadic Green's function and expand it with spherical vector waves. The spherical vector waves are the singular vectors of the dyadic Green's function operator in the sphere--shell bounding volume.  We present explicit expressions for these singular vectors and derive their corresponding singular values.  Results in this section supplement the arguments presented in Section~\ref{subsec:chap6-3D} of Chapter~\ref{chap:comm}.

\subsection{Dyadic Green's function in spherical vector waves representation}
The dyadic Green's function  $\vG(\vx, \vx')$ in Chapter~\ref{chap:comm} is defined as the solution of the following wave equation under a point source excitation:
\begin{equation}
    \frac{1}{k^2}\nabla\times{\nabla\times{\vG(\vx, \vx')}} - \vG(\vx, \vx') = \vI\delta(\vx - \vx'),
    \label{eq:appenI-waveEq}
\end{equation}
where $\vI$ is the unit dyad and $k$ is the magnitude of the free-space wavevector.
Slightly different from the conventional definition by a factor of $k^2$, the wave equation in Eq. (\ref{eq:appenI-waveEq}) has the advantage of giving dimensionless singular values of the Green's function operator since $\vG(\vx, \vx')$ now has dimensions of reciprocal volume.
The solution of Eq. (\ref{eq:appenI-waveEq}) is commonly written as~\cite{tsang2004scattering}
\begin{equation}
    \vG(\vx, \vx') = (k^2 \vI + \nabla\nabla)\frac{e^{ik|\vx-\vx'|}}{4\pi|\vx-\vx'|}.
    \label{eq:appenI-GFrealspace}
\end{equation}
Considering the spherical symmetry of the sphere--shell bounding volume, we express the dyadic Green's function in Eq. (\ref{eq:appenI-GFrealspace}) in its spherical vector waves representation~\cite{tsang2004scattering}:
\begin{equation}
    \vG(\vx, \vx') = ik^3 \sum_{n=0}^\infty \sum_{m=-n}^n \sum_{j=1,2} \vv_{{\rm out,} nmj}(\vx)\vv_{{\rm reg,} nmj}^*(\vx'),
    \label{eq:appenI-GFsvw}
\end{equation}
where $\vv_{{\rm out,} nmj}(\vx)$ and $\vv_{{\rm reg,} nmj}(\vx')$ are the outgoing and regular spherical vector waves. Their explicit expressions are discussed in the following subsection. The index $n$ and $m$ are the two indices of the underlying spherical harmonics, and $j=1,2$ denotes the two possible polarizations of the transverse vector field. 

\subsection{Spherical vector waves as the singular vectors of the Green's function operator}
In this section, we give explicit expressions for the spherical vector waves, $\vv_{{\rm out,} nmj}(\vx)$ and $\vv_{{\rm reg,} nmj}(\vx')$. We also present a crucial orthogonal relationship for these spherical vector waves, which allows us to identify them as the singular vectors of the Green's function operator in the sphere--shell bounding volume.
The center of our sphere--shell bounding volume is chosen as the origin for the coordinates $\vx$ and $\vx'$. 
In spherical coordinates $(r,\theta,\phi)$, the spherical vector waves can be separated into a radial dependency of a spherical Hankel/Bessel function and an angular dependency of vector spherical harmonics. Their spatial distributions depend the polarization state $j = \{1, 2\}$, which we spell out separately:
\begin{align}
    \vv_{{\rm out}, nm1}(\vx) &= \gamma_{n} h_n^{(1)}(kr)\vV^{(3)}_{nm}(\theta,\phi) \label{eq:appenI-out_vsw1} \\
    \vv_{{\rm out}, nm2}(\vx) &= \gamma_{n}\left\{ n(n+1)\frac{h_n^{(1)}(kr)}{kr}\vV^{(1)}_{nm}(\theta,\phi) + \frac{\left[krh_n^{(1)}(kr)\right]'}{kr} \vV^{(2)}_{nm}(\theta,\phi) \right\} \label{eq:appenI-out_vsw2} \\
    \vv_{{\rm reg}, nm1}(\vx) &= \gamma_{n} j_n(kr)\vV^{(3)}_{nm}(\theta,\phi) \label{eq:appenI-reg_vsw1} \\
    \vv_{{\rm reg}, nm2}(\vx) &= \gamma_{n}\left\{ n(n+1)\frac{j_n(kr)}{kr}\vV^{(1)}_{nm}(\theta,\phi) + \frac{\left[krj_n(kr)\right]'}{kr} \vV^{(2)}_{nm}(\theta,\phi) \right\}, \label{eq:appenI-reg_vsw2}
\end{align}
where the prefactor $\gamma_{n}=1/\sqrt{n(n+1)}$. 
The three vector spherical harmonics are an extension of the scalar spherical harmonics: $\vV^{(1)}(\theta,\phi) = \hat{r} Y^m_n(\theta,\phi)$, $\vV^{(2)}(\theta,\phi) = r\nabla \left[Y^m_n(\theta,\phi)\right]$, and $\vV^{(3)}(\theta,\phi) = \nabla\times\left[\hat{r} Y^m_n(\theta,\phi)\right]$, where $Y_n^m(\theta,\phi)=\sqrt{\frac{(2n+1)(n-m)!}{4\pi(n+m)!}} P^m_n(\cos\theta)e^{im\phi}$ are the scalar spherical harmonics defined by associated Legendre polynomials $P^m_n(x)$. The radial dependency of the outgoing spherical vector harmonics in Eqs. (\ref{eq:appenI-out_vsw1}, \ref{eq:appenI-out_vsw2}) are the spherical Hankel function of the first kind, $h^{(1)}_n(kr)$, with the domain of $r$ restricted to the region of the bounding shell. On the other hand, the regular  spherical vector harmonics $\vv_{{\rm reg,} nmj}(\vx)$ are defined in the region of the bounding sphere. Because of this, their radial dependency follows the spherical Bessel function $j_n(kr)$. The regular spherical vector waves $\vv_{{\rm reg,} nmj}(\vx)$ in Eqs. (\ref{eq:appenI-reg_vsw1}, \ref{eq:appenI-reg_vsw2}) take the same forms as their outgoing counterparts $\vv_{{\rm out,} nmj}(\vx)$ but with every $h^{(1)}_n(kr)$ replaced by $j_n(kr)$.

The three vector spherical harmonics, $\vV_{nm}^{(1)}(\theta,\phi)$, $\vV_{nm}^{(2)}(\theta,\phi)$, and $\vV_{nm}^{(3)}(\theta,\phi)$, satisfy the following orthogonal property:
\begin{equation}
    \int_0^\pi d\theta \sin\theta\int_0^{2\pi}d\phi\vV_{nm}^{(\alpha)}(\theta,\phi)\cdot\vV_{n'm'}^{(\beta)*}(\theta,\phi) = z_{\alpha n}\delta_{\alpha\beta}\delta_{mm'}\delta_{nn'},
    \label{eq:appenI-Vorth}
\end{equation}
where the prefactor $z_{1n} = 1$ and $z_{2n} = z_{3n} = n(n+1).$
This orthogonality is crucial because it ensures that 1. different outgoing spherical vector waves $\vv_{{\rm out}, nmj}(\vx)$ are orthogonal to each other in the outer bounding shell and 2. different regular spherical vector waves $\vv_{{\rm reg}, nmj}(\vx)$ are orthogonal to each other in the inner bounding sphere. 
Considering these two orthogonal conditions and the fact that the Green's function operator in Eq.~(\ref{eq:appenI-GFsvw}) can be expanded as the sum of the outer products between $\vv_{{\rm out}, nmj}(\vx)$ and $\vv_{{\rm reg}, nmj}(\vx)$, we identify these two types of spherical vector waves as the left and right singular vectors of the Green's function operator in the sphere--shell bounding volume.


The discussion of the spherical vector wave representation in this subsection mostly follows the presentation in Ref. \cite[chapter~2.1]{tsang2004scattering}, though with different notations for the spherical vector waves and an extra factor of $k^2$ in the Green's function. We also adopt a more conventional definition of the scalar spherical harmonics as in Jackson \cite{jackson1999classical}. This leads to different prefactors in Eqs. (\ref{eq:appenI-out_vsw1} -- \ref{eq:appenI-Vorth}) compared to the ones in Ref. \cite{tsang2004scattering}.

\subsection{Singular values of the Green's function operator in the sphere--shell bounding volume}
Given the spherical wave expansion of the Green's function in Eq. (\ref{eq:appenI-GFsvw}), the singular values of the Green's function operator in the sphere--shell bounding volume can be identified as the products between the norms of the unnormalized singular vectors $\vv_{nmj}(\vx)$ and $\text{Rg}\vv_{nmj}(\vx)$ in their respective domains:
\begin{equation}
    |s_{nmj}^{\text{(sphere--shell)}}|^2 = k^6  \int_{V_{\text{sphere}}} \left|\vv_{{\rm reg,} nmj}(\vx)\right|^2 d\vx
    \int_{V_{\text{shell}}} \left|\vv_{{\rm out,} nmj}(\vx)\right|^2 d\vx .
    \label{eq:appenI-s_3Dintegral}
\end{equation}
The angular part of the integrals in Eq. (\ref{eq:appenI-s_3Dintegral}) can be computed by plugging in the explicit expressions of $\vv_{{\rm reg,} nmj}(\vx)$ and $\vv_{{\rm out,} nmj}(\vx)$ in Eqs. (\ref{eq:appenI-out_vsw1} -- \ref{eq:appenI-reg_vsw2}), which are simplified under the orthogonality relation of the vector spherical harmonics $\vV^{(i)}(\theta,\phi)$ in Eq. (\ref{eq:appenI-Vorth}).
The result is several remaining one-dimensional integrals in the radial direction:
\begin{align}
    |s^{\text{(sphere--shell)}}_{nm1}|^2 &= \int^{kR_{\text{sphere}}}_0 x^2 |j_n(x)|^2 dx \int^{kR_{\text{outer}}}_{kR_{\text{inner}}} x^2 |h^{(1)}_n(x)|^2 dx \label{eq:appenI-s_1Dintegral1}\\
    |s^{\text{(sphere--shell)}}_{nm2}|^2 &= \int^{kR_{\text{sphere}}}_0 \left\{n(n+1)|j_n(x)|^2 + \left[xj_n(x)\right]'^{\hspace{1pt}2} \right\} dx \nonumber \\ 
    & \qquad\qquad \times \int^{kR_{\text{outer}}}_{kR_{\text{inner}}} \left\{n(n+1)|h^{(1)}_n(x)|^2 + |xh^{(1)}_n(x)|'^{\hspace{1pt}2} \right\} dx, \label{eq:appenI-s_1Dintegral2}
\end{align}
where $R_{\text{sphere}}$ is the radius of the bounding sphere. The variables $R_{\text{inner}}$ and $R_{\text{outer}}$ are the inner and outer radii of the bounding shell, respectively.
The one-dimensional integrals in Eqs. (\ref{eq:appenI-s_1Dintegral1}, \ref{eq:appenI-s_1Dintegral2}) can be analytically integrated with the aid of indefinite integrals of the spherical Bessel functions in Ref. \cite{bloomfield2017indefinite}, after which we obtain explicit expressions of the singular values in the sphere--shell bounding volume:
\begin{align}
    |s^{\text{(sphere--shell)}}_{nm1}|^2 &= 
    \frac{\pi^2}{16}x^2\left[ J^2_{n+\frac{1}{2}}(x) - J_{n+\frac{3}{2}}(x)J_{n-\frac{1}{2}}(x)  \right]\Biggr\rvert_{x=0}^{x=kR_{\text{sphere}}} \nonumber \\ & \qquad\qquad\qquad\qquad  \times y^2\Re\left[ |H^{(1)}_{n+\frac{1}{2}}(x)|^2 - H^{(1)}_{n+\frac{3}{2}}(y)H^{(2)}_{n-\frac{1}{2}}(y)  \right]\Biggr\rvert_{y=kR_{\text{inner}}}^{y=kR_{\text{outer}}}  \label{eq:appenI-s_explicit1}
    \\
    |s^{\text{(sphere--shell)}}_{nm2}|^2 &= 
    \frac{\pi^2}{16} x^2\left\{ \frac{n+1}{2n+1} \left[ J^2_{n-\frac{1}{2}}(x) - J_{n+\frac{1}{2}}(x)J_{n-\frac{3}{2}}(x) \right] \right. 
    \nonumber \\ & \qquad\qquad\qquad\qquad \left. + \frac{n}{2n+1} \left[ J^2_{n+\frac{3}{2}}(x) - J_{n+\frac{5}{2}}(x)J_{n+\frac{1}{2}}(x) \right] \right\} \Biggr\rvert_{x=0}^{x=kR_{\text{sphere}}} \nonumber
    \\
    & \quad \times y^2\Re\left\{ \frac{n+1}{2n+1} \left[ |H^{(1)}_{n-\frac{1}{2}}(y)|^2 - H^{(1)}_{n+\frac{1}{2}}(y)H^{(2)}_{n-\frac{3}{2}}(y) \right] \right. 
    \nonumber \\ & \qquad\qquad\qquad\qquad \left. + \frac{n}{2n+1} \left[ |H^{(1)}_{n+\frac{3}{2}}(y)|^2 - H^{(1)}_{n+\frac{5}{2}}(y)H^{(2)}_{n+\frac{1}{2}}(y) \right] \right\} \Biggr\rvert_{y=kR_{\text{inner}}}^{y=kR_{\text{outer}}},
    \label{eq:appenI-s_explicit2}
\end{align}
where the functions $J_n(x)$ and $H^{(1)}_n(x)$ denote the Bessel function and the Hankel function of the first kind. Equations (\ref{eq:appenI-s_explicit1}, \ref{eq:appenI-s_explicit2}) are the explicit expressions of $|s^{\text{(sphere--shell)}}_{nmj}|^2$ we use in the Chapter~\ref{chap:comm} to calculate the upper bounds of the coupling strengths between any two regions in the bounding volume.

\begin{figure}[b!]
\centering
\includegraphics[width=1\textwidth]{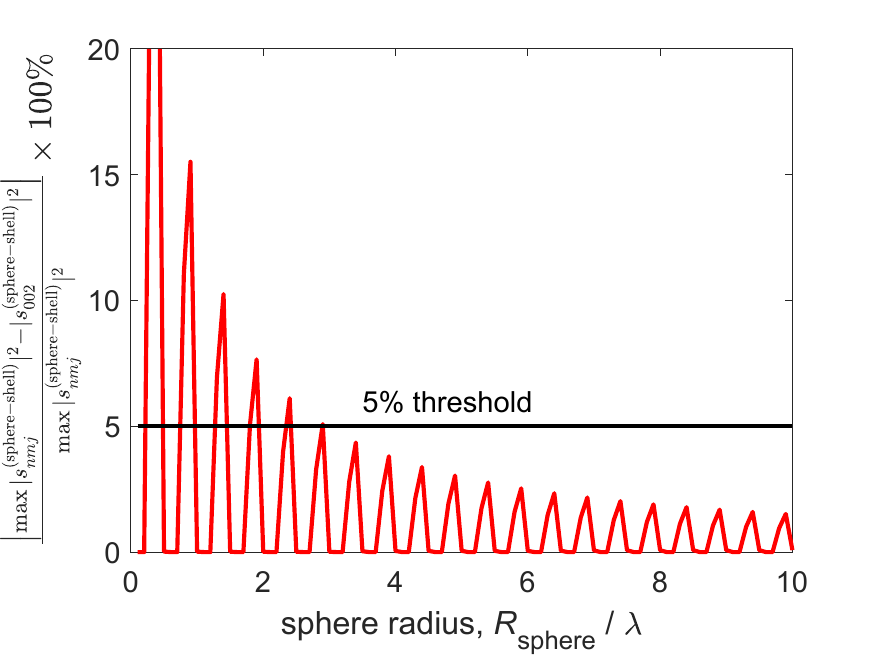}
\caption{Numerical evidence showing that the maximum channel strength, $\max{|s^{\text{(sphere--shell)}}_{nmj}|^2}$, is asymptotically attained by the first angular channel of the second polarization state, $|s^{\text{(sphere--shell)}}_{002}|^2$, in the limit of large sphere radius $R_s$ relative to the free-space wavelength $\lambda$.}
\label{fig:appenI-SM_Fig2}
\end{figure}

\hiddensection{Maximal channel strength in the limits of large bounding sphere and far-field bounding shells}
\label{sec:appenI-sec2}
We observe that the maximal channel strength in the sphere--shell bounding domain is asymptotically attained by the first angular channel of the second polarization state in the limit of large bounding sphere:
\begin{equation}
    \max{|s^{\text{(sphere--shell)}}_{nmj}|^2} = |s^{\text{(sphere--shell)}}_{002}|^2,\quad \text{for\ } R_{\rm sphere} \gg \lambda, 
    \label{eq:appenI-sv_max_eq1}
\end{equation}
This is evidenced by Fig.~\ref{fig:appenI-SM_Fig2}, which shows that the relative difference between $\max{|s^{\text{(sphere--shell)}}_{nmj}|^2}$ and $|s^{\text{(sphere--shell)}}_{002}|^2$ is smaller than 5\% for a bounding sphere with radius $R_{\rm sphere}$ larger than three times the wavelength $\lambda$ and the relative difference asymptotically tends to zero as the radius becomes much larger than the wavelength. The separation distance $d$ between the two bounding domains and the maximal thickness $2R_r$ of the spherical shell is assumed to be $10\lambda$ and $\lambda$, respectively, though our result does not appear to be sensitive to these two parameters.

The channel strength  $|s^{\text{(sphere--shell)}}_{002}|^2$ has a simple analytical form in the limits of large bounding sphere and far-field bounding shell. To show this, we first consider the limit of far-field bounding shell, where $|s^{\text{(sphere--shell)}}_{002}|^2$, according to Eq.~(\ref{eq:appenI-s_1Dintegral2}), simplifies to
\begin{equation}
    |s^{\text{(sphere--shell)}}_{002}|^2 = 2kR_r\int_0^{kR_{\rm sphere}}x^2|j_{-1}(x)|^2 dx,\quad \text{for\ } d \gg \lambda, 
    \label{eq:appenI-sv_max_eq2}
\end{equation}
where we approximate the spherical Hankel function as $h^{(1)}_n(x) \approx \frac{1}{x}e^{ix}i^{-n-1}$ under the condition of $d \gg \lambda$. The integral in Eq.~(\ref{eq:appenI-sv_max_eq2}) can be analytically evaluated considering that $j_{-1}(x) = \cos(x)/x$. Its result, under the limit of large bounding sphere, further reduces to 
\begin{equation}
    |s^{\text{(sphere--shell)}}_{002}|^2 = k^2R_rR_{\rm sphere}, \ \text{for\ } R_{\rm sphere} \gg \lambda \text{\ and \ } d \gg \lambda.
    \label{eq:appenI-sv_max_eq3}
\end{equation}

Combining Eqs.~(\ref{eq:appenI-sv_max_eq1}) and~(\ref{eq:appenI-sv_max_eq3}), we derive an analytical expression of the maximal channel strength in the limits of large bounding sphere and far-field spherical shell:
\begin{equation}
    \max{|s^{\text{(sphere--shell)}}_{nmj}|^2} = k^2R_rR_{\rm sphere},\ \text{for\ } R_{\rm sphere} \gg \lambda \text{\ and \ } d \gg \lambda,
    \label{eq:appenI-sv_max_eq4}
\end{equation}
which, we observe, scales linearly with the maximal radii of both the source and receiver domains.

\hiddensection{A lower bound on the sum rule}
\label{sec:appenI-sec3}
The sum rule $S=\sum_{nmj}|s_{nmj}|^2$ is conserved under a unitary transformation from the communication channel basis to the delta-function basis in real space. Conveniently, we express $S$ as a double integral of the Frobenius norm of the dyadic Green's function over both the source and receiver volumes:
\begin{equation}
     S = \int_{V_s}\int_{V_r} ||\vG(\vx,\vx')||_F^2 d\vx d\vx'.
\end{equation}
The Frobenius norm of the dyadic Green's function reads~\cite{miller_fundamental_2016}
\begin{equation}
    ||\vG(\vx,\vx')||_F^2 = \frac{k^6}{8\pi^2}\left[ \frac{1}{(k|\vx-\vx'|)^2} + \frac{1}{(k|\vx-\vx'|)^4} + \frac{3}{(k|\vx-\vx'|)^6}\right]
\end{equation}
which monotonically decays with respect to the separation distance, $|\vx-\vx'|$, between two points. This monotonic decay allows us to lower bound the sum rule by relaxing the separation distance to the largest possible separation distance, $\max{|\vx-\vx'|} = d+2R_s+2R_r$, between the source and receiver volumes:
\begin{equation}
    S \geq \frac{k^4V_sV_r}{8\pi^2(d+2R_s+2R_r)^2} + \mathcal{O}\left(\left[k(d+2R_s+2R_r)\right]^{-4}\right). \label{eq:appenI-sum}
\end{equation}
The variables $R_s$ and $R_r$ denote the maximal radii of the source and receiver domains.
For conciseness, we assume the furthest separated points are in the far field, i.e. $k(d+2R_s+2R_r) \gg 1$, so  that only the leading term in Eq. (\ref{eq:appenI-sum}) remains. This, of course, can be easily generalized by explicitly including two other higher-order terms with a slightly more complicated expression. 

\hiddensection{An upper bound on the relative coupling strengths in the large-channel limit}
\label{sec:appenI-sec4}
In this section, we derive large-channel asympotes of the coupling strengths in the sphere--shell bounding volume. 
Two differently polarized communication channels exhibit slightly different asymptotes, though both can be bounded above by a single expression. 
Together with the lower bound on the total sum rule, we derive an upper bound on the relative coupling strength for both polarizations. This section provides a theoretical basis for the bound we present at the end of Section~\ref{subsec:chap6-3D} of Chapter~\ref{chap:comm}.

The singular values $|s^{\text{(sphere--shell)}}_{nmj}|^2$ have simple analytical expressions in the large-channel limit when the index $n\rightarrow\infty$. They can be derived by substituting the large-$n$ asymptotes of the spherical Bessel and Hankel functions, $j_n(x) \sim \frac{1}{\sqrt{(4n+2)x}} \left(\frac{ex}{2n+1}\right)^{n+1/2}$ and $h^{(1)}_n(x) \sim \frac{-2i}{\sqrt{(4n+2)x}}\left(\frac{ex}{2n+1}\right)^{-n-1/2}$, into Eq. (\ref{eq:appenI-s_1Dintegral1}, \ref{eq:appenI-s_1Dintegral2}):
\begin{align}
    |s^{\text{(sphere--shell)}}_{nm1}|^2 &= \left( \frac{kR_{\text{sphere}}}{2n} \right)^4 \left( \frac{R_{\text{sphere}}}{R_{\text{inner}}} \right)^{2n-1} \quad \text{as\ } n\rightarrow\infty, \label{eq:appenI-s_asym1}
    \\
    |s^{\text{(sphere--shell)}}_{nm2}|^2 &= \frac{1}{4} \left( \frac{R_{\text{sphere}}}{R_{\text{inner}}} \right)^{2n+1} \quad \text{as\ } n\rightarrow\infty. \label{eq:appenI-s_asym2}
\end{align}
While both polarizations decay exponentially as a function of $n$, the first polarization channel is always smaller than the second one in the large $n$ limit due to the additional decay of the factor of $1/n^4$. The value of the second polarization thus serves as an upper bound for both:
\begin{equation}
    |s^{\text{(sphere--shell)}}_{nmj}|^2 \leq \frac{1}{4} \left( \frac{R_{\text{sphere}}}{R_{\text{inner}}} \right)^{2n+1} \quad \text{as\ } n\rightarrow\infty.
    \label{eq:appenI-s_asym_bd}
\end{equation}
This is an upper bound for the coupling strengths of both polarizations in a sphere--shell bounding volume in the large-channel limit. The bound only depends on the ratio between the radius of the bounding sphere, $R_{\text{sphere}}$, and the inner radius of the bounding shell, $R_{\text{inner}}$ --- the smaller the ratio, the faster the decay.

For any two domains that can be separated by a spherical surface, there are two possible sphere--shell bounding volumes: one that centers around the source region and one that centers around the receiver region. To obtain a tighter upper bound, we choose the one that centers around the smaller domain because it has the smaller ratio between $R_{\text{sphere}}$ and $R_{\text{inner}}$. Considering this and the fact that the number of channels with $n$-index less or equal to $n$ is $q=2(n+1)^2$, Eq. (\ref{eq:appenI-s_asym_bd}) can be written as
\begin{equation}
    |s^{\text{(sphere--shell)}}_{q}|^2 \leq \frac{1}{4} \left( 1 +  \frac{R_{\text{min}}}{d} \right)^{-\sqrt{2q}-1} \quad \text{as\ } q\rightarrow\infty,
    \label{eq:appenI-s_asym_bd_l}
\end{equation}
where $R_{\text{min}} = \text\{R_s, R_r\}$ is the smaller of the radius of the source domain $R_s$ and the radius of the receiver domain $R_r$, and $d$ is the distance between the two domains. This equation shows the coupling strengths between two regions always decay \textit{sub}-exponentially with the total channel index, $q$, in the large-channel limit.

Lastly, we invoke the domain-monotonicity theorem discussed in Section~\ref{sec:chap6-optimal} of Chapter~\ref{chap:comm} which implies that the coupling strengths $|s_q|^2$ between any two domains have to be smaller than their counterparts in a sphere--shell bounding volume:
\begin{equation}
    |s_{q}|^2 \leq |s_q^{\text{(sphere--shell)}}|^2,\quad  \text{for}\ q = 1, 2, ...
    \label{eq:appenI-sl_max}
\end{equation}
Combining this with the upper bound of $|s^{\text{(sphere--shell)}}_{q}|^2$ in Eq. (\ref{eq:appenI-s_asym_bd_l}) and the lower bound of the sum rule $S$ in Eq. (\ref{eq:appenI-sum}), we derive an upper bound on the relative channel strengths in the large-channel limit:
\begin{equation}
    \frac{|s_{q}|^2}{S} \leq \frac{2\pi^2(d+2R_s+2R_r)^2}{k^4V_sV_r(1+d/R_{\text{min}})^{\sqrt{2q}+1}}, \quad \text{as}\ q \rightarrow \infty.
    \label{eq:appenI-decay3D}
\end{equation}
This suggests that the relative coupling strength between any two domains decay at least sub-exponentially in the large-channel limit. Equation (\ref{eq:appenI-decay3D}) is a key result presented in Chapter~\ref{chap:comm} and we hereby provide a derivation in this section.

\chapter{Quick review of graph theory} 
\label{appen:graph}

Graphs distill real-world relations into abstract lines and vertices, such as social networks in sociology, genetic networks in biology, communication networks in computer science, lattice structures in physics, and molecular chains in chemistry. Complex dynamics untangle and straighten out on graphs.
In particular, chordal graphs represent a type of connections where many hard problems can be easily solved, including graph coloring, clique finding, and matrix factorization~\cite{Vandenberghe2015}. 
In this appendix, we review basics of graphs and chordal graphs, as well as theorems that foreground chordal graphs in sparse semidefinite programming.

A graph composes of a set of vertices $V = \{v_1, v_2, ..., v_n\}$ and their connecting edges $E \subseteq V \times V$. The latter two uniquely define the graph: $G(V,E)$.
We consider undirectional graph whose edges are unordered pairs, denoted by curly brackets such as $\{v_i, v_j\}$.
If a series of edges leads one vertex back to itself, then these edges form a cycle.
Two vertices $v_i$ and $v_j$ are adjacent if there is an edge between them, i.e., if $\{v_i, v_j\} \in E$.
Vertices that are all adjacent to each other form a clique. 
A clique can expand upon admitting new vertices that are adjacent to all its existing members. If no such new vertices exist, the clique is then called a maximal clique of the graph.

Many hard problems are easy to solve on a chordal graph. A chordal graph is a graph in which every cycle of length four and greater has a chord (an edge between nonconsecutive vertices of the cycle). In other words, if you trace a cycle of four (or more) edges without finding a shortcut, then the graph is not a chordal graph. 
Any graph can be made chordal by adding extra edges. In the extreme case, supplying every possible edges guarantees a chordal graph, but the resulting graph has no sparsity left. The art is to add as few edges as possible to make a graph chordal. This procedure is called chordal completion, often implemented via heuristic algorithms~\cite{amestoy1996approximate}.

Graphs characterize sparsity in matrices. Specifically, each missing edge in an undirectional graph $G(V, E)$ corresponds to two zeros in a symmetric matrix $\XX$: $\{v_i, v_j\}\notin E \iff \XX_{ij} = \XX_{ji} = 0$.
Two groups of symmetric matrices $\XX$ are of particular interests in sparse semidefinite programming. 
The first are matrices that are positive semidefinite and sparse, with sparsity pattern given by the graph $G(V, E)$:
\begin{equation}
    \SS_+^n(E,0) = \{\XX\in\SS^n_+\ |\ \XX_{ij}=\XX_{ji}=0\ \textrm{if}\ (i,j)\neq E\}.
\end{equation}
The second are matrices that are not necessarily positive semidefinite or sparse but can be ``completed'' into positive semidefinite matrices. To motivate this, consider an inner product $\Tr(\AA\XX)$ between the symmetric matrix $\XX$ and a sparse symmetric matrix $\AA$ whose sparsity is given by the graph $G(V, E)$.
The inner product multiplies matrices element-wise, so only the elements in $\XX$ that correspond to the edges in the graph $G(V, E)$ matter.  If the rest of the elements in $\XX$ can be altered to turn $\XX$ into a positive semidefinite matrix, say $\mathbb{M}$, then we say $\XX$ belongs to a group of completable partial symmetric matrices:
\begin{equation}
    \SS_+^n(E,?) = \{\XX\in\SS^n\ |\ \exists \mathbb{M}\geq 0, \mathbb{M}_{ij} = \XX_{ij}  \ \forall (i,j)\in E\}
\end{equation}
The two matrix spaces, $\SS_+^n(E,0)$ and $\SS_+^n(E,?)$, are both convex cones and, in fact, dual of each other~\cite{Vandenberghe2015}. Together, they constitute the basic matrix spaces in sparse semidefinite programming.

Crucially, if the graph $G(E, V)$ is a chordal graph, both types of matrices above can be decomposed into into smaller matrices defined on the maximal cliques of the graph: $\{C_1, C_2, ..., C_l\}$.
Aiding this decomposition is a projection matrix $\TT_l$ that projects a matrix $\XX$ into its principle submatrix $\XX_l = \TT_l \XX \TT_l^\top$, and reads $(\TT_l)_{ij}=1$ if $C_l(i)=j$ and zero otherwise, where $C_l(i)$ is the $i$-th vertex in $C_l$.
If the matrix $\XX$ is positive semidefinite completable, i.e., $\XX\in\SS_+^n(E,?)$, then all the submatrices $\XX_l$ are all positive semidefinite, as dictated by the Grone’s theorem~\cite{grone1984positive}:
\begin{equation}
    \XX \in \SS^n_+(E,?) \iff \XX_l = \TT_l \XX \TT_l^\top \in \SS^{|C_l|}_+.
    \label{eq:appenJ-Grone}
\end{equation}
If the matrix $\XX$ is sparse positive semidefinite, i.e., $\XX\in\SS_+^n(E,0)$, then all the submatrices $\XX_l$ are positive semidefinite and uniquely expand the matrix $\XX$, as dictated by Agler’s theorem~\cite{agler1988positive}:  
\begin{equation}
    \XX\in\SS_+^n(E,0) \iff \XX = \sum_{l=1}^{p} \TT_l^\top \XX_l \TT_l, \ 
    \XX_l \in \SS^{|C_l|}_+.
    \label{eq:appenJ-Agler}
\end{equation} 
Agler's theorem in Eq.~(\ref{eq:appenJ-Agler}) allows one to decompose sparse semidefinite programs in their dual forms, where the matrix variable $\XX$ directly inherits the sparsity of the problem; Grone's theorem in Eq.~(\ref{eq:appenJ-Grone}) allows one to decompose sparse semidefinite programs in their primal forms, where the the matrix variable $\XX$ does not inherit the sparsity but is multiplied by matrices that do.

\backmatter

\bibliography{theis_bib}

\end{document}